\documentclass[amsmath,amssymb,aps,prd,11pt,tightenlines,superscriptaddress,nofootinbib,preprintnumbers,notitlepage]{revtex4-2}

\input{header}

\begin{document}

\vspace*{-0.15in}
\title{{\LARGE  LETTER OF INTENT: \\ \vspace{5mm}  
\LARGE THE FORWARD PHYSICS FACILITY} }

\vspace*{0.6cm}
\begin{figure*}[h]
\centering
\includegraphics[width=0.8\textwidth]{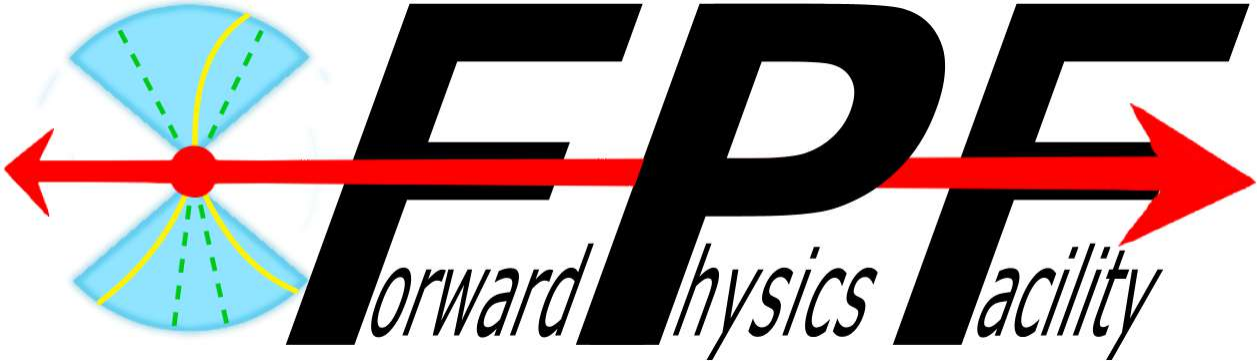}
\end{figure*}

\author{Luis~A.~Anchordoqui}
\affiliation{Department of Physics and Astronomy, Lehman College, City University of New York, Bronx, NY 10468, USA}

\author{John Kenneth Anders}
\affiliation{University of Liverpool, Liverpool L69 3BX, United Kingdom}

\author{Akitaka Ariga}
\affiliation{Albert Einstein Center for Fundamental Physics, Laboratory for High Energy Physics, University of Bern, CH-3012 Bern, Switzerland}
\affiliation{Department of Physics, Chiba University, Inage-ku, Chiba, 263-8522, Japan}

\author{Tomoko Ariga}
\affiliation{Kyushu University, Nishi-ku, Fukuoka, 819-0395, Japan}

\author{David Asner}
\affiliation{SNOLAB, Creighton Mine \#9, 1039 Regional Road 24, Sudbury, ON P3Y 1N2, Canada}

\author{Jeremy Atkinson}
\affiliation{Albert Einstein Center for Fundamental Physics, Laboratory for High Energy Physics, University of Bern, CH-3012 Bern, Switzerland}

\author{Alan~J.~Barr}
\affiliation{Department of Physics, University of Oxford, OX1 3RH, United Kingdom}

\author{Larry Bartoszek}
\affiliation{Bartoszek Engineering, Aurora, IL 60506, USA}

\author{Brian Batell}
\affiliation{Department of Physics and Astronomy, University of Pittsburgh, Pittsburgh, PA 15217, USA}

\author{Hans Peter Beck}
\affiliation{Albert Einstein Center for Fundamental Physics, Laboratory for High Energy Physics, University of Bern, CH-3012 Bern, Switzerland}
\affiliation{Department of Physics, University of Fribourg, CH–1700 Fribourg, Switzerland}

% ...

\begin{abstract}
\vspace*{0.2in} The Forward Physics Facility (FPF) is a proposed extension of the HL-LHC program designed to exploit the unique scientific opportunities offered by the intense flux of high energy neutrinos, and possibly new particles, in the far-forward direction. Located in a well-shielded cavern 627~m downstream of one of the LHC interaction points, the facility will support a broad and ambitious physics program that significantly expands the discovery potential of the HL-LHC. Equipped with four complementary detectors---FLArE, FASER$\nu$2, FASER2, and FORMOSA---the FPF will enable breakthrough measurements that will advance our understanding of neutrino physics, quantum chromodynamics, and astroparticle physics, and will search for dark matter and other new particles. With this Letter of Intent, we propose the construction of the FPF cavern and the construction, integration, and installation of its experiments. We summarize the physics case, the facility design, the layout and components of the detectors, as well as the envisioned collaboration structure, cost estimate, and implementation timeline. 
\end{abstract}

\maketitle
\clearpage

{
\setlength{\baselineskip}{11.9pt} 
\tableofcontents
}

\clearpage
\section{Executive Summary}
\label{sec:intro}
The Forward Physics Facility (FPF) is an underground cavern designed to house relatively small, far-forward experiments at the High-Luminosity LHC at CERN. The existing large experiments at the LHC are blind to particles produced in the far-forward direction.  History has shown that materially improving detector coverage at colliders frequently leads to new physics insights, and the LHC is currently missing many opportunities for discovery because of inadequate coverage of the forward direction.  The FPF fills this hole in the LHC program and enhances the potential for breakthroughs in particle physics on the time scale of years, not decades, with only a modest investment. \medskip

\noindent The FPF has many unique features that differentiate it from other proposed projects:  \medskip

\begin{wrapfigure}{R}{0.55\textwidth}
  \vspace{-6mm}
  \includegraphics[width=0.53\textwidth]{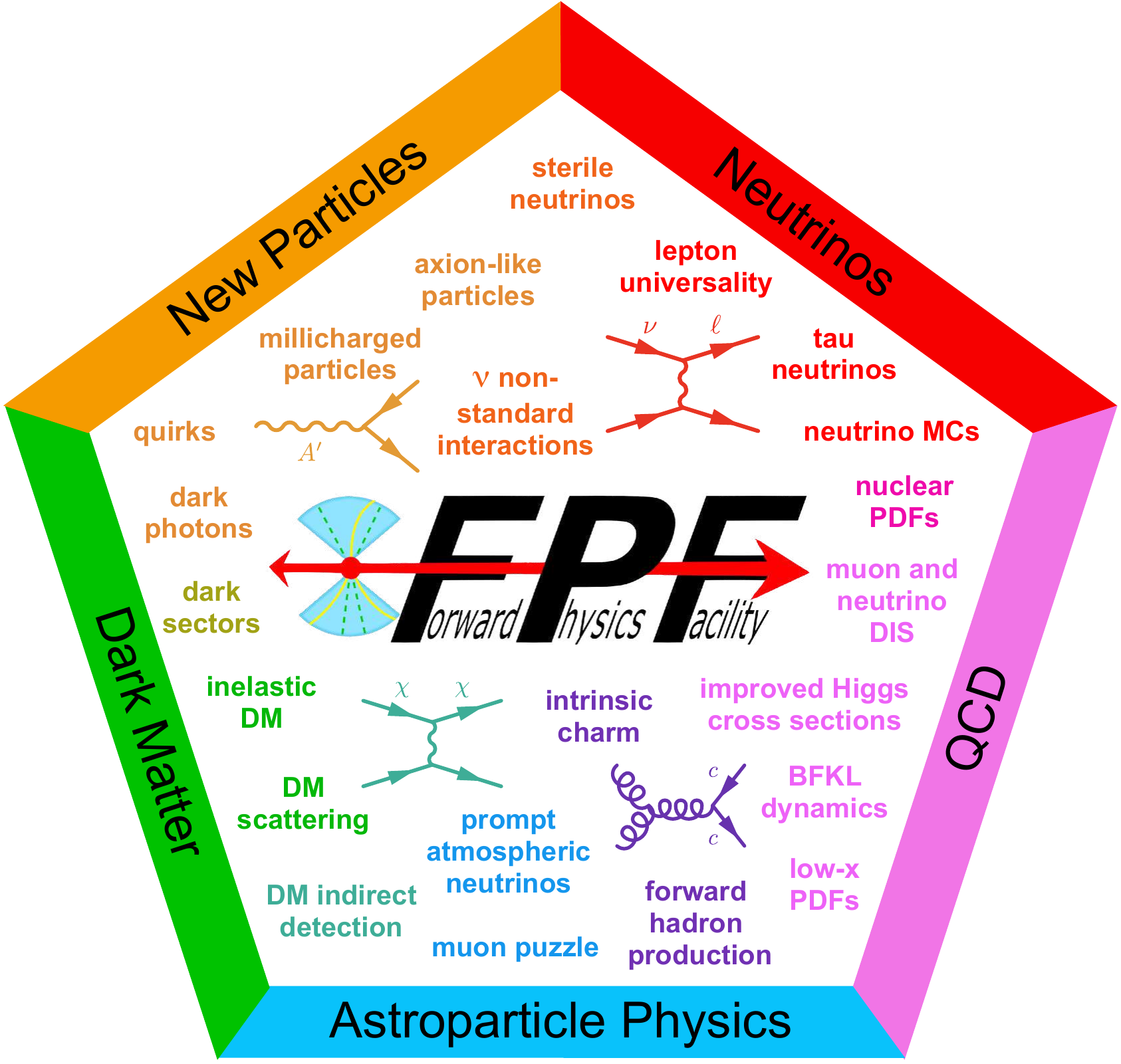}
  \caption{\textbf{Physics Overview.} The FPF will probe topics that span multiple frontiers, including new particles, neutrinos, dark matter, QCD, and astroparticle physics. }
  \label{fig:Phys_overviewIntro}
  \vspace{-5mm}
\end{wrapfigure}

\noindent \textbf{The guarantee of a qualitatively new view of particle physics at the high energy frontier.}  The FPF has a guaranteed physics case at the newly opened frontier of collider neutrino physics. As recently as three years ago, no neutrinos produced at a collider had been directly detected.  In 2023 FASER and SND@LHC observed collider neutrinos for the first time, opening a new window on the high-energy universe.  The study of TeV neutrinos, the most energetic ever produced by humans, is sensitive to their production mechanism, their propagation, and their interactions in detectors. Just as their extremely weak interactions make them ideal messengers from distant and dense astrophysical environments, neutrinos at colliders are ideal messengers of the inner structure of protons and nuclei.  The FPF neutrino program will refine our understanding of forward hadron production, uniquely probe QCD in uncharted regions of phase space, provide unique constraints on proton and nuclear structure, and sharpen theoretical predictions for Standard Model (SM) processes, enhancing the ability of other LHC detectors and future particle and astroparticle experiments to discover new physics. \medskip

\noindent \textbf{A rich new-physics case.} The FPF physics case touches on almost all areas of beyond-the-SM (BSM) physics, including neutrino and flavour physics, dark matter and dark sectors, and other searches for new fundamental particles and forces. The FPF also strengthens existing searches for BSM physics at existing LHC experiments, and builds new connections to nuclear physics, astroparticle physics, and cosmology (see \cref{fig:Phys_overviewIntro}).  The enormous fluxes of particles in the forward direction will enable the FPF to precisely study millions of neutrinos, including thousands of tau neutrinos, in the TeV energy range between fixed target and astroparticle experiments, and probe models of neutrino masses and violations of lepton flavour universality.  The FPF can also discover particles that are inaccessible at SHiP and other fixed target experiments and detect BSM particles with signal event rates that are two to four orders of magnitude beyond planned upgrades of FASER and SND@LHC. \medskip

\noindent \textbf{A diverse experimental program.}  The FPF is a facility, not an experiment. As noted above, the physics case in the far-forward region is exceptionally rich and cannot be fully explored by a single experiment. For this reason, the FPF will house a number of forward experiments, each with a different technology targeting a different kind of physics: 
\begin{itemize}
\setlength\itemsep{-0.05in}
\item FLArE: A liquid argon TPC with a fiducial mass of 10 tons and a magnetized hadronic calorimeter, which will detect hundreds of muon and electron neutrinos \textit{per day} over an unparalleled dynamic range of energy with excellent resolution and particle identification, and simultaneously search for light dark matter interactions with low-energy signatures. 
\item FASER$\nu$2: A 20-ton emulsion-tungsten detector with extraordinary spatial resolution, designed to detect neutrinos and anti-neutrinos of all three flavours, including thousands of tau neutrinos and anti-neutrinos, bringing them under the microscope of precision studies for the first time.  
\item FASER2: A large, high-field, magnetic tracking spectrometer with broad sensitivity to new particles, from light, long-lived particles predicted by models of dark matter and dark sectors, to exotic particles with masses up to the TeV-scale. 
\item FORMOSA: A scintillator detector, designed to discover milli-charged particles with world-leading sensitivity over four orders of magnitude in mass. 
\end{itemize}

\noindent \textbf{A firm foundation of existing pathfinder experiments.}  The FPF benefits from experience gained at existing pathfinder experiments, including FASER, FASER$\nu$, MilliQan, and small liquid argon neutrino detectors.  In contrast to proposals based on simulations and idealised assumptions, these detectors have proven, through complete, real-world analyses, that the far-forward region provides an extraordinarily quiet environment to study high-energy neutrinos and look for weakly-interacting particles with background-free searches and inherent advantages associated with high energies that cannot be found elsewhere.  \medskip

\noindent \textbf{The flexibility to respond to future developments.}  Because the particles targeted in the far-forward region are very weakly-interacting, the experiments do not ``block'' each other, and new detectors may be added either upstream or downstream of existing experiments without compromising their physics potential.  At the same time, both on-axis and off-axis locations have their virtues.  The FPF can accommodate new experiments at various locations, and therefore flexibly adapt to new and innovative ideas, new proposals for detectors, and future discoveries.  \medskip

\noindent \textbf{Modest cost and investment.}  
The cost of the FPF is modest compared to its physics potential. Mature cost estimates, based in part on experience with the HL-LHC, are 50 MCHF, and 40 MCHF for the Facility, including the integration of experiments and cryogenics, and the combined four baseline experiments, respectively. The required investment is very small compared to the amount already invested in the LHC, but it nevertheless opens up a large and complementary scientific program. \medskip

\noindent \textbf{A sustainable and responsible impact on the environment.}  The environmental impact of the FPF is minor compared to other proposals.  In contrast to other proposed experiments, the FPF requires no additional collider or beam modifications beyond what is already committed to the HL-LHC.  The total energy required by the FPF in its steady state is expected to be below 1 MW, less than the power provided by a single modern windmill, and small compared to the ${\cal O}(100~\text{MW})$ required by the LHC and other large accelerator facilities. \medskip

\noindent \textbf{Support for a new generation of particle physicists.}  The FPF is a qualitatively new idea that has attracted the interest of a new generation of particle theorists and experimentalists.  Junior physicists can contribute at every stage of an experiment’s life cycle, from detector design to construction to analysis, in the time it takes to obtain a graduate degree.  The FPF will develop future leaders of the field because the physics goals are exciting, and there is room for their creative and innovative ideas to have a real impact.  \medskip

\noindent \textbf{Synergy with FCC and other future colliders.}  The FPF is highly motivated by the need to preserve expertise at the high-energy frontier.  If the FCC or any other particle collider is to be built, near-term high-energy experiments are required to maintain the hard-won expertise that has built up over many decades---expertise that is preserved not by simulations, community planning exercises, and panel recommendations, as important as they are, but by actually building new experiments and analyzing their results.  In all scenarios, it is absolutely essential that the LHC be exploited to its full capacity.  Building the FPF in the coming few years is a requirement for the LHC to realise its promise and maximize its potential for new discoveries that will point us to a new golden age of particle physics.

\clearpage
\section{Physics Case}
\label{sec:physics}
\begin{figure}[bp]
\includegraphics[width=0.91\textwidth]{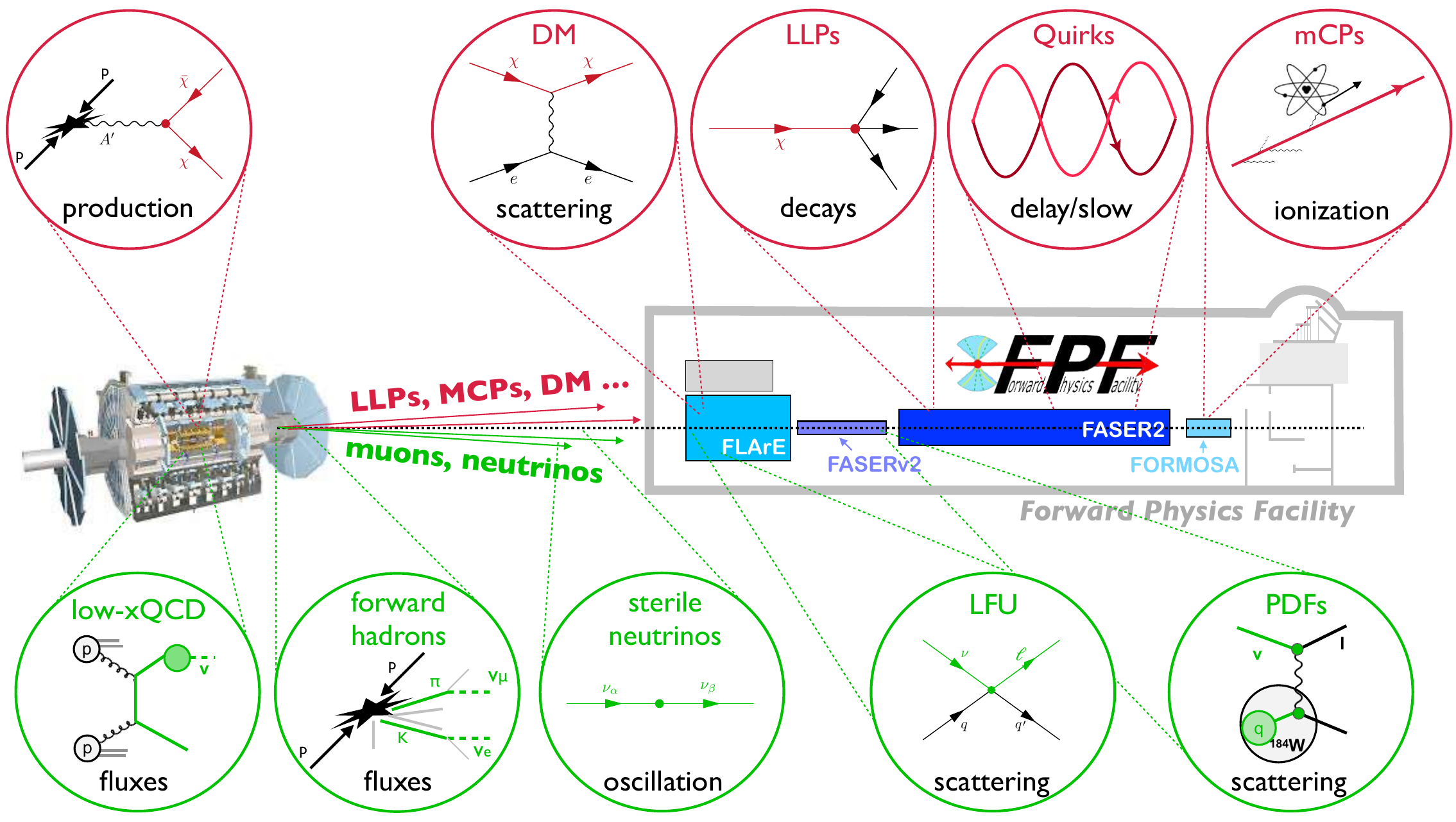}
  \caption{\textbf{New Particle Searches and Neutrino Measurements at the FPF.} Representative examples of DM and other new particles that can be discovered and studied at the FPF (top) and of some of the many topics that can be illuminated by TeV-energy neutrino measurements at the FPF (bottom). }
  \label{fig:Phys_overview}
\end{figure}

\noindent \textbf{Overview:} The FPF physics program encompasses a wide range of searches for BSM phenomena alongside unique and guaranteed SM measurements, as illustrated in \cref{fig:Phys_overview}. This is made possible by the complementary capabilities of the suite of FPF experiments. The SM program exploits the unprecedented flux of collider neutrinos observed by FLArE and FASER$\nu$2 to study lepton flavour universality and non-standard interactions in the neutrino sector, probe QCD in unexplored kinematic regimes, and address longstanding puzzles in astroparticle physics. On the BSM front, the program includes searches for long-lived particles (LLPs) decaying into visible final states at FASER2, dark matter (DM) scattering signatures detectable at FLArE, and unconventional ionisation signals from particles with fractional electric charge, observable at FORMOSA. \medskip

\noindent \textbf{Documentation:} The science case for the FPF has been developed in nine dedicated FPF meetings~\cite{FPF1Meeting, FPF2Meeting, FPF3Meeting, FPF4Meeting, FPF5Meeting, FPF6Meeting, FPF7Meeting, FPF8Meeting, FPFTheoryWS}. The physics opportunities have been summarised in an 80-page review~\cite{Anchordoqui:2021ghd} and a more comprehensive 430-page White Paper~\cite{Feng:2022inv}, written and endorsed by 400 physicists. While these documents reflect the state of knowledge and priorities at the time of their publication, the field has continued to evolve, with steady progress in the development of new ideas and the refinement of earlier studies. A recent overview of the FPF's scientific program was published in Ref.~\cite{Adhikary:2024nlv}, and a more concise version written for the 2024-2026 European Particle Physics Strategy Update can be found in Ref.~\cite{FPFWorkingGroups:2025rsc}.  

In the following, we summarize this broad program for neutrino physics, QCD, astroparticle physics, DM, and other new physics searches. We note that the associated physics sensitivity studies have been carried out with varying levels of maturity, for example in their treatment of backgrounds and efficiencies. However, it should be noted that for those assuming zero background and 100\% efficiency, it has been shown at the current FASER experiment that the actual sensitivity of published analyses are almost identical to truth level projections with these assumptions~\cite{FASER:2023tle}. We note that all quantitative results have been updated to an integrated luminosity of $2~\iab$. \medskip

\subsection{Neutrino Physics}

\noindent \textbf{The Dawn of Collider Neutrino Physics:}
The LHC is the highest-energy particle collider built to date, and it is therefore also the source of the most energetic neutrinos produced in a controlled laboratory environment. Indeed, the LHC generates intense, strongly-collimated, and highly-energetic beams of both neutrinos and antineutrinos of all three flavours in the forward direction. Although this high flux was noted in the 1980s~\cite{DeRujula:1984pg}, experimental efforts to exploit this potential started only in the beginning of the 2020s. During the LHC's Long Shutdown~2 (2019-2022), two experiment were installed to take advantage of this opportunity: FASER~\cite{FASER:2022hcn}, with its dedicated FASER$\nu$ neutrino detector, and SND@LHC~\cite{SNDLHC:2022ihg}. These experiments have been taking data since the summer of 2022, and they reported the first direct detection of collider neutrinos in 2023~\cite{FASER:2023zcr,SNDLHC:2023pun}.

To use the words of Elizabeth Worcester’s Viewpoint article~\cite{Worcester:2023njy}, the first observation of neutrinos at the LHC by FASER and SND@LHC in 2023 marks the \textit{dawn of collider neutrino physics}. Since then, the FASER Collaboration has performed the first measurement of the neutrino interaction cross section at TeV energies~\cite{FASER:2024hoe} or, alternatively, assuming the SM interaction cross section as known, the far-forward neutrino flux at the LHC~\cite{FASER:2024ref}. By the end of LHC Run 3 in 2026, the existing experiments are expected to detect approximately $10^3$ electron neutrino, $10^4$ muon neutrino, and $10^2$ tau neutrino charged current (CC) interactions.\medskip

\begin{figure}[tb]
  \centering
  \includegraphics[width=0.99\textwidth]{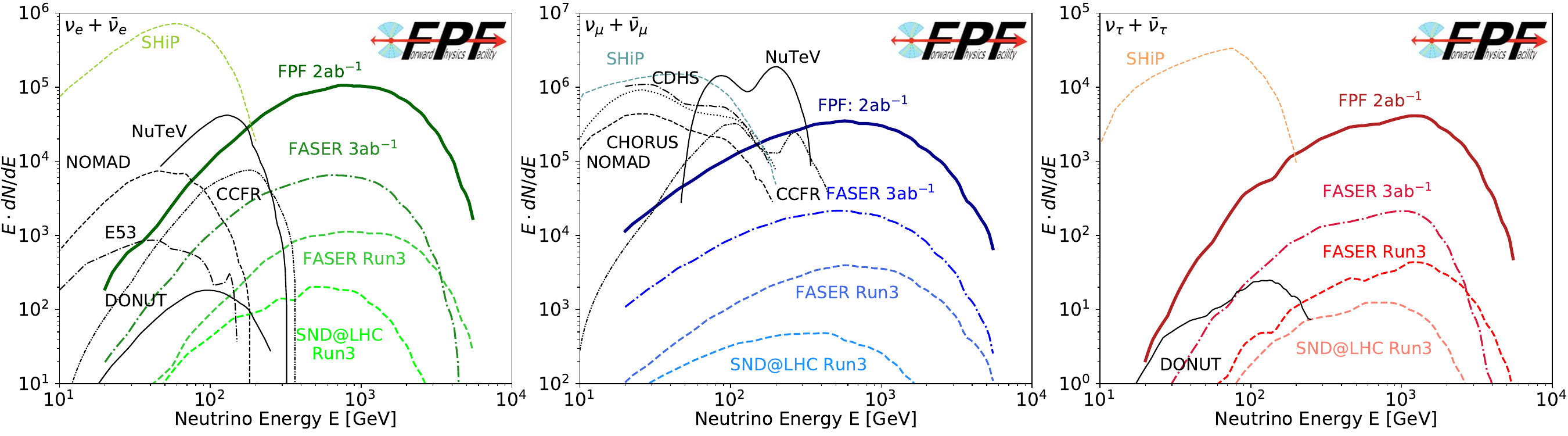}
  \caption{\textbf{Neutrino Yields at the FPF.} The panels show the expected event rates and energy spectra of neutrinos undergoing CC interactions at the existing FASER$\nu$ and SND@LHC detectors with about 1~ton target mass and 350~fb$^{-1}$ integrated luminosity (dashed); a detector at the FASER location with 1~ton target mass operating during the full HL-LHC era with 3~ab$^{-1}$ integrated luminosity (dash-dotted); and the FPF, with the 20~ton FASER$\nu$2 and 10~ton FLArE detectors, with 2~ab$^{-1}$ integrated luminosity (solid). The spectra shown are for electron (left), muon (middle), and tau (right) neutrinos. Spectra from previous accelerator experiments~\cite{ParticleDataGroup:2020ssz} and the planned SHiP experiment~\cite{Ahdida:2023okr} are also shown for comparison. Details of the simulation are discussed in the text.}
  \label{fig:Phys_NuSpectra}
\end{figure}

\noindent \textbf{Expected Neutrino Event Rates:} The FPF experiments, with larger detectors and higher luminosities, are projected to detect a hundred thousand electron neutrino, a million muon neutrino, and thousands of tau neutrino CC interactions, providing approximately 100 times more statistics than current experiments. We have estimated the expected neutrino fluxes and event rates in the detectors using a fast neutrino flux simulation~\cite{Kling:2021gos}. We use \texttt{EPOS-LHC}~\cite{Pierog:2013ria} to simulate the production of light hadrons and POWHEG matched with Pythia as obtained in Ref.~\cite{Buonocore:2023kna} to simulate the production of charm hadrons. The expected energy spectra of interacting neutrinos for the existing experiments operating in LHC Run~3, for a proposed 1~ton detector at the FASER location operating at the HL-LHC~\cite{FASER:2025myb}, and the FPF neutrino detectors, FLArE and FASER$\nu$2, are shown in \cref{fig:Phys_NuSpectra} for all three neutrino flavours. For comparison, we also show the spectra of previous neutrino experiments, such as NuTeV, CHARM and DONuT. The energy spectra of collider neutrino experiments peak at $\sim \tev$ energies, above the coverage of all previous accelerator-based neutrino detectors. We also display the expected neutrino event rates at SHiP, which peak at much lower energies $E \alt$ 100~GeV.\footnote{Additional neutrino detectors positioned at the surface exit points of the LHC neutrino beam---the closest being approximately 9~km from the corresponding interaction point (IP)---have also been considered~\cite{Ariga:2025gtj, Kamp:2025phs}. However, the neutrino flux is significantly diluted over such distances, leading to a substantial reduction in neutrino yield and necessitating large, coarse detectors. It was concluded that these factors considerably limit the physics potential of surface-level detectors compared to the FPF detectors, which are located much closer to the IP.}

\begin{table}[tb]
\setlength{\tabcolsep}{5.0pt}
  \centering
  \begin{tabular}{c|c|c|c||c|c|c}
  \hline\hline
  \multicolumn{4}{c||}{Detector} & 
  \multicolumn{3}{c}{CC Interactions} \\
  \hline
  Name & Mass & Luminosity & Rapidity 
  & $\nu_e\!\!+\!\bar{\nu}_e$ 
  & $\nu_\mu\!\!+\!\bar{\nu}_\mu$
  & $\nu_\tau\!\!+\!\bar{\nu}_\tau$
  \\
  \hline\hline
  FASER$\nu$ at Run~3
  & 1.1~t & 350~fb$^{-1}$ & $\eta>8.8$
  & 2.3k & 12k & 40 \\
  SND@LHC at Run~3
  & 0.8~t & 350~fb$^{-1}$ & $7.2<\eta<8.4$
  & 300  & 1.5k & 12 \\ 
  \hline
  FASER at HL-LHC
  & 1.1~t & 3~ab$^{-1}$ & $\eta>8.8$
  & 19k & 102k & 360 \\
  SND@HL-LHC 
  & 1.3~t & 3~ab$^{-1}$ & $6.9<\eta<7.6$
  & 2.9k & 15k & 143 \\
  \hline
  FASER$\nu$2 at FPF
  & 20~t & 2~ab$^{-1}$ & $\eta>8.5$
  & 127k & 647k & 2.3k\\
  FLArE at FPF
  & 10~t & 2~ab$^{-1}$ & $\eta>7.5$
  & 34.7k & 167k & 1.0k\\
  FLArE HCAL at FPF
  & 41~t & 2~ab$^{-1}$ & $\eta>6.5$
  & 34.0k & 180k & 1.5k\\
  FASER2 veto at FPF
  & 0.9~t & 2~ab$^{-1}$ & $\eta>6.7$
  & 1.6k & 6.8k & 54\\

\hline\hline
\end{tabular}
\caption{Expected numbers of CC neutrino interactions in LHC neutrino detectors, together with their mass, expected luminosities and rapidity coverage. Numbers for SND@HL-LHC were taken from Ref.~\cite{SNDLHC:2025qtx}.}
  \label{tab:Phys_NuNumber}
\end{table} 

In \cref{tab:Phys_NuNumber}, we present the number of CC neutrino interacions expected at the FLArE and FASER$\nu$2 experiments, along with the currently-operating detectors and their planned upgrades. The planned upgrades assume that detectors with similar size compared to the current ones continue to operate in roughly the same locations during the HL-LHC era~\cite{FASER:2025myb, SNDLHC:2025qtx}.\footnote{We note that the SND@HL-LHC proposal~\cite{SNDLHC:2025qtx, Abbaneo:2926288} presents expected neutrino event rates using both DPMJET and POWHEG to simulate the neutrino flux from charm hadron decay. Here we display their results obtained using POWHEG, consistent with the other numbers presented. We note, however, that DPMJET is well known to overestimate the neutrino flux by up to an order of magnitude since it neglects the charm quark mass~\cite{FASER:2024ykc} and its predictions are strongly disfavoured by data~\cite{FASER:2024ref}. We caution the reader that many results presented in the SND@HL-LHC proposal and previous documents by the SND@LHC collaboration rely on DPMJET and may hence overestimate the physics sensitivity.} The table also summarises the respective target masses, luminosities and rapidity coverage. In addition to the main FPF neutrino targets, we also present numbers for the hadronic calorimeter of FLArE, a roughly 40-ton steel detector, which can act as a neutrino target on its own. Although it exceeds the other detectors in total mass, its transverse area of $3.5~\m \times 2~\m$ extends further off-axis to lower rapidities, where the neutrino flux is smaller, leading to an overall rate that is comparable to FLArE. Also shown are numbers for a plastic target in the veto system at FASER2~\cite{Kling:2025lnt}. While small in mass, this target has the unique feature that electrons can escape and reach the spectrometer before interacting, allowing separate $\nu_e$ and $\bar\nu_e$ measurements. Although \cref{tab:Phys_NuNumber} and \cref{fig:Phys_NuSpectra} highlight the total number of neutrino interactions, we note that detector capabilities are equally important. We expect FLArE and FASER$\nu$2 to offer better performance than SND@HL-LHC, FASER at Run-4, and the FLArE HCAL, the latter of which features much less granular readout than FLArE and FASER$\nu$2.

\begin{figure}[bth]
\centering
\vspace{3mm}
\includegraphics[width=1.00\textwidth]{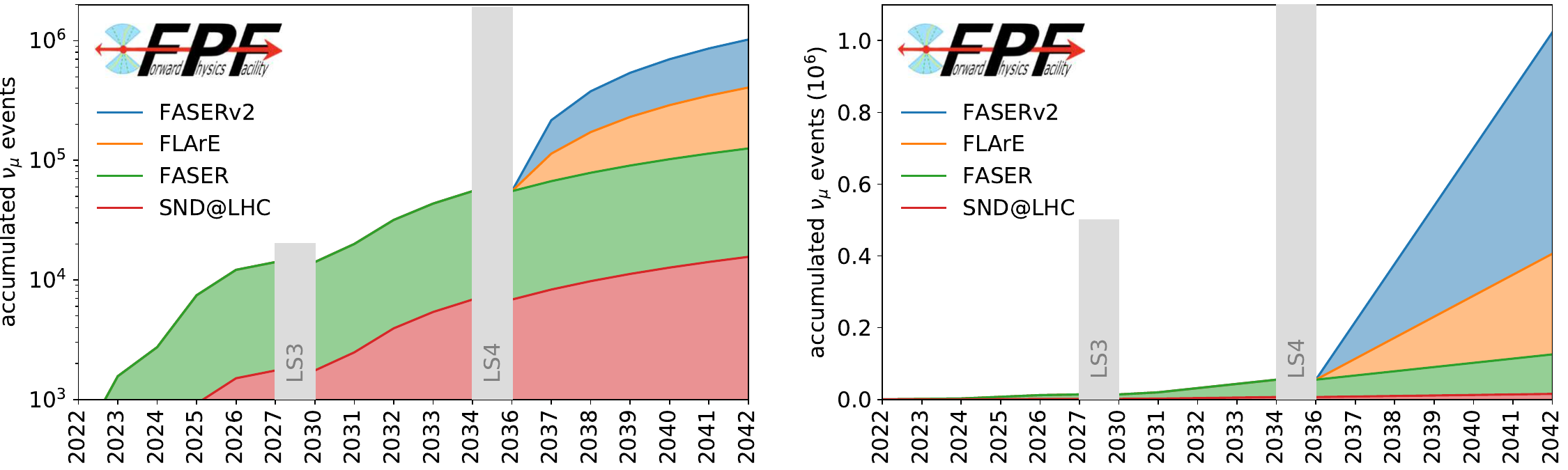}
\caption{\textbf{Neutrino Event Rate Timeline.} Accumulated number of CC neutrino interaction events recorded by existing and proposed detectors at the end of each year in logarithmic (left) and linear (right) scales assuming $150~\ifb$ of data collected in 2031 and $300~\ifb$ in all later years. The projection includes the currently operating FASER and SND@LHC experiments, assuming --- for simplicity --- that similarly-sized detectors continue to operate at the same locations during the HL-LHC era. It also includes the proposed FASER$\nu$2 and FLArE detectors at the FPF, including the FLArE HCAL, which are assumed to begin full data-taking at he beginning of LHC Run~5.}
\vspace{1mm}
\label{fig:Phys_NuTimeline}
\end{figure}

In \cref{fig:Phys_NuTimeline}, we show the expected number of neutrino events as a function of time. The figure includes both the existing experiments, FASER and SND@LHC, as well as the FASER$\nu$2 and FLArE experiments at the FPF. Since there is some  uncertainty on the timeline for approval the figure conservatively assumes physics data taking starting at the beginning of Run~5 in 2036.\medskip

\begin{figure}[tb]
  \centering
  \includegraphics[width=0.99\textwidth]{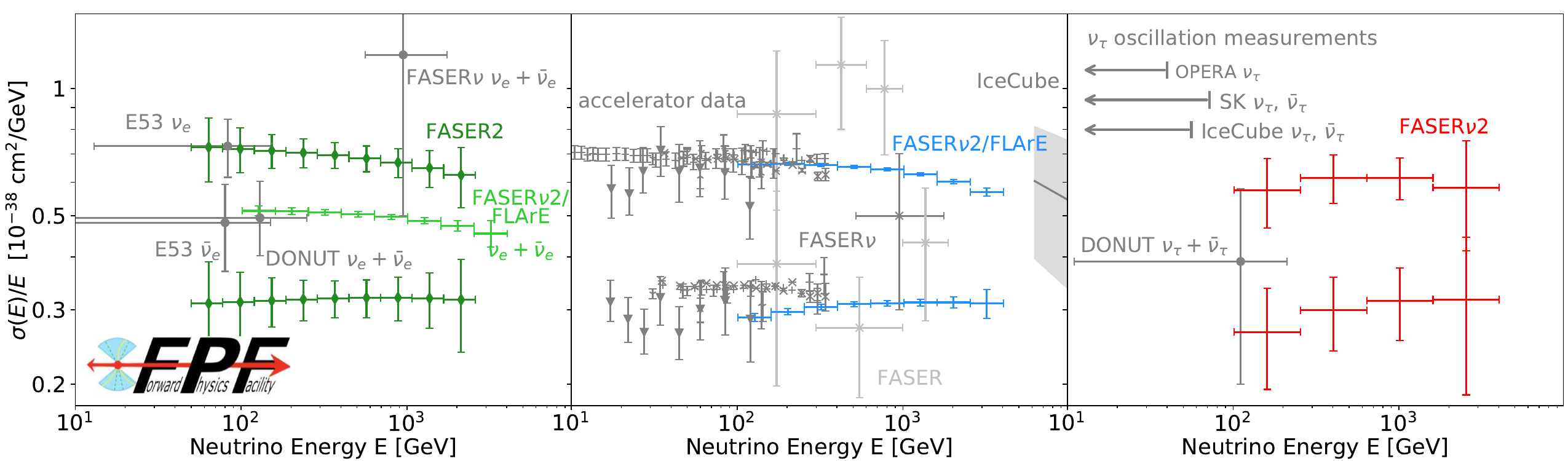}
  \caption{\textbf{Neutrino Cross Sections at the FPF.} The expected precision of FPF measurements of neutrino CC interaction cross sections (statistical errors only) as a function of energy for electron (left), muon (middle), and tau (right) neutrinos with an integrated luminosity of $2~\iab$. In the case of muon and tau neutrinos, separate measurements of the neutrino and antineutrino measurement can be performed at FASER$\nu$2 and FLArE using muons passing through the FASER2 spectrometer, where a 17\% branching fraction of taus into muons was considered. For electron neutrinos, FASER$\nu$2 and FLArE measure a flux-averaged cross section, while a dedicated 0.6~ton plastic target placed in the veto system of the FASER2 spectrometer will make it possible to perform separate measurements~\cite{Kling:2025lnt}. Existing data from accelerator experiments~\cite{ParticleDataGroup:2020ssz}, IceCube~\cite{IceCube:2017roe}, and FASER~\cite{FASER:2024hoe, FASER:2024ref} are also shown.}
  \label{fig:Phys_NuXS}
\end{figure}

\noindent \textbf{Neutrino Cross Sections at TeV Energies:} Given the large energies available at the LHC, neutrino experiments at the LHC currently provide the unique opportunity to probe neutrino interactions at TeV energies in a controlled laboratory environment. \cref{fig:Phys_NuXS} displays the expected statistical precision of the FPF experiments to measure the inclusive neutrino-nucleon CC scattering cross sections for all three neutrino flavours, along with existing measurements. For muon neutrinos, the low-energy region has been well-constrained by previous accelerator neutrino experiments~\cite{NOMAD:2007krq, Berge:1987zw, Seligman:1997fe, NuTeV:2005wsg, Baltay:1988au, DONuT:2007bsg}. In contrast, fewer and less precise measurements are available for electron and tau neutrinos. IceCube \cite{IceCube:2016zyt} has also measured the muon neutrino cross section at very high energies using atmospheric neutrinos, although with relatively large uncertainties~\cite{IceCube:2017roe, Bustamante:2017xuy, IceCube:2020rnc}. First measurements from FASER are also shown~\cite{FASER:2024hoe, FASER:2024ref}. Measurements at the FPF will provide an unprecedented precision for cross section measurements at TeV energies. \medskip

\noindent \textbf{Heavy Flavour-Associated Neutrino Interactions:}
Beyond inclusive neutrino interaction cross sections, the FPF experiments will enable detailed studies of exclusive processes, particularly those involving heavy-flavour production. Charm-associated muon neutrino interactions, $\nu_\mu N\!\to\!\mu X_c \!+\! X$, which account for approximately 15\% of neutrino interactions at TeV energies, have previously been observed by CCFR, NuTeV, and CHORUS~\cite{Rabinowitz:1993xx, NuTeV:2001dfo, Kayis-Topaksu:2011ols}. FASER$\nu$2 with $2~\ifb$ of integrated luminosity is expected to collect around 100k such events, which will significantly increase the available statistics for this process. In addition, FASER$\nu$2 will record approximately 19k charm-associated electron neutrino interactions and 330 charm-associated tau neutrino interactions, which will make it possible to observe and study these processes for the first time. Unlike CCFR and NuTeV, which identified charm via muonic decays leading to dimuon final states, FASER$\nu$2 will use its emulsion detector to directly reconstruct charm decays via displaced secondary vertices. This approach offers roughly an order-of-magnitude improvement in signal efficiency. These measurements will provide valuable probes of the nucleon's strangeness content, the CKM matrix element $V_{cd}$, and charm fragmentation fractions. 

Additionally, FASER$\nu$2 will be able to see bottom-associated neutrino interactions $\nu N\!\to\!X_b\!+\! X$. This process requires a large neutrino energy to exceed the kinematic threshold of $b$-quark production. It is further strongly CKM suppressed in the SM, resulting in a suppression of $V_{ub}^2 \sim 10^{-5}$. Simulations show that FASER$\nu$2 is expected to record about 26 such events, providing a unique opportunity to observe this rare process for the first time.\medskip

\noindent \textbf{Tau Antineutrino Observation and Tau Neutrino Precision Physics:} Of the seventeen particles in the SM, the tau neutrino remains the least well measured. In particular, no separate observations of tau neutrinos and tau antineutrinos have been made to date. Detecting a $\nu_\tau$ directly requires a sufficiently energetic neutrino beam to produce a $\tau$ lepton, followed by the successful identification of that $\tau$. This is challenging due to the $\tau$ lepton’s short lifetime of $c\tau = 87~\mu\text{m}$ and the presence of an undetectable $\nu_\tau$ in all $\tau$ decays. To date, the most direct information on tau neutrinos comes from the DONuT and OPERA experiments, which each observed approximately ten $\nu_\tau$ events~\cite{DONuT:2007bsg, OPERA:2018nar}. In addition, Super-Kamiokande and IceCube have reported higher-statistics $\nu_\tau$ appearance through atmospheric oscillations~\cite{Super-Kamiokande:2017edb, IceCube:2019dqi}. Their results, however, rely on statistical identification rather than direct observation. As a consequence, their constraints on the $\nu_\tau$ cross section remain at the $\sim$30\% uncertainty level, comparable to those from OPERA. In the future, the neutrino detector at SHiP may provide high statistics $\nu_\tau$ measurements up to $E_\nu \sim 150~\gev$~\cite{SHiP:2015vad}.

Tau neutrinos constitute approximately 0.3\% of the neutrino flux produced at the LHC. Despite their relatively small fraction, the high intensity of the LHC beam ensures that thousands of tau neutrinos and antineutrinos will interact in the detectors of the FPF. The FASER$\nu$2 detector has been designed to identify these interactions, leveraging its emulsion-based technology and its integration with the FASER2 spectrometer for charge and momentum measurement. This capability will enable FASER$\nu$2 to achieve a major milestone in particle physics: \textbf{the first observation of the tau antineutrino}, the only remaining SM particle yet to be directly detected. Similar efforts are pursued by the SHiP experiment, likely on a similar timescale.\medskip

\begin{wrapfigure}{R}{0.5\textwidth}
  \centering  
  \vspace{-3mm}
  \includegraphics[width=0.50\textwidth]{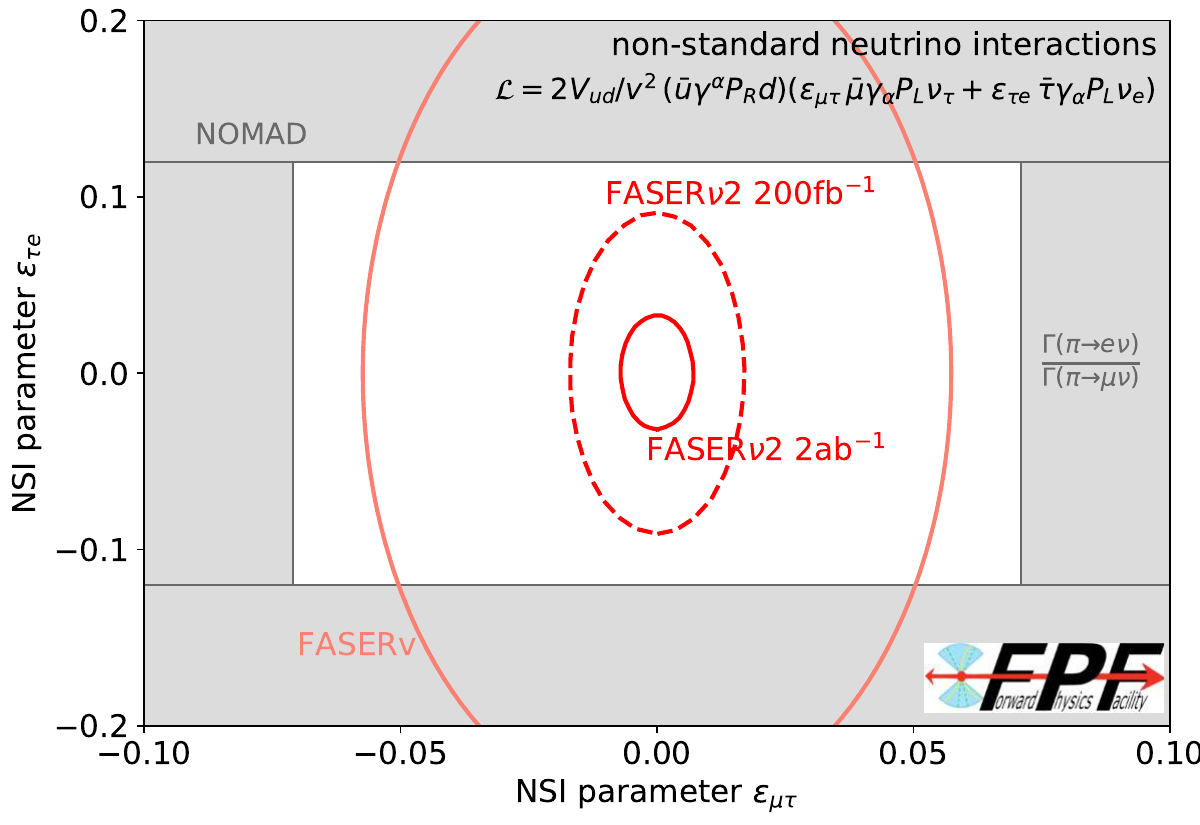}
\caption{\textbf{Precision Tau Neutrino Studies at the FPF.} The projected sensitivities of  FASER$\nu$ and FASER$\nu$2 to neutrino NSI parameters that violate lepton flavour universality~\cite{Kling:2023tgr}. Past NOMAD bounds~\cite{NOMAD:2001xxt, NOMAD:2003mqg} are presented based on Ref.~\cite{Biggio:2009nt}. Constraints from the measured ratio of pion decay widths to the electron and muon~\cite{ParticleDataGroup:2024cfk} are obtained based on Ref.~\cite{Falkowski:2021bkq}.}
  \label{fig:tau_neutrino}
\end{wrapfigure}

\noindent \textbf{Tests of Lepton Flavour Universality:} The large expected number of tau neutrino interactions will also open up a new window to an era of tau neutrino precision studies at TeV energies. These measurements will provide a powerful platform for testing lepton flavour universality in the neutrino sector. Deviations from universality can, for example, be parameterised by neutrino non-standard interactions (NSI), which are four-fermion effective field theory operators involving neutrinos. The potential of LHC neutrino experiments to constrain CC NSI using neutrino interaction measurements $\nu_\ell q \to \ell' q'$ has been studied in Ref.~\cite{Falkowski:2021bkq}. Although many NSI operators are already well constrained by either precision measurements of meson decays at flavour factories or LHC measurements probing the related processes $q q' \to \nu_\ell \ell'$, collider neutrino experiments have the potential to obtain world-leading constraints on operators associated with tau neutrinos by utilising the sizable $\nu_\tau$ flux at the LHC. 

An example of FASER$\nu$2's sensitivity to probe two such NSI operators is shown in \cref{fig:tau_neutrino}, as studied in Ref.~\cite{Kling:2023tgr}. Here, two operators associated with the right-handed quark current are considered, $\mathcal{L} = 2\,V_{ud}/v^2 \times (\bar{u}\gamma^\alpha P_Rd) \times \left[\epsilon_{\mu\tau}\,(\bar{\mu} \gamma_\alpha P_L \nu_\tau) + \epsilon_{\tau e}\,(\bar{\tau} \gamma_\alpha P_L \nu_e)\right]$. The first term, proportional to $\epsilon_{\mu \tau}$, leads to an enhanced $\nu_\tau$ flux from the new pion decay mode $\pi^+ \to \mu \nu_\tau$, while the second term, proportional to $\epsilon_{\tau e}$, induces the interaction $\nu_e d \to \tau u$. In both cases, an excess of tau lepton events would be observed in the detector. In \cref{fig:tau_neutrino} we see that FASER$\nu$2 at the FPF will significantly improve the constraints on these NSI operators compared to previous results. We also note that the $\epsilon_{\mu \tau}$ term is difficult to probe at beam dump experiments, including SHiP, where pions are absorbed before decaying. These results underscore the unique role of collider neutrino experiments in advancing our understanding of tau neutrino interactions and their potential connection to new physics. In addition, the FPF program also has strong sensitivity to neutral-current (NC) NSI operators, as discussed in Refs.~\cite{Ismail:2020yqc, Kling:2025lnt}.\medskip

\noindent \textbf{Resonances in Neutrino Electron Scattering:} Despite the relatively small cross section, about 2000 CC neutrino–electron scattering events are expected to occur in  FASER$\nu$2 alone. This will provide a valuable standard candle for neutrino flux measurements. In addition, the high energy of the LHC neutrino beam also opens a unique opportunity to observe $s$-channel resonant vector-meson production in neutrino scattering for the first time. The lightest viable resonance is the $\rho^-$ meson, requiring neutrino energies $E_\nu \sim m_\rho^2 / 2m_e \simeq 570$~GeV which are accessible only via LHC neutrinos. Ref.~\cite{Brdar:2021hpy} estimated that approximately 33 $\nu_e e^- \to \rho^- \to \pi^- \pi^0$ events are expected to occur in the FPF experiments and presented a background mitigation strategy to enhance the signal-to-background ratio sufficiently to permit a definitive discovery at FASER$\nu$2. \medskip

\noindent \textbf{Scattering with Low Energy Transfer:} Although LHC neutrinos typically carry energies well above 10~GeV, their energy transfer to nuclei in FPF detectors can be much lower, especially in elastic, quasi-elastic (QE), or resonant interactions. Several approaches to low-Q extrapolations of weak structure functions in the non-perturbative regime, such as the Bodek-Yang~\cite{Bodek:2002vp, Bodek:2004pc, Bodek:2010km}, NNSF$\nu$~\cite{Candido:2023utz} and CKMT+PCAC-NT~\cite{Jeong:2023hwe} models, can be tested and refined with FPF experiments. Identifying such events in FASER$\nu$2 and FLArE, which are characterised by low hadronic activity but potentially large electromagnetic deposits or high-energy muons, will provide an additional handle to constrain neutrino fluxes. These and other channels with low energy transfer allow for a more precise reconstruction of the incident neutrino energy and, due to their well-known cross sections, can serve as standard candles for flux measurements -- a technique commonly referred to as the low-$\nu$ method~\cite{Wilkinson:2023vvu}. Moreover, studying these exclusive processes at the FPF will extend previous measurements into the high-energy regime, where data remain sparse. The results will advance efforts to develop a universal model of QE and QE-like scattering across a broad energy range, one of the field’s key open challenges identified by the Neutrino Scattering Theory Experiment Collaboration (NuSTEC)~\cite{NuSTEC:2017hzk}.\medskip

\noindent \textbf{Neutrino Tridents:} The production of dilepton pairs in neutrino trident scattering off a nucleus $N$, $\nu N\to\nu^{(\prime)} \ell^-\ell^{(\prime)+} N$, is well recognised as a sensitive probe of electroweak and BSM physics. Although previous claims of the observation of dimuon neutrino tridents have been made at the 3$\sigma$ level by the CHARM-II~\cite{CHARM-II:1990dvf} and CCFR~\cite{CCFR:1991lpl} Collaborations, further backgrounds, previously unaccounted for, were identified in a later analysis by the NuTeV Collaboration~\cite{NuTeV:1999wlw}. It has been shown in Ref.~\cite{Altmannshofer:2024hqd} that the forward neutrino program at the FPF provides a promising opportunity to measure dimuon neutrino tridents with a statistical significance exceeding 5$\sigma$ for the first time, based on the reverse tracking of dimuon pairs in the FASER$\nu$2 detector, mitigating all backgrounds. Moreover, the unique energy range of the FPF experiments enables probes of BSM contributions to vector and axial vector couplings~\cite{Altmannshofer:2024hqd} that are complementary to those projected for DUNE~\cite{Altmannshofer:2019zhy}.
\medskip

\noindent \textbf{Neutrino Electromagnetic Properties:} Although neutrinos are electrically neutral in the SM, electromagnetic interactions can arise through quantum loops or new physics, leading to properties such as nonzero charge, magnetic dipole moments, and charge radii. These effects could help explain various anomalies~\cite{Sakstein:2020axg, MiniBooNE:2007uho}, and a measured dipole moment would offer insight into the Dirac or Majorana nature of neutrinos~\cite{Shrock:1982sc, Frere:2015pma}. Ref.~\cite{MammenAbraham:2023psg} studied the FPF's potential to probe these properties and found that searches for low-recoil electron scattering at FLArE will set the strongest laboratory-based limits on the tau neutrino magnetic moment of $\mu_{\nu_\tau} \leq 7 \cdot 10^{-8}~\mu_B$ and an electric charge of $|Q_{\nu_\tau}| \leq 4.1 \cdot 10^{-5}~e$. Furthermore, precise measurements of NC deep inelastic scattering (DIS) in FLArE and FASER$\nu$2 could yield leading constraints on the charge radii of electron and muon neutrinos, almost reaching their predicted SM value~\cite{MammenAbraham:2023psg}. \medskip

\noindent \textbf{Weak Mixing Angle Measurements:} The measurement of neutrino interactions at the energies accessible at the FPF provide an opportunity to precisely measure electroweak parameters. This includes the  weak mixing angle, $\sin^2 \theta_W$, which governs the relative strength of NC and CC interactions in the SM. Various measurements have been carried out across a wide energy range, including neutrino scattering measurements at NuTeV, which measured a value about $3\sigma$ above the SM prediction at a scale of $\sim 4~\gev$~\cite{NuTeV:2001whx}. Ref.~\cite{MammenAbraham:2023psg} studied the precision to similarly constrain the weak mixing angle via neutrino scattering at the FPF. The study found that $\sin^2 \theta_W$ could be measured to about 3\% precision at a scale of $\sim 10~\gev$, providing a data point complementary to existing measurements at other energy scales. 

\subsection{Quantum Chromodynamics}
\label{sec:qcd}

\noindent \textbf{A New Window into QCD:} In addition to offering unique insights into neutrino physics, the intense beam of TeV-energy neutrinos and muons reaching the FPF will serve as powerful tools for studying QCD. 
The interactions of these leptons with the detector target will probe proton~\cite{Gao:2017yyd} and nuclear~\cite{Klasen:2023uqj} parton distribution functions (PDFs) in poorly understood regimes, in particular the large-$x$ region, and scrutinise quark/antiquark flavour separation including the proton's heavy quark content. 
Furthermore, the neutrino fluxes measured at the FPF neutrino detectors will provide additional information on forward particle production and the structure of the colliding protons in the small-$x$ regime that is inaccessible to the main LHC experiments. 
Both types of measurements (scattering and production) will have wide-ranging implications refining our understanding of perturbative and non-peturbative QCD, allowing us to improve predictions for the ATLAS and CMS experiments in the search for new physics and the characterisation of the Higgs boson during the HL-LHC era, and providing a better understanding of cosmic-ray collisions in the atmosphere as described in \cref{sec:astroparticle}.\medskip 

\noindent \textbf{The FPF as a Neutrino-Ion Collider:} The FPF will extend the legacy of previous neutrino DIS experiments such as NuTeV~\cite{NuTeV:2005wsg}, CDHS~\cite{Berge:1987zw}, and CHORUS~\cite{CHORUS:2005cpn} by pushing into significantly higher energies and thus broader kinematic regimes.
With center-of-mass energies for neutrino-nucleus interactions in the $10 - 70$~GeV range, the FPF enables a CC \textit{neutrino-ion collider}~\cite{Cruz-Martinez:2023sdv} program at the LHC, analogous to the NC program at the upcoming Electron-Ion Collider (EIC) which probes $\sqrt{s} = 30 - 80$~GeV~\cite{Accardi:2012qut}. These complementary DIS measurements probe different linear combinations of quark and antiquark PDFs, and thus are critical for breaking degeneracies and disentangling the underlying structure of protons and nuclei.

\begin{wrapfigure}{R}{0.62\textwidth}
  \centering
  \vspace{-7mm}
  \includegraphics[width=0.62\textwidth]{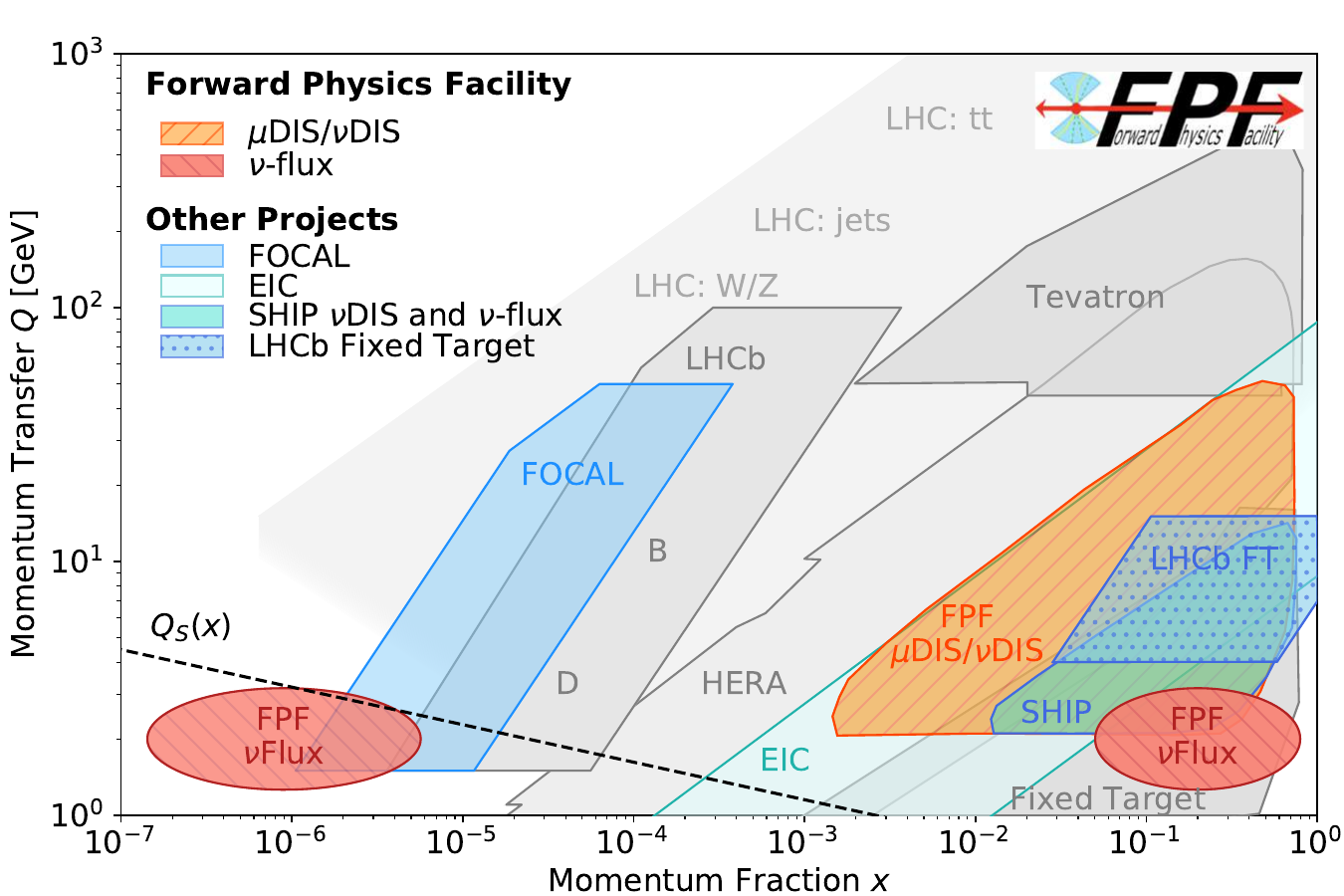}
\caption{\textbf{Kinematic Landscape of QCD Measurements.} Kinematic coverage of measurements at the FPF, both those related to neutrino production (red) and neutrino/muon scattering (orange), compared to past/ongoing experiments (gray), and other proposed or planned experiments (blue) in the partonic momentum fraction $x$ versus momentum transfer $Q$ plane. Also indicated is the saturation scale $Q_s(x)$ in the proton (dashed line)~\cite{ALICE:2023fov}. Adapted from Ref.~\cite{PBC:2025sny}.}
  \vspace{-3mm}
  \label{fig:QCD_Landscape}
\end{wrapfigure}

The kinematics of these neutrino DIS events are parameterised in terms of the partonic momentum fraction $x= Q^2/(2m_p E_\nu)$ and momentum transfer $Q^2=4E_\nu E_\ell\sin^2(\theta_\ell/2)$.
The factor-of-10 larger neutrino energies at the LHC compared to previous fixed target experiments make it possible to probe roughly 10-times-smaller values of $x$ and 10-times-larger values of $Q$. 
This extension is reflected in \cref{fig:QCD_Landscape}, which shows the kinematic coverage of past, ongoing, and future experiments in the $(x, Q)$ plane. As indicated by the orange-shaded region, neutrino DIS measurements at the FPF can access regions of $x$ and $Q^2$ that no other previous or ongoing fixed target experiment, nor the planned SHiP experiment, is capable of. 
In particular, the FPF will have statistically-significant event rates down to $x$ $\gtrsim 10^{-3}$ and $Q\lesssim 50~{\rm GeV}$~\cite{Cruz-Martinez:2023sdv}.  This sensitivity is driven by muon-neutrinos, which dominate the CC event rate. 
\cref{fig:QCD_Landscape} also highlights the excellent sensitivity of the FPF, via forward neutrino production, to the gluon PDF down to $x\sim 10^{-7}$ and at very large-$x$ (red ellipses).

Beyond reconstructing the relevant kinematic quantities, the FPF experiments are able to make important measurements that allow for further scrutiny of  DIS events.
As mentioned before, about 15\% of neutrino interactions will produce a charm hadron, which can be tagged through its decay topology in FASER$\nu$2. The charm tagging effectively indicates that the neutrino interacted with a down-type quark, predominantly the strange quark, which provides a strong handle to probe the relatively less-constrained strange quark PDF~\cite{Faura:2020oom}. %
Further information on the DIS event provided by the magnet in FASER2, which allows for the incident particle to be identified as a neutrino or antineutrino and further distinguish contributions from initial-state quarks and anti-quarks.
Finally, semi-inclusive DIS measurements with identified pions and kaons can be compared with predictions evaluated at NNLO in the QCD expansion~\cite{Bonino:2025tnf}, providing constraints on complementary quark flavour combinations, which can be used to extract non-perturbative QCD fragmentation functions~\cite{Metz:2016swz}. 
\medskip 

\noindent \textbf{Proton Structure from Neutrino DIS:}
By simulating the DIS event distribution at forward experiments, the constraining power of these measurements to reduce PDF uncertainties in state-of-the-art global PDF analyses was demonstrated in Ref.~\cite{Cruz-Martinez:2023sdv}. The impact of FPF data on the global PDF4LHC21~\cite{PDF4LHCWorkingGroup:2022cjn} PDFs was estimated via a Hessian PDF profiling procedure~\cite{Paukkunen:2014zia, Schmidt:2018hvu, AbdulKhalek:2018rok, HERAFitterdevelopersTeam:2015cre}, implemented in the \textsc{xFitter} open-source QCD analysis framework~\cite{Alekhin:2014irh, Bertone:2017tig, xFitter:2022zjb, xFitter:web} and is shown in \cref{fig:FPF_PDFs}. Despite the wealth of previously used data, FPF data can reduce uncertainties in the up valence and down valence quark PDFs by up to a factor of 2 or more, with even larger improvements for the strange quark for $x\gtrsim10^{-3}$ and $Q^2\lesssim10^3~{\rm GeV}^2$. The large improvement in the strange quark content is largely due to the FASER$\nu$2 charm tagging ability, which significantly increases the event statistics with respect to previous experiments relying on a di-muon signature. These results assumed that the best-fit central PDFs remain unchanged, while the observation of a deviation would have an even larger impact on the PDFs. Similar results were obtained by also including the simulated data in a global fit in the NNPDF framework. Although these studies were performed using a proton PDF as the baseline, the improvement is also verified to be in good correspondence~\cite{Cruz-Martinez:2023sdv} after accounting for nuclear corrections using a nuclear PDF set for tungsten~\cite{Eskola:2021nhw}. The impact of upcoming measurements at current neutrino experiments FASER and SND@LHC was also estimated: it was found that the yield after Run~3 would be insufficient to have any impact, while measurements at upgraded detectors operating during the HL-LHC era would only mildly improve the PDFs as also shown in \cref{fig:FPF_PDFs}. 
\medskip

\begin{figure}
  \centering
  \includegraphics[width=0.32\textwidth]{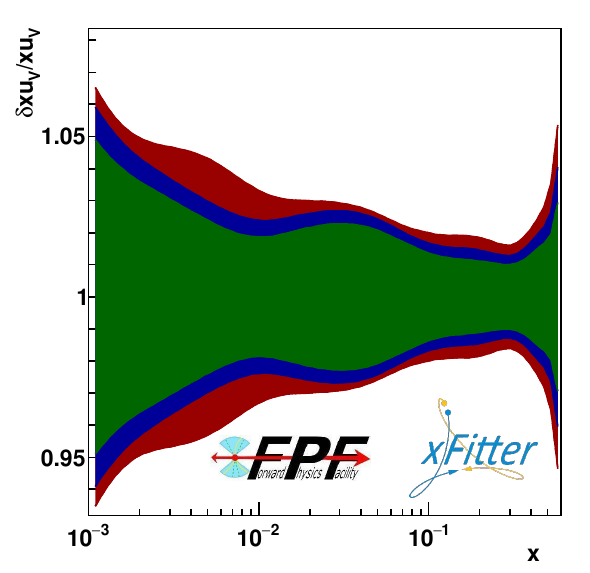}
  \includegraphics[width=0.32\textwidth]{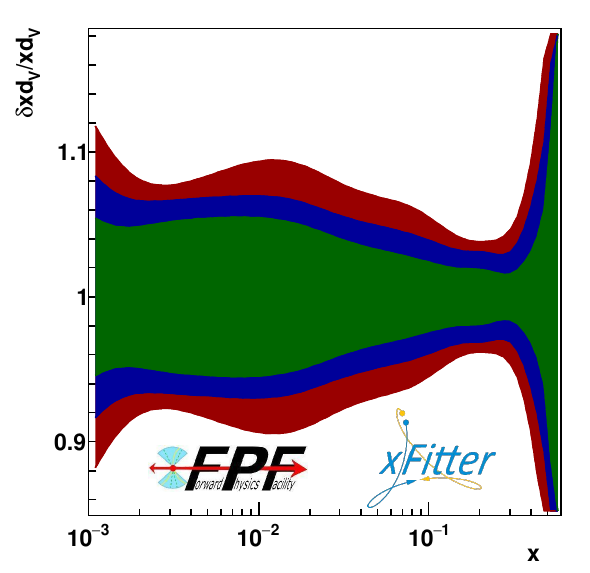}
  \includegraphics[width=0.32\textwidth]{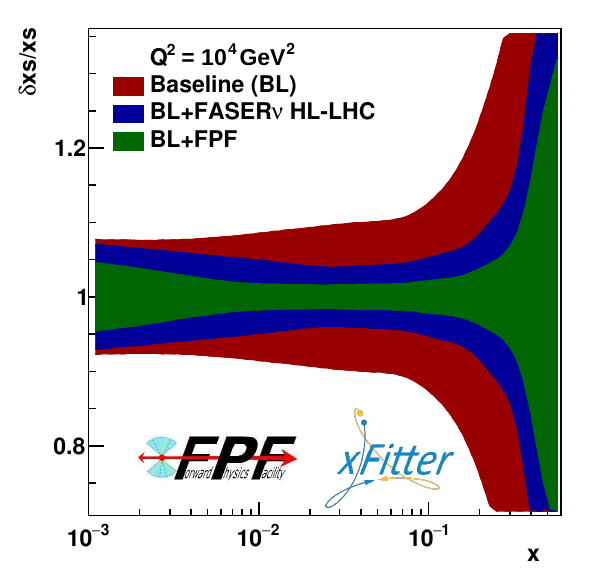}
  \caption{\textbf{Constraining Proton Structure via $\nu$DIS.} Fractional uncertainties (68\% CL) at $Q^2 = 10^4$~GeV$^2$ for the valence up (left), the valence down (middle) and the strange (right) quark PDF in the PDF4LHC21~\cite{PDF4LHCWorkingGroup:2022cjn} baseline scenario (red), compared to the results with FASER$\nu$ (blue) and FPF (green) pseudodata corresponding to the LHC Run~5 with $2~\iab$. The projections account for estimated statistical uncertainties~\cite{Cruz-Martinez:2023sdv}.}
\label{fig:FPF_PDFs}
\end{figure}

\noindent \textbf{Nuclear Structure from Neutrino DIS:}  In addition to probing proton PDFs, neutrino DIS measurements on heavy target nuclei will provide complementary information on nuclear PDFs. Characteristic nuclear modifications include shadowing at low $x$, anti-shadowing at intermediate $x$, and the EMC effect at high $x$, whose origin is currently not fully understood~\cite{Arneodo:1992wf, Norton:2003cb}. This picture is further complicated by discrepancies between charged-lepton and neutrino scattering data, with the latter exhibiting a much weaker shadowing and anti-shadowing behavior than observed in electron and muon scattering~\cite{Schienbein:2009kk,  Kovarik:2010uv, Muzakka:2022wey,  SajjadAthar:2020nvy}. In addition, models predict that shadowing disappears in the large $Q^2$ limit, but no consensus has been reached regarding the size of shadowing at intermediate values $Q^2 \sim 10^2-10^3$ GeV$^2$ similar to those at the FPF. Differences in heavy quark mass effects for neutrino DIS and charged-lepton DIS structure functions further complicate scattering data comparisons~\cite{Risse:2025smp}. The NuSTEC consortium identified the study in different regions of $x$ and $Q^2$ for neutrino and anti- neutrino interactions with nuclei a high priority goal~\cite{NuSTEC:2017hzk}.  Neutrino DIS measurements at the FPF will provide data over a wide kinematic range and for various nuclear targets, notably argon in FLArE, iron in the FLArE HCAL, and tungsten in FASER$\nu$2, whose combination allows the extraction of nuclear effects. Moreover, it may be possible to include additional target materials in either FASER$\nu$2 or the FLArE HCAL if deemed necessary. \medskip

\noindent \textbf{Structure Functions F4 and F5:} Neutrino DIS on nucleons is generically described in a model-independent framework involving intermediate vector boson exchange between lepton and hadron currents, leading to five independent structure functions~\cite{Albright:1974ts, Kretzer:2003iu}. The commonly studied structure functions, $F_1$, $F_2$,  and $F_3$, have been measured for the proton. In contrast, $F_4$ and $F_5$ are suppressed for small charged lepton masses, making charged-current $\nu_\tau$ interactions the only viable way to access them. Ref.~\cite{Francener:2024euo} showed that differential $\nu_\tau$ measurements at FASER$\nu$2 will be sensitive to the magnitude of these structure functions. \medskip

\begin{figure}
  \centering
  \includegraphics[width=0.49\textwidth]{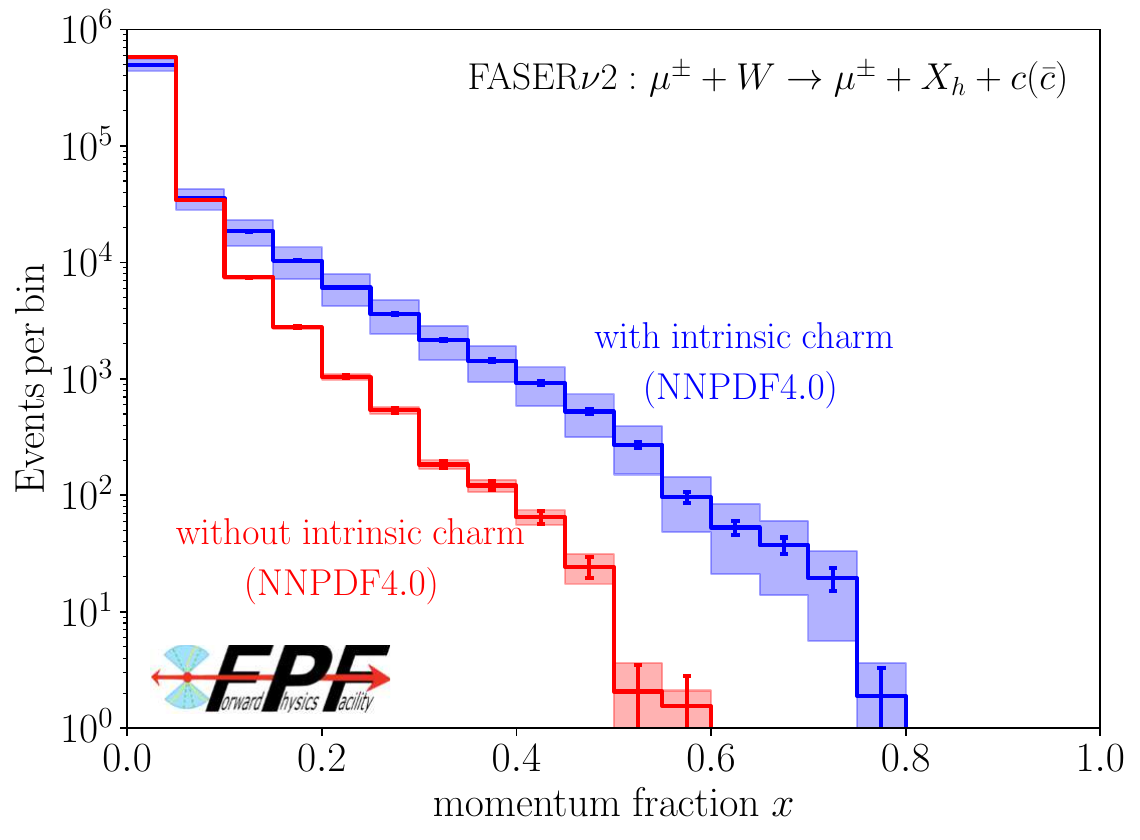}
  \includegraphics[width=0.49\textwidth]{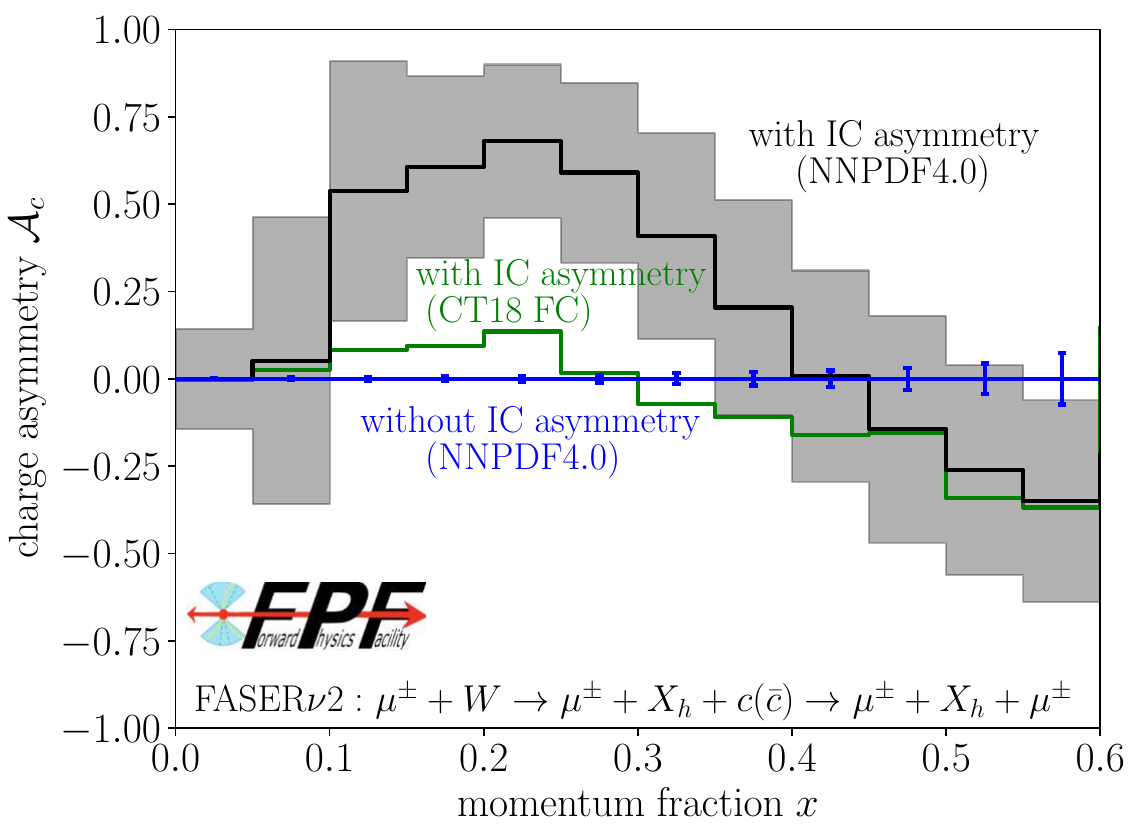}
\caption{\textbf{Constraining Inrinsic Charm via Muon DIS.} Left: the event rate at FASER$\nu$2 in bins of $x$ for $Q>1.65~{\rm GeV}$ for semi-inclusive charm production through muon DIS for a PDF set with fitted charm, including IC (blue), or with a perturbatively-generated charm PDF (red) for an integrated luminosity of $2~\iab$. Right: Charm production in muon DIS can be split into $\mu^\pm\mu^+$ and $\mu^\pm\mu^-$, from which an asymmetry, ${\cal A}_c(x)$, in the (anti-)charm quark that the muon interacted with can be extracted. In Ref.~\cite{Francener:2025pnr} it was found that muon DIS measurements at FASER$\nu$2 provide sufficient statistics to determine if there is an IC component (left) and also whether this IC is symmetric or asymmetric (right). Indeed, PDF sets with $c\neq\bar{c}$ from NNPDF4.0 and CT18IC have values of ${\cal A}_c(x)$ that are non-zero with magnitudes that are much larger than the expected statistical uncertainties.} 
  \label{fig:muon_DIS}
\end{figure}

\noindent \textbf{Proton Structure via Muon DIS:} The FPF is also sensitive to NC DIS by using the forward-going muons~\cite{Francener:2025pnr} in a kinematic region closely overlapping with the NC measurements at the EIC.
Along with neutrinos, muons are the only other SM particle capable of penetrating through the $\sim 250$ meters of rock separating the FPF from the ATLAS IP.
These muons lose relatively little energy as they travel through the rock, and, as in the case of neutrinos, their high energy provides sensitivity to smaller $x$ and larger $Q^2$ in comparison to previous fixed target experiments, as shown in \cref{fig:QCD_Landscape}.
Therefore, muon DIS measurements at the FPF allow for a unique independent corroboration of EIC results up to nuclear effects.  
Ref.~\cite{Francener:2025pnr} estimated that FASER$\nu$2 would be able to observe about $2.5\cdot 10^7$ muon DIS interactions, allowing for multi-differential cross section measurements. 
Hence, the FPF can constrain structure functions similarly to the neutrino case, including their material dependence, allowing a study of nuclear effects via neutrino and muon DIS in the same experiment. \medskip

\noindent \textbf{Observation of Intrinsic Charm via Muon DIS:} 
Notably, muon DIS at the FPF is able to cleanly probe the intrinsic charm (IC) hypothesis~\cite{Brodsky:1980pb, Ball:2022qks, Guzzi:2022rca}, which states that there is a non-perturbative charm quark component in the proton wave function. 
This will provide closure on the decades-long controversy on whether the EMC $F_2^c$ measurements in the early 1980s~\cite{EuropeanMuon:1982fow} did or did not reveal IC.
The left panel of \cref{fig:muon_DIS} presents the number of charm-production events in muon DIS at FASER$\nu$2, highlighting the large statistics (up to $10^6$ events) and the strong differences between predictions based on PDF sets with and without IC.

Furthermore, the FPF can establish whether there is an asymmetry between the charm and anti-charm content of the nucleus, which is the ultimate smoking gun for IC.
For this, the FPF exploits the fact that charm will decay to muons about 10\% of the time, and the charge of the muon can be used to identify at FASER2 whether a charm or anti-charm quark interacted in the event. 
Thus, muon DIS at the FPF can measure the asymmetry of the resulting charm production spectra from $\mu^\pm +N \rightarrow \mu^\pm +X+c(\bar{c})\rightarrow \mu^\pm +X+\mu^+(\mu^-)$.
The right panel of \cref{fig:muon_DIS} shows a projection for the measured asymmetry distribution, ${\cal A}_c(x) = [N_{\rm DIS}(c)-N_{\rm DIS}(\bar{c})]/[N_{\rm DIS}(c)+N_{\rm DIS}(\bar{c})]$, where $N_{\rm DIS}(c)$ and $N_{\rm DIS}(\bar{c})$ are the DIS rate for a muon interacting with a charm and anti-charm quark, respectively.
The figure shows predictions for two PDF sets that allow for an asymmetric IC component, compared to the case of a PDF set where IC is symmetric.
With measurements at FASER$\nu$2, the statistical uncertainties are small enough that a nonzero IC asymmetry can be resolved, and one can also discriminate between different IC models available in the literature.
Similar IC studies could also be performed by FASER and SND@LHC at the HL-LHC, albeit with smaller statistics. \medskip

\begin{figure}
  \centering
  \includegraphics[width=0.47\linewidth]{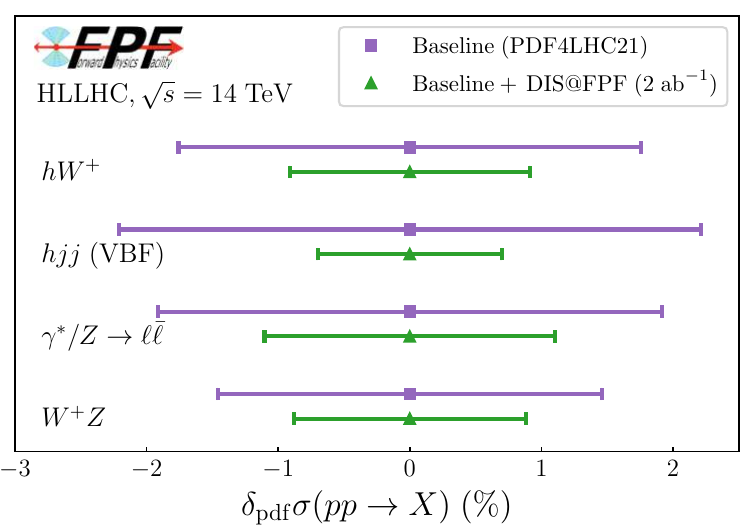}
  \includegraphics[width=0.49\linewidth]{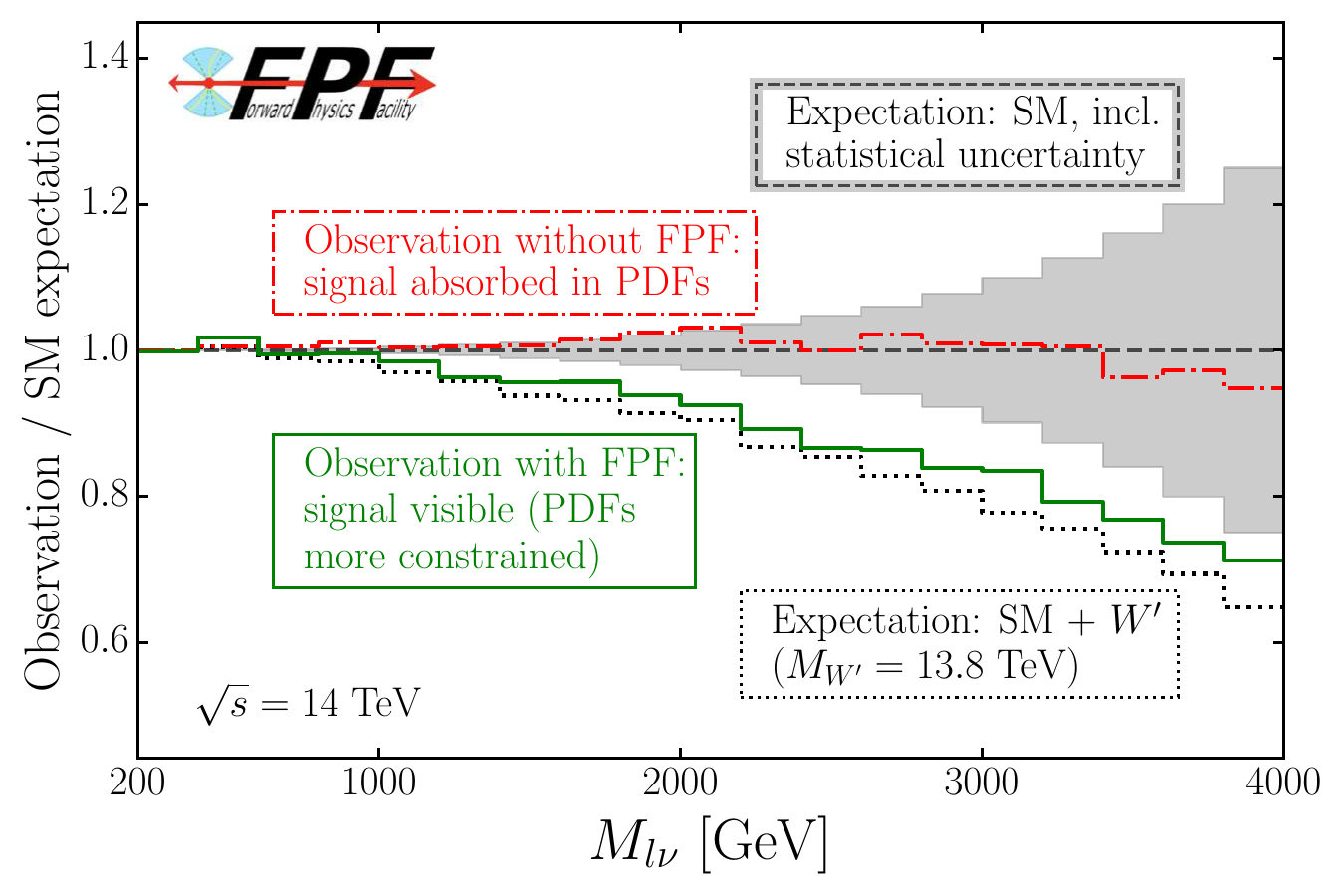}
  \caption{\textbf{Impact of FPF on HL-LHC Measurements.} Left: Proton PDF fits with FPF neutrino DIS data enable a reduction of PDF uncertainties in HL-LHC cross sections predictions for a number of Higgs and electroweak processes, with cross sections with the baseline PDFs shown in purple, and cross sections with PDFs constrained by FPF data in green~\cite{Cruz-Martinez:2023sdv}.
  Right: Constraints from the FPF can prevent BSM signals, from being absorbed into the PDFs~\cite{Hammou:2023heg}. We consider the production of a 13.8 TeV $W'$ boson producing a signal in high-energy Drell-Yan tails (black dotted line).
  Here two
  PDFs fits are carried out in the  presence of the $W'$ signal: one without FPF constraints included (red curve) and one with FPF constraints included (green curve). 
  Without the FPF constraints, the fitted PDFs reabsorb the BSM signal, leading to an  observation that is compatible with the SM within uncertainties. With the FPF constraints, the PDFs are more robustly constrained, and the observation of the BSM signature becomes possible.}
\label{fig:PDF_applications}
\end{figure}

\noindent \textbf{Impact on High-Luminosity LHC Measurements:}
By using TeV-scale neutrinos and muons to probe the quark/gluon content of the nucleon, not only do we make progress towards answering foundational questions on the deep structure of matter, but we also sharpen theoretical predictions for Higgs, electroweak, and BSM particle production at ATLAS and CMS.
Indeed, better PDFs from FPF measurements enable more reliable predictions for quark-initiated processes at the LHC, such as the Drell-Yan production of electroweak bosons and the Higgs boson in vector-boson-fusion and in associated production~\cite{Cruz-Martinez:2023sdv}.
This reduced uncertainty on the PDFs translates into a more precise prediction of high-$p_T$ cross sections, paving the way for more robust measurements of Higgs boson couplings~\cite{LHCHiggsCrossSectionWorkingGroup:2016ypw}, the $W$-boson mass~\cite{CMS:2024lrd}, the strong coupling $\alpha_s(m_Z)$~\cite{ATLAS:2023lhg}, and the weak mixing angle $\sin^2\theta_W$ at the LHC, all of which are already limited by PDF-related systematics and will be more so at the HL-LHC. 
To illustrate the significant FPF impact, in the left panel of~\cref{fig:PDF_applications} we show the reduction of PDF uncertainties on representative HL-LHC cross sections for a number of Higgs and electroweak processes that will be measured by ATLAS and CMS.

PDFs constrained from FPF data will also allow for improvements in the high-$p_T$ BSM search program of ATLAS and CMS.
In Ref.~\cite{Hammou:2023heg}, it was shown that if a global PDF fit is carried out in the presence of high-mass Drell-Yan cross sections from ATLAS and CMS where a BSM signal from an off-shell 13.8 TeV $W'$-boson has been injected, this signal would be completely reabsorbed by a change in the large-$x$ quark and antiquark PDFs~\cite{Greljo:2021kvv} without worsening the fit quality.
The resulting observation would then appear as consistent with the SM, due to the fact that the limited constraints on the large-$x$ antiquark PDFs allow them to mimic the new physics signature. 
However, as demonstrated in the right panel of \cref{fig:PDF_applications}, once the FPF data are included in the global PDF fit, this degeneracy is broken by the additional information on the large-$x$ quark and antiquark PDFs that neutrino DIS at the FPF provides~\cite{Hammou:2024xuj}. Thus, the PDF-induced bias is removed, and theoretical predictions for high-mass Drell-Yan cross sections can be calculated accurately, such that if a 13.8 TeV $W'$ exists, it would clearly be observed through a destructive interference and a marked reduction in the event yields. 
Moreover, the same study compared the impact of the FPF projections with those of the EIC and found that the FPF had a greater ability to disentangle the $W'$ effects from the PDFs.\medskip 

\noindent \textbf{Neutrino Interactions as a Probe of Cold Nuclear Matter:} The FPF detectors use heavy target material, e.g tungsten in FASER$\nu$2. Hence, the primary neutrino interaction vertex is surrounded by a dense nuclear medium and newly created coloured objects must be transported through this medium.  How the development of a parton shower is affected by this medium and how eventual hadronisation takes place are still open questions. 
The possibility of probing this physics has attracted a lot of attention in the context of electron-ion collisions. In fact, investigations of this phenomenon form an important component of the physics program for the EIC. As noted in the EIC White Paper~\cite{Accardi:2012qut}, “cold QCD matter could be an excellent femtometer-scale detector of the hadronisation process from its controllable interaction with the produced quark (or gluon)”. In this sense, the neutrino detector at the FPF would be a neutrino-ion collider, which would be able to access the same physics, in a complementary way.
An important characteristic of the planned neutrino detectors at the FPF is their precision tracking capability. This capability will make it possible to record not only the energy and direction of the outgoing lepton, but also the detailed properties of the final-state hadronic system that emerges from the large nucleus of the target, such as the hadron multiplicity or hadron momenta. By carefully analyzing this data, the FPF experiments can gain unique insights into the fundamental physics of QCD in the regime of cold nuclear matter. \medskip 

\noindent \textbf{Probe of Forward Particle Production:} The neutrino beam at the LHC originates from the decays of the lightest hadrons of a given flavour, primarily charged pions, charged and neutral kaons, and charm hadrons. Notably, the production of these particles has not been previously measured in the forward region at LHC energies. Therefore, neutrino flux measurements provide a novel method to probe and constrain forward hadron production, providing insights into the underlying physics that are otherwise inaccessible. \medskip

\noindent \textbf{Forward Light Hadrons and Hadronic Interaction Models:} The production of forward light hadrons lies beyond the scope of perturbative QCD. Instead, their production is described by phenomenological hadronic interaction models that were designed to match the available data~\cite{Pierog:2013ria, Ostapchenko:2010vb, Riehn:2019jet, Fieg:2023kld}. Until now, forward particle measurements at the LHC have been limited to neutral pions, $\eta$ mesons and neutrons, as measured by the LHCf experiment~\cite{LHCf:2017fnw, LHCf:2018gbv, Piparo:2023yam}. Collider neutrino measurements will provide complementary information on the forward production of charged pions, charged kaons, neutral kaons, and hyperons~\cite{Kling:2023tgr}. Tightening constraints on forward particle production at high energies and refining hadronic interaction models is crucial for simulating minimum bias events at the LHC and has important implications for astroparticle physics~\cite{FASER:2021pkt}, where such models are used to describe particle production in extreme astrophysical environments as well as in cosmic ray interactions with the Earth’s atmosphere as discussed in \cref{sec:astroparticle}. \medskip

\noindent \textbf{Forward Charm Production as a Window into Uncharted Regions of QCD:} An additional component of the neutrino flux stems from the decay of charm hadrons. In contrast to light hadron production, forward charm production can, in principle, be modeled using perturbative QCD methods~\cite{Bai:2020ukz, Maciula:2022lzk, Bhattacharya:2023zei, Buonocore:2023kna}. 

Forward charm quarks offer a unique probe of gluon dynamics, as they are mainly produced via the gluon fusion process that generates a $c\bar{c}$ pair. A simple kinematic estimate shows that, to obtain TeV-energy forward neutrinos from charm decays, one of the initial state gluons needs to carry a large momentum fraction $x \sim E_\nu/E_\text{proton} \sim 1$, while the other carries a very small momentum fraction $x \sim 4 m_c^2 / s \sim 10^{-7}$. Measuring the flux of neutrinos from charm decay will constrain parton structure in both regimes, as indicated by the red ellipses in \cref{fig:QCD_Landscape}. The former regime would allow one to constrain high-$x$ PDFs, complementing measurements performed at low-energy beam dump experiments as well as muon and neutrino DIS measurements at the FPF. The latter regime is sensitive to the gluon PDF at very low-$x$, well beyond the coverage of other experiments~\cite{Anchordoqui:2021ghd, Bai:2021ira}. 

Neutrinos from charm decay are the dominant source of electron neutrinos at high energies as well as tau neutrinos. In addition, their relative contribution to the overall flux increases when moving away from center of the beam, so to lower rapidities. Ref.~\cite{Kling:2023tgr} demonstrated that, by leveraging these features, the neutrino flux from forward charm decays can be extracted and constrained at the percent level. More recently, it was shown that machine learning techniques applied to neutrino event yields at the FPF can enable a precise, data-driven determination of the LHC forward neutrino fluxes, further improving the connection to theoretical predictions~\cite{John:2025qlm}. Additionally, it may be possible to tag neutrinos from charm decays via their coincidence with charm hadrons detected in ATLAS, as discussed in Refs.~\cite{SNDLHC:2025qtx, Abbaneo:2926288}. \medskip

\begin{wrapfigure}{R}{0.50\textwidth}
  \centering
  \vspace{-8mm}
  \includegraphics[width=.5\textwidth]{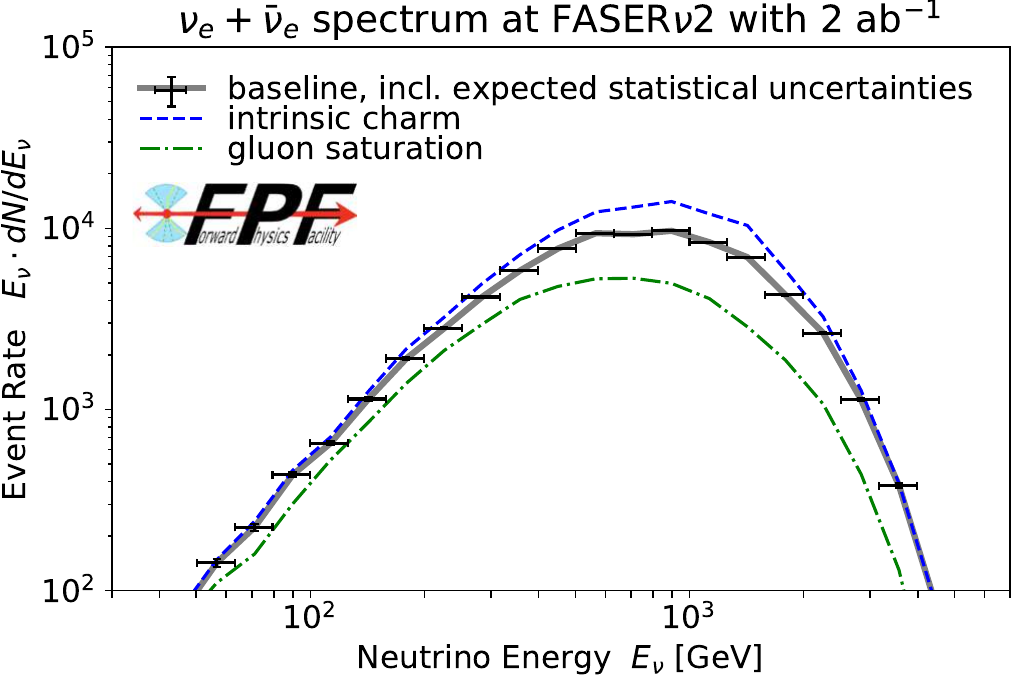}
  \caption{\textbf{Intrinsic charm and gluon saturation at the FPF.} The energy distribution of interacting electron neutrinos in FASER$\nu$2. The baseline prediction, shown as gray solid line, uses SIBYLL 2.3d~\cite{Riehn:2019jet} to model light hadron production and $k_T$-factorisation approach discussed in Ref.~\cite{Bhattacharya:2023zei} for charm production. The coloured lines illustrate an intrinsic charm component in the proton, following the BHPS model~\cite{Brodsky:1980pb} implemented in the CT14 PDF~\cite{Hou:2017khm} as estimated in Ref.~\cite{Maciula:2022lzk} (blue dashed) and gluon saturation at low-$x$~\cite{Kutak:2012rf}, as estimated in Ref.~\cite{Bhattacharya:2023zei} (green dash-dotted).}
  \label{fig:QCD_IC}
  \vspace{-5mm}
\end{wrapfigure}

\noindent \textbf{Intrinsic Charm:} Due to their sensitivity to high-$x$ PDFs, collider neutrino flux measurements may also shed light on intrinsic charm~\cite{FASER:2019dxq}. According to both in the original BHPS model~\cite{Brodsky:1980pb}, as well as the recent PDF fits performed by the NNPDF~\cite{Ball:2022qks} and CTEQ~\cite{Hou:2017khm} collaborations, the possibility of the intrinsic charm component in the proton, discussed above, would lead to a significant increase in the charm PDF at $x \sim 0.3$. In this case, also the $c g  \to c g$ process becomes a notable contribution to forward hadron production besides gluon fusion. In particular, this leads to an enhancement in forward charm production at high energies, increasing the neutrino flux at the FPF~\cite{Feng:2022inv, Maciula:2022lzk}. The resulting changes in the shape and normalisation of the observed neutrino spectra, illustrated in \cref{fig:QCD_IC}, will hence allow the IC component to be constrained as discussed in Refs.~\cite{Maciula:2022lzk, John:2025qlm}.
\medskip

\noindent \textbf{Proton Structure at Low-$x$ and BFKL dynamics:} As discussed above, forward charm production provides a powerful probe of novel QCD dynamics at small $x$. The potential of the FPF to constrain the gluon PDF in this regime is illustrated in \cref{fig:smallxQCD}. The left panel shows the predicted energy spectrum of electron neutrinos, including expected statistical uncertainties represented by the black error bars. This prediction is based on NNPDF3.1, with the corresponding gluon PDF shown in the central panel. As illustrated by the light shaded regions, the PDF uncertainty grows significantly at small $x$, inducing large neutrino flux uncertainties. Using the Hessian profiling method and carefully chosen observables where other theory uncertainties associated to the choice of scale and charm mass cancel out, in this case the ratio of electron to tau neutrino event rates, we have demonstrated that collider neutrino flux measurements can substantially tighten constraints on the gluon PDF down to $x \sim 10^{-7}$, as indicated by the dark shaded region~\cite{Rojo:2024tho}.

Such measurements inform the study of novel QCD dynamics at small~$x$, a region where BFKL-like and non-linear effects are expected to become relevant. The standard collinear framework, based on on-shell matrix elements and integrated parton densities, is well-suited for processes involving hard scales. In this approach, the Dokshitzer-Gribov-Lipatov-Altarelli-Parisi (DGLAP) evolution resums logarithms enhanced by the momentum transfer scale. However, at very high energies, an additional class of logarithmic enhancements -- associated with the center-of-mass energy or, equivalently, with $1/x$ -- becomes important. In this regime, the evolution of the gluon density is expected to follow the Balitsky-Fadin-Kuraev-Lipatov (BFKL) equation~\cite{Fadin:1975cb, Kuraev:1976ge, Kuraev:1977fs, Balitsky:1978ic}, which resums these large logarithms. This leads to a rapid growth of the gluon density as $x$ decreases. Measurements at the FPF provide a unique opportunity to probe this behaviour and test BFKL dynamics in a previously unexplored region of phase space. In particular, this will be essential for informing cross sections at the FCC-hh, since, at $\sqrt{s} = 100$~TeV, even Higgs and gauge boson production become small-$x$ processes, with potentially large corrections from BFKL resummation~\cite{Bonvini:2018vzv, Rojo:2016kwu}. \medskip

\begin{figure}[thb]
  \centering
  \includegraphics[width=.99\textwidth]{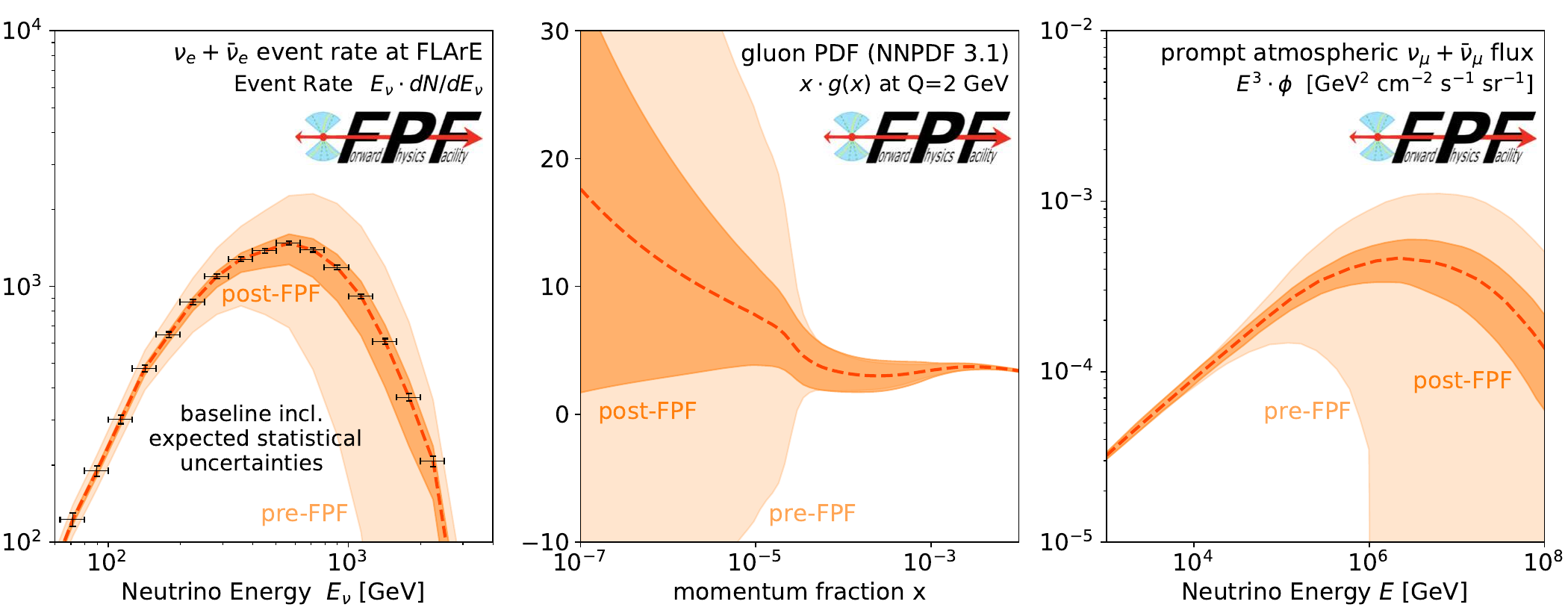}
  \caption{\textbf{Small-$x$ QCD at the FPF.} Constraints on various quantities with (dark orange) and without (light orange) FPF constraints on PDFs at small $x$. Left: The impact of FPF data on theoretical predictions for the spectra of interacting electron neutrinos in the FLArE detector at the FPF.  The spectra are obtained using \texttt{SIBYLL-2.3d}~\cite{Riehn:2019jet} and \texttt{POWHEG + Pythia}~\cite{Buonocore:2023kna} with \texttt{NNPDF~3.1}.  Also shown is the expected measurement of the spectrum at the FPF, with the black crosses indicating the expected energy uncertainty and the statistical uncertainty in event rates. Central: The impact of FPF data on constraining the proton's gluon PDF at small $x$, as modeled in NNPDF~3.1. Right: The impact of FPF data on reducing the uncertainty in the prompt atmospheric muon neutrino flux.}
  \label{fig:smallxQCD}
\end{figure}

\noindent \textbf{Gluon Saturation:} Another phenomenon expected to occur at low $x$ is called parton saturation~\cite{Gribov:1983ivg}. When the density of gluons becomes very high, the splitting processes leading to the increase of the gluon density have to be accompanied by recombination processes, which in turn lead to the taming of the fast growth of the gluon density and the corresponding cross sections. This regime can be described by an effective theory at high energy and density, the Colour Glass Condensate (CGC)~\cite{McLerran:1993ka,McLerran:1993ni}. In the presence of parton recombination, the evolution equations become modified by additional non-linear terms in the density which becomes relevant below the saturation scale, which divides the dilute and dense regimes. The saturation scale $Q_s$ scales as $Q_s^2\propto A^{1/3}x^{-\lambda}$ with an empirical parameter $\lambda\sim 0.3$ as illustrated via the dashed line in \cref{fig:QCD_Landscape}. This indicates that forward charm measurements will probe the kinematic region sensitive to gluon saturation in protons. This is hence complementary to planned measurements at the EIC, which aim to probe gluon saturation in heavy nuclei. The impact of saturation on the FPF neutrino spectra was studied in Ref.~\cite{Bhattacharya:2023zei} and the result is shown in \cref{fig:QCD_IC}. The onset of gluon saturation will sizably reduce the neutrino flux from charm decays as measured by the FPF neutrino detectors. 

\subsection{Astroparticle Physics}
\label{sec:astroparticle}

\noindent {\bf Connection to Astroparticle Physics:} Cosmic rays with energies as large as $10^{11}\,\mathrm{GeV}$ enter Earth's atmosphere, where they interact with air nuclei, generating extensive air showers (EASs) that are routinely detected by large ground-based detector arrays~\cite{Anchordoqui:2018qom}. While the initial cosmic-ray interaction energies are often beyond the reach of current colliders, secondary interactions of hadrons during the shower development typically occur at energies approximately equivalent to LHC energies, but in the laboratory frame and in the far-forward region. Modeling of these interactions is critical for a large variety of astrophysical measurements, for example, to infer the properties of cosmic rays, such as their energy and mass, from indirect measurements at ground level~\cite{Albrecht:2021cxw,Kampert:2012mx}, to understand charm production in EASs, which gives rise to the main backgrounds for high-energy astrophysical neutrino searches~\cite{IceCube:2014stg}, or to understand particle production in the most extreme astrophysical environments in the Universe~\cite{Dorner:2025egk,Koldobskiy:2021nld}. A key challenge is accurately modeling hadronic interactions in the forward region, a domain that current collider facilities are unable to directly probe.

Figure~\ref{fig:eta} shows the distribution of particles in pseudorapidity $\eta$ from 13 TeV $pp$ collisions (dashed lines), as obtained by \texttt{SIBYLL~2.3d}~\cite{Riehn:2019jet}, along with the pseudorapidity coverage of existing LHC experiments~\cite{Albrecht:2021cxw, SoldinICRC2025}. Solid lines represent the distribution of the number of muons $N_\mu$ produced by these particles in EASs, with the assumption $N_\mu \propto E_{\text{lab}}^{0.9}$, where $E_{\text{lab}}$ is the laboratory energy of the secondary particles. The majority of particles in LHC collisions are produced at central rapidity in the center-of-mass frame, within the main detector coverage. However, the vast majority of energy is carried by hadrons emitted in the forward direction which naturally have a profound influence on the EAS development and also decay to the most energetic atmospheric neutrinos. Most important is the production of pions and kaons with $\eta>6$, and charm hadrons with $\eta>4$, a region which is largely outside of the LHC main detector coverage and not fully covered by FASER ($\eta > 8.5$). 

The experiments at the FPF will cover a larger pseudorapidity range and will thus be able to probe the region most relevant for astroparticle physics via high-energy neutrinos produced in collisions at the LHC. These measurements will be critical to improve the modeling of high-energy hadronic interactions, which will help reduce uncertainties in air shower observations, atmospheric neutrino fluxes, as well as particle interactions in extreme astrophysical environments. Hence, the FPF will be crucial for refining our understanding of the properties of the highest-energy cosmic rays and neutrinos, as well as their sources.  \medskip

\begin{figure}[tb]
  \centering
  \includegraphics[width=1\textwidth]{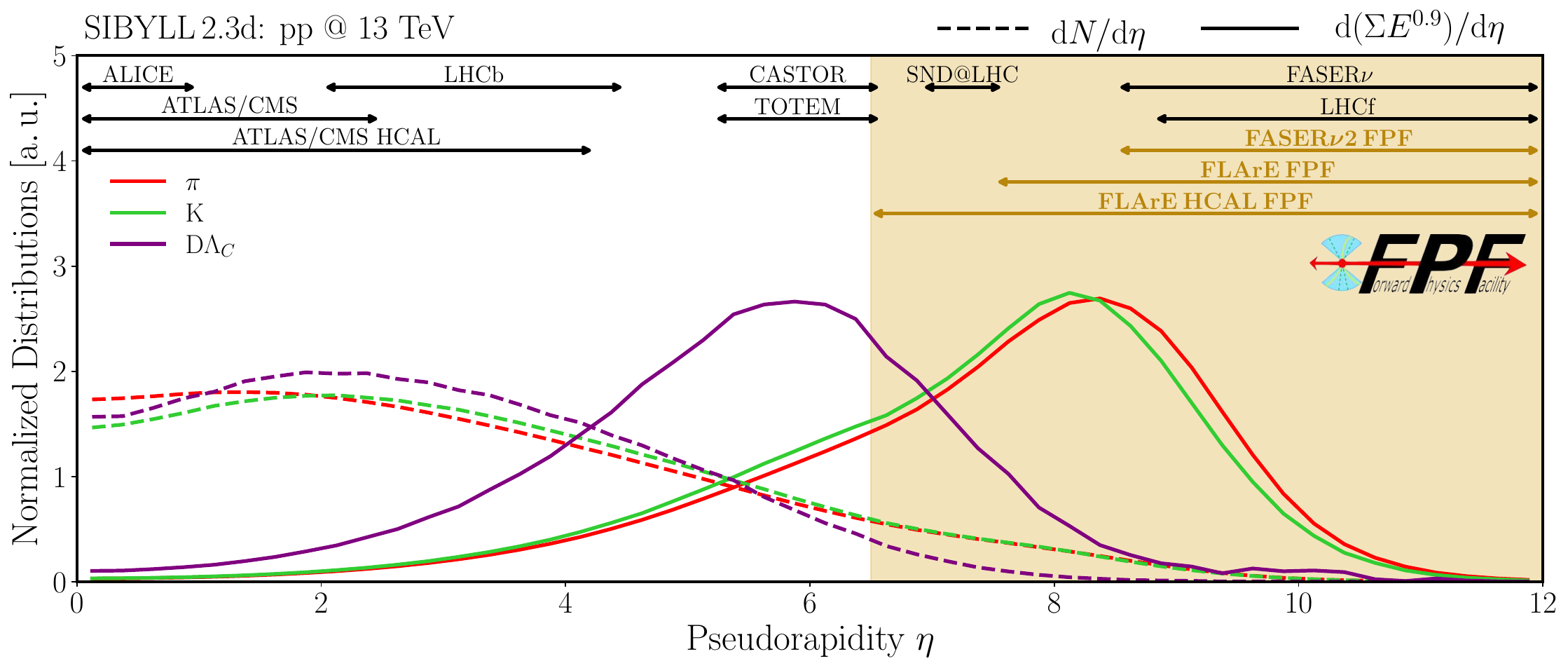}
  \caption{\textbf{Pseudorapidity Distributions and Experimental Coverage.} Dashed lines show the distribution of particles in pseudorapidity $\eta$~\protect\cite{SoldinICRC2025}, in arbitrary units, in 13 TeV $pp$ collisions, as obtained from \texttt{SIBYLL-2.3d}~\cite{Riehn:2019jet}. Solid lines show the distribution of muons produced by these particles, in arbitrary units, assuming an equivalent energy for the fixed target collisions in the laboratory frame, $E_{\rm{lab}}$, and $N_\mu \propto E_{\rm{lab}}^{0.9}$.}
  \label{fig:eta}
\end{figure}

\noindent \textbf{Astrophysical Neutrinos:} High-energy neutrinos of astrophysical origin have long been observed in the IceCube Neutrino Observatory~\cite{IceCube:2014stg} and more recently by other large-scale neutrino telescopes KM3NeT~\cite{KM3Net:2016zxf} and Baikal-GVD~\cite{GVD:2025lya}. Atmospheric neutrinos from decays of hadrons produced in cosmic-ray interactions with air nuclei are an irreducible background to these measurements. At energies above 1~PeV, approximately equivalent to center-of-mass energies at the LHC, they are produced primarily from the decays of charm hadrons. This component, known as the ``prompt'' atmospheric neutrino flux, constitutes the dominant background for astrophysical neutrino searches and remains poorly constrained due to the lack of data. The experiments at the FPF will provide crucial measurements to constrain the prompt atmospheric neutrino flux.

As highlighted above in the discussion on small-$x$ physics, the production of charm quarks is dominated by the gluon fusion process and can be described using perturbative QCD~\cite{Gauld:2015kvh}. Measurements of the neutrino flux at the FPF therefore provide access to both the very high-$x$ and the very low-$x$ regions of the parton densities in the colliding protons. These measurements yield information about high-$x$ PDFs, in particular IC, as well as novel QCD production mechanisms at very low-$x$, such as BFKL effects and non-linear dynamics~\cite{Duwentaster:2023mbk}, well beyond the coverage of any existing experiments and providing key inputs for astroparticle physics, particularly for large-scale neutrino telescopes. 

The FPF measurements will provide stringent constraints on the prompt atmospheric neutrino flux, contributing to the scientific program of large-scale neutrino telescopes. This is quantified in \cref{fig:smallxQCD} (right), showing theoretical predictions for the prompt muon-neutrino flux~\cite{Bai:2022xad} based on the formalism of Refs.~\cite{Bai:2021ira, Bai:2022xad}, considering only PDF uncertainties (error bands), before and after FPF constraints are included. Although other sources of theory uncertainty also contribute to the total uncertainties, \cref{fig:smallxQCD} demonstrates the strong sensitivity of the FPF to the mechanisms governing atmospheric neutrino production from charm decays. \medskip

\begin{figure}[tb]
  \centering
  \includegraphics[width=1.\textwidth]{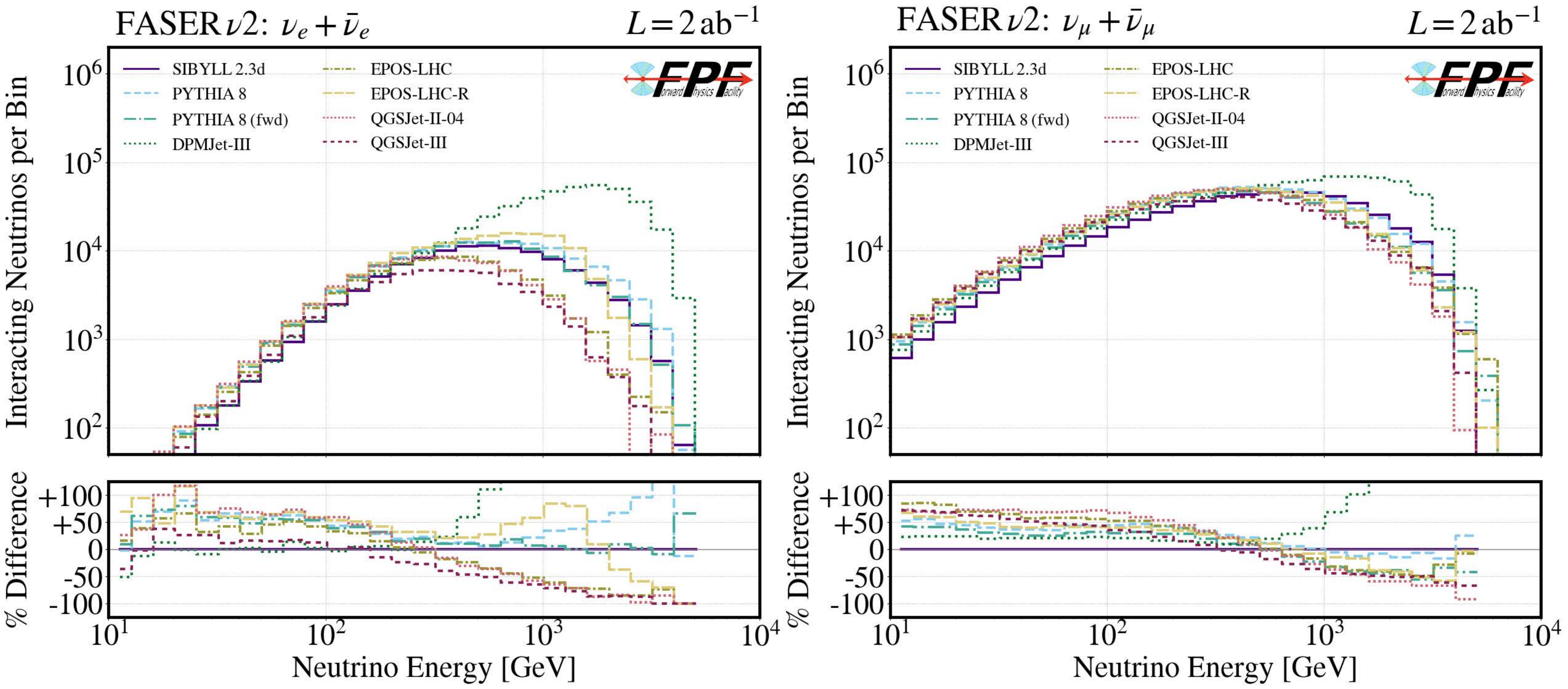}
  \caption{\textbf{Neutrino Flux.} Simulated neutrino energy spectra~\cite{Kling:2021gos,SoldinICRC2025} for electron neutrinos (left) and muon neutrinos (right) interacting in FASER$\nu$2 for an integrated luminosity of $2\,\rm{ab}^{-1}$. The predictions are obtained from \texttt{SIBYLL-2.3d}, \texttt{DPMJET-III}, \texttt{EPOS-LHC}, \texttt{EPOS-LHC-R}, \texttt{QGSJET-II.04}, \texttt{QGSJET-III}, and \texttt{Pythia~8.2}, including a dedicated forward tune (fwd). The bottom panel shows the corresponding model differences with respect to \texttt{SIBYLL-2.3d}. 
  }
  \label{fig:AstroNuRate}
\end{figure}

\noindent \textbf{Cosmic Ray Monte Carlo Generators and the Muon Puzzle:} Muons act as tracers of hadronic interactions, making their measurement in EASs essential for testing hadronic interaction models. For many years, several EAS experiments have reported discrepancies between model predictions and experimental data, a phenomenon known as the muon puzzle in EASs. Specifically, analyses from the Pierre Auger Observatory have shown an excess in the number of muons compared to simulations~\cite{PierreAuger:2014ucz, PierreAuger:2016nfk,PierreAuger:2024neu,PierreAuger:2025kym}, and a systematic meta-analysis of data from nine different air-shower experiments has revealed an energy-dependent trend in these discrepancies with high statistical significance~\cite{Dembinski:2019uta,Soldin:2021wyv,Cazon:2020zhx}. Recent studies suggest that the origin of these discrepancies may be complex and indicate severe deficits in our understanding of soft QCD~\cite{Albrecht:2021cxw,ArteagaVelazquez:2023fda}.

Measurements at the FPF will provide key insights into light hadron production in the far-forward region. Neutrinos are produced in decays of pions, kaons, and heavy hadrons and hence encode information on their production. Measurements of neutrinos at the FPF will provide a variety of information to disentangle the individual hadron components, such as energy spectra for different neutrino flavours or rapidity distributions. The ratio of electron neutrino to muon neutrino fluxes, for example, offers an indirect method for determining the ratio of charged kaons to pions. Electron neutrinos are predominantly produced from kaon decays, while muon neutrinos originate from both pion and kaon decays. These neutrinos have distinct energy spectra, which allows them to be differentiated. Additionally, neutrinos from pion decays are more concentrated around the LOS compared to those from kaon decays, due to the lower mass of the pion, meaning pions impart less transverse momentum. As a result, the proximity of the neutrinos to the LOS, or their rapidity distribution, can help disentangle the origins of the neutrinos and provide an estimate of the pion-to-kaon ratio. In fact, Ref.~\cite{Kling:2023tgr} demonstrated that flux measurements at the FPF will be able to constrain the individual flux components with percent level precision. In addition, the large muon fluxes at the FPF may encode complementary information to further constrain hadronic models. 

Figure~\ref{fig:AstroNuRate} shows predictions of the neutrino energy spectra for $\nu_e$ and $\nu_\mu$ interactions in FASER$\nu$2, assuming an integrated luminosity of $2\,\mathrm{ab}^{-1}$. These predictions, based on various commonly used models including \texttt{SIBYLL-2.3d}~\cite{Riehn:2019jet}, \texttt{DPMJET-III}~\cite{Roesler:2000he}, \texttt{EPOS-LHC}~\cite{Pierog:2013ria}, \texttt{EPOS-LHC-R}~\cite{Pierog:2023ahq}, \texttt{QGSJET-II-04}~\cite{Ostapchenko:2013pia,Ostapchenko:2005nj}, \texttt{QGSJet-III}~\cite{Ostapchenko:2024myl}, and \texttt{Pythia 8.2}~\cite{Sjostrand:2014zea}, including a dedicated forward tune~\cite{Fieg:2023kld}, show flux differences exceeding a factor of two at high energies, much larger than the expected statistical uncertainties at the FPF~\cite{Kling:2021gos,Cruz-Martinez:2023sdv}. Differences can also be observed for a series of phenomenologically-modified versions of \texttt{Sibyll-2.3d} (referred to as \texttt{SIBYLL}$^\bigstar$~\cite{Riehn:2024prp}, not shown in the figure), as demonstrated in Ref.~\cite{SoldinICRC2025}. This model increases muon production from hadronic multi-particle production processes and thereby provides a possible solution to the muon puzzle. Figure~\ref{fig:AstroNuRateComposion} shows the model predictions separately for pion, kaon, charm, and hyperon decays. Since the muon puzzle is thought to have its origins in soft QCD processes~\cite{Albrecht:2021cxw}, this directly ties it to the QCD program at the FPF. Measurements at the FPF will therefore be crucial for improving our understanding of particle production in EASs. \medskip

\begin{figure}[tb]
  \centering
  \includegraphics[width=1.\textwidth]{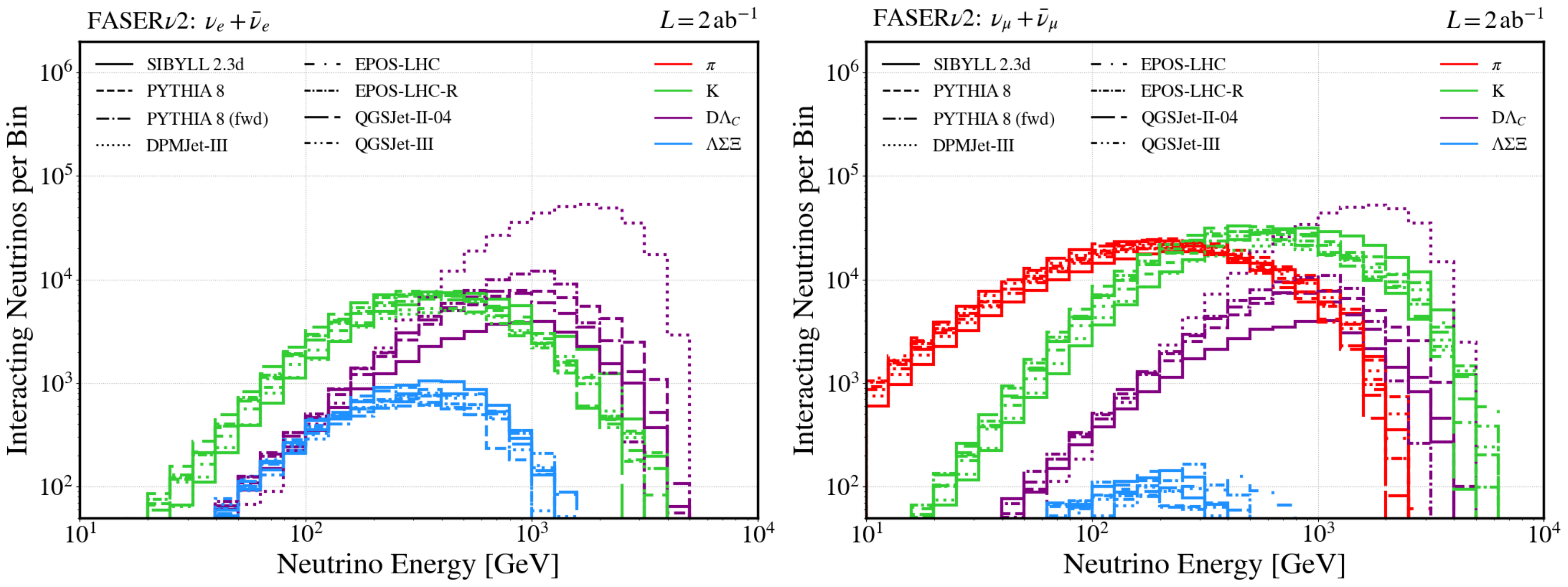}
  \caption{\textbf{Neutrino Flux Composition.} Simulated neutrino energy spectra~\cite{Kling:2021gos,SoldinICRC2025} for electron neutrinos (left) and muon neutrinos (right) interacting in FASER$\nu$2 for an integrated luminosity of $2\,\rm{ab}^{-1}$ for different production modes: pion (red), kaon (green), charm (purple), and hyperon (blue) decays.}
  \label{fig:AstroNuRateComposion}
\end{figure}

\begin{figure}[!b]
  \centering
  \vspace{-1em}
  \includegraphics[width=1.\textwidth]{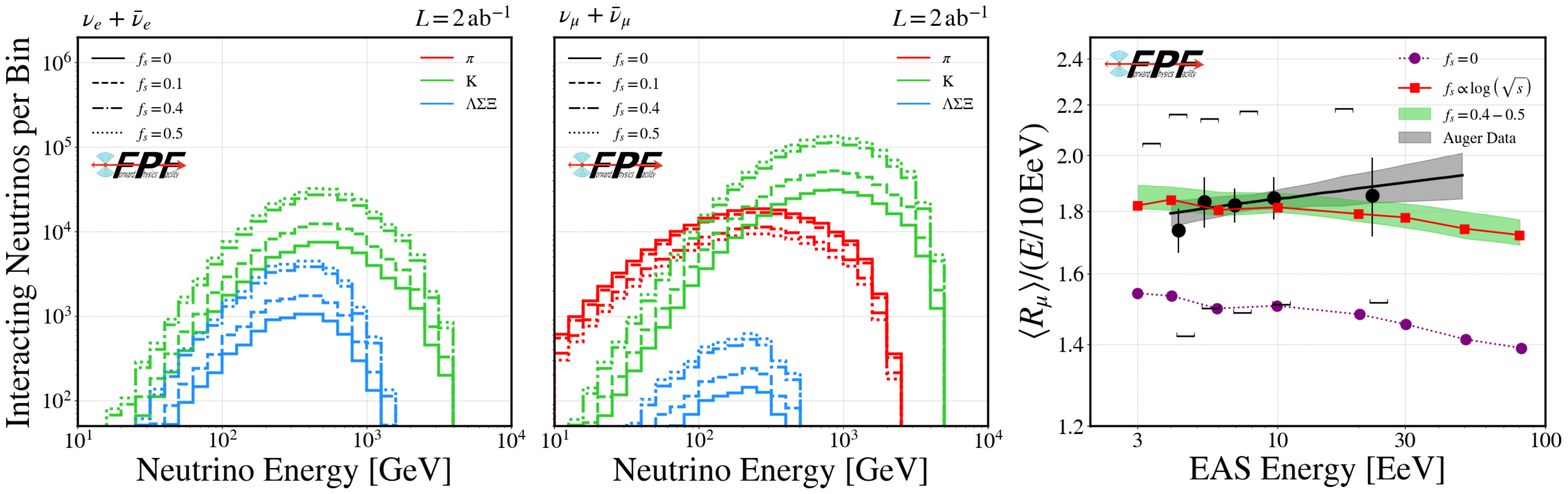}
  \caption{\textbf{Neutrino Flux with Strangeness Enhancement.} Simulated neutrino energy spectra~\protect\cite{Sciutto:2023zuz,SoldinICRC2025} for electron neutrinos (left) and muon neutrinos (center) interacting in FASER$\nu$2 assuming the {\tt piKswap} model, for different production modes: pion (red), kaon (green), and hyperon (blue) decays. The different line styles correspond to predictions obtained from \texttt{SIBYLL-2.3d} by varying $f_s$; see text for details. Also shown is the dimensionless muon shower content, $R_\mu$, as a function of the EAS energy, as predicted by the {\tt piKswap} model through simulations with \texttt{SIBYLL-2.3d} and \texttt{AIRES}~\cite{Sciutto:1999jh} (right), compared to data from the Pierre Auger Observatory~\cite{PierreAuger:2014ucz}. For details, see Ref.~\protect\cite{Sciutto:2023zuz}.}
  \label{fig:NuSpectraFs}
\end{figure}

\noindent \textbf{Testing Strangeness Enhancement:} A potential key to resolving the muon puzzle may lie in the universal strangeness and baryon enhancement observed by the ALICE experiment in mid-rapidity collisions~\cite{ALICE:2016fzo}. This enhancement, which depends solely on particle multiplicity and not on specific details of the collision system, could allow for predictions of hadron composition in EASs in a phase space beyond current collider capabilities. If this enhancement increases in the forward region, it would affect muon production in EASs and could be traced by measuring the ratio of charged kaons to pions at the FPF. The phenomenological {\tt piKswap} model provides a specific example that accounts for the enhancement of strangeness production~\cite{Anchordoqui:2022fpn}. In the simplest one-parameter model, strangeness enhancement is introduced by allowing the substitution of pions with kaons in \texttt{SIBYLL-2.3d} using a constant probability $f_s$ in shower collisions above a threshold energy ($s \sim 2~{\rm TeV}^2$) and large pseudorapidities ($\eta > 4$)~\cite{Anchordoqui:2022fpn}.  For values of $f_s$ between $0.4$ and $0.6$, this simple model can explain the observed data, suggesting a potential solution to the muon puzzle. However,  if $f_s$ is taken along with a Heaviside step function of energy, then the predicted electron neutrino flux at LHC energies is $1.6$ ($2.2$) times higher at its peak than the baseline prediction, already for $f_s = 0.1$ ($f_s = 0.2$), as shown in \cref{fig:NuSpectraFs}. While simple scenarios like this can be constrained by FASER~\cite{FASER:2024ref}, additional measurements at the FPF are required to constrain or exclude more sophisticated scenarios with high statistical significance; see also \cref{fig:NuRatio}. A {\tt piKswap} extension, in which $f_s$ grows logarithmically with $\sqrt{s}$, is shown in the right panel of \cref{fig:NuSpectraFs} to also accommodate Auger data, providing an equivalent solution to the muon puzzle~\cite{Sciutto:2023zuz}.  The FPF experiments will be able to probe $f_s \sim 0.005$~\cite{Kling:2023tgr}, providing a final verdict for models addressing the muon puzzle via strangeness enhancement. This example demonstrates how the FPF will be uniquely positioned to test and constrain hadronic interaction models, providing a significant advancement in our understanding of multi-particle production in EASs. Furthermore, if data collected by the FPF experiments were to return only null results on increments of strange-particle yields relative to pions, this would provide strong motivation to explore alternative schemes (perhaps rooted in BSM physics~\cite{Anchordoqui:2025nmb}) to address the muon puzzle. \medskip

\begin{figure}[tb]
  \centering
  \includegraphics[trim={0 4mm 0 0},clip, width=0.85\textwidth]{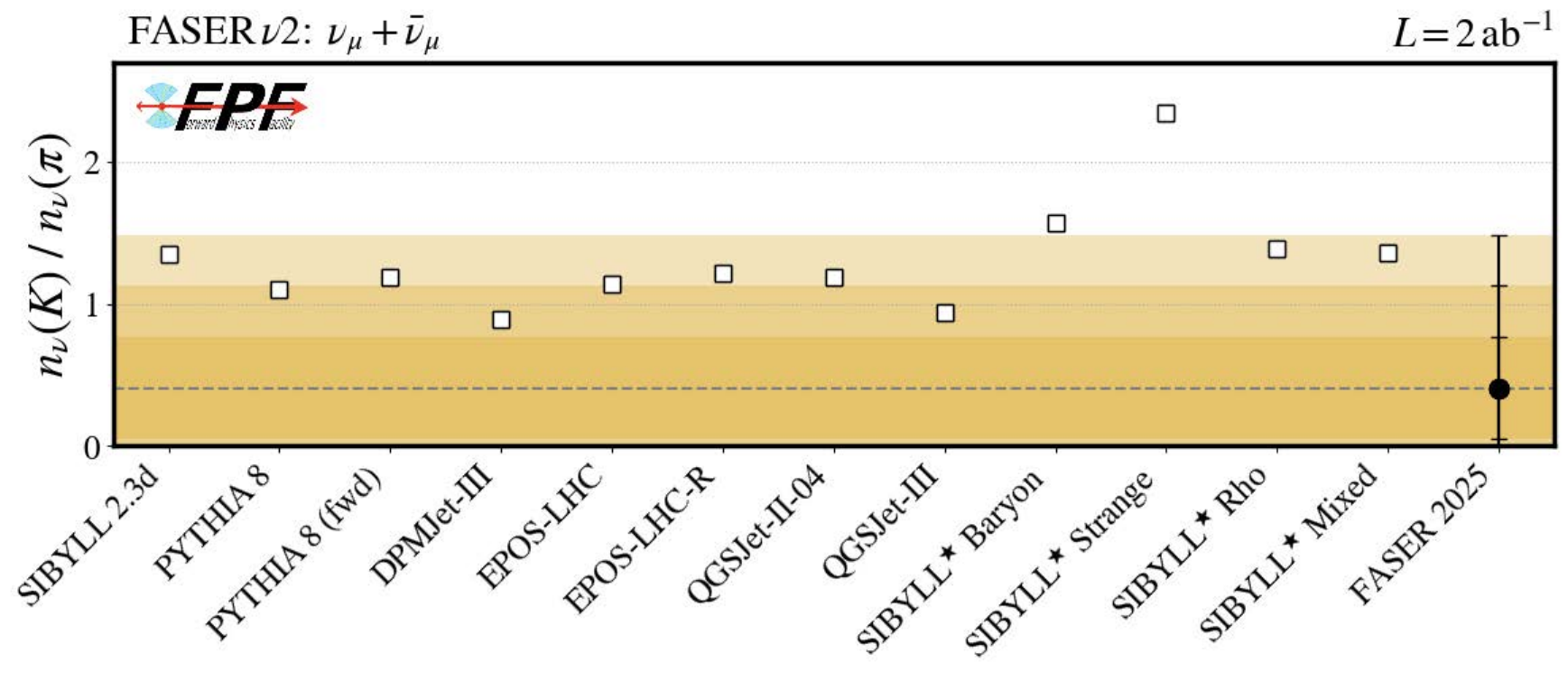}
\caption{ \textbf{Neutrinos from Pion and Kaon Decays.} Ratio of the number of muon neutrinos from kaon decays ($n_\nu(K)$) and pion decays ($n_\nu(\pi)$) in FASER$\nu$2, obtained from the hadronic models indicated~\cite{SoldinICRC2025}. Also shown is a recent measurement by FASER~\cite{FASER:2024ref} with 1$\sigma$, 2$\sigma$, and 3$\sigma$ uncertainty bands.}
\vspace{-4mm}
  \label{fig:NuRatio}
\end{figure}

Figure~\ref{fig:NuRatio} shows the expected ratio of the number of neutrinos from kaons to pions in FASER$\nu$2, obtained from the different hadronic interaction models, including multiple variants of \texttt{SIBYLL}$^\bigstar$; for details, see Refs.~\cite{Riehn:2024prp,SoldinICRC2025}. The predictions are compared to a recent measurement by FASER~\cite{FASER:2024ref}. All hadronic model predictions are in tension with the experimental data, which indicates an overestimation of the ratio of kaons to pions in all models. Predictions from the simple one-parameter {\tt piKswap} model yield ratios of more than $2$ for values of $f_s>0.1$ and are not shown. However, it is important to note that FASER probes a narrower pseudorapidity region ($\eta>8.8$) compared to FASER$\nu$2 ($\eta>8.4$) or FLArE ($\eta>7.5$). A measurement of the individual flux components at the FPF with high statistics is therefore of particular importance to constrain current model predictions with high statistical significance.  \medskip

\noindent {\bf High Energy Neutrino Event Generators and Cross Sections:} The FPF will also provide a decisive testbed to validate and improve the models and tools of high-energy neutrino interactions that are widely used in astroparticle physics. In particular, FPF measurements will improve our descriptions of neutrino structure functions, nuclear modifications, and nuclear final-state effects, as described \cref{sec:qcd}. By confronting state-of-the-art predictions, for example those based on the widely used cross-section models of CSMS~\cite{Cooper-Sarkar:2011jtt} and BGR~\cite{Bertone:2018dse,Garcia:2020jwr}, as well as neutrino event generators such as \texttt{GENIE}~\cite{Andreopoulos:2009rq} and modern \texttt{POWHEG}-based frameworks~\cite{vanBeekveld:2024ziz, FerrarioRavasio:2024kem, Buonocore:2024pdv}, with TeV-scale data from FLArE and FASER$\nu$2, the FPF will be able to quantify and reduce modeling systematics. Current models typically differ at the $\mathcal{O}(5\text{–}10\%)$ level across relevant kinematics, whereas the statistical uncertainties of FLArE and FASER$\nu$2 will be well below this threshold. Hence, the input of FPF measurements will sharpen both collider-neutrino and neutrino-telescope interpretations. In addition, FPF measurements of neutrino production of charm will help to characterise double cascade signatures from electron neutrino CC production of charm followed by charm decays, a background to neutrino telescope measurements of the astrophysical tau neutrino flux \cite{IceCube:2024nhk}. 

These efforts will also provide valuable input to geophysics. It has been shown that the attenuation of high-energy atmospheric neutrinos measured by IceCube can be used to carry out a statistically-significant tomography of the Earth~\cite{Donini:2018tsg, Chattopadhyay:2025cgt}. Such measurements typically have a degeneracy between the assumed neutrino cross section and density profile of the Earth. A measurement of the neutrino-nucleon cross section at TeV energies at the FPF will break this degeneracy and thus naturally reduce the systematic uncertainties in modeling the Earth’s interior.   

\subsection{Dark Matter 
\label{sec:darkmatter}}

\noindent \textbf{Dark Matter and Light New Physics:} The quest for new physics is strongly linked to understanding the matter-energy budget of the universe, with dark matter (DM) playing a prominent role in motivating numerous experimental programs. The particle nature and origin of DM could be closely tied to a broader dark sector. Consequently, related experimental efforts include both direct searches for DM candidates and more comprehensive investigations of possible mediators between the SM and the dark sector. 

A multitude of BSM scenarios have been proposed to incorporate DM. One guiding principle for navigating this vast landscape is the postulated DM production mechanism in the early universe. Specifically, thermal production---which extends the concept of the WIMP miracle to a broader range of dark sector masses, including the sub-GeV scale~\cite{Boehm:2003hm, Pospelov:2007mp, Feng:2008ya}---provides clear theoretical targets for DM searches that have only been partially constrained so far. Experiments at the FPF will be capable of probing many of these targets in searches for both DM and mediator particles. \medskip

\noindent \textbf{Dark Sector Mediators as Long-Lived Particles (LLPs):}
If the mediator is the lightest particle in the dark sector, then it may be naturally long-lived due to its feeble portal coupling to the SM. This general scenario provides strong motivation for the search for highly-displaced decays of light LLPs, a primary goal of the currently-operating FASER experiment. The FPF experiments, and FASER2 in particular, will not only extend this program to the HL-LHC era, but will also significantly expand its reach to probe compelling and widely-discussed BSM scenarios. These include popular benchmark models from the Physics Beyond Colliders (PBC) initiative~\cite{Alemany:2019vsk, PBC:2025sny}, such as the vector (dark photon), scalar (dark Higgs boson), and fermion (heavy neutral lepton (HNL)) portals, as well as axion-like particles (ALPs)~\cite{FASER:2024bbl}. These capabilities have been described in numerous studies by the FPF community and are summarised in Ref.~\cite{Feng:2022inv}.

An extensive program to search for light and feebly-interacting particles (FIPs) will position CERN as the leading facility for discovering such particles in the coming years. Establishing the FPF is a major step toward this goal, complementing other experimental efforts. Characterizing any new FIP state requires a broad program that employs diverse beams to probe different production modes and mass ranges, along with a variety of detectors spanning multiple baselines and capabilities, to determine properties such as mass, spin, and couplings to the SM. The FPF will crucially support these FIP searches while also extending its reach to direct light DM searches. We discuss selected examples of this complementarity below. \medskip

\begin{figure}
  \centering
  \includegraphics[width=0.49\linewidth]{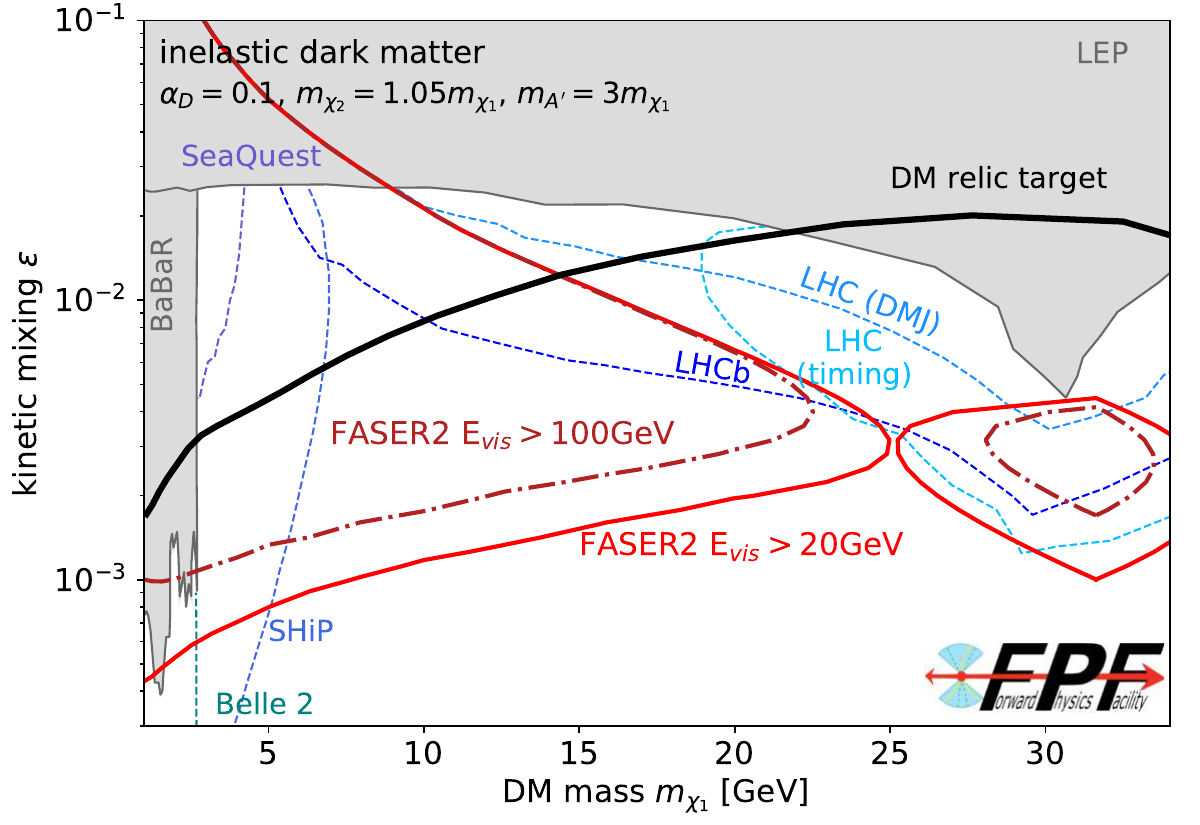}
  \includegraphics[width=0.49\linewidth]{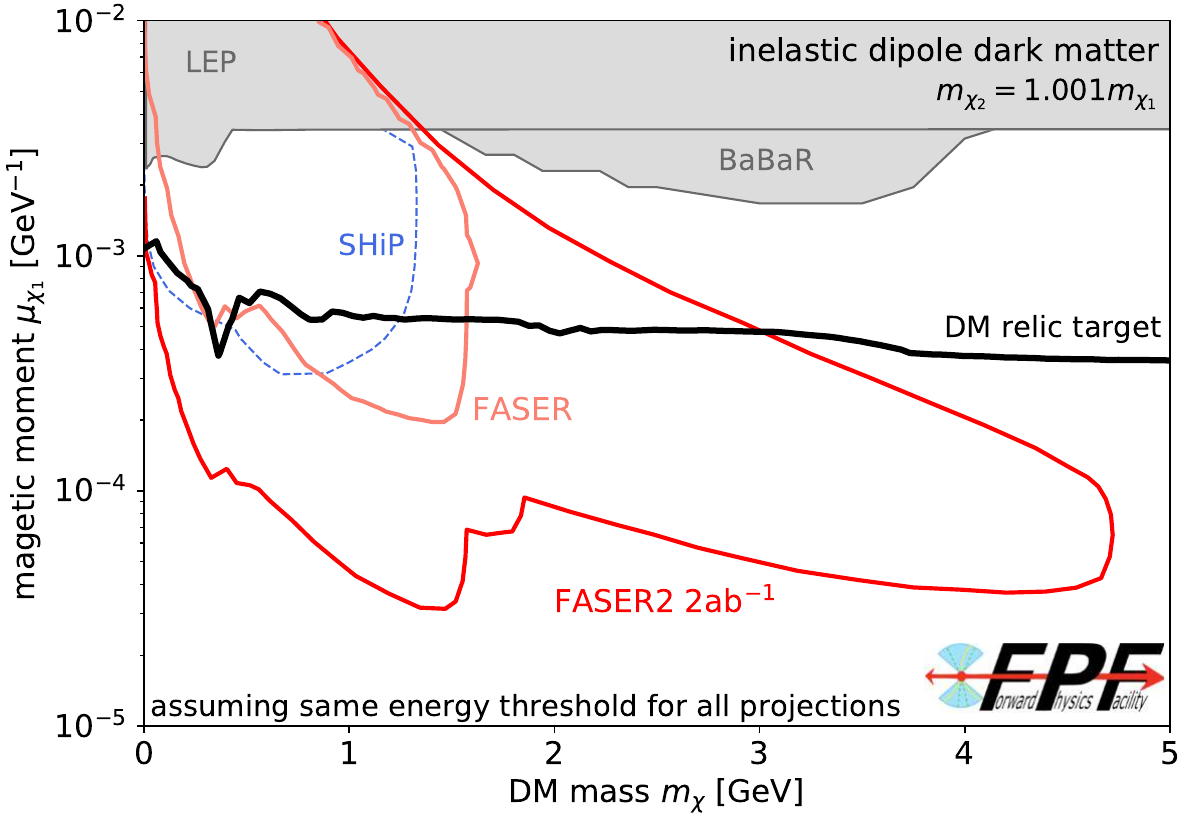}
\caption{\textbf{Inelastic Dark Matter Searches at the FPF}. The discovery potential of FASER2 with $2~\iab$ and other experiments for two different realisations of iDM. The left panel considers the case of heavy inelastic DM interacting via a dark photon portal, as discussed in Ref.~\cite{Berlin:2018jbm}, where the high energy of the LHC allows FASER2 to probe DM masses up to tens of GeV.  The right panel considers the case of light inelastic DM with very small mass splittings that is mediated by a dipole portal, as analyzed in Ref.~\cite{Dienes:2023uve}, where the large LHC energy boosts the signal to observable energies. In both scenarios, the reach of FASER2 extends beyond all other experiments, including direct and indirect DM searches, LHC main detectors, and beam dump experiments, such as SHiP. It also covers the thermal DM relic target (solid black lines), that is, the cosmologically-favoured parameter space where the model predicts the observed DM relic abundance as produced through thermal freeze-out. For comparison, we have shown the leading constraints provided by  BaBaR~\cite{BaBar:2017tiz} and LEP~\cite{Hook:2010tw,Curtin:2014cca}, as well as projections for a number of other proposed searches, including those for displaced muon jets (DMJ) and delayed particles (timing) at the main LHC experiments~\cite{Izaguirre:2015zva, Liu:2018wte}, as well as displaced particle searches at LHCb~\cite{LHCb:2017trq,Ilten:2016tkc,Pierce:2017taw}, SHiP~\cite{SHiP:2021nfo}, Belle 2~\cite{Belle-II:2018jsg}, and SeaQuest~\cite{Berlin:2018pwi}.}
  \label{fig:inelasticDarkMatter}
\end{figure}

\noindent \textbf{Searches for Heavy LLPs:} In comparison to other probes such as fixed-target experiments, the FPF experiments naturally excel in the search for relatively heavy LLPs with masses in the tens of GeV range. This is a simple consequence of the much larger center-of-mass energy available in the LHC $pp$ collisions. As a well-motivated DM scenario illustrating the power of the FPF to probe heavy LLPs, consider inelastic dark matter (iDM)~\cite{Tucker-Smith:2001myb, Berlin:2018jbm, Izaguirre:2015zva}. The simplest iDM models feature a kinetically-mixed dark photon mediator with an off-diagonal coupling between DM, $\chi_1$, and a heavier excited state, $\chi_2$. Interestingly, this construction renders direct and indirect detection strategies ineffective for sufficiently large mass splittings between $\chi_1$ and $\chi_2$, making iDM models a prime target for accelerator experiments. At the LHC, the 14 TeV collision energy results in a highly-boosted and collimated beam of $\chi_1$ and $\chi_2$ particles in the forward direction. The heavier state $\chi_2$ decays to the somewhat lighter DM $\chi_1$ and a pair of SM fermions. Notably, as a consequence of the small mass splitting, $\chi_2$ can be naturally long-lived and reach the detectors housed in the FPF cavern, leaving a signature of visible SM particles. FASER2 can probe large regions of the relic target parameter space~\cite{Berlin:2018jbm}, with masses reaching ${\cal O} (10)$ GeV, greatly exceeding the mass reach that is achievable at beam-dump experiments, as shown in the left panel of \cref{fig:inelasticDarkMatter}. The sensitivity is shown for two requirements on the minimum calorimeter energy: the 20 GeV threshold (solid) shows the best sensitivity but raising this to a more conservative 100 GeV threshold does not have a large impact on the sensitivity (dash-dotted). \medskip

\noindent \textbf{Strongly-Boosted LLPs:} 
Another broad scenario where the FPF experiments shine is when the visible decay products of the LLP carry only a small amount of energy.
To illustrate this possibility, consider the iDM model with an an effective dipole interaction, $\mu\bar{\chi}_2\sigma^{\mu\nu}\chi_1 F_{\mu\nu}$, where $\mu$ is a dimensionful coupling~\cite{Izaguirre:2015zva,Dienes:2023uve}. 
For nearly-degenerate $\chi_1$ and $\chi_2$, the photon from $\chi_2$ decay will have a very low energy, 
$E_{\text{vis}} \sim E_{\chi_2} \Delta$ where $\Delta \equiv (m_{\chi_2} - m_{\chi_1})/m_{\chi_2}$ is the fractional mass splitting. Such low visible energy is likely to fall below the detection threshold at lower energy beam-dump experiments that require ${\cal O}({\rm GeV})$ energy deposits~\cite{Dienes:2023uve,Batell:2014mga}. In contrast, the GeV-scale $\chi_2$ produced  in the forward region of the LHC pp collisions are highly boosted, with Lorentz boost parameters $\gamma \sim {\rm TeV}/m_{\chi_2}=1000 \ ({\rm GeV}/m_{\chi_2})$, leading to significant visible energy from the photon in the final state. Thus, the FPF is well-suited to search for such a model with a small mass splitting, as the signal would be more likely to fall below threshold at other experiments with identical thresholds. This is illustrated in the right panel of \cref{fig:inelasticDarkMatter}, which shows that FASER2 can cover large regions of parameter space for $m_{\chi_1}\lesssim 5$~GeV that cannot be probed by other experiments~\cite{Dienes:2023uve}. The plot shown here, together with the sensitivity for larger mass splittings from Fig. 3 in \cite{Dienes:2023uve}, illustrates that the FPF provides complementary coverage of the relic target when combined with lower-energy experiments, which probe larger couplings and larger mass splittings. This provides one illustration of a broad class of signatures, where the FPF can leverage the significant boost of LLPs produced at the LHC to obtain unique sensitivity to new physics. \medskip

\begin{wrapfigure}{R}{0.5\textwidth}
\centering
\vspace{-7mm}
\includegraphics[width=0.49\textwidth]{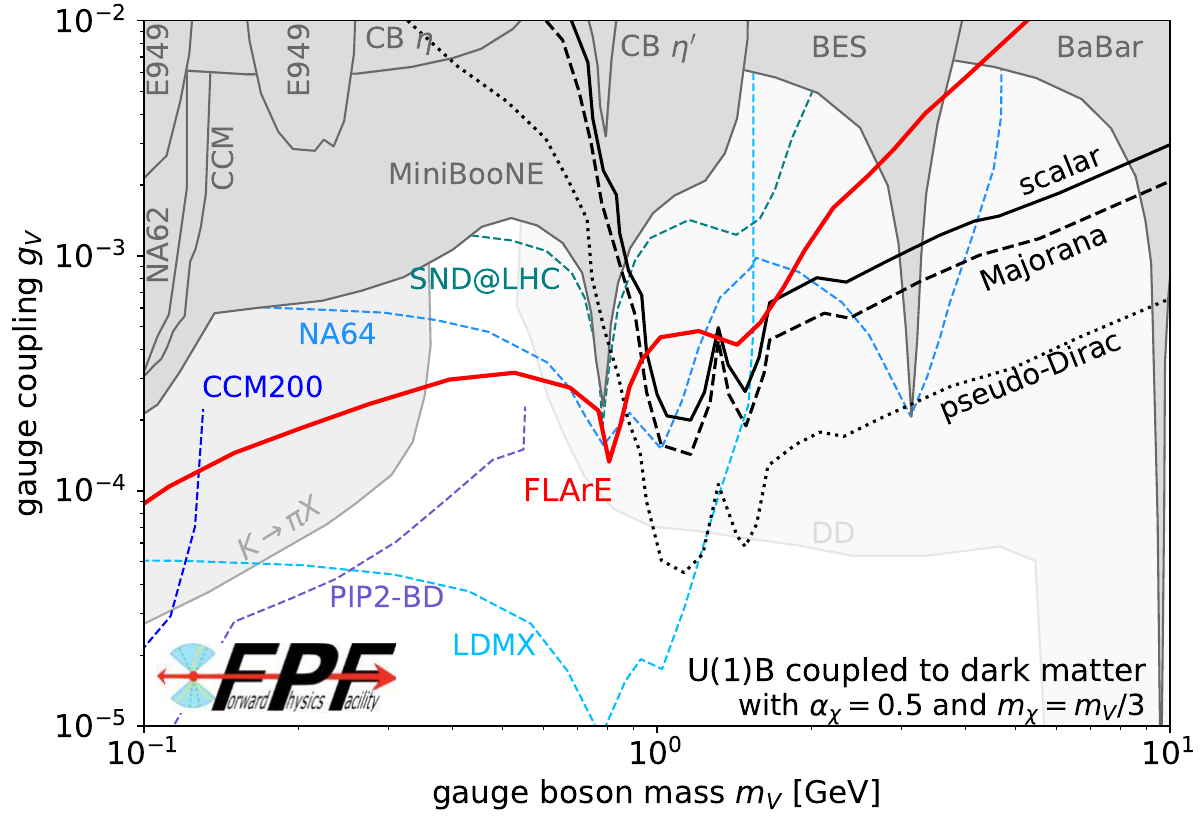}
\caption{\textbf{Scattering Signatures at the FPF}. The parameter space of a hadrophilic model with a $U(1)_B$ baryon number gauge boson mediator, spanned by its mass $m_V$ and gauge coupling $g_V$, coupling to DM with strength $\alpha_D =0.5$ and mass $m_\chi = m_V/3$. The black contours are the thermal relic targets for different types of DM; DM is thermally overproduced below these contours. The solid red contour shows the projected reach of FLArE at the FPF with $2~\iab$, while dashed contours show the projected reach of other proposed experiments and the dark gray shaded regions are excluded by current bounds~\cite{Batell:2021snh, Krnjaic:2022ozp}. The light gray shaded regions correspond to bounds from anomaly-induced kaon decays. The very light gray shaded regions are constraints from DM direct detection; these do not apply to Majorana and inelastic scalar DM. }
\vspace{-4mm}
\label{fig:bsm_dm}
\end{wrapfigure}

\noindent \textbf{Dark Sector Scattering Signatures:} The detection of DM particle scatterings at the FPF will rely on different experimental techniques, primarily those utilised in the proposed FLArE detector. With its low energy threshold and precise time resolution, FLArE can enable successful searches for light DM scattering off electrons~\cite{Batell:2021blf} or nuclei~\cite{Batell:2021aja}. This relies on the expected ability to effectively suppress neutrino- and muon-induced backgrounds. The former are suppressed because DM interactions with light mediators are predicted to have lower energy depositions in the detector than high-energy neutrino interactions. On the other hand, muon-induced backgrounds can be rejected using precise event timing, a capability that could also be implemented in other experimental setups in the FPF with sufficient time resolution.

Scattering signatures at the FPF can also probe other BSM particles, such as those in hadrophilic dark sectors~\cite{Batell:2021snh}, with expected sensitivity in a mass regime that is complementary to the reach of electron fixed target experiments~\cite{Schuster:2021mlr}. This is illustrated in \cref{fig:bsm_dm} for a dark matter candidate mediated via a $U(1)_B$ gauge boson. Furthermore, this experimental setup can also be used to study new neutrino interactions, as discussed in the context of the neutrino dipole portal to HNLs~\cite{Jodlowski:2020vhr, Ismail:2021dyp}.

\subsection{New Particles and Phenomena
\label{sec:newparticles}}

\begin{wrapfigure}{R}{0.6\textwidth}
\centering
\includegraphics[width=0.59\textwidth]{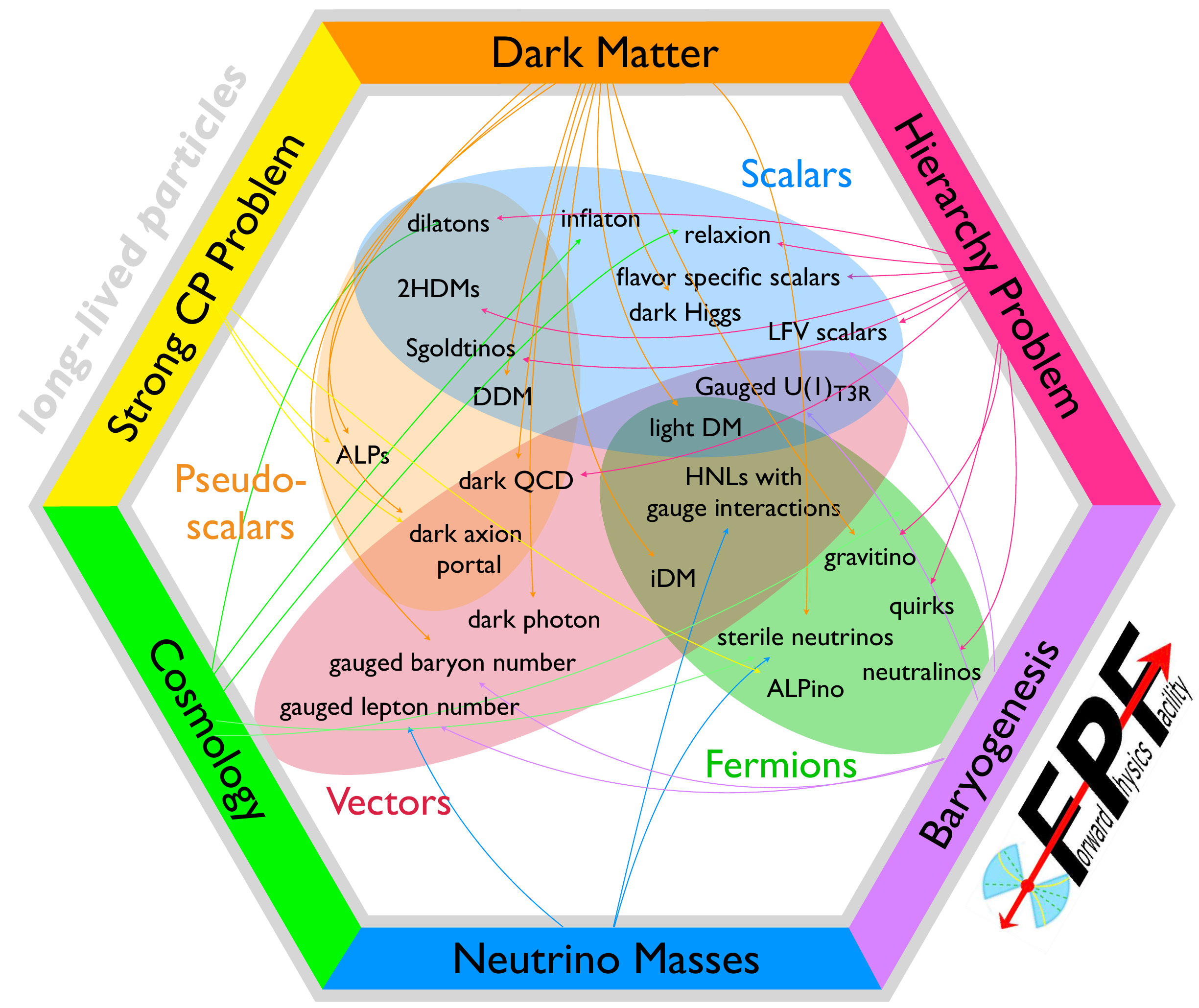}
\caption{\textbf{LLPs at the FPF}. Models of LLPs that have been studied in the context of the FPF, grouped by their quantum numbers. Connections are shown to the multitude of new physics motivations on the outside of the figure.}
\vspace{-2mm}
\label{fig:bsm_llp}
\end{wrapfigure}

\noindent \textbf{LLP Solutions to Fundamental Problems:} The FPF experiments offer exceptional sensitivity to a variety of theoretically-motivated LLPs linked to fundamental open questions in particle physics and cosmology, as depicted in~\cref{fig:bsm_llp} and studied extensively in the FPF White Paper~\cite{Feng:2022inv}.  These include LLPs associated with the origin of neutrino masses and baryogenesis (such as HNLs, sterile neutrinos, and new gauge bosons of baryon or lepton number), the strong CP problem (axions and ALPs), the hierarchy problem (light neutralinos and the relaxion), and inflation (light inflatons).  Furthermore, as already highlighted above, a variety of mediators to DM can naturally manifest as LLPs and are accessible at the FPF. This underscores the FPF's broad BSM search program, with sensitivity to LLPs across diverse theoretical scenarios. Given the wide range of possible LLP properties, including masses, spins, and SM couplings, the FPF is well positioned to play a leading role in both the discovery and diagnostic phases of LLP exploration.
\medskip

\noindent \textbf{Millicharged Particles:} Millicharged particles (mCPs) are intriguing candidates in searches for BSM physics, motivated by several compelling theoretical considerations and experimental anomalies. The discovery of mCPs could have important implications for the principle of charge quantisation~\cite{Dirac:1931kp}, as well as motivated UV extensions of the SM, such as grand unified theories~\cite{Pati:1973uk,Georgi:1974my} and string theory~\cite{Wen:1985qj,Shiu:2013wxa}. Additionally, mCPs emerge naturally within simple models featuring a massless dark photon that kinetically mixes with the photon~\cite{Holdom:1985ag}. From a cosmological perspective, mCPs are attractive due to their potential to constitute a fraction of the DM abundance~\cite{Brahm:1989jh,Feng:2009mn,Cline:2012is}, and furthermore, mCPs have been proposed as a possible explanation for observational puzzles, such as the EDGES anomaly~\cite{Bowman:2018yin,Barkana:2018lgd,Berlin:2018sjs,Liu:2019knx}. In addition, accelerator searches for mCP can be interpreted as a probe of reheating in cosmology~\cite{Gan:2023jbs}

Experiments at the FPF will offer unique and unprecedented sensitivity to mCPs. Compared to other ongoing and proposed efforts, the FPF provides some key advantages: it exploits the high-energy $pp$ collisions at the LHC to copiously produce comparatively heavy mCPs up to ${\cal O}(100\, {\rm GeV})$, and it leverages the enhanced production rates of mCPs in the forward region.  The proposed FORMOSA experiment~\cite{Foroughi-Abari:2020qar}, a scintillator-based detector, will build on the demonstrated success of the milliQan experiment~\cite{Haas:2014dda}, which is currently operating at the LHC. FORMOSA detects mCPs via coincident signals across multiple plastic scintillator bars induced by the small ionisation deposited as the mCPs traverse the material. This approach enables FORMOSA to achieve world-leading sensitivity to mCPs over a broad mass range, spanning from approximately $10\, {\rm MeV}$ to $100\, {\rm GeV}$, as can be seen in the left panel of~\cref{fig:BSM_newparticles}.  Alongside FORMOSA, the FLArE experiment will also be able to explore new regions of mCP parameter space~\cite{Kling:2022ykt}, as seen in the figure.  The FLArE search strategy targets single mCP-electron scattering events with energy deposits above $30\, {\rm MeV}$. Collectively, these experiments will significantly advance the experimental frontier in mCP searches.\medskip

\begin{figure}[bt]
  \centering
  \includegraphics[width=0.49\textwidth]{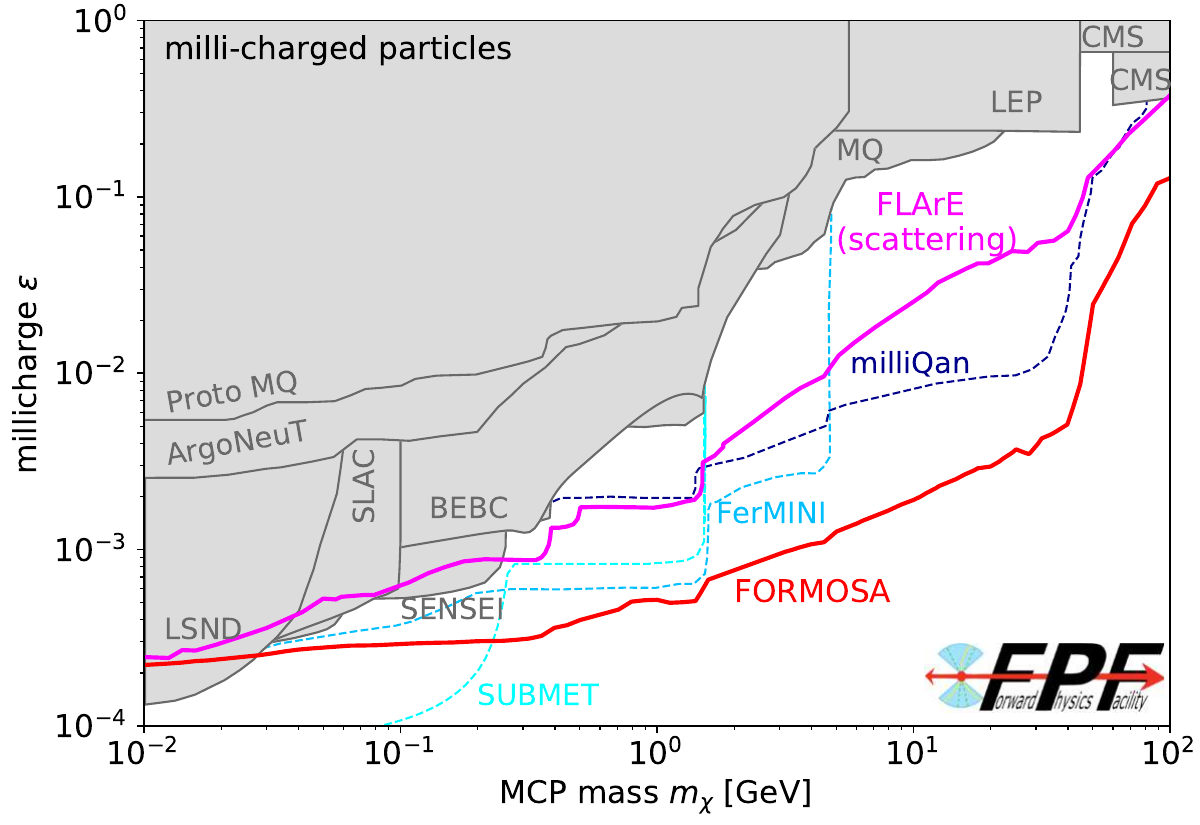}
  \includegraphics[width=0.49\textwidth]{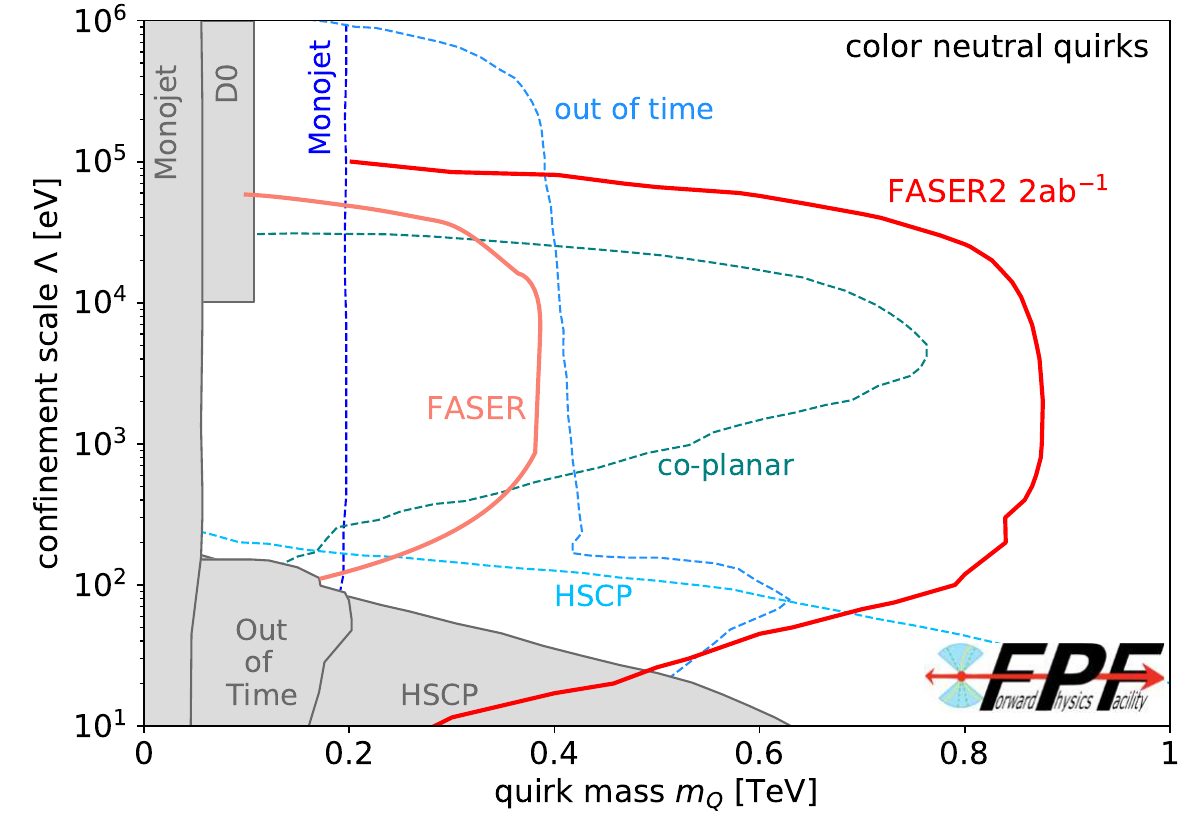}
  \caption{\textbf{New Particle Searches at the FPF.} Left: The discovery reach of FORMOSA and FLArE with $2~\iab$ for millicharged particles~\cite{Foroughi-Abari:2020qar, Kling:2022ykt}. Right: The discovery reach of FASER and FASER2 for colour-neutral quirks~\cite{Feng:2024zgp}.  In both panels, we also show existing bounds (gray shaded regions) and projected sensitivities of other experiments (dashed contours), including millicharged particle searches at BEBC~\cite{Marocco:2020dqu}, SLAC~\cite{Prinz:1998ua}, LEP~\cite{Davidson:2000hf,OPAL:1995uwx}, CMS~\cite{CMS:2012xi, Jaeckel:2012yz, CMS:2024eyx}, LSND~\cite{Magill:2018tbb}, ArgoNeuT~\cite{ArgoNeuT:2019ckq}, milliQan~\cite{Ball:2016zrp, Ball:2020dnx, Alcott:2025rxn}, FerMINI~\cite{Kelly:2018brz}, SUBMET~\cite{Choi:2020mbk}, and SENSEI~\cite{SENSEI:2023gie}, as well as quirk searches at D0~\cite{D0:2010kkd},  monojet searches~\cite{Farina:2017cts,CMS-PAS-EXO-16-037,ATLAS:2016bek}, heavy stable charged particle searches (HSCP)~\cite{Farina:2017cts,CMS-PAS-EXO-16-036,ATLAS:2016onr}, co-planar hits searches~\cite{Knapen:2017kly}, and out-of-time searches~\cite{Evans:2018jmd,ATLAS:2013whh}.}
  \label{fig:BSM_newparticles}
\end{figure}

\noindent \textbf{Quirks:} Many compelling models of new physics predict the existence of quirks, which are LLPs that are charged under both the SM and an additional strongly-interacting gauge force. For example, colour neutral quirks arise in many models of neutral naturalness~\cite{Curtin:2015bka}, which are built to address the gauge hierarchy problem~\cite{BasteroGil:2000bw, Bazzocchi:2012de}.  These include folded supersymmetry~\cite{Burdman:2006tz, Burdman:2008ek}, twin Higgs models~\cite{Chacko:2005pe, Craig:2015pha, Serra:2019omd, Ahmed:2020hiw}, quirky little Higgs models~\cite{Cai:2008au}, and minimal neutral naturalness models~\cite{Xu:2018ofw}. Quirks can also have SM colour in more general cases.  If the confinement scale $\Lambda$ of the hidden gauge group is much smaller than the quirk mass, a pair-produced quirk and anti-quirk remain bound by a macroscopic gauge flux tube~\cite{Kang:2008ea}. When $\Lambda \sim 100~\text{eV} - \text{keV}$, the bound state exhibits oscillations around its center-of-mass with macroscopic amplitudes ranging from millimeters to meters. These systems appear as pairs of minimum-ionizing particles with unusual, correlated trajectories. 

At colliders, colour-neutral quirks are typically produced in pairs via Drell–Yan processes. Because of their small transverse momentum, these pairs are preferentially emitted in the forward direction and travel together along the beamline~\cite{Li:2021tsy}. Discovery prospects at the FPF are illustrated in the right panel of \cref{fig:BSM_newparticles}. For hidden confinement scales around 100~eV, current bounds from searches for heavy stable charged particles, monojets, and other exotic signatures~\cite{Feng:2024zgp} do not even exclude quirk masses of 100~GeV. Remarkably, FASER2 can probe masses up to 1~TeV---the weak-scale range favoured by models that address the gauge hierarchy problem~\cite{Batell:2022pzc}---simply by identifying pairs of slow or delayed charged tracks~\cite{Feng:2024zgp}. An additional search strategy based on the helical nature of the quirk tracks leads to similar sensitivity~\cite{Li:2021tsy}. Notably, these high-mass quirks cannot be accessed in fixed-target experiments, underscoring the unique search capabilities of forward detectors at high-energy colliders. \medskip

\noindent \textbf{The LHC and the FPF as a Light-Shining-Through-Wall Experiment:} Light axions and ALPs with masses at the eV scale and below are the focus of an expanding experimental program. While cosmological and astrophysical observations have placed strong constraints on these particles, they often rely on additional assumptions and carry significant uncertainties.  This led to a growing community interest in purely laboratory-based ALP searches, as advocated in Ref.~\cite{Jaeckel:2006xm}, and motivated the construction of Light-Shining-Through-Wall experiments. The most powerful realisation of this idea using optical lasers is the ALP-II experiment at DESY~\cite{Spector:2019ooq}, which probes ALP masses up to 10~meV. Ref.~\cite{Kling:2022ehv} noted that the LHC and forward detectors create a similar setup at higher energies, allowing sensitivity to heavier ALPs. LHC collisions produce an intense flux of high-energy photons in the forward region, which can convert into ALPs in the LHC magnets. These ALPs can then traverse about 250 meters of rock and reconvert into photons in the magnetic fields of FASER2, where they may be detected by an electromagnetic calorimeter. Ref.~\cite{Kling:2022ehv} showed that FASER2 can set the strongest purely laboratory-based bounds on ALPs with masses between 10 meV and 1 keV.\medskip

\begin{figure}[tb]
  \centering \vspace{-5mm}
  \includegraphics[trim={0 -1.5mm 0 0},clip,width=0.49\textwidth]{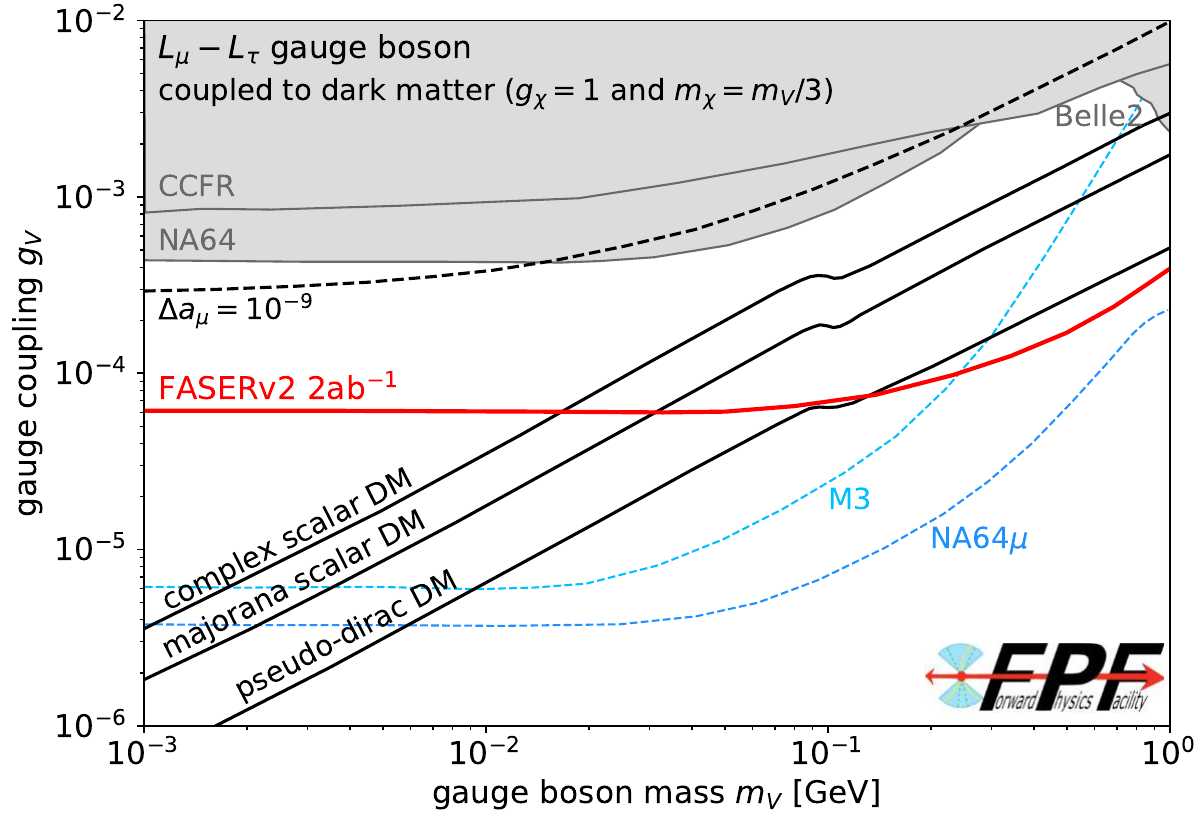}
  \includegraphics[trim={0 2mm 0 0},clip, width=0.47\textwidth]{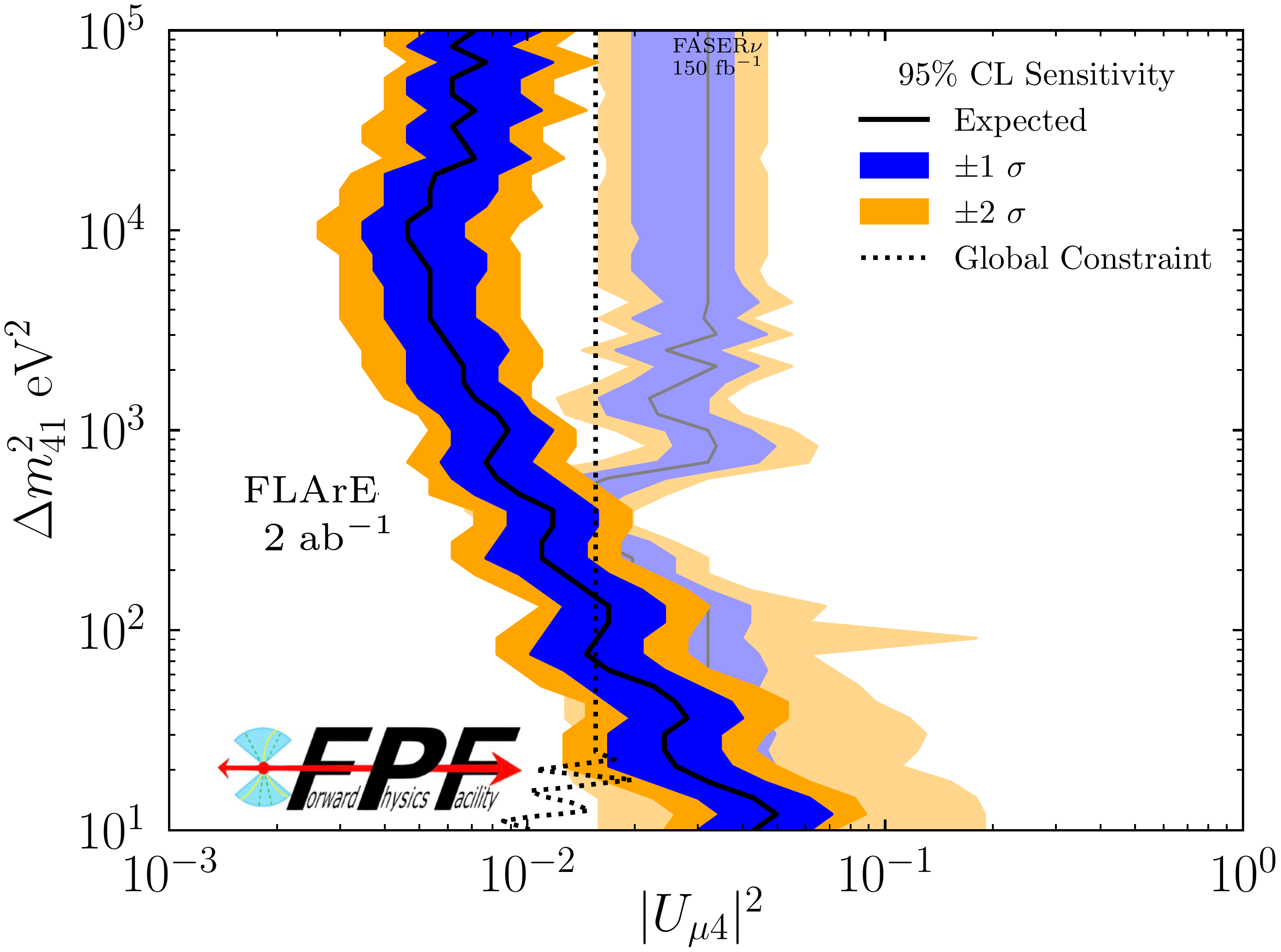}
\caption{\textbf{Left: Muon-philic Force Carriers.} Sensitivity projections at FASER$\nu$2 with $2~\iab$ for an invisibly decaying $L_\mu - L_\tau$ gauge boson in the plane spanned by the gauge boson's mass $m_V$ and coupling $g_V$. DM relic target lines are shown in the solid black contours for several models and assume fixed DM coupling $g_\chi=1$ and a DM mass $m_\chi = m_V / 3$. Existing exclusion limits come from NA64~\cite{NA64:2024klw}, Belle~II~\cite{Belle-II:2022yaw} and CCFR~\cite{Altmannshofer:2014pba} and other projected sensitivities are presented for  NA64$\mu$~\cite{NA64:2024klw}, and $M^3$~\cite{Kahn:2018cqs}. (Figure adapted from Ref.~\cite{PBC:2025sny}.) \textbf{Right: Sterile Neutrino Oscillations.} The expected sensitivity to neutrino oscillations in the $\nu_\mu$ disappearance channel at the LHC using Feldman-Cousins and Asimov for FASER$\nu$ at LHC Run~3 and FLArE at the FPF. The existing oscillation averaged constraints coming mostly from MINOS+ and MiniBooNE are also shown. (Figure taken from Ref.~\cite{Anchordoqui:2021ghd}.) }
  \label{fig:BSM_muons}
\end{figure}

\noindent \textbf{Muons as Probes of New Physics:}  The FPF detectors are exposed to a large flux of highly energetic muons---about $10^{11}$ muons per $\m^2$ per $\iab$---thereby effectively providing a muon fixed target experiment with TeV-scale muons. This setup can be used to search for a variety of new particles that primarily couple to muons, which would be produced in muon interactions, while evading constraints from electron or proton beam dump experiments. Ref.~\cite{Ariga:2023fjg} studied the scenario in which the new particle would decay invisibly, for example into DM. This provides a signature where the muon is deflected and loses a significant amount of its energy.  This scenario could be searched for in FASER$\nu$2 at the FPF, utilising the large target mass of the detector, as well as its ability to measure muon momenta and small deflection angles. The sensitivity is shown in \cref{fig:BSM_muons} for the example of an $L_\mu-L_\tau$ gauge boson as a mediator to DM, which was a showcase model in the previous PBC report~\cite{PBC:2025sny}. We can see that FASER$\nu$2 can probe a large region of unexplored parameter space that is motivated by the observed DM relic abundance~\cite{Berlin:2018bsc} and would also give a sizable contribution to the muon anomalous magnetic moment~\cite{Chen:2017awl}. Additional studies also demonstrated the leading potential of the FPF to constrain decaying muon-philic scalars~\cite{MammenAbraham:2025gai}, as well as to look for the axion-mediated lepton flavour violating scattering process $\mu N \to \tau N a$~\cite{Batell:2024cdl}. 
\medskip

\noindent \textbf{Neutrinos as Probes of New Physics:} The high energy of the neutrino beam and the potentially large number of interactions make collider neutrino experiments an appealing setting for exploring new physics in the neutrino sector. A variety of signatures and scenarios have been explored in the literature. One example is sterile neutrino oscillations: while no oscillations are expected to occur among the active neutrino states alone, the presence of additional heavy sterile neutrinos can induce visible oscillation patterns. The strongest effects are expected for masses $m \sim (E_\nu / L)^{1/2} \sim 50$~eV, assuming neutrino energies of $E_\nu \sim \tev$ and a baseline of $L \sim 620$~m. The combination of the LHC and FPF thus constitutes a short-baseline experiment with a smaller $L/E_{\nu}$ than any other neutrino experiment. The sensitivity to sterile neutrino oscillations has been studied in Refs.~\cite{FASER:2019dxq, Bai:2020ukz, Feng:2022inv}, which found that the FPF could place world-leading constraints on muon neutrino disappearance due to heavy sterile neutrinos. As example, we show the sensitivity obtained by searching for modulation in the energy spectra in the muon neutrinos appearance channel at FLArE in the right panel of \cref{fig:BSM_muons}. Similarly, the FPF experiments can search for signals of neutrino-modulino oscillations to probe models with string scale in the grand unification region and SUSY breaking driven by sequestered gravity in gauge mediation~\cite{Anchordoqui:2023qxv, Anchordoqui:2024ynb}.

In addition, precision neutrino measurements at the FPF can also be used to search for neutrino-philic particles. One scenario is force carriers decaying primarily to tau neutrinos, such as a $B - 3L_\tau$ gauge boson, thereby affecting the tau-neutrino flux. Refs.~\cite{Kling:2020iar, Batell:2021snh} found that FASER$\nu$2 will have world-leading sensitivity for such particles in the GeV-mass region. In addition, FLArE will be able to probe neutrino self-interactions and neutrino-philic mediators to DM produced in neutrino interactions by searching for a missing transverse momentum signal in CC neutrino interactions. Refs.~\cite{Kelly:2021mcd, Berryman:2022hds} found that FLArE will be able to place world-leading constraints on such mediators with masses between 0.3 GeV and 20 GeV and small couplings not yet probed by existing searches.

\clearpage 
\section{The Facility }
\label{sec:facility}


The FPF facility has been studied, in the context of the PBC study group, by CERN experts over the last four years, with technical studies detailed in Refs.~\cite{Feng:2022inv, PBCnote, PBCnote2, vibration-note}. 
The work has benefited from the vast experience at CERN in designing and implementing many similar large underground facilities.  A particularly relevant example was the design/construction of the HL-LHC underground galleries at the ATLAS and CMS sites, where the civil engineering (CE) work was carried out in 2018-2022 and the outfitting is ongoing. Many of the same technical solutions from this project can be adopted for the FPF, and lessons learned can also be applied. 


\begin{figure}[b]
  \centering
  \includegraphics[width=0.9\textwidth]{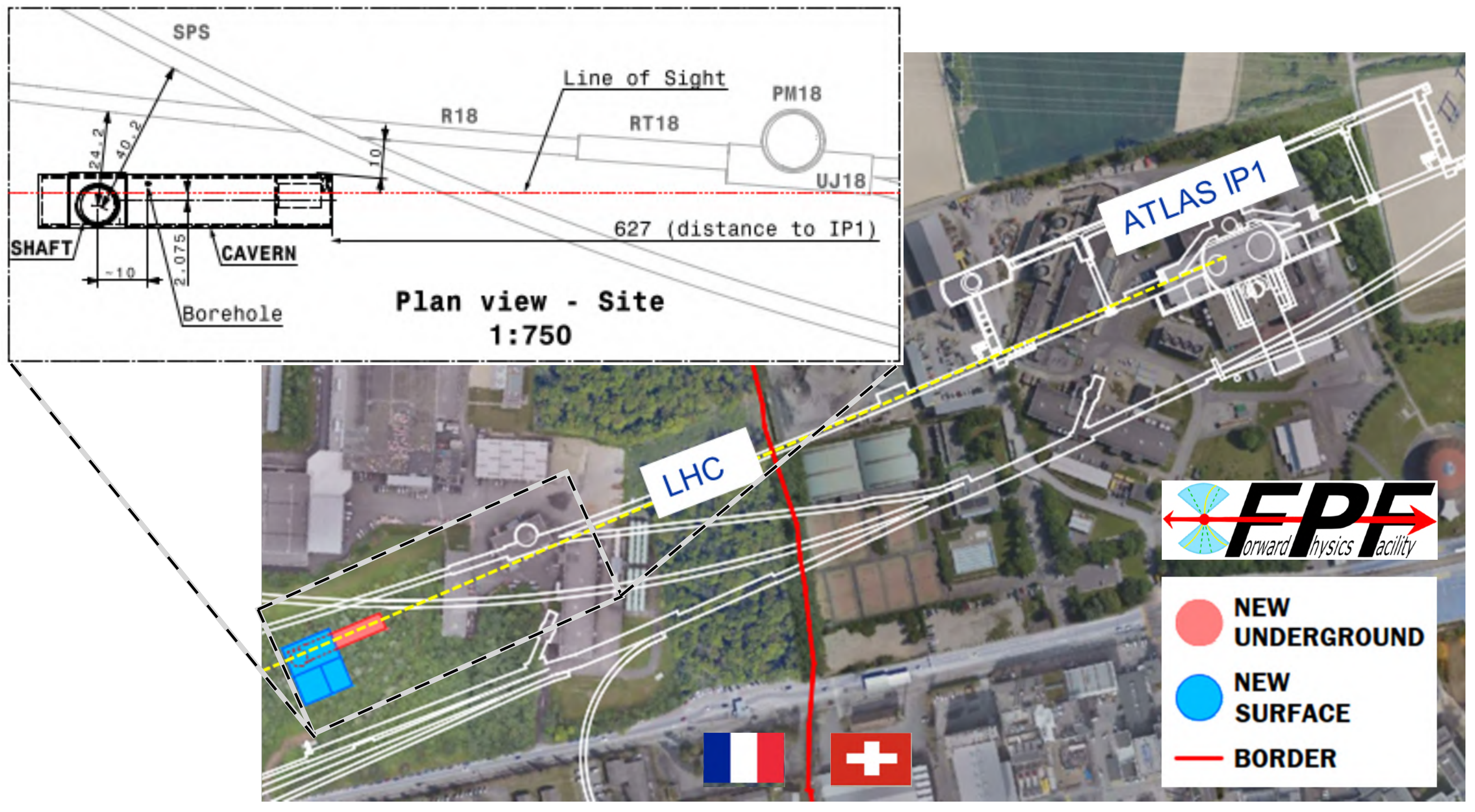}
  \caption{\textbf{FPF Location.} The location of the FPF shown on a map of the area around IP1. The LOS is shown as a yellow dashed line. The inset shows a more detailed view of the area around the FPF. All distances are given in meters. }
  \label{fig:plan}
\end{figure}

\subsection{Site Selection and Cavern Design} 

A site optimisation to find the best location for the FPF facility was carried out, with the goal of allowing several large detectors to be placed on the collision axis line-of-sight (LOS). Initially, this included the possibility of widening the UJ12 junction cavern close to where the FASER experiment is currently situated. However, this was found to be too disruptive and was dismissed early on, leaving the possibility of a new purpose-built facility at the depth of the LHC beamline. The four locations on either side of the two high luminosity interactions points (IP1 and IP5) were then studied, with the requirement that the new cavern should be as close to the IP as possible, while being 10~m from the LHC tunnel for structural stability and radiation protection considerations. It quickly became clear that the optimal choice is the site to the west of IP1. This has the advantages that (i) it is in France, which is cheaper for construction works, (ii) it is within the CERN fenced area, (iii) it is expected that the geological conditions in this area are good for the excavation, and (iv) it is close to one of the CERN surface sites (SM18) and would be easy to connect to existing service networks. The location can be seen in \cref{fig:plan}.  The cavern covers distances along the LOS between 627 m and 702 m from the ATLAS IP, and particles produced at the ATLAS IP must pass through $\sim 250$~m of concrete and rock, before entering the FPF cavern.

\begin{sidewaysfigure}[thbp!]
  \centering
  \includegraphics[width=0.90\textwidth]{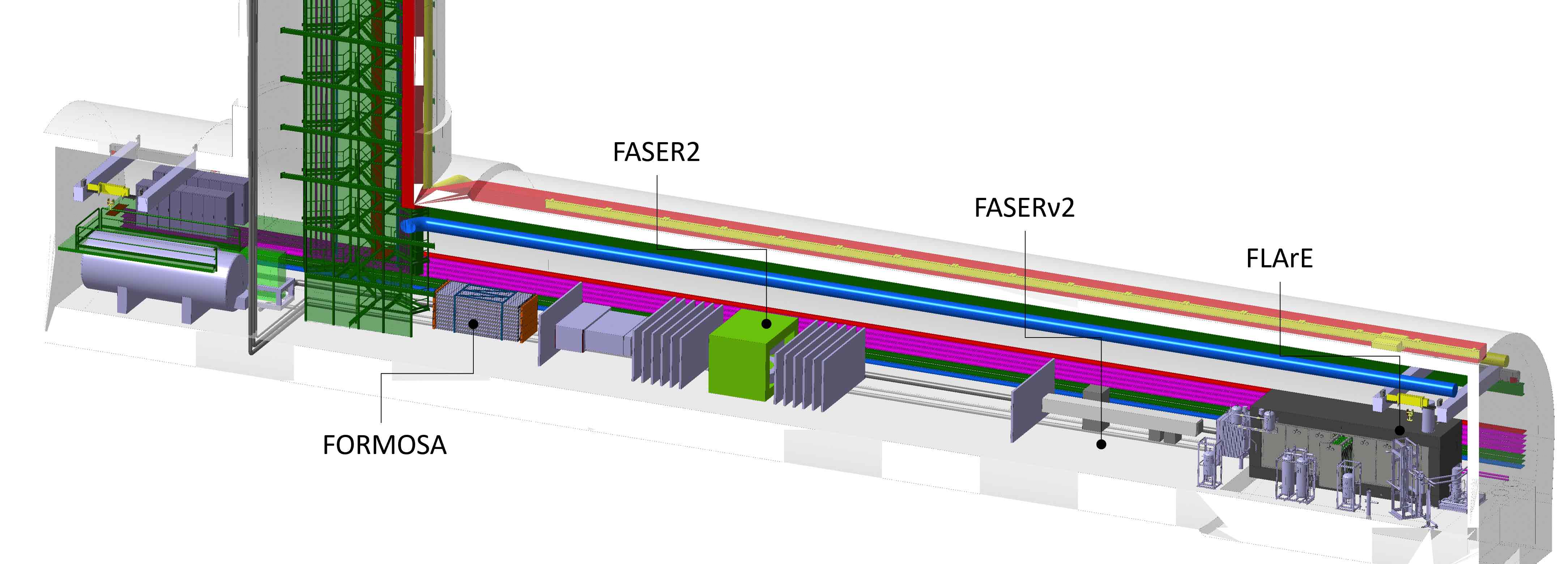}
  \caption{\textbf{FPF Cavern Design.} The baseline layout of the FPF facility, showing the four proposed experiments and the large infrastructure. }
  \label{fig:facility-layout}
\end{sidewaysfigure}

The facility design has been through several iterations to optimise the layout for the proposed detectors, along with the needed technical infrastructure. The overarching design principle has been to minimise the size of the cavern to reduce the cost, while ensuring there is sufficient space for all the needed infrastructure and taking into account that the detailed designs of the detectors are still evolving. The current baseline facility design is shown in \cref{fig:facility-layout}. This includes a 75~m-long, 12~m-wide underground cavern, with a dedicated experimental area (65~m long) and a service cavern (5~m long), as well as an 88~m-deep shaft and the associated surface building for access and services. 

The left panel of \cref{fig:facility-layout2} shows a cross section (transverse to the LOS direction) of the cavern showing the relevant dimensions. The position of the nominal LOS is 1.5~m above the floor and 3.8~m from the nearest cavern wall. The actual LOS is affected by the LHC crossing angle, which for the current baseline HL-LHC configuration (250~$\mu$rad half crossing in the horizontal plane) will move the LOS by around 16.5~cm in the horizontal plane.\footnote{There is a possibility that the crossing plane would change to vertical after several years of HL-LHC running, which would move the LOS by 16.5~cm from the nominal LOS but in an up or down direction (this would be a 23~cm shift with respect to the previous LOS alignment). Since this change is small compared to the size of the FPF detectors, it will not require realigning the detectors with the LOS, except perhaps in the case of FASER$\nu$2, where it may make sense to re-align the detector, which would be relatively easy to do.} 
To mitigate the risk of people being trapped in the FPF cavern in case of a fire or other safety incident that blocks escape via the shaft, a fire-proof over-pressure safety corridor will be installed along the cavern wall. The 1.2~m side and 2.5~m high safety corridor is shown in the figure. 
The right panel of \cref{fig:facility-layout2} also shows a 3D model of the full FPF facility, including the cavern, shaft and surface buildings.

\begin{figure}[tb]
  \centering
  \includegraphics[width=0.49\linewidth]{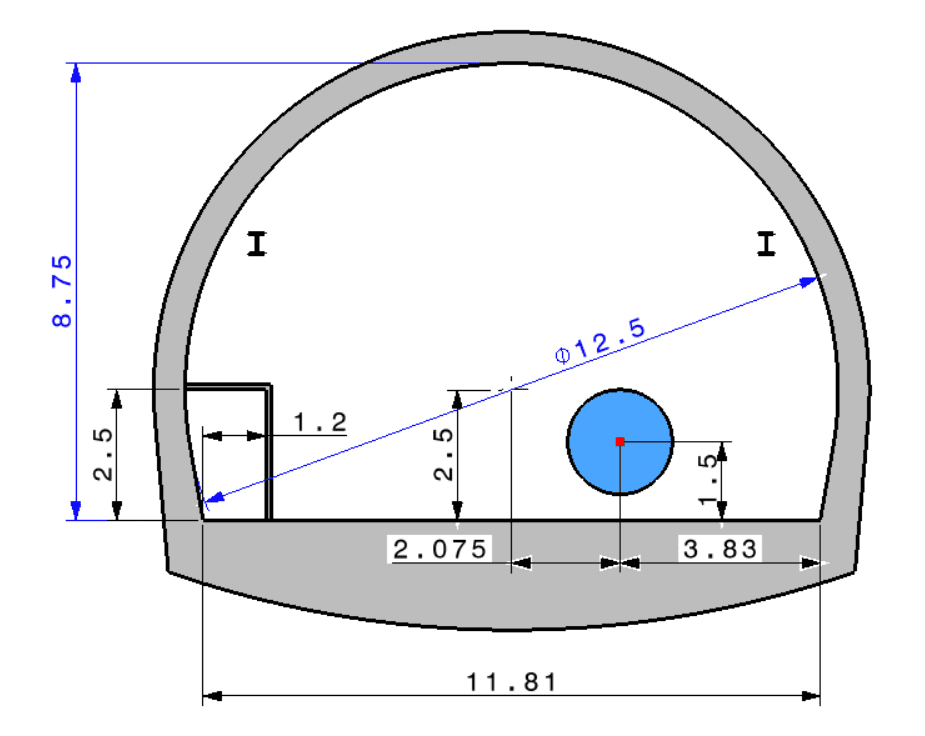}
  \hspace{1.5cm}
  \includegraphics[width=0.3\textwidth]{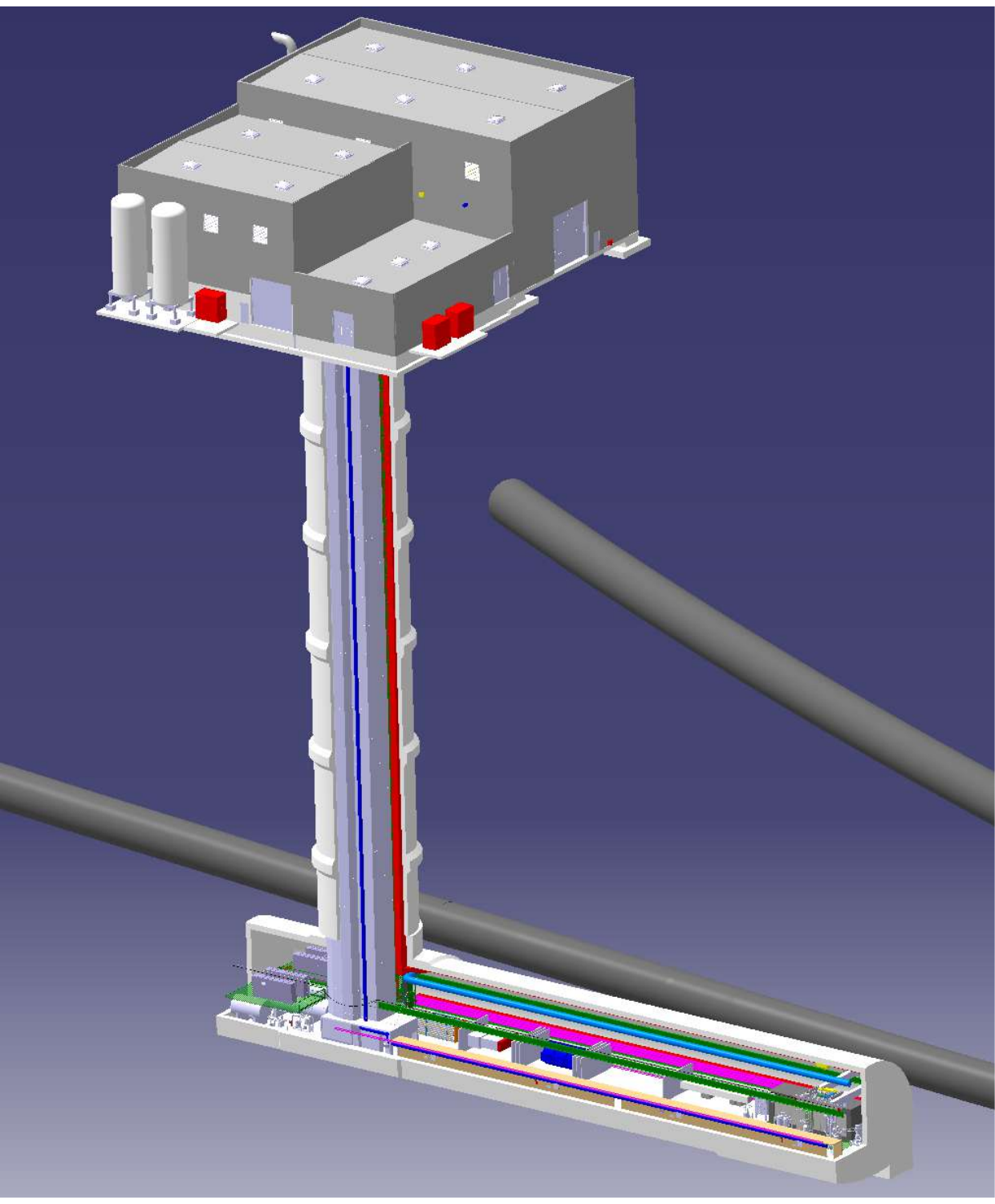}
  \caption{\textbf{FPF Cavern Cross Section and 3D View.} The left panel 
  shows the FPF cavern cross section. The nominal LOS is shown as a red dot. The right panel shows the 3D view of the Facility. }
  \label{fig:facility-layout2}
\end{figure}

The design of the surface buildings and the access to the site is shown in \cref{fig:CE:Surfacebuldings1}. This has been optimised to minimise the interference with existing infrastructure and underground service networks.

\begin{figure}[!ht]
  \centering
  \includegraphics[width=0.65\linewidth]{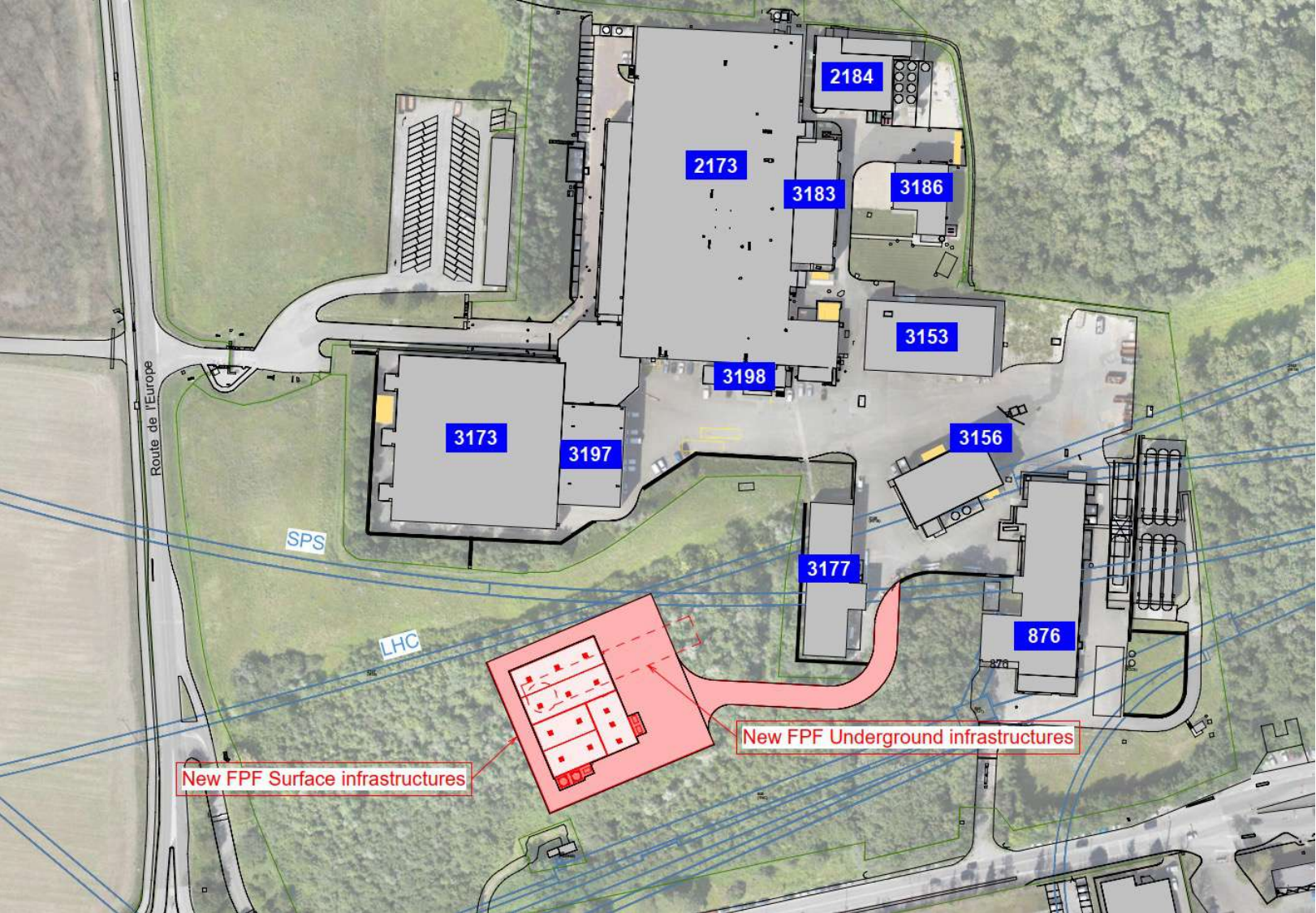}
  \caption{\textbf{Surface Building.} General layout of the proposed FPF surface area. This includes the surface buildings and road access route.}
  \label{fig:CE:Surfacebuldings1}
\end{figure}

\subsection{Site Investigation and Geological Conditions} 

To establish the subsurface conditions in which the facility will sit, site investigation works were carried out in Spring 2023. 
A single core (20~cm in diameter) was drilled to the full depth of the proposed shaft, 100~m deep, at the estimated location for the new shaft, 24~m from the LHC and 40~m from the SPS.

The location of the core drilling was defined by the CERN survey team, and 
the works were carried out using a single drilling machine. Once the core was drilled, it was divided into segments for transportation and analysis of the different ground types shown within it. 
The phases of this ground investigation are shown in \cref{fig:CoreSamples}.

\begin{figure}[tbhp]
  \centering
  \begin{minipage}[t]{0.32\textwidth}
    \centering
    \includegraphics[width=0.9\linewidth, height=6cm]{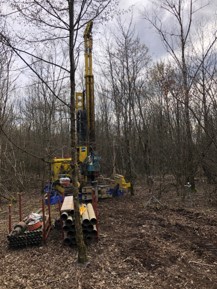}
    \par\vspace{0.5em}
    \textbf{(a)} Drilling machine in place.
  \end{minipage}
  \hfill
  \begin{minipage}[t]{0.32\textwidth}
    \centering
    \includegraphics[width=0.9\linewidth, height=6cm]{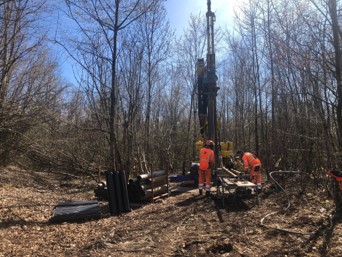}
    \par\vspace{0.5em}
    \textbf{(b)} Works started.
  \end{minipage}
  \hfill
  \begin{minipage}[t]{0.32\textwidth}
    \centering
    \includegraphics[width=0.9\linewidth, height=6cm]{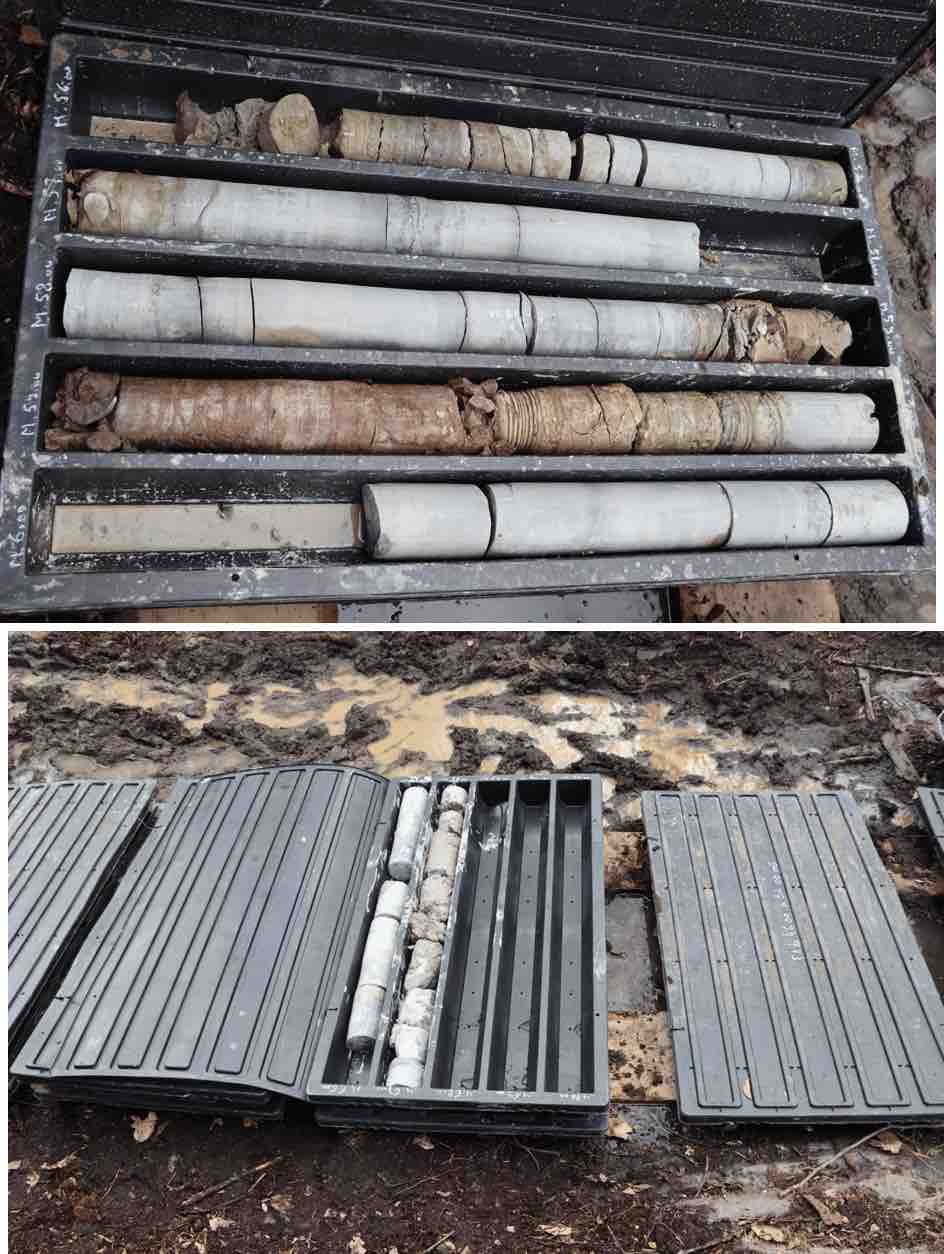}
    \par\vspace{0.5em}
    \textbf{(c)} Core samples.
  \end{minipage}
  \caption{\textbf{Site Investigation.} Photos showing the extraction of core samples at the FPF site~\cite{PBCnote2}.}
  \label{fig:CoreSamples}
\end{figure}

The results of the site investigation were broadly positive, with favourable ground conditions noted and no water table identified. Some attention will still be needed to correctly manage the presence of hydrocarbons, fluoride, and swelling potential on the site, but these can be addressed during the design phase. A full outline and interpretation of the site investigation works were given in a report by the Geotechnical consultants, GADZ~\cite{FPF_GADZ_SI_report}. These results were then contextualised into the rest of the FPF CE proposed works in a report by Arup~\cite{FPF_Arup_SI_report}. 

A simplified overview of the results of the site investigation study can be seen in \cref{tab:Site_Investigation_Table}.

\begin{table}[tbhp]
  \centering
  \begin{tabular}{c|c}
    \hline\hline
     \textbf{Results} &
     \textbf{Recommendations} \\ 
    \hline\hline
     Ground found mostly competent for tunneling & 
     N/A \\
     purposes & \\
    \hline
     Signs of hydrocarbons were found in the soft &
     1) Excavation material contaminated with liquid \\
     sandstone at depths between 84 and 90~m. &
     hydrocarbons will require specific spoil management. \\
     & 2) Underground tunnels and works in contact with \\
     & soils contaminated with hydrocarbons will require \\
     & specialised waterproofing membrane. \\ 
    \hline
     Foundations of the surface buildings will sit within & 
     N/A \\
     ompetent moraine.& \\  
    \hline 
     No water table has been identified. Overall the & 
     N/A \\
     ground is not very permeable.& \\  
    \hline 
     Vertical swelling test carried out showed a high &
     Swelling pressures to be considered during the design \\
     swelling potential.&  of the final lining. \\ 
    \hline 
     Slight elevation of fluoride levels shown in the &
     Existing backfill material will need to be disposed of\\
     existing backfill material. &at appropriate facilities.\\ 
    \hline \hline
  \end{tabular}
  \caption{\textbf{Results and recommendations of the site investigation.}}
  \label{tab:Site_Investigation_Table}
\end{table}

\subsection{CE Costing and Schedule}
A Class 4 costing~\cite{CEcosting} for the CE work was carried out in September 2024, based on similar work carried out at CERN in the last decade and taking into account the findings of the site investigation. 
The Class 4 estimate has an expected range around the quoted point estimate from $-30\%$ to $+50\%$. 
The costing methodology has been cross-checked by an external CE consultant. The cost estimate is 35~MCHF for the underground works, shaft, and surface buildings.  
A detailed breakdown of these costs is given in \cref{tab:CE-costs}. 

Given the unknown time for approval of the project, a technically feasible timeline for the implementation of the FPF is shown in~\cref{fig:CE-schedule}. The schedule is based on the extensive experience at CERN of implementing similar projects. It shows work that has already been completed - the feasibility work and concept design, as well as the site investigation works. It assumes that the technical design would start in Q4 2025, and that this is then followed by the detailed design and tendering process. The expected time for the CE works is 3~years. Work is ongoing to optimise these steps, and to see if different activities can be done in parallel, to minimise the time from approval to the start of construction.

\begin{figure}[th]
  \centering
  \includegraphics[width=0.95\textwidth]{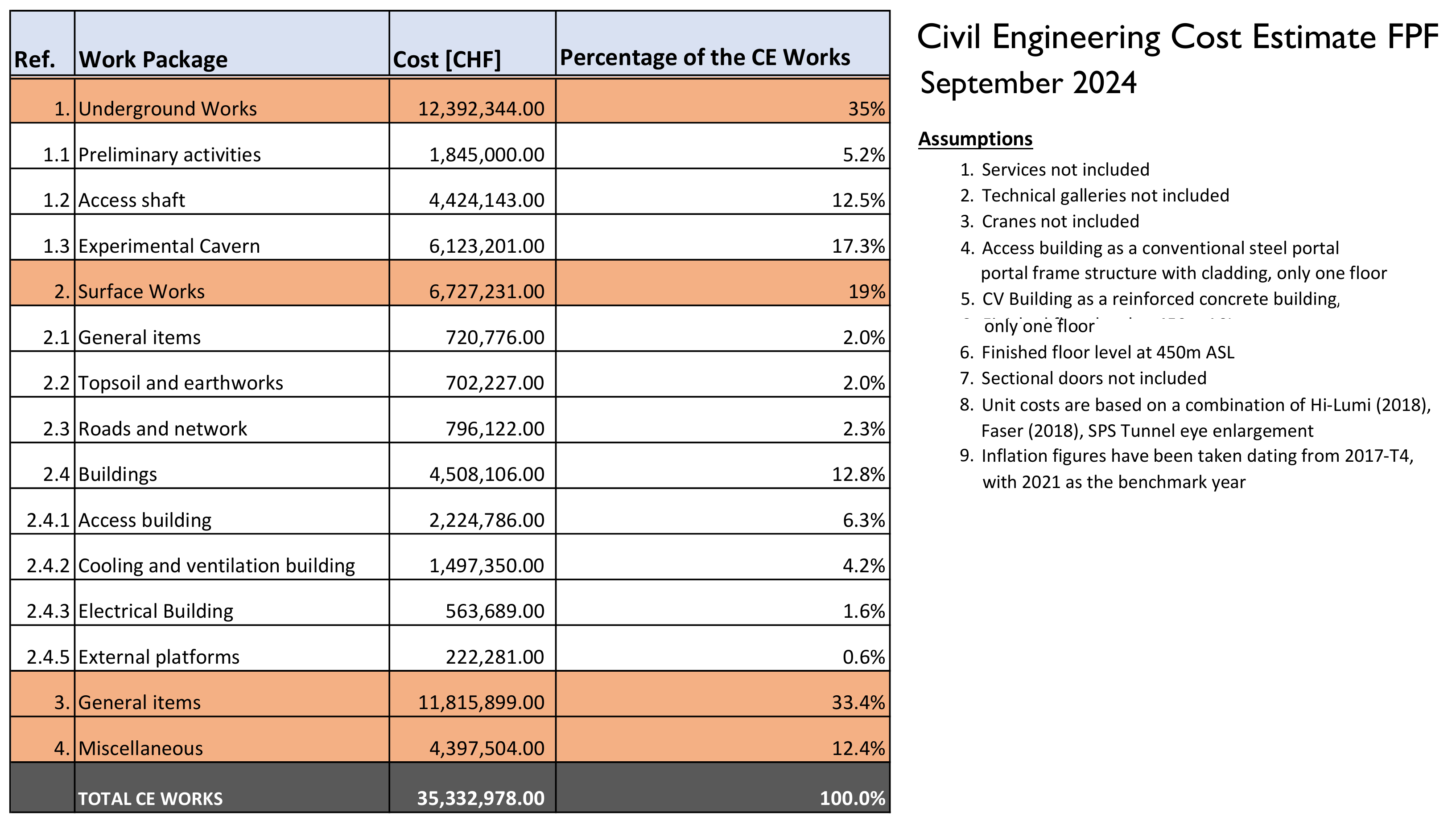}
  \caption{\textbf{CE Cost Estimate.} A detailed breakdown of the costs of the baseline CE works. }
  \label{tab:CE-costs}
\end{figure}

\begin{figure}[th]
  \centering
  \includegraphics[width=0.99\textwidth]{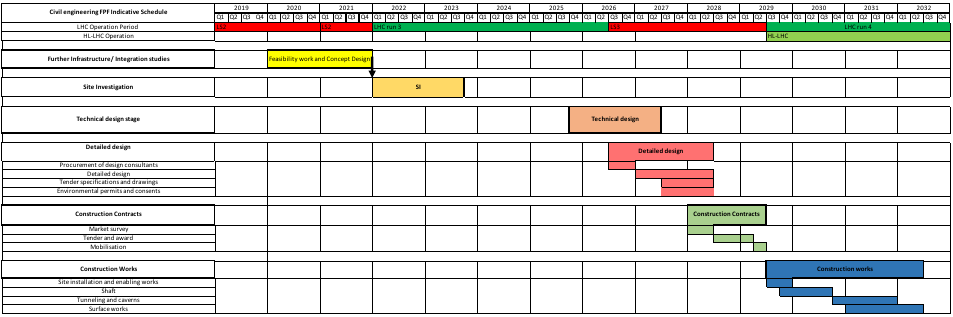}
  \caption{A technically feasible timeline for the FPF CE works, including the preparation and implementation times.  }
  \label{fig:CE-schedule}
\end{figure}

\subsection{Excavation Work and Vibrations} 

The possibility of carrying out the FPF excavation work during beam operation will allow much more flexibility in the FPF implementation schedule. However, concerns have been raised that the excavation works could impact beam operations of the LHC or SPS, leading to beam losses and possible beam dumps. The CERN beam physics group has carried out detailed studies~\cite{vibration-note} of the effect of the expected vibration level and possible static tunnel movements or deformations from the excavation on beam operation performance.

The studies are based on a similar analysis that was done for the HL-LHC works at IP1, but also include the analysis of measurements of the vibration level in the LHC tunnel during these works. 
Given that the FPF works would be significantly further from the IP (which is where the beam is most sensitive to vibrations), although closer to the LHC tunnel, the effect from the FPF is expected to be lower than that seen during the HL-LHC works and at an acceptable level for beam operations. Mitigation measures that can be put in place to minimise the risk include (i) carrying out ground compactification works (for the surface area preparation) during times with no operations (technical stops etc.), (ii) compactification of the spoil removed during the excavation should be done offsite, and (iii) the possibility to switch from using rock-breaker to road-header excavators for the works if excessive vibrations are observed during the works. In terms of possible deformation of the LHC or SPS tunnels, this is expected to be at less than the 1~mm level and can be corrected for by orbit corrector magnets. The overall conclusion of these studies is that no problems are foreseen, and the excavation can be carried out during beam operations. 

\subsection{Muon and Hadron Fluxes
\label{sec:muon_fluxes}}

Given that the FPF is shielded from the LHC collisions by 250~m of rock, and charged particles are swept away by strong LHC magnetic fields, the facility is exposed to relatively low-background conditions, considering the 40~MHz of high-energy proton collisions occurring at IP1. Muons and neutrinos are the only SM particles produced at, or near, the IP which can make it to the FPF. Muons typically lose 150~GeV of energy as they traverse the rock in front of the FPF.

\begin{figure}[bt]
  \centering
  \includegraphics[trim=1.3cm 0cm 1cm 2.cm,clip,width=0.49\textwidth]{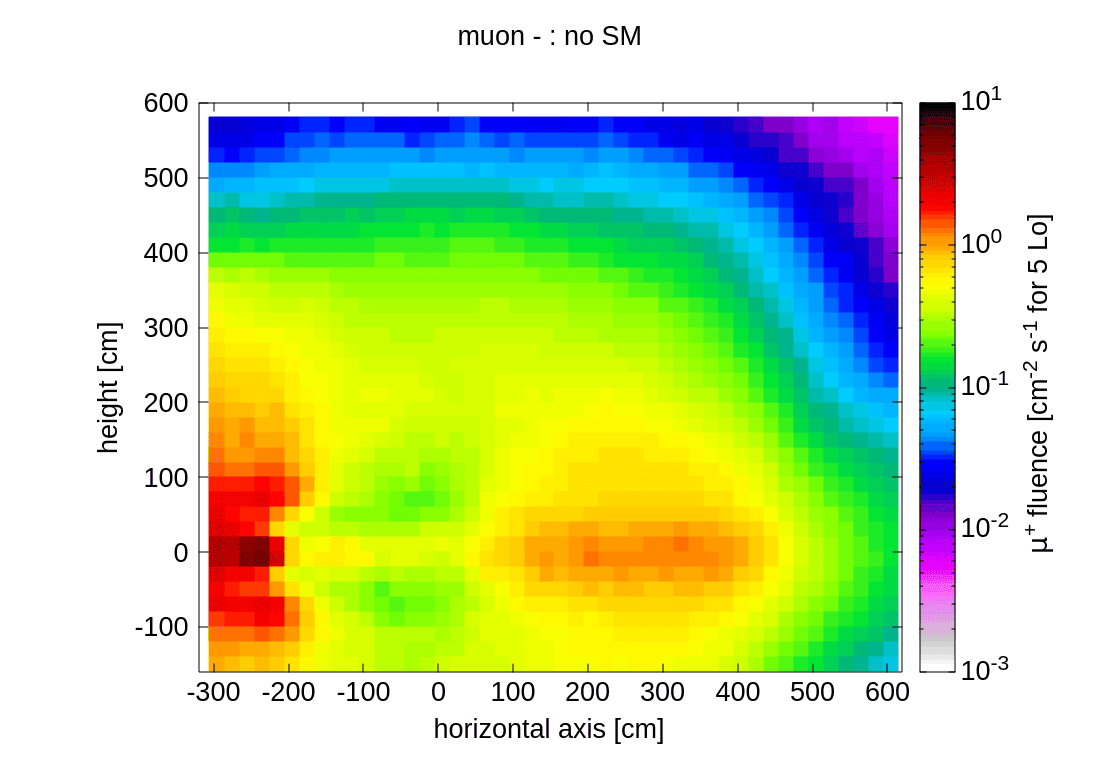}
  \includegraphics[trim=1.3cm 0cm 1cm 2.cm,clip,width=0.49\textwidth]{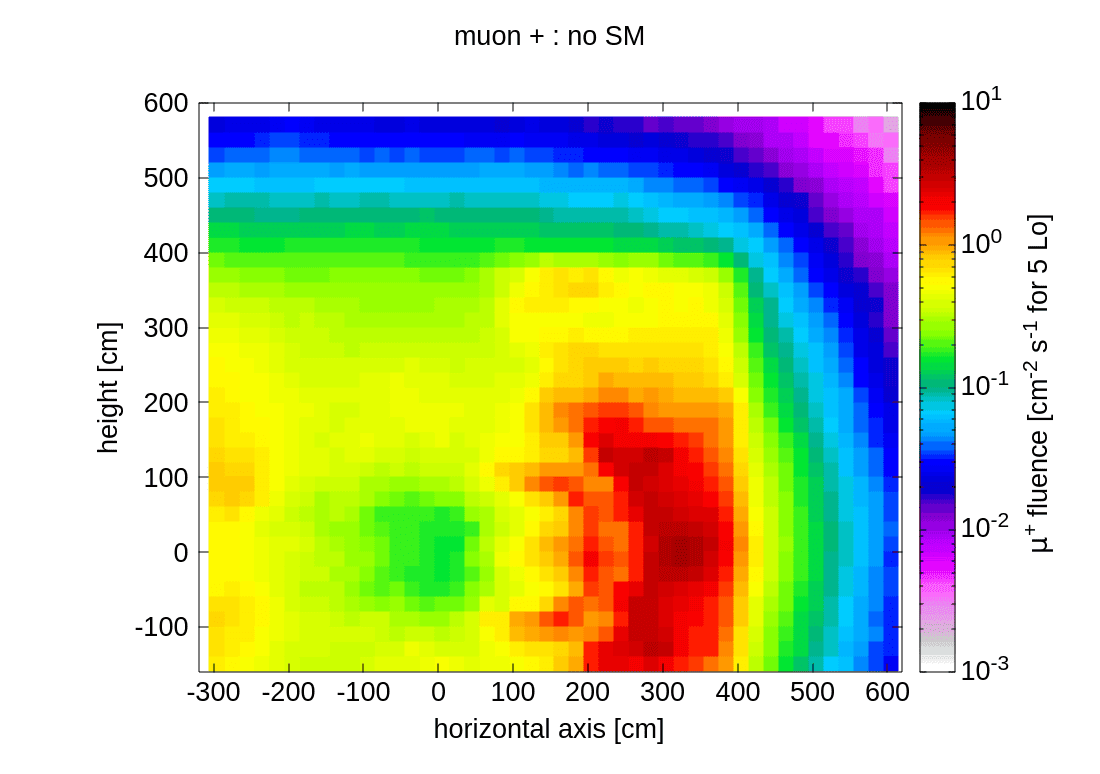}
  \caption{\textbf{Muon Fluence.} The muon fluence rate for $\mu^-$ (left) and $\mu^+$ (right) in the transverse plane in the FPF cavern for the HL-LHC baseline luminosity of $5 \times 10^{34}$~cm$^{-2}$~s$^{-1}$. The coordinate system is defined such that $(0,0)$ is the LOS. }
  \label{fig:mu-flux}
\end{figure}

The expected muon background rate in the FPF has been estimated using \texttt{FLUKA} simulations~\cite{fluka}. These simulations include a detailed description of the infrastructure between IP1 and the FPF. For the LHC Run 3 setup, the simulations have been validated at the $\mathcal{O}$(25\%) level with FASER~\cite{FASER:2018bac} and SND@LHC~\cite{SNDLHC:2023mib} data.  However, for the HL-LHC, much of the accelerator infrastructure (magnets, absorbers, etc.) in the relevant region will change. As shown in \cref{fig:mu-flux}, for the baseline HL-LHC luminosity of $5 \times 10^{34}$ cm$^{-2}$ s$^{-1}$, \texttt{FLUKA} simulations predict a muon flux of 0.6~cm$^{-2}$~s$^{-1}$ within 50~cm of the LOS, with the flux substantially higher when going to 2~m from the LOS in the horizontal plane. In general, the expected muon rate is acceptable for the proposed experiments; however, reducing the rate would be beneficial. Here, studies on the effectiveness of installing a sweeper magnet in the LHC tunnel or using the beam corrector magnets to reduce the flux are ongoing; see \cref{sec:sweeper_magnet}.

The incoming muon beam can undergo photonuclear interactions in the rock close to the FPF, producing hadrons which can enter the FPF cavern. Long-lived hadrons produced in this way could be problematic for detector electronics (for example, triggering single event upsets) and could cause radiation damage to detectors. The \texttt{FLUKA} simulation was used to simulate the expected high-energy hadron fluence in the FPF, as well as the neutron field.

In \cref{fig:FLUKA:R2E}, the transverse distributions of Silicon 1~MeV neutron equivalent fluence\footnote{Virtual 1~MeV neutron fluence that produces the same damage in Silicon as the actual radiation field.} and high-energy hadron equivalent fluence at the entrance of the cavern are presented for an integrated luminosity corresponding to one year of ultimate HL-LHC operation. The former is many orders of magnitude lower than that in the main LHC experiments, and  not problematic for the expected detector technologies in the FPF. The annual high energy hadron equivalent fluence, determining the single event error rate in electronics, does not exceed 3 $\times$ $10^6$ cm$^{-2}$ y$^{-1}$, which is the threshold adopted in the LHC for declaring an area safe from the radiation to electronics (R2E) point of view~\cite{R2E}. 

\begin{figure}[tb]
  \centering
  \includegraphics[trim=2cm 0cm 1cm 1.5cm,clip,width=0.49\textwidth]{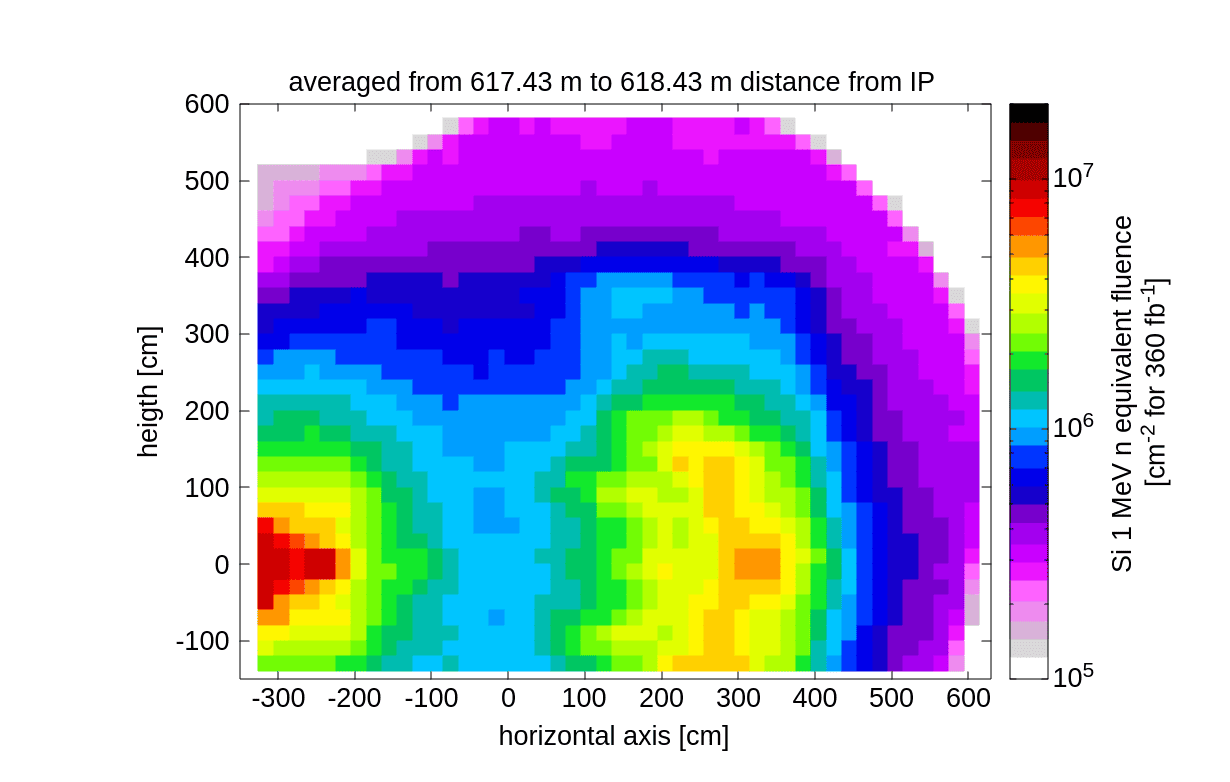} \includegraphics[trim=2cm 0cm 1cm 1.5cm,clip,width=0.49\textwidth]{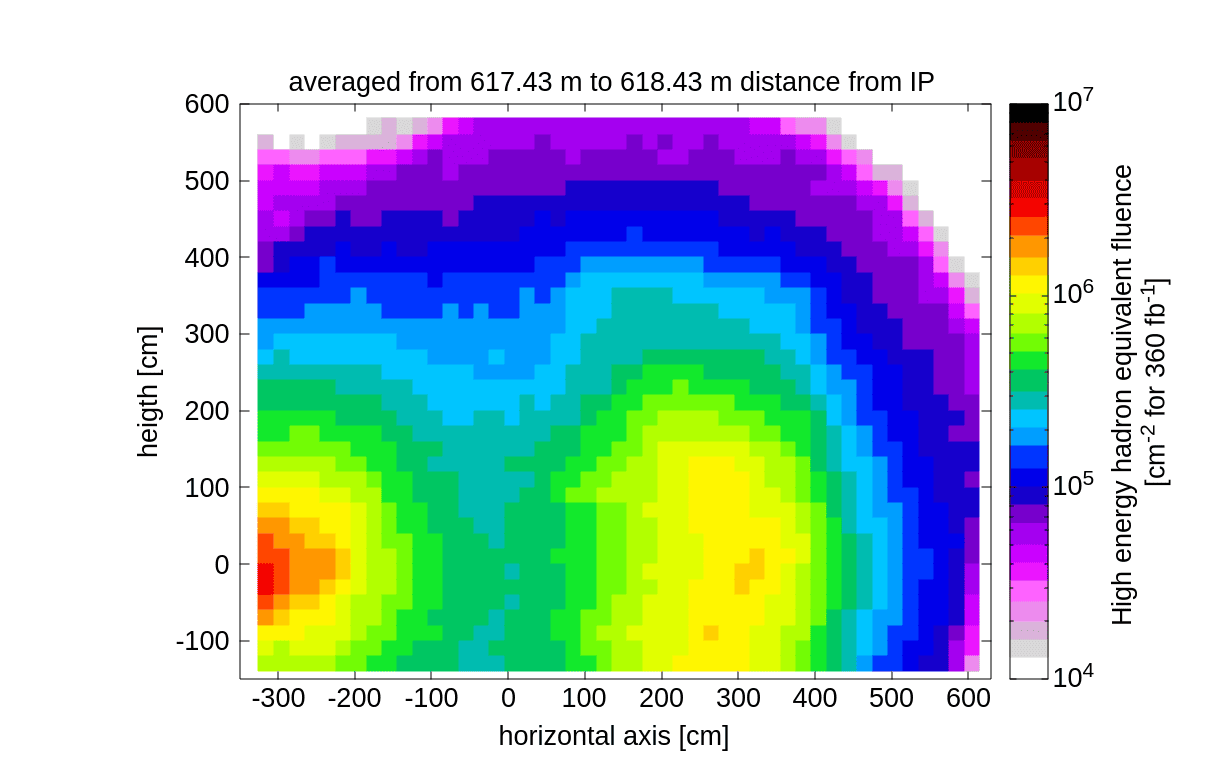}
  \caption{\textbf{Hadron Fluence.} Silicon 1~MeV neutron equivalent fluence (left) and high energy hadron equivalent fluence (right) in the FPF cavern for 360 fb$^{-1}$ integrated luminosity.}
  \label{fig:FLUKA:R2E}
\end{figure}

\subsection{Radiation Levels and Safety}
Being able to access the cavern during beam operations will be extremely valuable for detector installation, commissioning, and maintenance tasks. It will also allow the experiments to be upgraded or even replaced, as may be desirable to respond to the evolution of the physics landscape over the time period of the HL-LHC. \texttt{FLUKA} simulations, carried out by the CERN Radio-Protection (RP) group, have been used to assess the radiation level in the FPF cavern during beam operation.\footnote{Note, these studies were carried out for an older design of the FPF. however the differences are small and are not expected to effect the overall conclusions.} 
These studies are conservatively carried out for the ultimate HL-LHC luminosity of $7.5 \times 10^{34}$~cm$^{-2}$~s$^{-1}$, and consider possible radiation sources from (i) beam gas interactions in the LHC close to the FPF, (ii) accidental loss of the full LHC beam close to the FPF, and (iii) the radiation induced by the through-going muons from the IP and the Long Straight Section (LSS). These studies concluded that the dose from the first two sources considered are below the relevant level, whereas, in some local hot spots in the cavern, the radiation induced by incoming muons could be above 6~mSv/year, the threshold for classifying a location as a supervised radiation areas at CERN~\cite{SafetyCodeF}. Figure~\ref{fig:muon_dose_rate_XY_1m} shows the prompt ambient dose in $\mu$Sv/h in the transverse plane for three slices along the FPF cavern (left is the upstream end of the cavern, middle is the center of the cavern, and right is the downstream end). The hot spots related to the peaks in the muon flux can clearly be seen, as well as a larger dose at the front of the cavern from short-lived hadrons produced by muon interactions in the rock before the FPF. Most of this dose falls off over the first 5~m of the cavern. 

Based on these results the RP group concludes that access to the FPF cavern during beam operations will be possible with the following limitations:
\begin{itemize}
\setlength\itemsep{-0.05in}
\item The access should be limited to 20\% of time over a year (this means that no permanent control rooms should be foreseen in the cavern);
\item People entering the cavern during beam operations must be trained as radiation workers, and wear a dosimeter;\footnote{Note, depending on the timeline for the CE works, this may also apply to construction workers carrying out the works. This has not been considered in the cost estimate given in~\cref{tab:CE-costs}.}
\item Some parts of the cavern could be further restricted as local short stay areas if more detailed simulations and measurements show the dose is too high in these areas.
\end{itemize}

\begin{figure}[ht]
\centering
\includegraphics[width=0.99\linewidth]{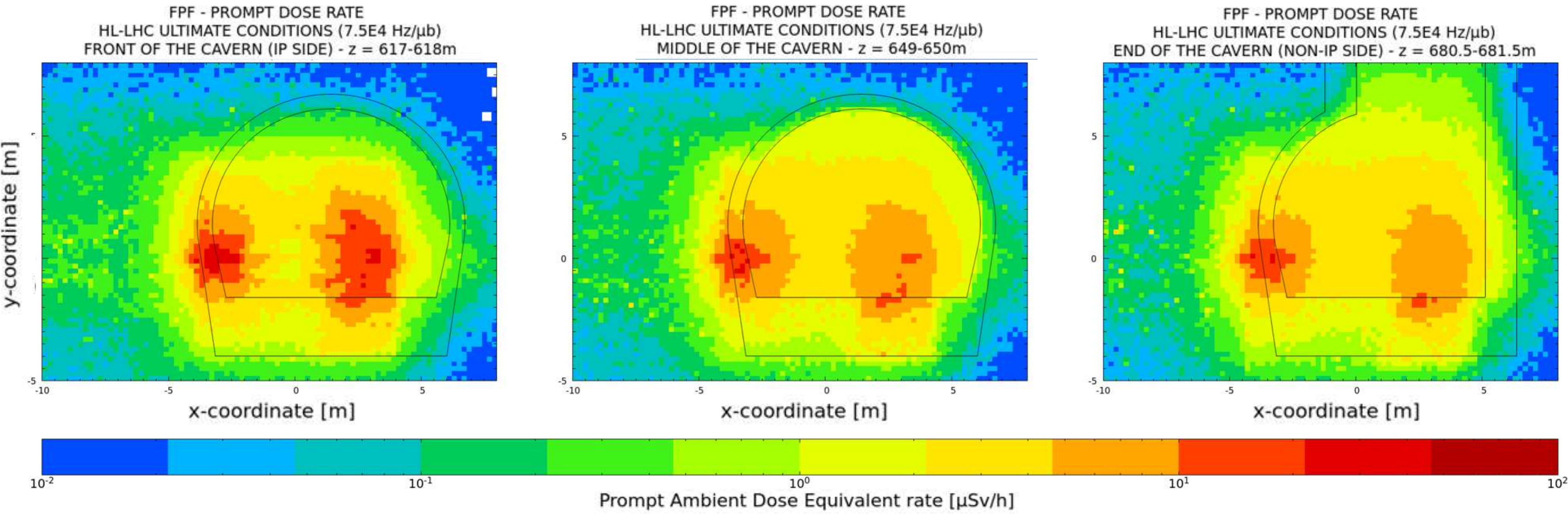}
\caption{\textbf{Radiation Dose Induced by Muons from the IP.} Prompt ambient dose equivalent rate in the transverse $(x, y)$ plane at the front, middle, and back of the FPF cavern, as indicated.  The dose is averaged over $\Delta z=\pm 50$~cm.}
\label{fig:muon_dose_rate_XY_1m}
\end{figure}

\subsection{Infrastructure, Technical Services, Transport, and Detector Integration} 

Integration studies have shown that the proposed experiments (in their current form) can be installed and fit into the baseline cavern, including their main associated infrastructure. 
Standard infrastructure and services that have been considered so far include cranes and handling infrastructure, electrical power, ventilation systems, fire/smoke safety, access, and evacuation systems. A very preliminary costing of these services (based on existing CERN standard solutions) is shown in \cref{tab:services}, giving an approximate total of less than 10~MCHF.\footnote{Most of these items were costed in mid-2021 and so may not reflect the costing today. The ventilation costing was updated based on a more thorough study in November 2022~\cite{EDMS2801032}.} The cryogenic infrastructure required for the FLArE detector and the FASER2 magnet is discussed and costed in section~\cref{sec:experiments}.

\begin{table}[tbhp]
  \centering
  \begin{tabular}{l|l|c} 
  \hline\hline
  {\bf Item} & {\bf Details} & {\bf Approximate cost} {\bf (MCHF)} \\
  \hline\hline
  Electrical Installation~\cite{EDMS2588617} & 2MVA electrical power  & 1.5 \\
  \hline
  Ventilation~\cite{EDMS2801032} & Fresh air  & 2.5 \\
  & Pressurisation & \\
  & Smoke extraction & \\
  & Ar extraction & \\
  \hline
  Access/Safety Systems & Access system & 2.5 \\
  & Oxygen Deficiency Hazard & \\ 
  & Fire Safety & \\
  & Evacuation & \\
  \hline
  Transport/Handling Infrastructure & Shaft crane (25 t) & 1.9 \\
  & Cavern crane (25 t)  & \\
  & Lift & \\
  \hline
  {\bf Total}  & & {\bf 8.4} \\
  \hline \hline
  \end{tabular}
\caption{\textbf{Integration Cost Estimate.} Breakdown of the main services and infrastructure for the FPF with very preliminary cost estimates.}
  \label{tab:services}
\end{table}


In the FPF baseline design, the cavern will house four proposed detectors: FLArE, FASER$\nu$2, FASER2, and FORMOSA. To maximise the physics reach, the detectors are centered on the LOS.
Integration studies have been carried out to demonstrate the detectors and needed large infrastructure can fit together in the cavern, and that the largest and most complicated pieces can be transported to their final location.
Additionally, more detailed equipment has been integrated into the global FPF model, for example including cable trays, electrical racks and safety and access equipment.

\clearpage 
\section{Experiments }
\label{sec:experiments}

To carry out the physics program highlighted in \cref{sec:physics}, four complementary experiments are proposed for the FPF. With locations from the most upstream to downstream these are:
\begin{itemize}
\item \textbf{FLArE}, a liquid argon TPC detector with 10~ton fiducial mass, optimised for neutrino physics and BSM searches with rare scattering signatures
\item \textbf{FASER$\nu$2}, a 20~ton tungsten/emulsion detector optimized for neutrino physics
\item \textbf{FASER2}, a large tracking spectrometer designed to search for decaying LLPs, and to precisely measure muons from neutrino interactions in the upstream experiments
\item \textbf{FORMOSA}, a scintillator based experiment with world leading sensitivity to milli-charged particles
\end{itemize}

\cref{fig:facility-layout} shows the location of the different experiments integrated into the FPF cavern. The different detectors have been optimized to be complementary and, where possible, for the information from each to be combined to maximise the physics output from the facility. The conceptual design, expected performance and current status (including possible R\&D) for each is detailed in the subsections below.

\subsection{FLARE }
\label{sec:FLArE}

FLArE is a modularised, liquid argon, time-projection chamber (LArTPC), designed as a multi-purpose detector for a wide range of energies. Arranged in a $3 \times 7$ module configuration, the active TPC volume spans $1.8 \times1.8 \times 7.0$ m$^3$, with an inner fiducial region of about $1.0 \times 1.0 \times 7.0$ m$^3$, corresponding to a fiducial (active) mass of approximately $10~\text{t}$ ($30~\text{t}$). 
FLArE is motivated by the requirements of neutrino detection~\cite{Anchordoqui:2021ghd} and light DM searches~\cite{Batell:2021blf} and builds on the considerable investment in liquid noble gas detectors over the last decade, including ICARUS~\cite{ICARUS-T600:2020ajz}, MicroBooNE~\cite{MicroBooNE:2016pwy}, SBND~\cite{SBND:2020scp}, and various components of DUNE~\cite{DUNE:2020txw,DUNE:2023nqi,DUNE:2024wvj}. In particular, the design of the FLArE detector has been informed by the design of the DUNE near detector~\cite{DUNE:2021tad} and the demonstrated performance of the ProtoDUNE detectors at CERN~\cite{DUNE:2021hwx}. Liquid argon as a detector medium allows one to precisely track and identify particles and measure track angle and kinetic energy from tens of MeV to many hundreds of GeV, thus covering both DM scattering and high-energy LHC neutrinos. As a fully active detector with both ionisation and scintillation capability, FLArE has a unique scientific reach that is complementary to other FPF detectors, such as the FASER$\nu$2 emulsion-based detector or the FASER2 magnetic spectrometer. While the other three FPF experiments are each designed to excel at a narrower range of measurements, FLArE offers access to a broader portfolio of physics topics within a single detector. In addition, FLArE is complemented downstream by a magnetised $40~\text{t}$ steel/scintillator hadronic calorimeter (HCAL), which can also act as a neutrino target.

\subsubsection{Scientific Portfolio and Requirements} 

FLArE is motivated by several physics opportunities within the FPF program, spanning from precision neutrino measurements to searches for new physics.
Alongside FASER$\nu$2, FLArE is one of the primary neutrino detectors in the FPF and it is expected to collect hundreds of thousands of charged-current (CC) neutrino interactions within its fiducial volume (\cref{tab:Phys_NuNumber}).
The detector is designed to reconstruct and identify all three neutrino flavours, measuring their kinematics with good energy containment and vertex resolution.
This will enable precise measurements of neutrino fluxes and cross sections in the multi-hundreds of GeV range. These measurements will be complementary to FASER$\nu$2 since they will be affected by different systematics, thus strengthening the FPF program.
Moreover, FLArE will have a wider pseudorapidity coverage than FASER$\nu$2 ($\eta>7.5$) and will measure the flux in an otherwise uncovered region of phase space where the fraction of neutrinos from charm decay is enhanced. This is of particular importance for informing low-$x$ QCD studies and constraining hadron production models relevant for astroparticle physics. 
The key requirement for neutrino detection in FLArE is the ability to suppress the large muon and muon-induced backgrounds coming from the ATLAS IP.
Operating a LArTPC in such a high-rate environment poses several challenges given the slow drift velocity of electrons in liquid argon ($1.6$ mm/$\mu$s @ 500 V/m). 
The current design mitigates them by segmenting the liquid argon volume in small, optically-independent modules with shorter drift length and relying on precise timing from the photo-detection system for triggering and event separation.
Assuming strict requirements on their charge readout, FLArE can also offer $\mathcal{O}(1\text{mm})$ spatial resolution and down to $\mathcal{O}(20~\text{mrad})$ angular resolution.
While this performance is poorer than that of an emulsion detector like FASER$\nu$2, it could be sufficient to statistically identify tau neutrinos through the short-lived tau decay, thus enabling detection of all three neutrino flavours in liquid argon. 
Since the decay length for charm is similar to that for taus, it may also be possible to statistically identify charm-production, which contributes roughly 15\% to the total charged current deep inelastic scattering, but this requires further study.
FLArE will also be complemented by a hadron calorimeter to enhance energy containment for the downstream region of the fiducial volume.
Although with coarser readout, the HCAL iron/steel absorbers can act as an additional neutrino target, and its transverse size further extends coverage towards lower psuedorapidity ($\eta>6.5$) thus adding to the uniqueness of FLArE in exploring the charm-enhanced region of the neutrino flux.
In addition, the HCAL measurements of neutrino interactions on iron targets will complement those on argon from FLArE itself and tungsten from FASER$\nu$2. This will help constrain initial-state nuclear effects, and other targets could be implemented as well to expand the study.
The abundant neutrino data will allow to probe for non-standard interactions (NSI). In particular, FLArE can cleanly identify neutral-current (NC) events, enabling precision tests of the weak mixing angle $\theta_W$ and neutrino charge-radii measurements that are more challenging for FASER$\nu$2~\cite{MammenAbraham:2023psg}.
FLArE will also contribute to the detection of neutrino tridents~\cite{Altmannshofer:2024hqd}, as these are rare events that would require the combined masses of FLArE, its HCAL, and FASER$\nu$2 to achieve sufficient statistics. 
Furthermore, the fully active nature of FLArE allows a lower threshold, down to approximately 30 MeV depending on the nature of the $\gamma$-ray background. The large kinematic range allows to see very small energy deposits. 
This capability will enable searches for light dark matter scattering on electrons or argon nuclei~\cite{Batell:2021blf, Batell:2021aja}, millicharged particles (mCPs) in synergy with FORMOSA, and neutrino magnetic moments~\cite{MammenAbraham:2023psg}. 
This is also essential for sensitivity to low-rate standard candle processes like inverse muon decay and neutrino-electron scattering for unbiased flux normalisation~\cite{Wilkinson:2023vvu}.

\subsubsection{Backgrounds and Space Charge}
The FPF cavern is shielded from the ATLAS IP by $250~\mathrm{m}$ of concrete and rock, however a significant background is still expected from the proton-proton collisions.
The most abundant component is high-energy muons which are produced directly at the IP; see \cref{sec:muon_fluxes}.
\texttt{FLUKA} simulations predict a flux of roughly 0.6~cm$^{-2}$~s$^{-1}$ of muons along the LOS, with hotter spots off-axis~\cite{PBCnote}.
The effects of the muon background on a liquid argon TPC are two-fold: first, the free charge ionised by muons lingers in the TPC, affecting the drift field and the charge transportation towards the anode plane (space charge effect); second, signal events can be confused by spurious tracks or be mimicked by secondary interactions of neutral particles produced by the crossing muons. Depending on the background rate, mitigations can be put in place for both of these effects.
In particular, the liquid argon volume needs to be segmented into multiple smaller TPC elements to keep the drift length small and reduce the space charge effects.
The 1D space charge model presented by Palestini in Ref.~\cite{Palestini:2020dhv} can be used to estimate the space charge effect in FLArE. In the case of a single drift volume with length 1.8 m, the resulting electric field distortion from the muon flux is about 25\% at the anode, increasing to 40\% at the cathode. However, with a modularised design and a 30 cm drift length, the distortion is reduced to $\sim1\%$ and can be safely handled. Details of this calculation can be found in Ref.~\cite{Linden:2927376}.

\subsubsection{Cryostat and Cryogenics}

\begin{figure}[htb]
  \centering
  \includegraphics[width=0.45\linewidth]{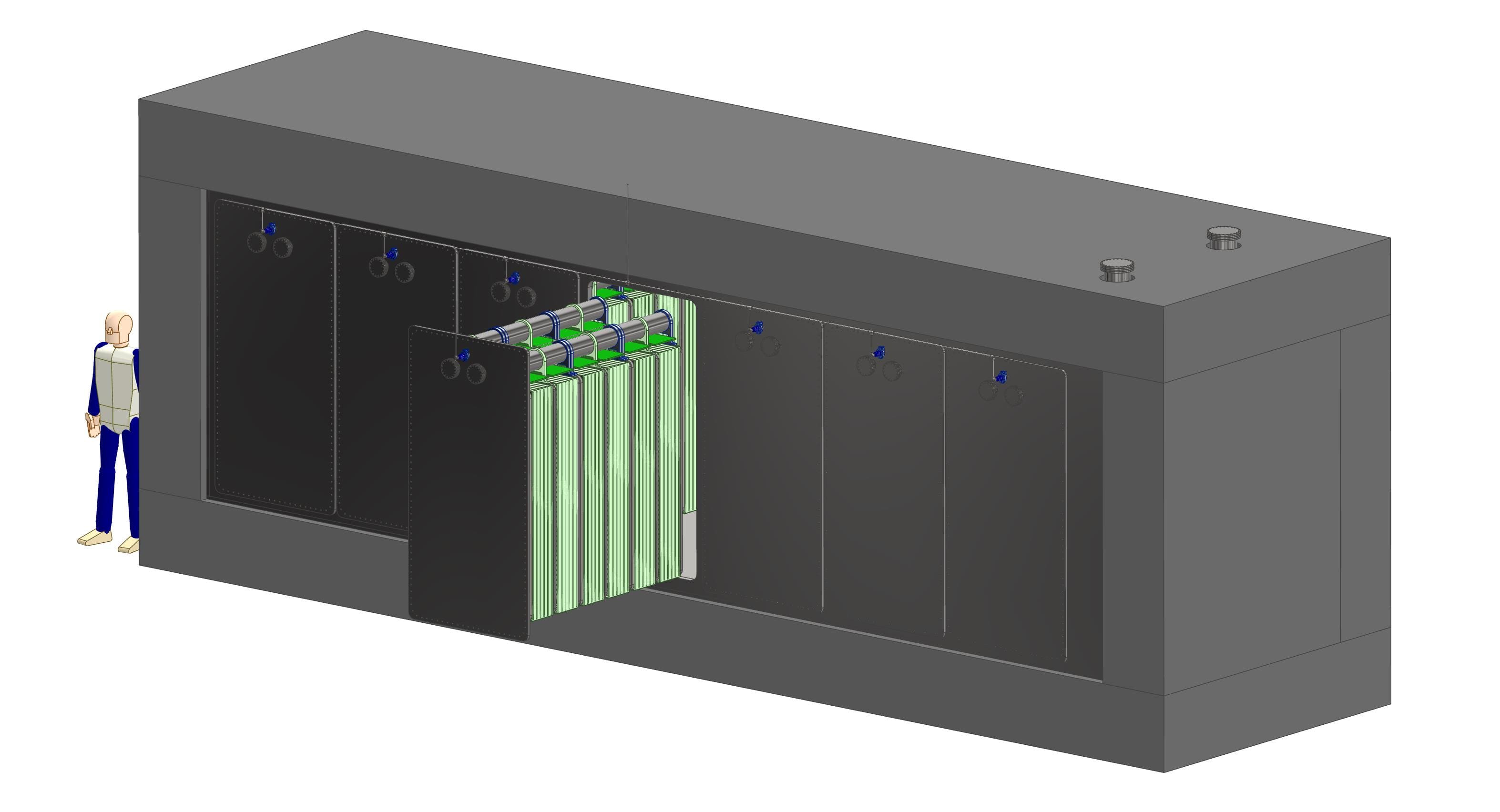}\,
  \includegraphics[width=0.45\linewidth]{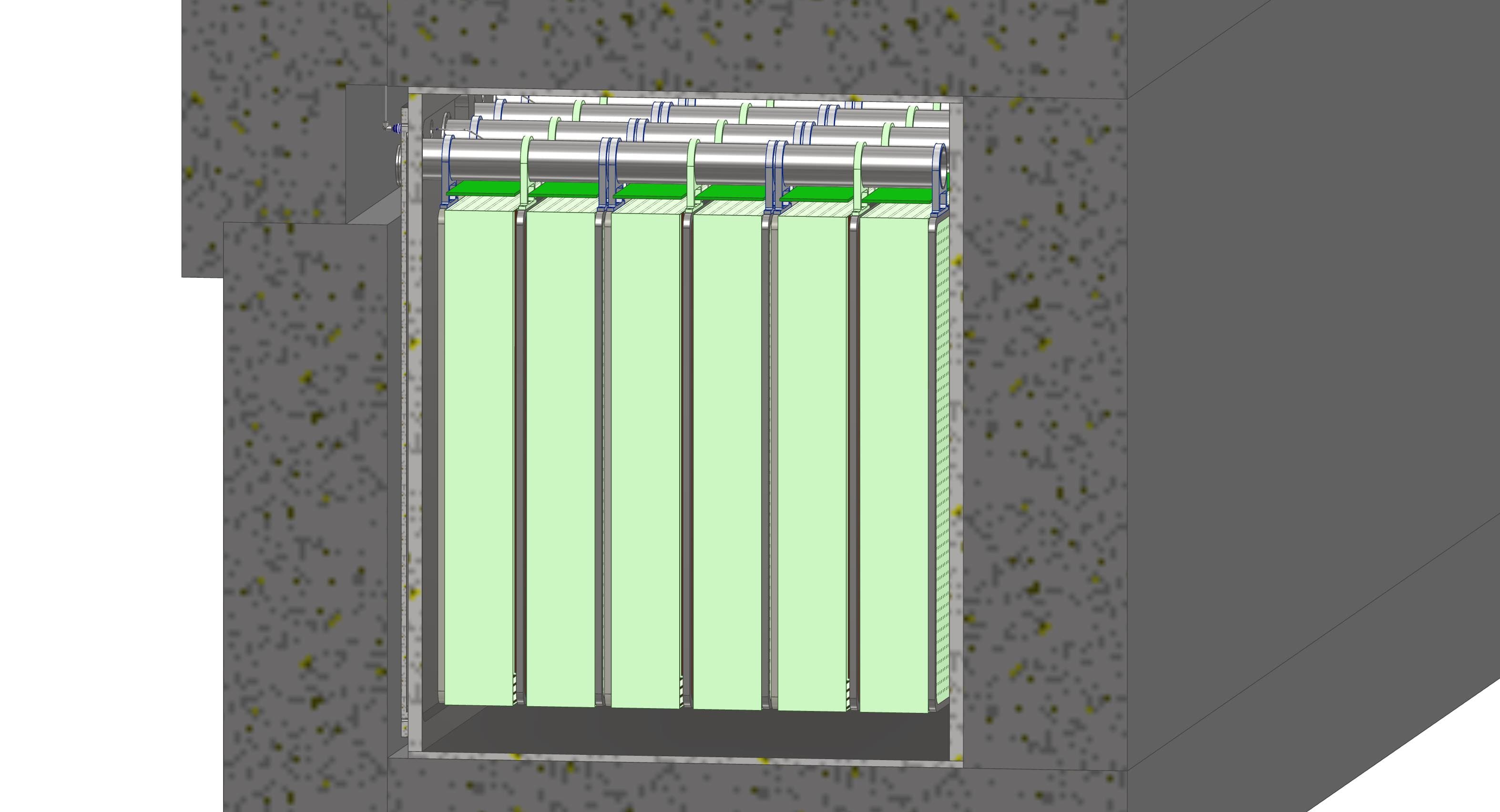}
  \caption{Left: The foam-insulated cryostat concept with side installation. The horizontal installation concept avoids difficulties with crane height presented by vertical installation. Right: Cross-section of the cryostat showing the foam insulation and the inner TPC modules.}
  \label{fig:foam_cryostat}
\end{figure}

The space available for the FLArE cryostat in the FPF hall is approximately 3.5 (W) x 3.5 (H) x 9.6 (L) m$^3$.  Options for the cryostat include a GTT membrane, a double-walled vacuum-insulated cryostat, or a single-walled foam-insulated cryostat. 
The current baseline design uses the single-walled, foam-insulated option and it is shown in \cref{fig:foam_cryostat}.
The dimensions and design of the cryostat and cryogenics must fulfill several requirements. First, the cryostat must be large enough to accommodate a TPC with the desired active argon volume.  The TPC must have an adequate active buffer mass to contain events in the transverse direction.  Simulation studies~\cite{Vicenzi:2927282} have determined the transverse TPC dimensions of $1.8~\text{m} \times 1.8~\text{m}$, providing a 40 cm buffer around the fiducial $1~\text{m} \times 1~\text{m}$ volume region.  The cryostat must allow additional space between the TPC field cage and the grounded cryostat wall to prevent high voltage breakdown.  This sets a minimum size for the cryostat.  On the other hand, it must be small enough in width to allow installation into the hall through the FPF shaft.  The designed size of the cryostat is a result of these considerations.

Another set of requirements on the cryostat derive from the needed temperature and purity of the liquid argon.  This means that the expected rate of heat transfer through the cryostat walls, combined with the heat load from electronics and other sources, must not exceed the cooling capacity of the cryocooler.  Cryostat materials and seals must be selected for ultra-high purity operation, and the cryogenic system must be capable of achieving and maintaining ultra-high purity for the liquid argon within one week of recirculation. Finally, it is desirable to have a cryostat design that allows for easy access to the TPC modules in the event that a module requires repair or replacement.

The single-walled, foam-insulated cryostat is expected to adapt the same technique used for the construction of the MicroBooNE experiment, which used a spray-on polyurethane insulation with acceptable thermal insulation quality demonstrated in Ref.~\cite{MicroBooNE:2016pwy}. This insulation technique has slightly higher heat load than the membrane insulation, but the more flexible thickness and application of the insulation materials are preferable, given the limited space and access in the FPF cavern. As an alternative option, the membrane cryostat could be manufactured by GTT with the membrane cryostat technique widely adopted in LArTPC experiments such as ProtoDUNE, under the exclusive contract between CERN and GTT. Given the significantly smaller size of the FLArE cryostat, a revised version without the corrugation on the inner wall of the cryostat may be pursued with the manufacturer to save space. 

The ultra-high purity requirements are crucial for the success of the experiment. Purities of less than 1.0 ppb electronegative concentrations, e.g., water and oxygen, and less than 1.0 ppm nitrogen concentration are required. A dedicated purification system with the standard purification scheme widely used in the LArTPC experiments, using a molecular sieve and copper catalyst, is expected to be implemented. Given that the purifier materials are fully saturated upon delivery, a regeneration system with a controlled heating system will be constructed to conduct the initial reactivation of the purifier materials and future regeneration when the scrubber materials are fully consumed by the absorption of impurities.

The cryogenic system plays a vital role by maintaining the LAr temperature for stable detector performance and ensuring efficient condensing and circulation for purity and operational stability. Excess heat is removed by a cooling mechanism that uses liquid nitrogen (LN2) as the medium with the main cooling power provided by a cryogenic refrigerator, absorbing any external heat and condensing gaseous argon back into a liquid state.
There are a number of constraints from CERN and the Facility that affect the cryogenic operation of the FLArE detector.  Foremost is that all operations regarding filling, recirculating, and emptying the cryostat must adhere to the oxygen deficiency standards at CERN.  These constraints have influenced our choices after discussions with CERN cryogenic experts.  The FLArE cryostat will be at a depth of 85 meters, and will be placed in a shallow pit designed to contain any leakage.  Furthermore, it must be filled from the surface through vacuum-insulated pipes, and it must be kept at a stable temperature by the use of LN2, which also must come down from the surface.  Given the difficulty of providing cryogenic liquids from the surface and extracting them from depth, it is necessary to have a secondary LAr storage in the FPF cavern. A local cryogenic refrigerator will also be provided in the cavern to condense and circulate LN2 to keep the LAr cold. The current choice calls for a Turbo-Brayton unit TB175 (made by Air Liquide Inc.) to be placed underground~\cite{turbo_brayton}. Secondary LAr and LN2 storage tanks have been sized for proper operation and placed on the opposite side of the FPF shaft, as shown in \cref{fig:facility-layout}. The Turbo-Brayton unit is expected to produce acoustic noise, and so it is vital to keep it separate from the detector hall.  The space created for this cryogenic equipment is expected to also be used for detector readout electronics racks for all of the FPF experiments.  Lastly, the Turbo-Brayton unit, as well as the storage tanks, have been sized so that they can be installed through the shaft before the shaft is fully outfitted with staircases and other equipment.  
The proposed Turbo-Brayton unit TB175 has a maximum cooling power of 16.5 kW at 77K with a cooling efficiency of 8.4\% and requires only cooling water supply and AC electrical power supply.  Cryogen supply is not required. 

The total capacity of the FLArE cryostat is approximately 40,000 liters with an instrumented volume of approximately 23,000 liters. The weight of the cryostat, not including foam insulation and seals, is 19,780 kg.  The heat load of the cryogenic system due to heat leaks is estimated to be approximately 2.7 kW with a breakdown of the heat budget shown in \cref{tab:heat_load}, excluding the electronics heat load. 
\begin{table}[htb]
  \centering
  \begin{tabular}{ l|c }
  \hline \hline
  Heat Source & Heat Load (watts)\\
  \hline
  Leak in cryogenic piping& 1000 \\
  Radiation from the top& 40\\
  Conduction through gas& 30\\
  Conduction from top flange through wall& 100\\
  Penetration components& 500 \\
  Conduction from liquid through wall& 832\\
  Convection & 200\\
  \hline
  Total & 2702\\
  \hline \hline
  \end{tabular}
  \caption{Heat load estimation of the FLArE cryostat and cryogenic system exluding the load from electronics.}
  \label{tab:heat_load}
\end{table}
The electronics system heat load is expected to be a significant source of heat for the system, depending on the readout electronics selection. For a configuration with 42 drift regions (cf. \cref{tpc_section}) with a charge readout arrangement of 360 by 200 channels, we have a total of about 3 million channels (see details in \cref{tpc_section}). With the standard front cold ASIC used in DUNE, the heat load is $<$ 50 mW per channel, which for 3 million channels results in an unacceptably high heat load of $<$ 150 kW.  However, if we consider a conventional two-dimensional $x$-$y$ coordinate readout scheme for this case, with channels reduced to $\approx$ 24k (42$\times$(360+200)), the heat load is expected to be 1.2 kW. With Q-Pix electronics \cite{Hoch:2024mce}, the per channel heat load can be reduced by about 50\%, to $<$ 25 mW per channel, bringing the total down to about 0.6 kW. An alternative and more realistic solution is to use the LArPix pixelated readout electronics \cite{Dwyer:2018phu} used in the DUNE ND-LAr detector, with $<$ 100 $\mu$W power per channel, for a total heat load of 0.3 kW with a fully pixelised readout (360$\times$200 channels per anode). The LArPix scheme is therefore considered our default choice, unless further R\&D shows that lower heat loads are possible with Q-Pix.  Assuming the heat load from the photon detection system with SiPM is less than the charge readout electronics heat load, a conservative estimate is to attribute the same heat load to the photon detection system.
To summarise, the total heat load of the cryogenics depends the selection of readout electronics and can range from 3.0 kW to 5.1 kW. This head load is well below the cooling capacity of the Turbo-Brayton Unit of 16.5 kW.

Due to safety regulations at CERN, in an emergency, there must be sufficient storage volume for the total amount of liquid argon in the FLArE cryostat to be transferred out into storage immediately. An industrial standard cryogenic storage tank with a 35000-liter volume is planned be installed at the other end of the FPF cavern to satisfy the safety requirement. This tank must be kept at LAr temperature so that if used to empty the detector cryostat there is no evaporative loss. A heat exchange coil in the storage tank with constant LAr circulation is implemented for this purpose.
\begin{figure}[htb]
 \centering
 \includegraphics[width=0.70\linewidth]{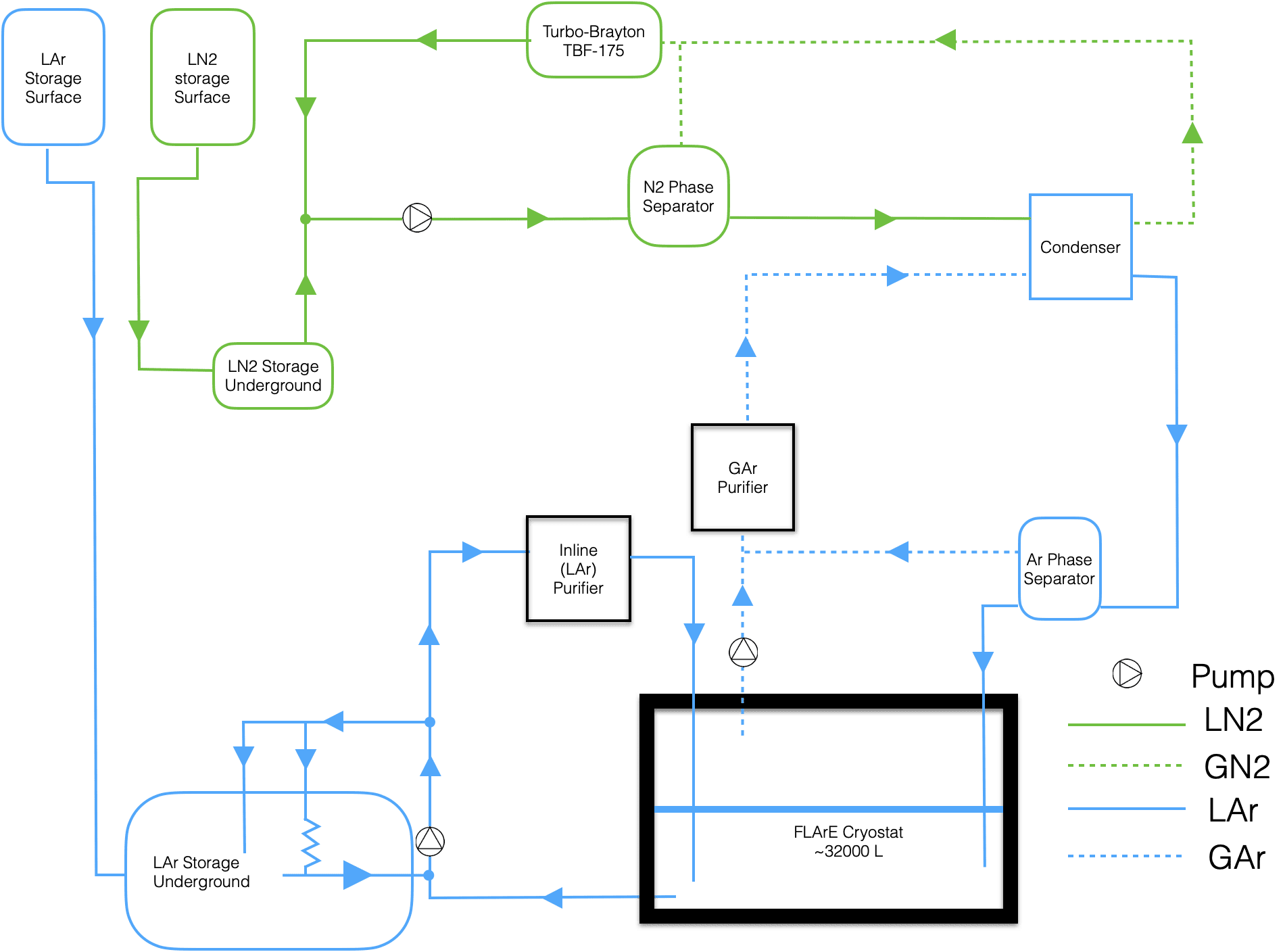}
 \caption{The cryogenic system schematic of FLArE Cryostat.} 
 \label{fig:flare_cryo}
\end{figure}
The schematic diagram of the cryogenic system is shown in \cref{fig:flare_cryo}. A gas purification scheme is designed with the capability of circulating liquid argon. The purity performance of such a design has been demonstrated in a local test stand at a smaller scale of hundreds of liters, as well as in DarkSide 20k, with a similar scale~\cite{Thorpe:2022dhw}. 

A summary of the cryostat and cryogenics for the current design is provided in \cref{tab:cryostat}.

\begin{table}[htb]
  \centering
  \begin{tabular}{ l|l|l }
  \hline   \hline
  Parameter & Value & Comment \\  
  \hline
  Total capacity & 40,000 l  &  Includes inactive region  \\ 
  Inner dimensions & 1930 mm $\times$ 2394 mm $\times$ 8788 mm  & Cryostat internal dimensions  \\ 
  Outer dimensions &  3645 mm $\times$ 3461 mm $\times$ 10,039 mm  &    \\ 
  Avg.  Insulation thickness & 600 mm  &  \\ 
  Material & Rigid polyurethane foam  &  \\ 
  Maximum Design Pressure & 34.5 kPa &  \\ 
  Weight of the Stainless Vessel & $\sim$ 20 tons & \\  
  Estimated heat load from electronics & $<$ 2400 watts & \\
  Estimated cryogenics heat load & 2702 watts & \\ 
  \hline  \hline
  \end{tabular} 
  \caption{Initial parameters for the FLArE cryostat and cryogenics. Maximum internal pressure is a preliminary evaluation.}\label{tab:cryostat}
\end{table}

\subsubsection{Detector and TPC Design} \label{tpc_section}

The reference design for the FLArE detector calls for a modular TPC, with 21 modules arranged in seven rows (looking from above, perpendicular to the beam) by three columns (along the beam).  Each module has nominal dimensions of 180 cm vertically by 60 cm horizontally by 100 cm in the beam direction. The drift direction is in the $x$-direction, that is, horizontal and perpendicular to the beam.  Each module is divided into two volumes by a central cathode, with an anode at either end.  The 21 TPC modules thus comprise 42 separate drift volumes, read out by 42 anodes.  The resulting maximum drift length for electrons is 30 cm.
The anodes of the TPCs will be pixelated.  Each anode will host an array of readout pads with a size of 5 mm (baseline design).  For a 180 by 100 cm anode, this means a 360 by 200 array, for a total of 72000 channels per anode.  With 42 anodes, this results in over 3 million readout channels.  The DUNE near detector design has roughly the same channel density; we will adapt the design for our purposes with considerations of specific changes. 

Figure~\ref{fig:oneTPC-man} shows the scale of a person next to a TPC assembly with three modules hung below two stainless steel tubes.  The tubes are cantilevered from the outer flange.  The mass of the module shown is approximately 1,110 kg. About half of that mass (540 kg) is in the outer flange, with the cantilever arms contributing roughly 90 kg and the TPC anodes, cathodes, and field cage contributing about 480 kg. The assembly is mechanically supported from three lifting points on the outer face of the flange: two near the bases of the cantilevered tubes and the other near the bottom of the flange.  Detailed lift points are not shown in the figure, but are to be designed.  \cref{fig:foam_cryostat} shows the horizontal installation of a TPC assembly in the foam-insulated cryostat.
\begin{figure}[htb]
  \centering
  \includegraphics[width=0.75\linewidth]{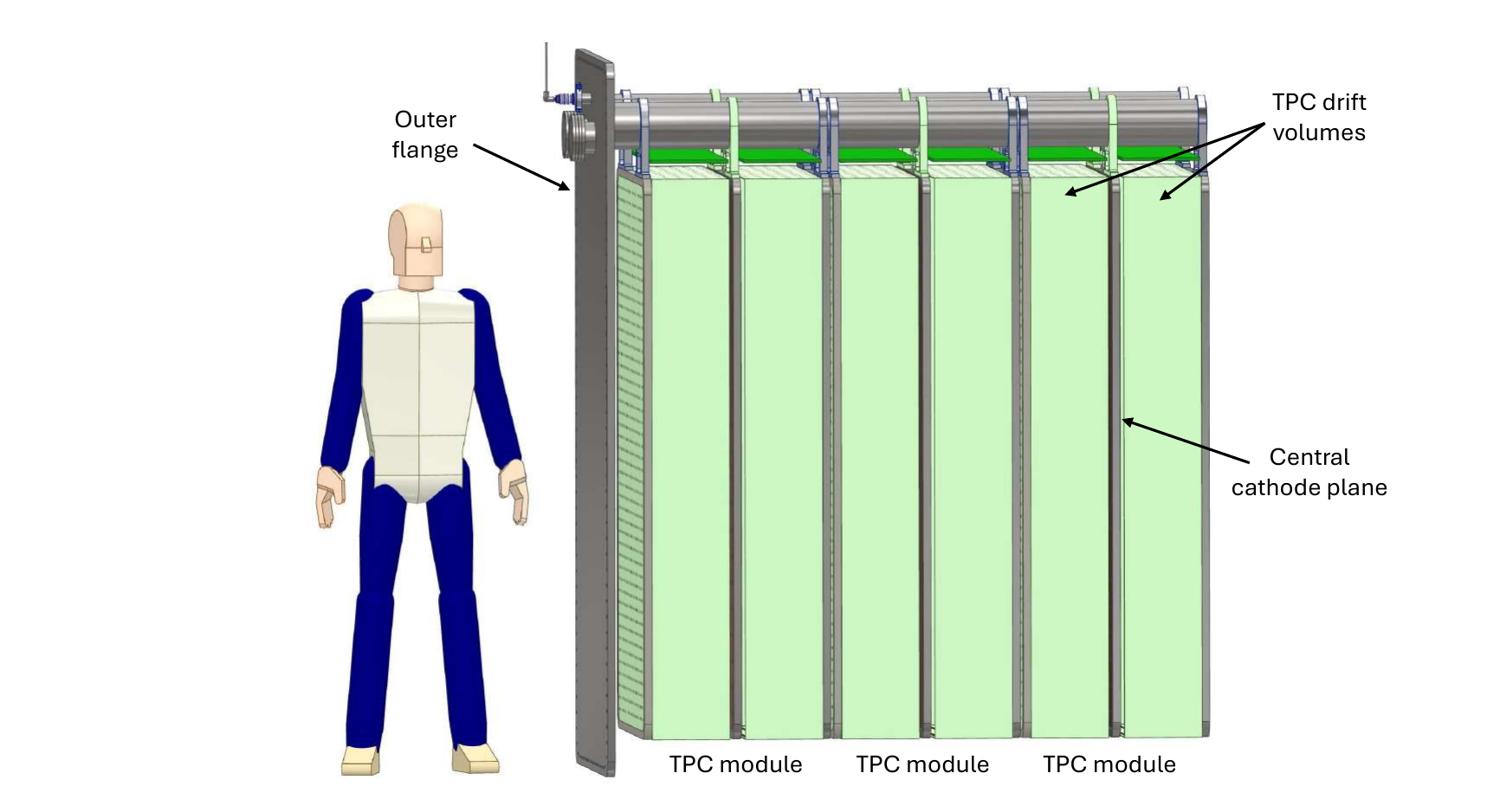}
  \caption{A single TPC assembly with 3 modules and a 6-foot person for scale. Each module is divided in two drift volumes by their central cathode.}
  \label{fig:oneTPC-man}
\end{figure}

\cref{fig:explodeddrift} illustrates how each TPC module and drift elements will be assembled onto the cantilever tubes.  Yet to be designed are cable routing paths for the anodes. Figure~\ref{fig:HVclose-up} shows a conceptual design for the high-voltage route to the cathodes from a high-voltage feedthrough near the top of the outer flange.  The cathodes are supported from the cantilevered tubes by insulating G-10 rings, while the anodes are supported by stainless steel rings.  The darker green plates shown in the figure represent the envelope of electronic board positions.  
\begin{figure}[htb]
  \centering
  \includegraphics[width=0.75\linewidth]{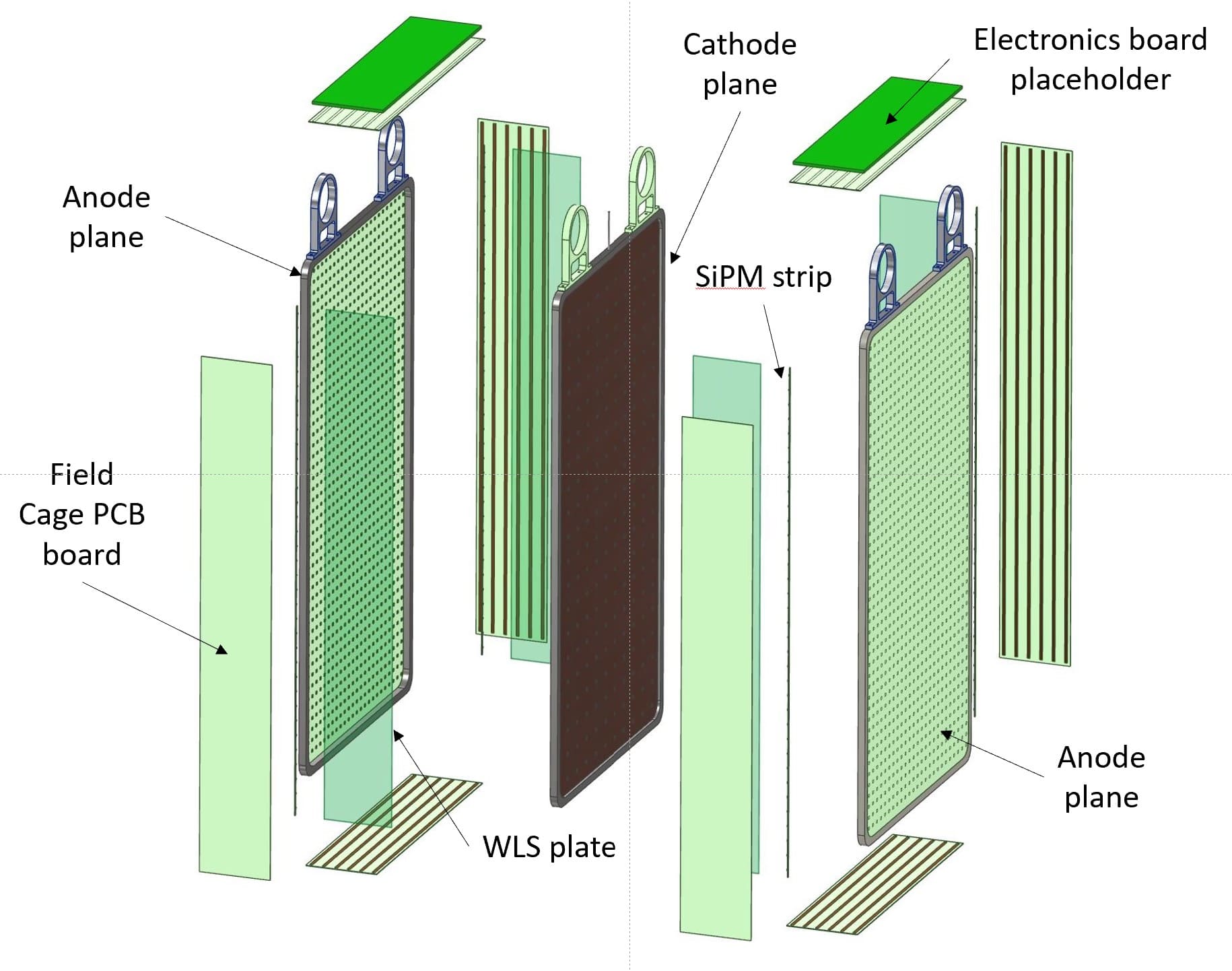}
  \caption{Exploded view of a single TPC module with one cathode plane, two anode planes, and optical detector planes}
  \label{fig:explodeddrift}
\end{figure}
\begin{figure}[htb]
    \centering
    \includegraphics[width=0.75\linewidth]{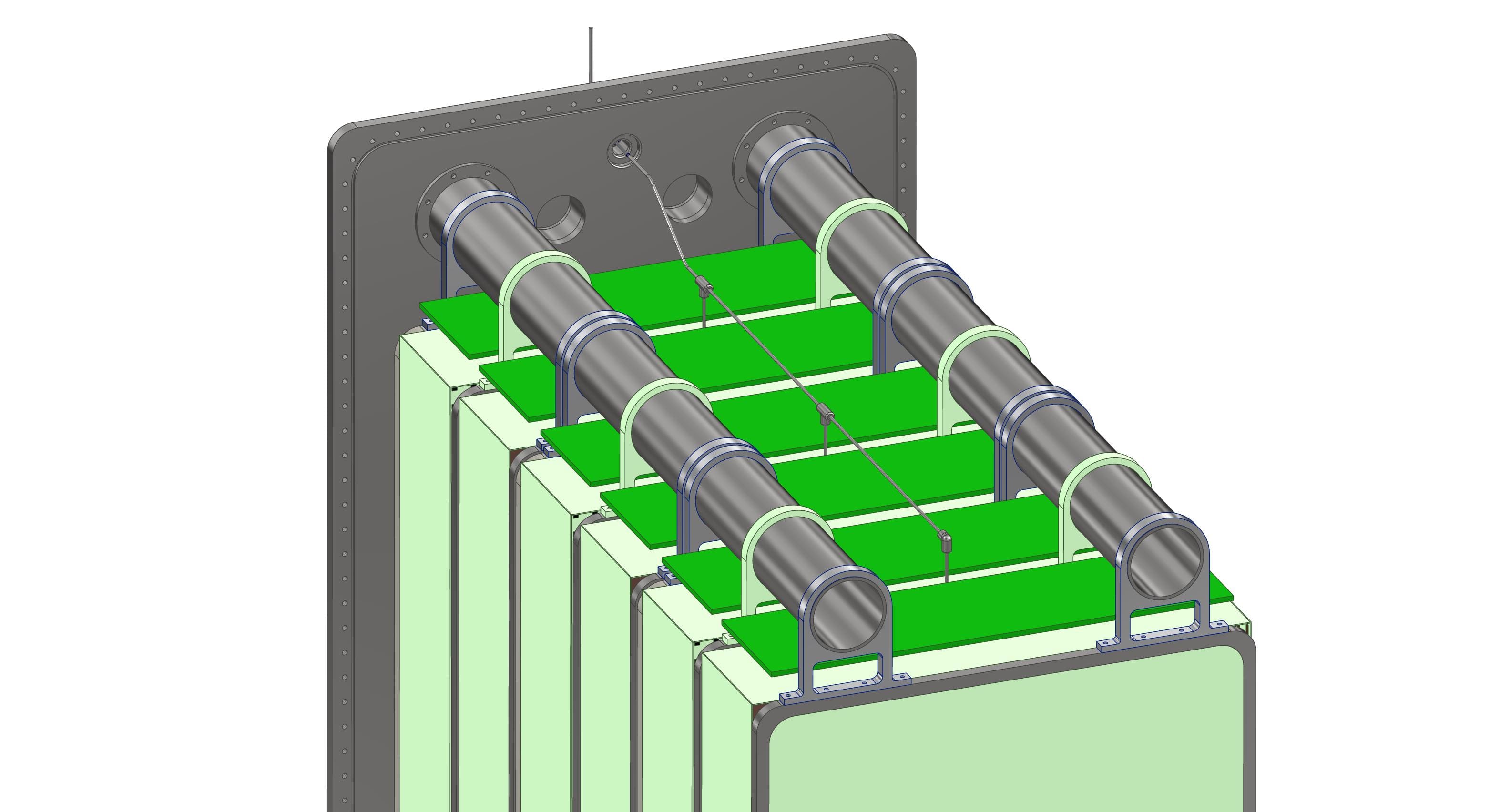}
    \caption{Close-up of the top of a TPC assembly showing a possible high voltage routing scheme for the cathodes.}
    \label{fig:HVclose-up}
\end{figure}

Figures~\ref{fig:explodeddrift} and  \ref{fig:SiPM-WLSclose-up} illustrate an implementation of the photon counter, consisting of a wavelength shifting (WLS) plate with SiPM readout on the anode side. Liquid argon emits $128~\text{nm}$ scintillation light, so a wavelength shifter is coated on the plate to convert it to the visible range ($430-450~\text{nm}$). Once converted, light is trapped inside by internal reflection until collected by the SiPMs. This plate would need to be placed inside the field cage and subjected to the electric field. Such a scheme has been used in the DUNE near detector design; in the case of FLArE we anticipate a few changes, which are detailed in Ref.~\cite{Linden:2927376}.  
\begin{figure}
  \centering
  \includegraphics[width=0.5\linewidth]{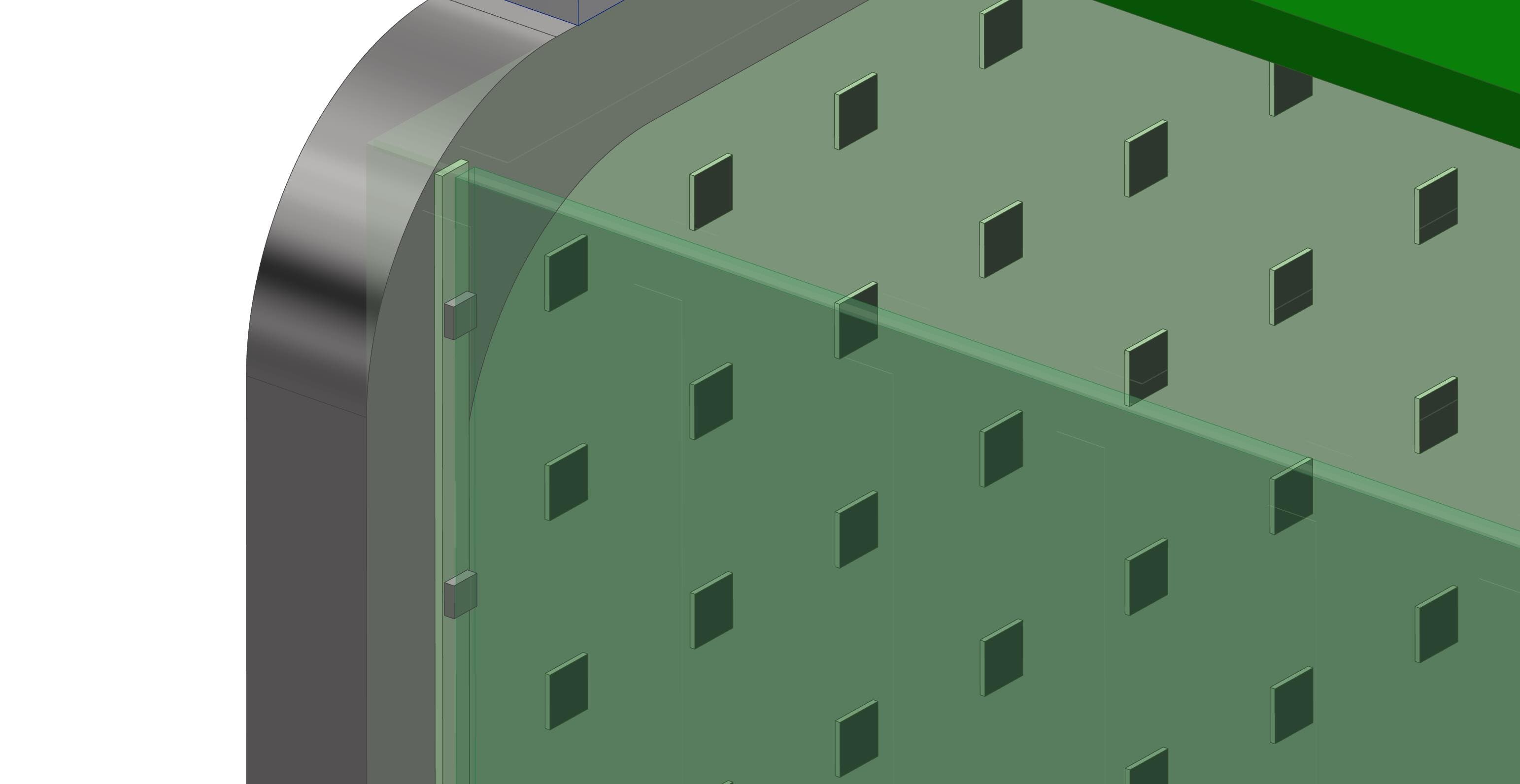}
  \caption{Closeup of the SiPMs that read out the WLS plates. The WLS coating is envisioned to p-terphenyl (pTP) or tetraphenyl-butadiene (TPB). Once converted, light is trapped inside by internal reflection until collected by the SiPMs. The outside and top field cage plates have been made transparent for visualisation. }
  \label{fig:SiPM-WLSclose-up}
\end{figure}

Readout cables will be routed through the two 6-inch conflat flanges on either side of the high-voltage feedthrough, and through the gap between the cryostat and the foam insulation.  A penetration for the cables through the foam is not shown in the figures.

A summary of the preliminary parameters for the FLArE TPC and its readout is given in \cref{tab:tpc}.

\begin{table}[htb!]
  \centering
  \begin{tabular}{ l|l|l}
  \hline  \hline
  Parameter & Value & Comment \\  
  \hline
  TPC liquid fill & LAr & radiation length 14 cm \\ 
  Modules & 3 (W) $\times$ 7 (L)   &  21 modules  \\ 
  Module dimension & 60 cm (W) $\times$ 100 cm (L) $\times$ 180 cm (H)     & approximate  \\ 
  Gap length  &  30 cm  &  Cathode in center  \\ 
  Drift field  & 500 V/cm   &  \\ 
  Max voltage & 15000 V &  \\ 
  Anode pixel size & 4 mm x 4 mm     & 5 mm spacing \\
  Charge channels/anode & 72000 &  two anodes per mod \\ 
  Gap between TPC modules & 19 cm & \\
  Photon system & WLS plate with TPB &   \\ 
  SiPM channels/anode  &  50   &    \\ 
  \hline  \hline
  \end{tabular} 
  \caption{Preliminary parameters for the FLArE TPC. }
  \label{tab:tpc} 
\end{table}

\subsubsection{Hadron Calorimeter}
A calorimeter located downstream of the FLArE TPC is desirable from the point of view of energy containment to measure the energy of particles and showers that escape the TPC.  The proposed design for this calorimeter is based on Baby MIND (Magnetised Iron Neutrino Detector), a design developed by the T2K collaboration for deployment at J-PARC~\cite{BabyMIND:2017mys}.  

As the name suggests, Baby MIND consists of plates of magnetised iron interspersed with scintillator planes.  In the case of the detector at J-PARC, there are 33 magnetised modules and 18 scintillator planes, but these numbers will be optimised for use of this design as a FLArE calorimeter.  The current assumption is a calorimeter with $3.5~\text{m} \times 2~\text{m}$ transverse size and a mass of 40~tons, but the exact dimensions, and thus also the momentum and energy resolution, are still be to optimised.  The magnet modules are individually magnetised by conducting coils wound on their surface.  Each consists of ARMCO steel sheets with two horizontal slits machined in the center, and wrapped by an aluminum coil.  They are designed to provide a minimum field of 1.5 T over the central tracking region, with a minimum 10\% field uniformity. No cooling system is required.
The power consumption is approximately $200$ W per module with an operating current of $120-140$ A~\cite{Rolando:2017rpx}. 
A 3D model of the calorimeter, with a person for scale, is shown in \cref{fig:babymind}.

\begin{figure}[htb]
    \centering
    \includegraphics[width=0.70\linewidth]{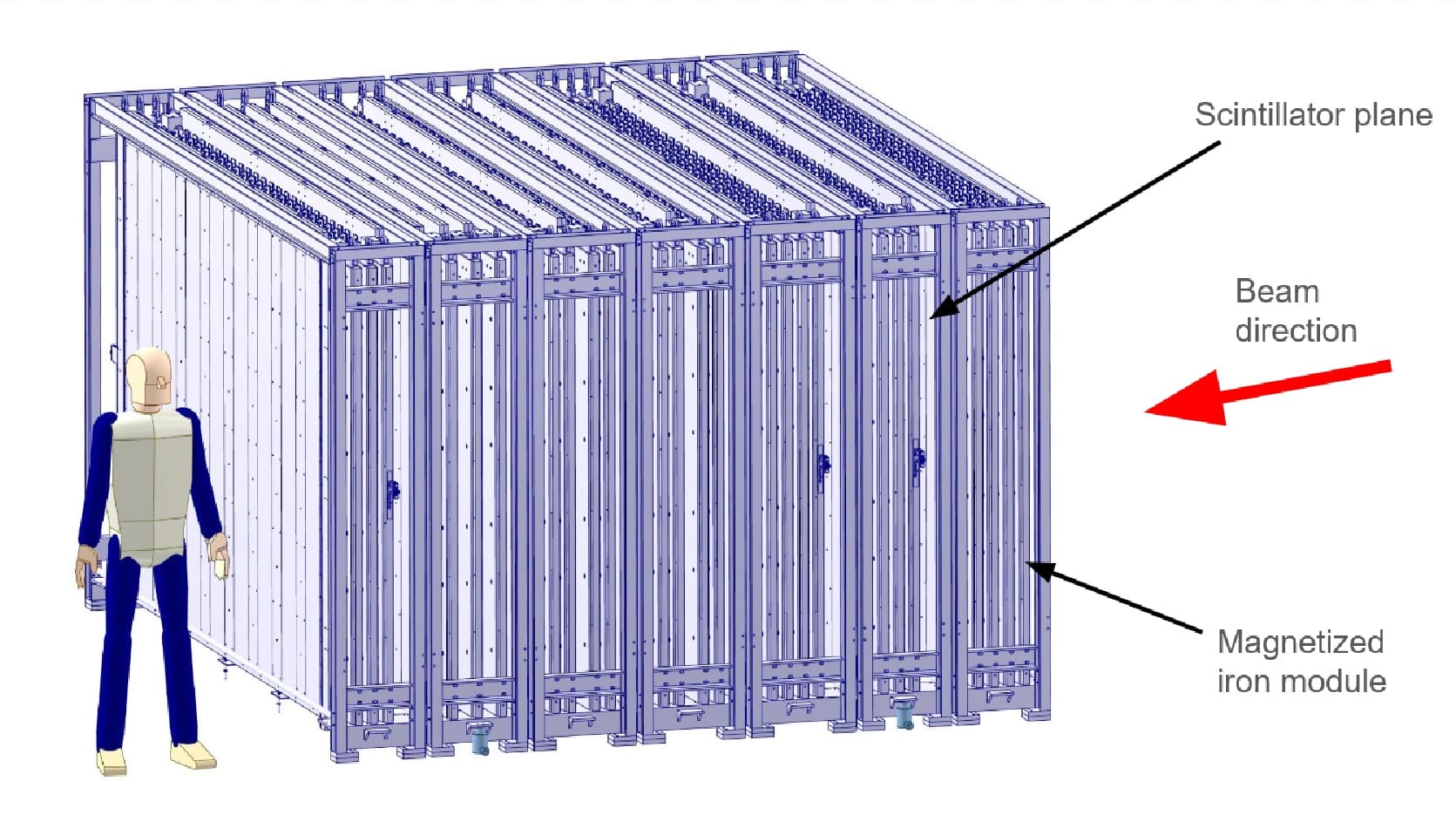}
    \caption{The BabyMIND concept for the FLArE downstream calorimeter, which consists of plates of magnetised iron interspersed with scintillator planes.}
    \label{fig:babymind}
\end{figure}

Each scintillator module consists of 95 horizontal and 16 vertical scintillator bars read out by SiPMs via a wavelength shifting fiber.  The horizontal bars have dimensions of 0.7 $\times$ 3 $\times$ 288 cm, while the vertical bars have dimensions of 0.7 $\times$ 21 $\times$ 195 cm.  The scintillator consists of polystyrene doped with 1.5\% paraterphenyl (PTP) and 0.01\%  1,4-bis(5-phenyloxazol-2-yl) benzene (POPOP).  These are read out with 1~mm diameter WLS Kuraray Y11 fibers covering the scintillator in a U-shape.
As mentioned above, the hadron calorimeter can also be used as a physics target, providing an additional interaction medium.  The details of its energy resolution, spatial resolution, and particle ID depend on the specific configuration and dimensions of the calorimeter, which have yet to be optimised for physics performance.

\subsubsection{Alternative Design: 3D Optical TPC Readout}

An alternative TPC readout design is based on a 3D optical TPC similar to that developed within the ARIADNE programme. The ARIADNE approach utilises the 1.6~ns timing resolution and native 3D raw data of a Timepix3 camera to image the wavelength-shifted secondary scintillation light generated by a novel glass THGEM (THick Gaseous Electron Multiplier) within the gas phase of the dual-phase LArTPC~\cite{Roberts:2018sww, Lowe:2020wiq}. 

The main detection principle sees incoming particles ionising LAr and creating prompt scintillation light (known as S1). The ionisation electrons then drift towards an extraction grid situated below the liquid level, where they are transferred to the gas phase and subsequently amplified using a THGEM. The drift charge multiplication produces secondary scintillation light (S2), which is wavelength-shifted before imaging with a Timepix3 camera. With no need for thousands of internal charge TPC readout channels, pre-amps, etc., reduction in construction costs is one of a number of advantages of the ARIADNE technology. Furthermore, since this is a monolithic TPC, there would be further associated cost reductions. It is also relevant to note that the Timepix3 cameras and associated image intensifiers and optics are commercially available, allowing for the immediate deployment to a large optical TPC.  

Within the Timepix3 camera, each pixel operates independently. When a pixel detects a light signal above a predefined threshold, the pixel readout process is triggered. Following the triggering of a pixel, several quantities are measured. First, the pixel's $(x,y)$ coordinates are recorded, allowing for 2D reconstruction, akin to a traditional photograph. The time at which the pixel was triggered, known as the Time of Arrival (ToA), is timestamped with 1.6~ns resolution. Given the drift velocity of free electrons in LAr of approximately 1.6~mm per microsecond (for a 0.5~kV/cm field), the 1.6~ns ToA timestamp resolution allows for complete 3D track reconstruction. Finally, through use of an integrated pre-amplifier within each pixel, the Timepix3 camera measures the total integrated light signal detected at the pixel. This integrated signal, known as Time over Threshold (ToT), is recorded with 10-bit resolution and enables calorimetry within each pixel. These four pieces of information ($x$, $y$, ToA, ToT) are continuously streaming from the Timepix3 camera for any detected ``hits'' that produce signals above the predefined threshold. This approach enables a ``data-driven'' readout with native zero suppression and highly efficient readout for the expected sparse data. 

 The cameras, which can be mounted far from the TPC, even outside of the cryostat, are readily accessible, allowing for easy maintenance and future upgrades throughout the life of an experiment. The separation of the cameras also provides a total decoupling from the TPC, eliminating any parasitic noise that may couple between the TPC and the readout electronics. Through the selection and optimisation of the lenses used, it is possible to read out a large area or to obtain very high resolution by increasing the number of cameras (each camera has 256 $\times$ 256 pixels). Furthermore, the next generation Timepix4 camera will have a larger sensor providing even higher resolution, an improved ToT dynamic range, and 200~ps timing resolution~\cite{Llopart:2022hyz}.
 
\begin{figure}[htbp]
    \centering
    \includegraphics[width=0.9\linewidth]{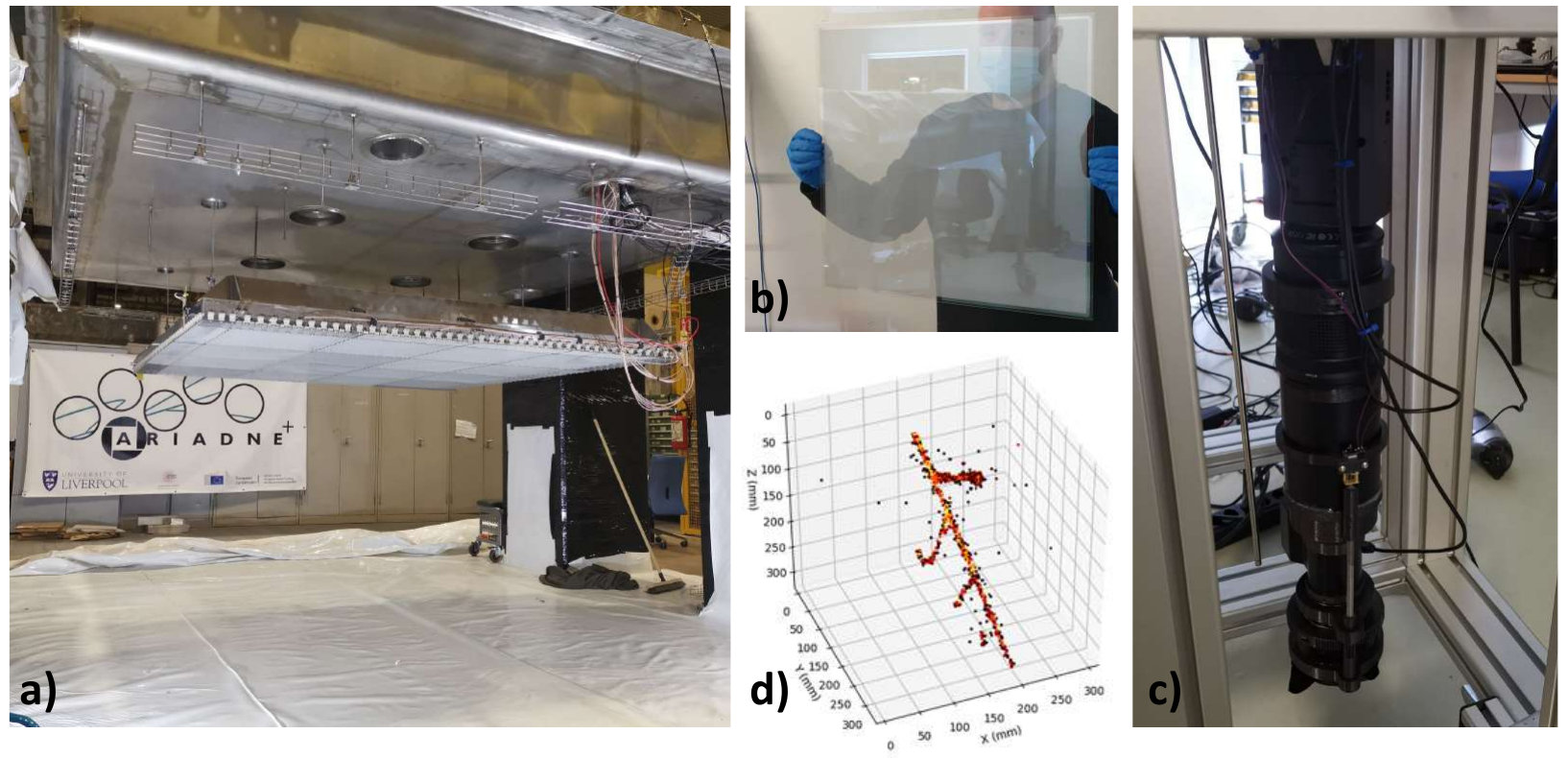}
    \caption{(a) The 2~m~$\times$~2~m glass THGEM LRP during commissioning of the ARIADNE$^{+}$ experiment at CERN. (b) An individual 50~cm~$\times$~50~cm glass THGEM. (c) A Timepix3 camera with associated image intensifier and lenses. (d) A LAr cosmic event. }
    \label{fig:ARIADNE}
\end{figure}

 The new glass THGEMs developed in the ARIADNE programme, when compared to previous FR4 PCB-based designs, offer better mechanical stability, higher purity and robustness against discharges~\cite{Lowe:2021nkz}. There has been a successful large-scale demonstration of the Timepix3 and glass THGEM technologies within the ARIADNE$^{+}$ experiment at the CERN Neutrino Platform~\cite{Lowe:2023pfk}. Four cameras were used to image a 2~m~$\times$~2~m glass THGEM light readout plane (LRP) and cosmic muon data were collected (see \cref{fig:ARIADNE}). As detailed below, similar size LRPs are being considered for the FLArE TPC.

A conceptual model of a FLArE TPC with 1.8~m electron drift and instrumented with 56 TimePix3 cameras is shown in \cref{fig:TPCTPX3}. This design configuration will provide a 1.8 mm per pixel resolution. The cameras and optics will be installed externally at cryostat view-ports, where one camera views one THGEM. In this design there are two 3.3~m~$\times$~2~m LRPs, each containing 28, 45~cm~$\times$~45~cm glass THGEMs.

\begin{figure}[htbp]
    \centering
    \includegraphics[width=0.8\linewidth]{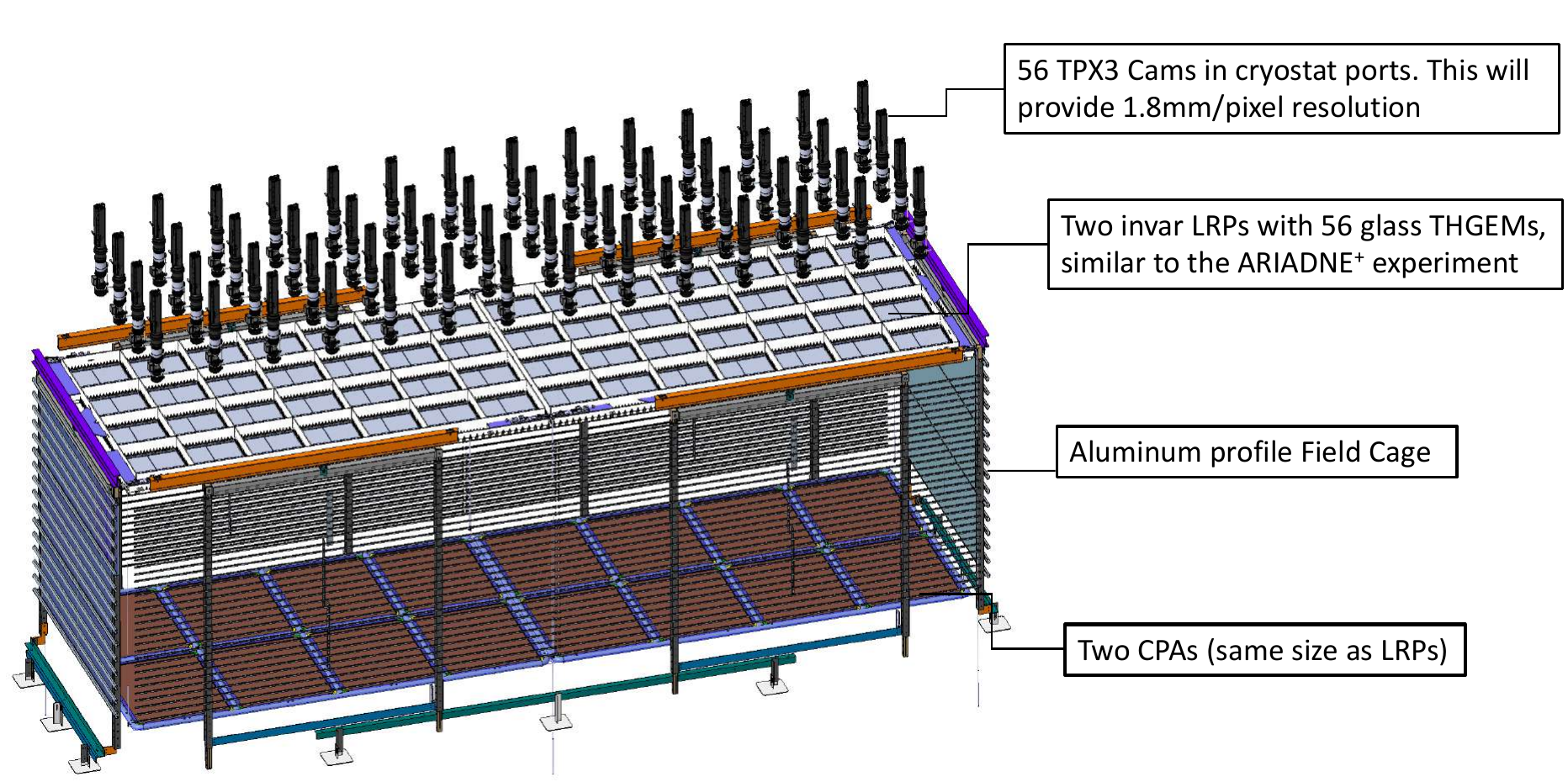}
    \caption{A conceptual model of a 3D optical dual-phase TPC option for FLArE.}
    \label{fig:TPCTPX3}
\end{figure}

This optical readout design offers some advantages over the baseline modular design.  It is potentially cheaper, using off-the-shelf cameras, and the technology has been successfully demonstrated.  On the other hand, a challenge with this approach is presented by the longer drift dimension (1.8~m, as opposed to 30 cm for the horizontal drift).  For a given electric field, this increases the diffusion of electrons as well as the space charge effect limiting spatial resolution. A higher electric field can partially mitigate these effects, and an offline correction may need to be applied for any non-uniform electric field regions. A detailed simulation taking account of the true event rate, as well as the position of the interactions, is needed to understand this properly.

\subsubsection{Cryostat Alternative for 3D Optical TPC}

The current baseline design uses a single-walled, foam-insulated cryostat.  Other options for the cryostat include a double-walled, vacuum-insulated vessel, or a GTT membrane cryostat.

An alternative design of a cryostat to accommodate the dual-phase camera readout TPC option is shown in \cref{fig:cryostat_optical}. This cryostat is vacuum-jacketed and can be manufactured with standard commercial techniques. The cryostat lid will be re-openable, and the entire TPC will hang under the lid. Hydraulic arms are envisaged to lift the cryostat lid, mitigating the need for a high-clearance crane. Preliminary studies show that the cryostat without the lid can be lowered down the shaft at a 60 degrees angle, similar to the Turbo-Brayton cryo-cooler unit. The benefit of this cryostat is that it avoids any cold seals, utilises standard cryostat manufacturing techniques, and provides flexibility and easy TPC installation/access in the case of the optical TPC. There is an ongoing FEA analysis to optimise the design considering various operating pressure scenarios.

\begin{figure}[h]
    \centering
    \includegraphics[width=0.8\linewidth]{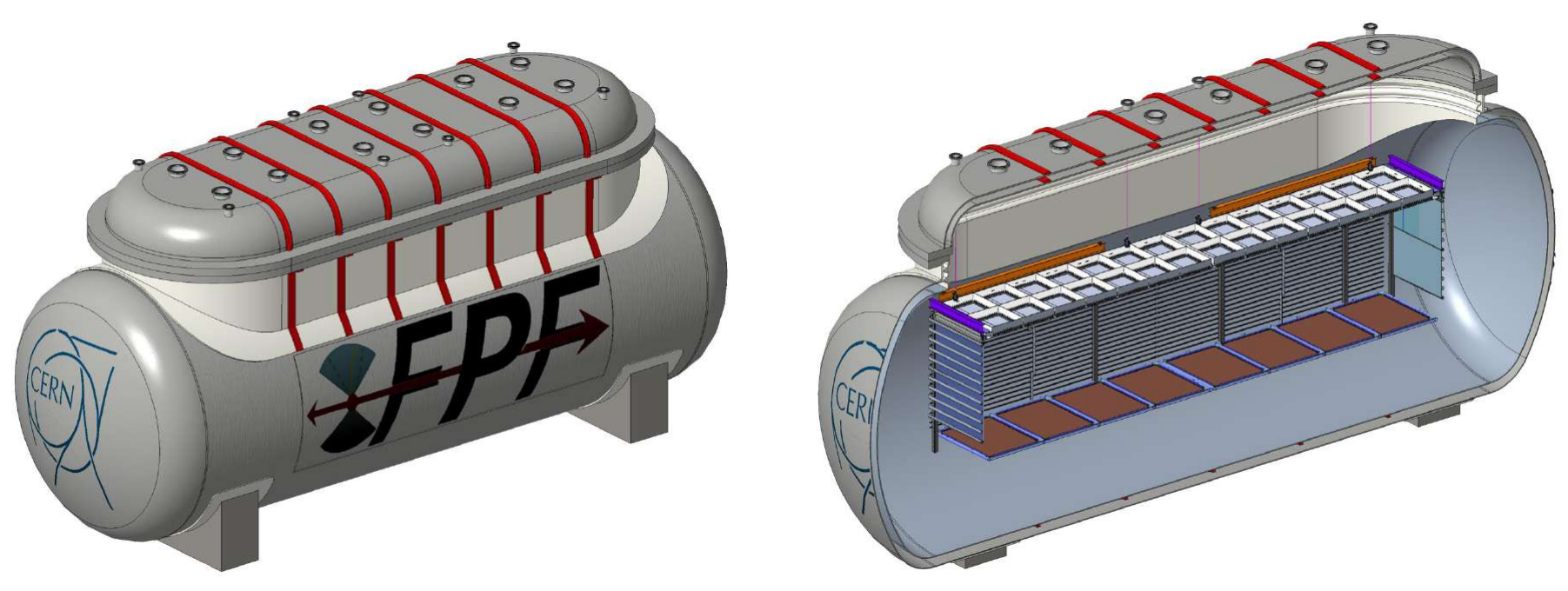}
\caption{A conceptual model of a vacuum-jacketed commercial cryostat with a re-openable lid for the dual-phase fast optical readout option. The cryostat lid will open with hydraulic arms instead of a crane, since there is not enough height clearance.  The cryostat is 10 m in length and 4.2 m in height to the top of the lid.}
    \label{fig:cryostat_optical}
\end{figure}

The vacuum-jacketed cryostat will have significantly lower heat losses than a foam-insulated design.  However, it should be noted that the vacuum-insulated cryostat would require some additional infrastructure for the vacuum system.

\subsection{FASER$\nu$2 }

\subsubsection{Design of the Experiment}

FASER$\nu$2 is a 20-ton neutrino detector located on the LOS, a much larger successor to the FASER$\nu$~\cite{FASER:2019dxq} detector in the FASER experiment. With the FASER$\nu$ detector, the first evidence for neutrino interaction candidates produced at the LHC was reported in 2021~\cite{FASER:2021mtu}, and the first measurements of the $\nu_e$ and $\nu_\mu$ interaction cross sections at TeV energies were reported in 2024~\cite{FASER:2024hoe}. These results confirm the FASER$\nu$ emulsion detector's ability to deliver physics measurements in the LHC environment. 

An emulsion-based detector will enable the localisation of neutrino interaction vertices, identification of muons and electrons, the measurement of charged particle momenta by the multiple Coulomb scattering method, and the energy measurement of EM showers~\cite{FASER:2024hoe}. It will also identify heavy flavour particles produced in neutrino interactions, including tau leptons and charm and beauty particles. 
FASER$\nu$2 can perform precision tau neutrino measurements and heavy flavour physics studies, testing lepton universality in neutrino scattering and new physics effects, as well as providing important input to QCD and astroparticle physics, as described in \cref{sec:physics}. 

In addition to neutrino measurements, FASER$\nu$2 can measure muon DIS and the associated charm production, providing important input to QCD, as described in \cref{sec:physics}.

Figure~\ref{fig:FASERnu2_schematic1} shows a schematic of the proposed FASER$\nu$2 detector, which is composed of 3300 emulsion layers interleaved with 2 mm-thick tungsten plates. The total volume of the tungsten target is 25~cm $\times$ 64~cm $\times$ 6.6~m, with a mass of 20~tons. The emulsion detectors will be placed in two cooling boxes and kept at around $10^\circ$C to avoid the fading of the recorded signal. 
Although the baseline target material is tungsten, part of the detector could be equipped with a different material, such as iron, to study the nuclear target dependence of interactions. This would, for example, help to constrain nuclear PDFs and to test shadowing effects.

\begin{figure}[htb]
  \centering
  \includegraphics[width=\linewidth]{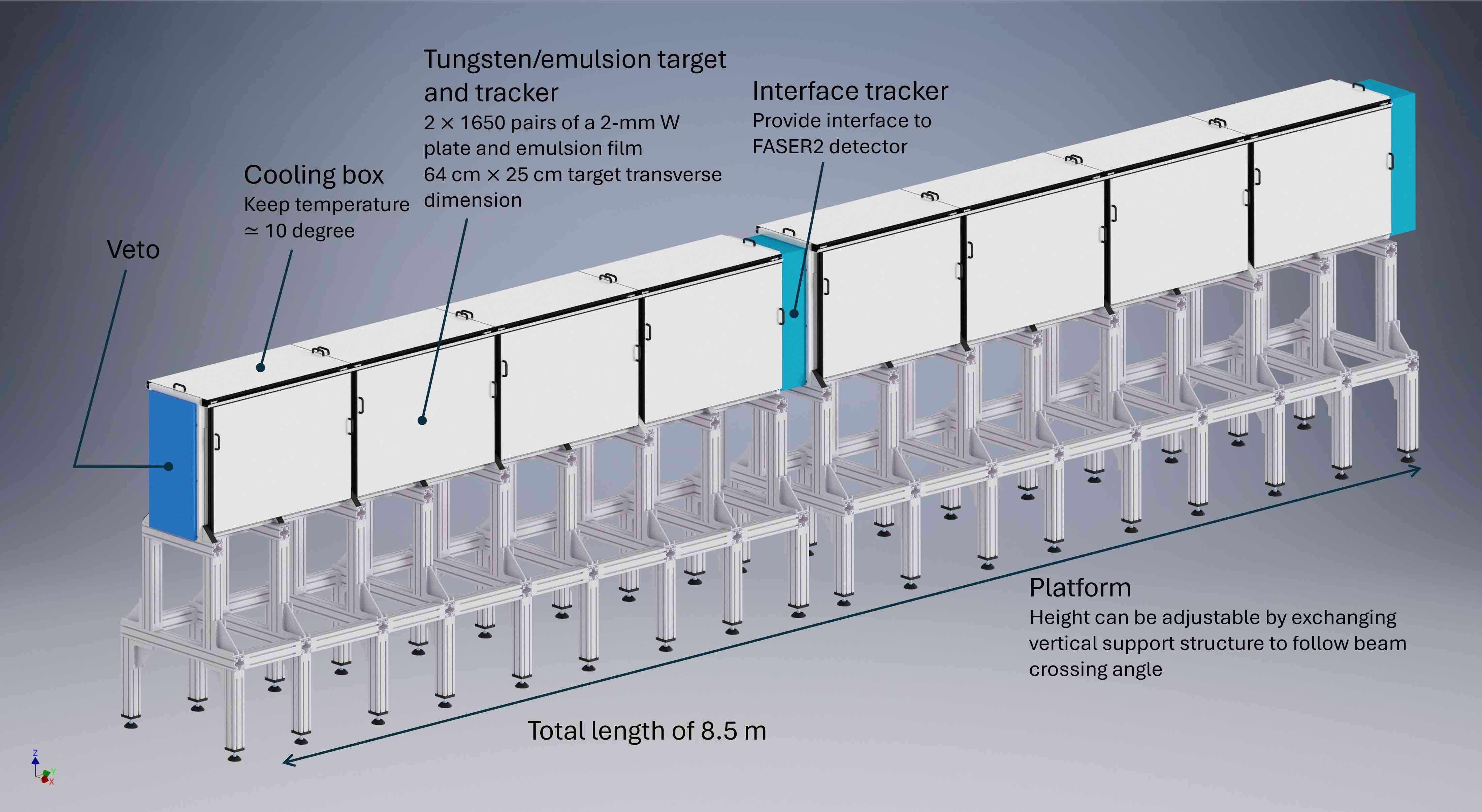}
  \caption{Schematic of FASER$\nu$2.}
  \label{fig:FASERnu2_schematic1}
\end{figure}

The detector will be placed directly in front of the FASER2 spectrometer along the LOS. The FASER$\nu$2 detector will also include a veto system and interface detectors to the FASER2 spectrometer, with one interface detector in the middle of the emulsion modules and the other detector downstream of the emulsion modules. These additional systems will enable a FASER2-FASER$\nu$2 global analysis and make measurement of the muon charge possible, a prerequisite for $\nu_\tau$/$\bar\nu_\tau$ and also $\nu_\mu$/$\bar\nu_\mu$ separation. The veto system will be scintillator-based, and the interface detectors could be based on silicon strip sensors or scintillating fiber tracker technology. The detector length, including the emulsion films and interface detectors, will be approximately 8.5 m. 

\subsubsection{Detector Assembly and Emulsion Development}

The assembly of each of the 1.1-ton FASER$\nu$ detectors in LHC Run 3 takes about two weeks at the surface emulsion facility, after which they are transported to the LHC tunnel. As shown in \cref{tab:detector_specifications}, to scale up to the 20-ton FASER$\nu$2 detector, both the assembly and transportation will become significantly more challenging. An updated detector assembly scheme is therefore essential to enable the FASER$\nu$2 data-taking campaigns.

\begin{table}[htbp]
  \centering
  \begin{tabular}{l|c|c|c|c|c}
  \hline\hline
  Name & Mass & Film size & \# of films & Exchanges/year & Assembling site\\
  \hline
  FASER$\nu$ & 1.1 tons & 25 cm $\times$ 30 cm & 730 & 3 & Surface lab\\
  \hline
  FASER$\nu$2 & 20 tons & 25 cm $\times$ 64 cm & 3300 & 1 & FPF cavern\\
  \hline \hline
  \end{tabular}
  \caption{Detector specifications.}
  \label{tab:detector_specifications}
\end{table}

The proposed strategy for FASER$\nu$2 is as follows:
\begin{itemize}
\setlength\itemsep{-0.05in}
\item A single film exchange per year.
\item Assembly in the FPF cavern under non-darkroom conditions.
\item A new mechanical architecture to apply pressure to ensure good alignment.
\end{itemize}

The emulsion film exchange will be performed in the FPF cavern under ambient light. Therefore, each emulsion film will be individually sealed in a light- and humidity-tight bag. As illustrated in \cref{fig:fasernu2_schematic2}, T-shirt-shaped tungsten plates are suspended from a rail mounted at the top of the cooling box. Emulsion films are inserted between the tungsten plates, and the entire stack is compressed by air actuators using compressed air. A pressure of one atmosphere will be applied to maintain sub-micron alignment between the films.

\begin{figure}[htb]
  \centering
  \includegraphics[height=5.1cm]{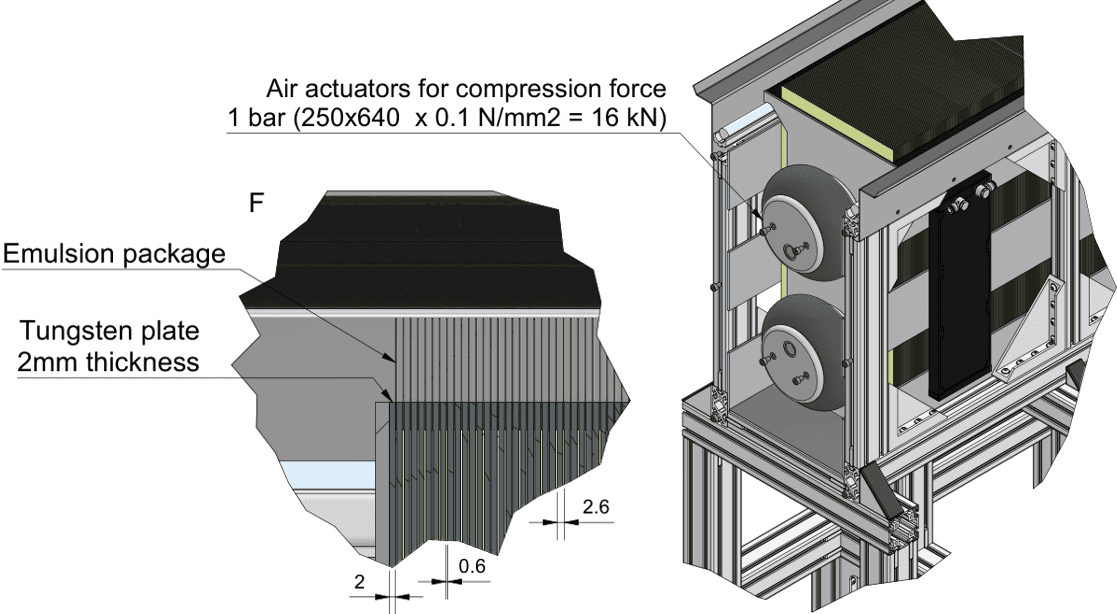}
  \includegraphics[height=5.1cm, trim=8cm 0 4cm 0, clip]{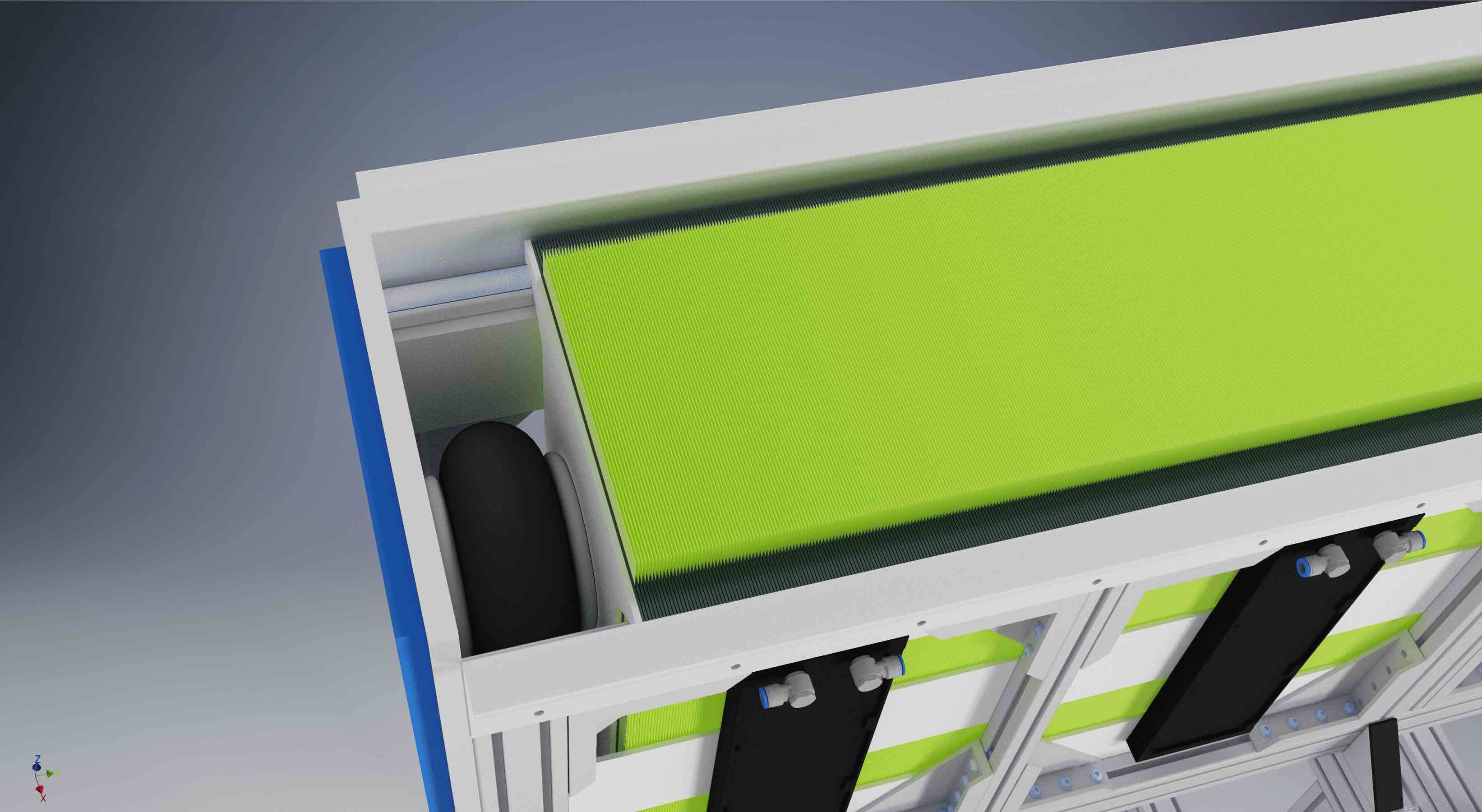}
  \caption{Inner design of the FASER$\nu$2 detector.}
  \label{fig:fasernu2_schematic2}
\end{figure}

A mechanical prototype has been produced to test critical technical challenges, namely applying pressure to fix sub-micrometer alignment and assembly under room light in the FPF experimental hall. As shown in \cref{fig:fasernu2_prototype}, a test beam experiment was performed in July 2024 at the SPS-H8 beamline, confirming the concept of the techniques. 

\begin{figure}[htb]
  \centering
  \includegraphics[height=5.5cm]{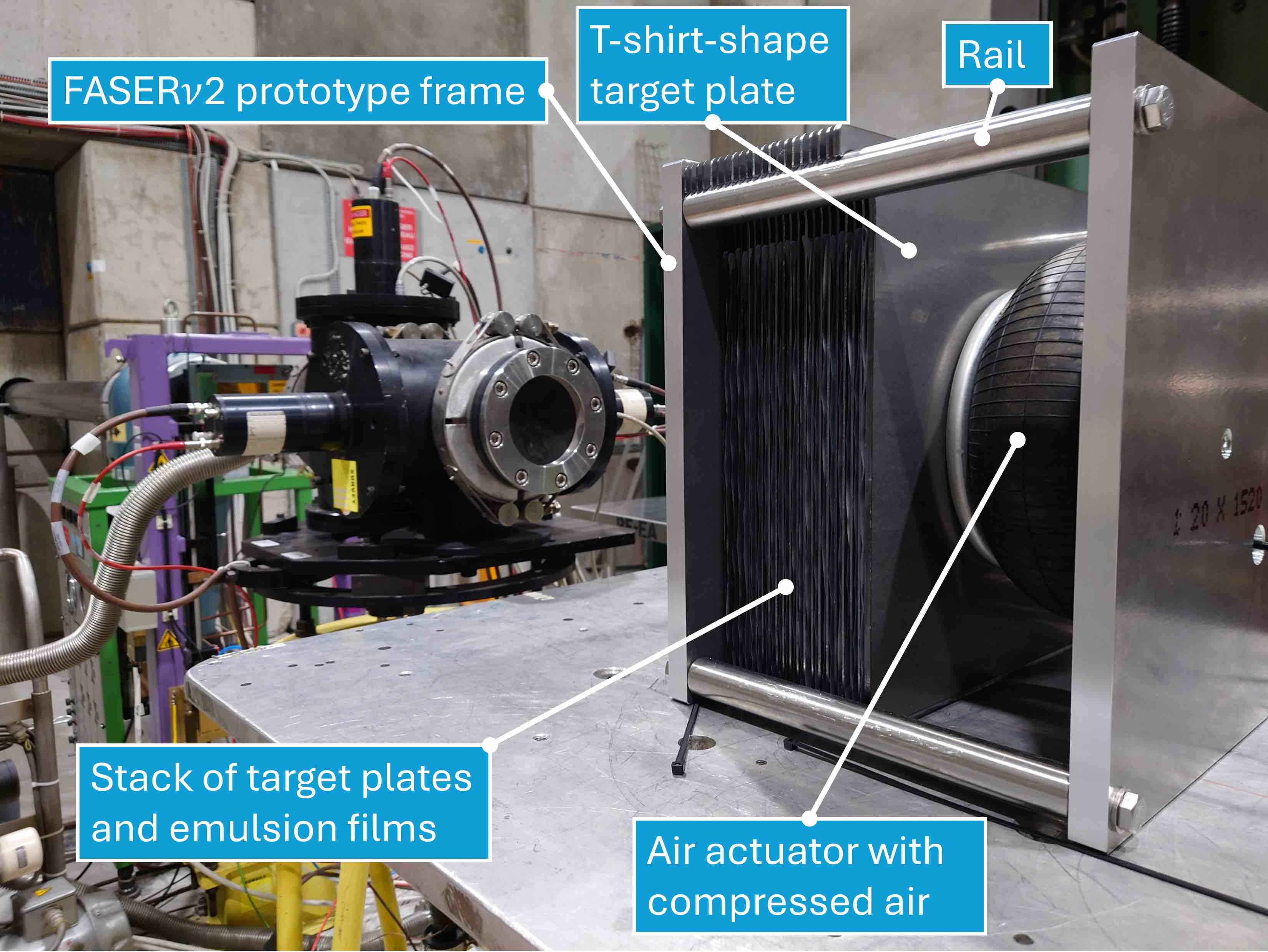}
  \includegraphics[height=5.5cm]{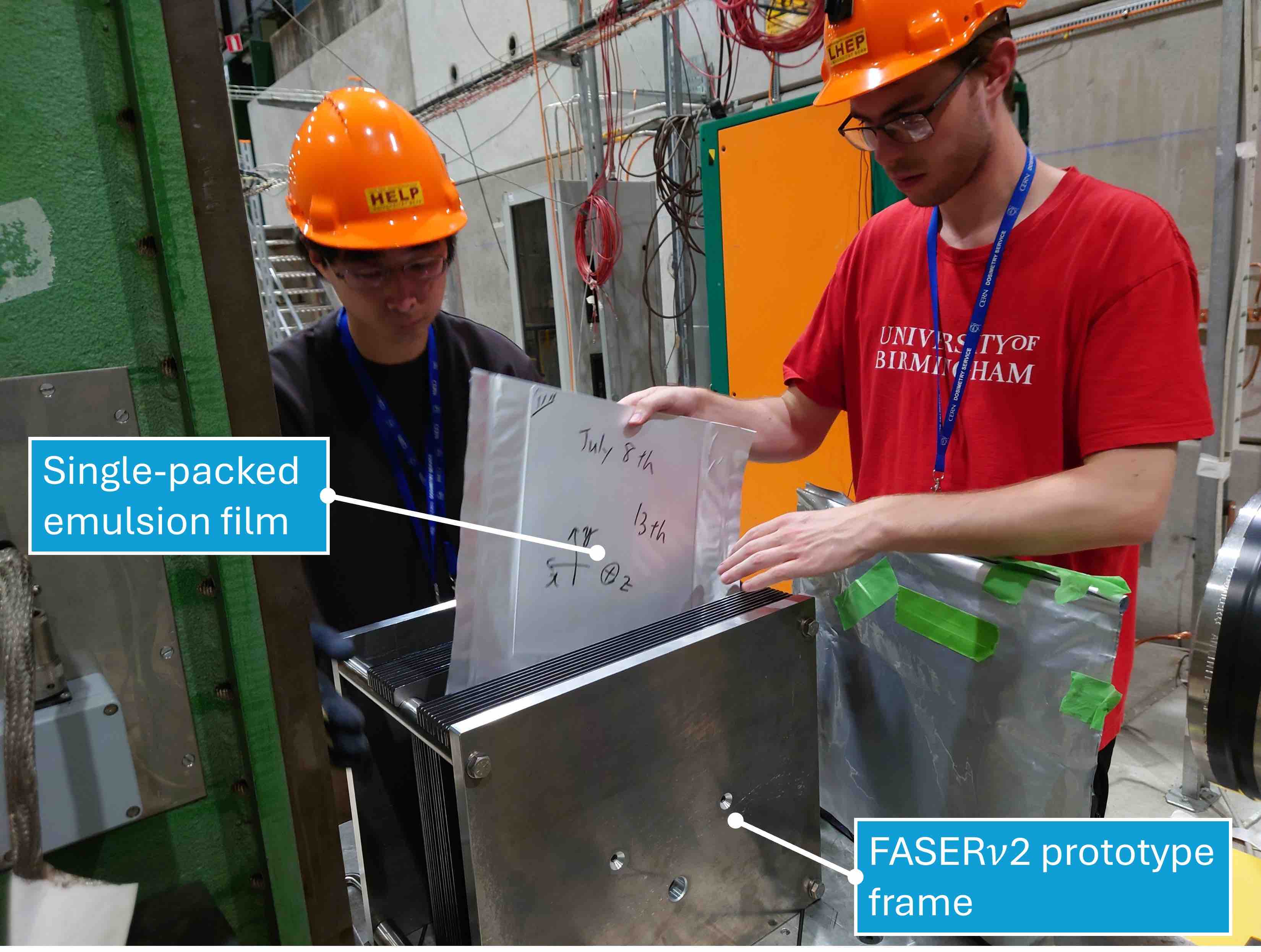}
  \caption{Left: FASER$\nu$2 prototype module on the SPS-H8 beamline. Right: Assembly of the FASER$\nu$2 prototype module.}
  \label{fig:fasernu2_prototype}
\end{figure}

In addition to testing the mechanical prototype, other test samples were produced and exposed to the beam. These samples allow to evaluate the long-term performance of emulsion films intended to operate for a full year without replacement. They also enable testing of a new type of photo-development solution that enhances chemical amplification gain, potentially contributing to faster readout of the emulsion detector. The analysis of these samples is currently ongoing.

For the emulsion development after exposure to the beam, our current plan is to use the same darkroom facility as FASER$\nu$, but with updated tools. It currently takes about 1.5 weeks for a FASER$\nu$ module to be developed, so we expect it will take about 6 weeks for a FASER$\nu$2 module. This time may be shortened with the updated tools.

\subsubsection{Detection of Short-Lived Particle Decays}

As the tau lepton has a decay length of $c\tau \simeq 87~\mu\text{m}$, a high-precision emulsion detector~\cite{Ariga:2020lbq} is essential to detect tau decays topologically. After optimisations of the detector performance in terms of precision, sensitivity, and long-term stability, emulsion gel with silver bromide crystals with a diameter of 200~nm will be used, which provides an intrinsic position resolution of 50~nm. 
The alignment accuracy and angular resolution are proven in FASER$\nu$ as shown in \cref{fig:deltaX} and \cref{fig:dtx}.

\begin{figure}[htb]
\centering
\includegraphics[width=0.3\linewidth]{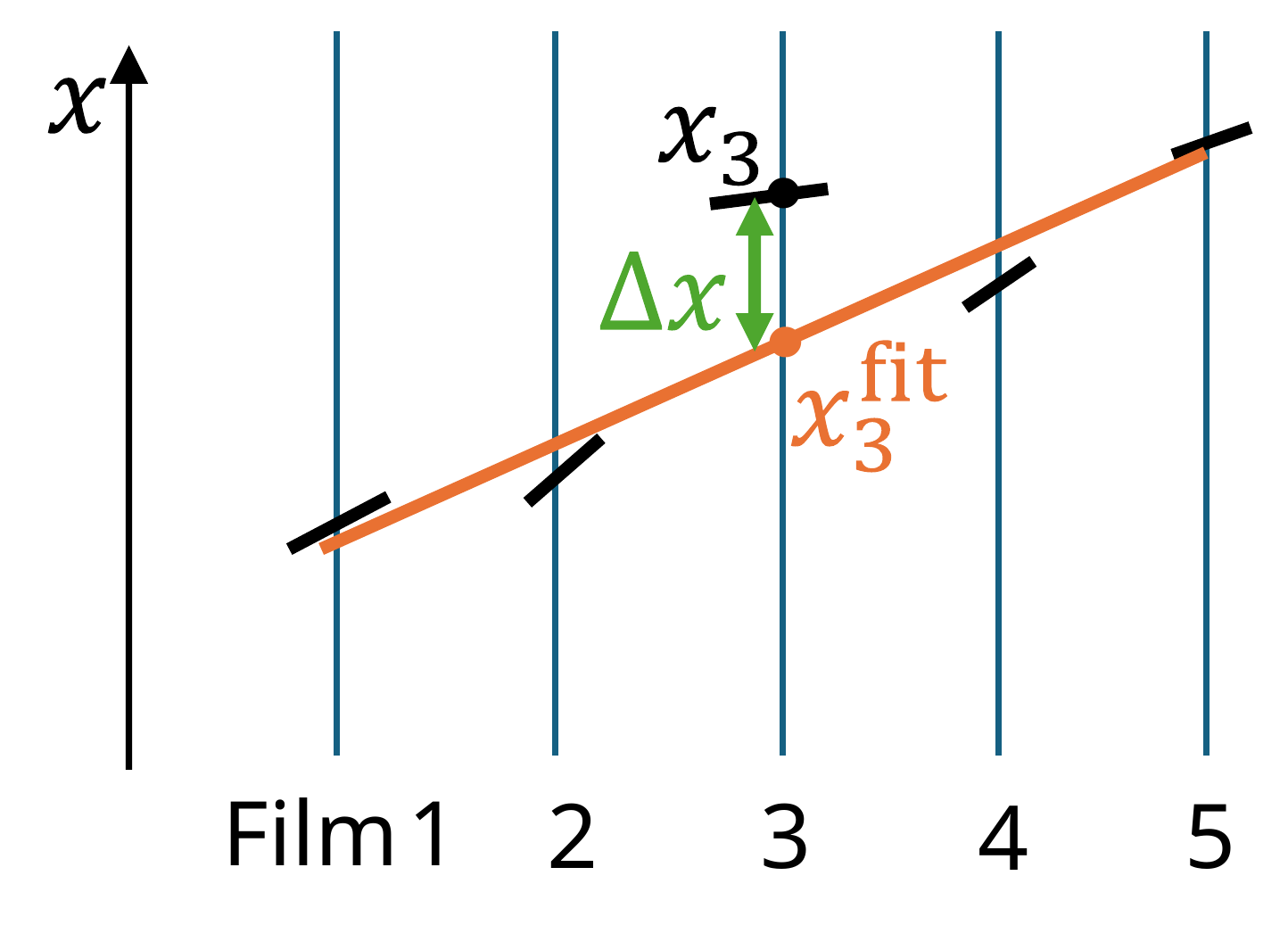}
\includegraphics[width=0.3\linewidth]{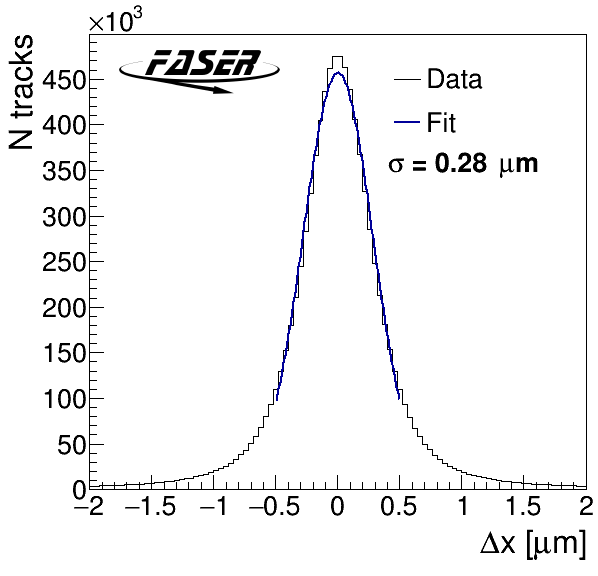}
\includegraphics[width=0.3\linewidth]{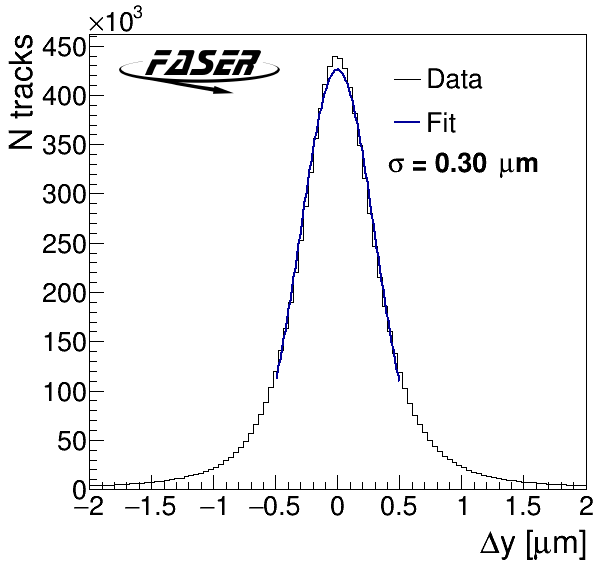}
\caption{Evaluation of the alignment accuracy in FASER$\nu$~\cite{FASER:2025qaf}. Left: a schematic of the method. Centre and right: distributions of the position deviation of the base-tracks from the fitted line.
}
\label{fig:deltaX}
\end{figure}

\begin{figure}[htb]
    \centering
    \includegraphics[width=0.35\linewidth]{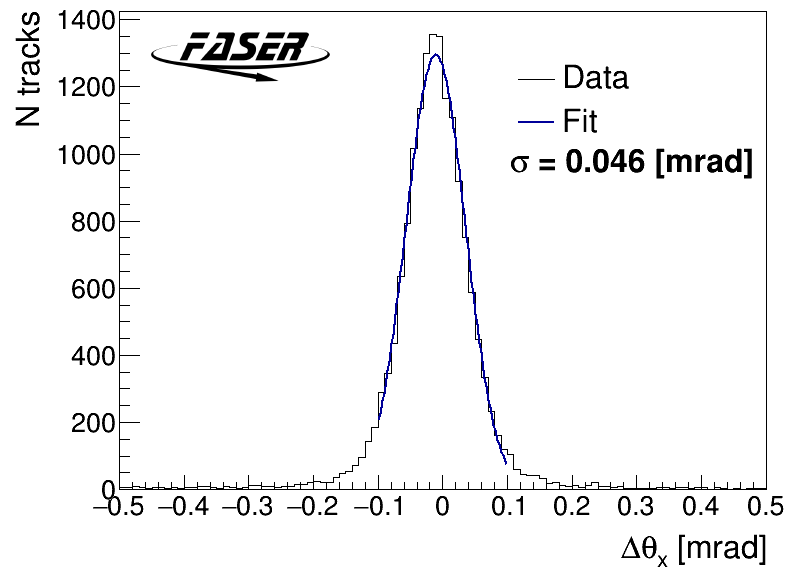}
    \includegraphics[width=0.35\linewidth]{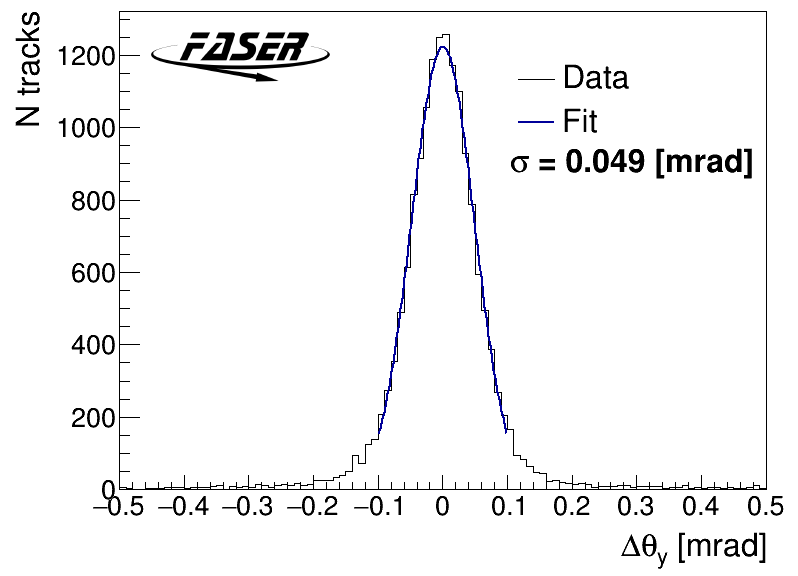}
    \caption{Angular resolution for an arm length of about 1.4 cm measured in FAESR$\nu$~\cite{FASER:2025qaf}.}
    \label{fig:dtx}
\end{figure}

The left panel of \cref{fig:tau_to_mu} shows a tau decay topology in the emulsion detector. As shown in the right panel of \cref{fig:tau_to_mu}, a global analysis that links information from FASER$\nu$2 with the FASER2 spectrometer will make it possible to measure the charge of muons from tau decays and thereby enable the definitive detection of $\bar\nu_\tau$ for the first time. 

Figure~\ref{fig:charm_and_tau_decays} shows event displays of simulated neutrino interaction vertices for a charm-associated $\nu_\mu$ interaction and a $\nu_\tau$ interaction in the emulsion detector. Algorithms to detect the small angle differences between short-lived particles and their daughters have been developed, and those decays in \cref{fig:charm_and_tau_decays} are successfully reconstructed.

\begin{figure}[htb]
  \centering
  \includegraphics[width=\textwidth]{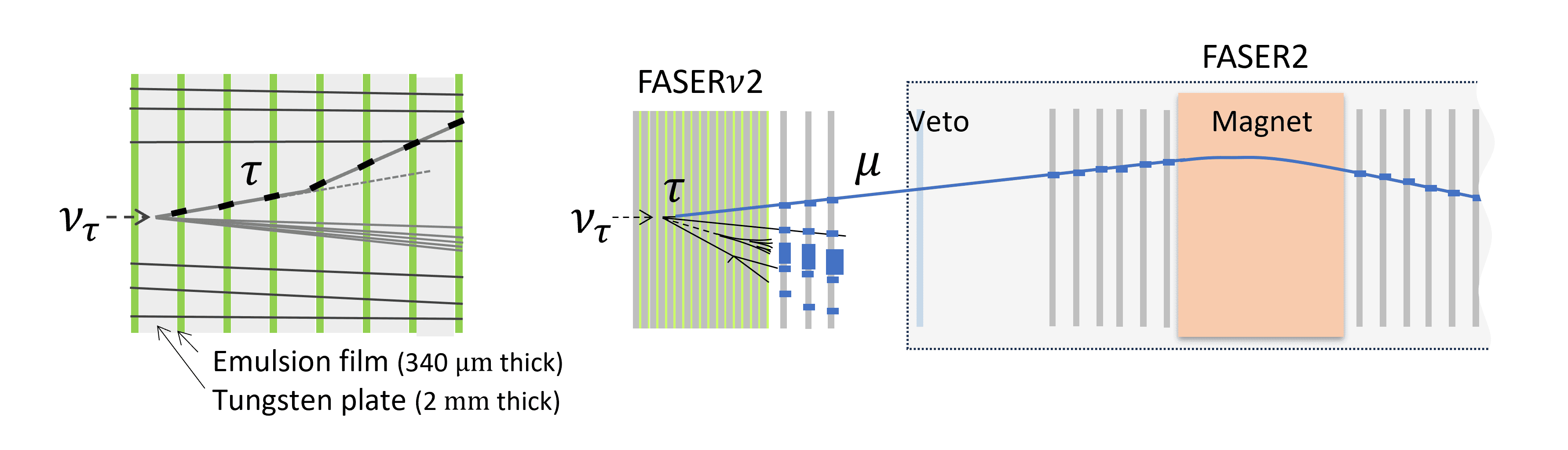}
  \caption{Left: Tau decay topology in the emulsion detector. Right: Charge measurement of a muon from tau decay.}
  \label{fig:tau_to_mu}
\end{figure}

\begin{figure}[htb]
  \centering
  \includegraphics[width=\textwidth]{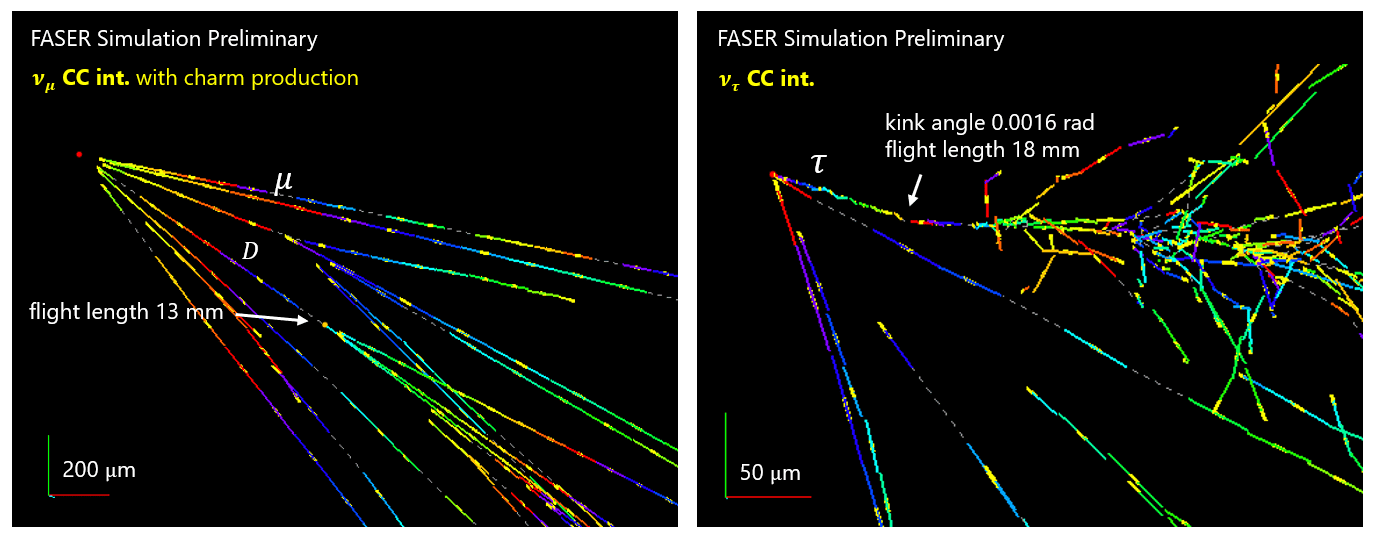}
  \caption{Event displays of simulated neutrino interaction vertices for a charm-associated $\nu_\mu$ interaction (left) and a $\nu_\tau$ interaction with the $\tau$ decaying into an electron (right) in the emulsion detector.}
  \label{fig:charm_and_tau_decays}
\end{figure}

\subsubsection{Interface Detector}\label{sec:fasernu2interface}
In the FASER$\nu$2 experiment, an Interface Tracker (IFT) is used to connect the high-resolution emulsion detector with the FASER2 tracking system placed downstream. The IFT is made of three layers of tracking detectors and is located just after the emulsion so that it can match particle tracks recorded in the emulsion to those seen downstream in the tracker. Possible technologies for the IFT include silicon strip detectors (SCT) and scintillating fiber trackers, which offer different levels of position and angle resolution.
To check the track matching performance, the positions $(x, y)$ and angles $(\theta_x, \theta_y)$ of tracks measured by the emulsion and IFT are compared. A match is confirmed if the differences between the two sets of measurements are small enough. The allowed differences depend on the measurement accuracy (resolution) of each detector and the 3.5~cm gap between them. The emulsion detector is assumed to have a resolution of 1~$\mu$m in position and 0.5~mrad in angle. The IFT is studied with three resolutions: 16~$\mu$m (for SCT), and 50~$\mu$m and 100~$\mu$m (for scintillating fibers), which give angle resolutions of about 0.4~mrad, 1.2~mrad, and 2.4~mrad.
Simulations were carried out using realistic track directions for muon neutrino and antineutrino interactions (about 35~mrad and 27~mrad). Background tracks were also added, with a smaller spread of about 2~mrad. The study measured how often the correct track from the emulsion was matched to the right one in the IFT (signal--signal), and how often it was wrongly matched to a background track (signal--background). The results show that better position resolution in the IFT gives better matching accuracy. Therefore, to keep good matching efficiency and reduce wrong matches, the IFT should have a position resolution better than 50~$\mu$m. 

\subsubsection{R\&D Efforts}
Emulsion detector analysis is limited by the accumulated track density and becomes increasingly difficult above $10^6$ tracks/cm$^2$ with the current tracking algorithms. To address the high track density caused by muon backgrounds in emulsion trackers, ongoing R\&D efforts focus on improvements in hardware, image processing, and reconstruction algorithms, together with the possible implementation of the sweeper magnet described in \cref{sec:sweeper_magnet}.

At the hardware level, optimizing the silver bromide crystal size and revising photo-development chemicals aim to reduce hit spread. The optical readout resolution, currently limited to approximately 300 nm due to distortions, is being improved through the application of 3D image deconvolution techniques. On the reconstruction side, the integration of machine learning methods shows strong potential for resolving ambiguities caused by overlapping tracks in three-dimensional space.

FASER$\nu$ employs a dedicated tracking algorithm optimised for reconstruction in high track density environments~\cite{DsTau:2019wjb}. Nevertheless, the FASER environment, with a flux of $10^5$ to $10^6$ muons/cm$^2$ accompanied by electromagnetic showers from bremsstrahlung of TeV muons, presents significant challenges for reliable track reconstruction. An updated version of the tracking algorithm, which actively identifies and corrects mis-reconstructions, is currently under development for FASER$\nu$. 99\% of muons are successfully reconstructed through 100 films with the current algorithm, for a muon track density of 2$\times$10$^5$ tracks/cm$^2$.

To reduce the emulsion cost, studies are ongoing to reconstruct emulsion data with 2-mm-thick tungsten plates, instead of the 1.1-mm-thick plates used in FASER$\nu$. As a test of such reconstruction, neutrino candidate events in the FASER$\nu$ data were reconstructed using only odd plates. \cref{fig:FASERnu2_FASERnu_nue_odd_plates} shows event displays of a neutrino interaction vertex in FASER$\nu$ reconstructed with all plates and with only odd plates. The results demonstrate good performance with 2.2-mm-thick tungsten plates, indicating the feasibility of reducing both cost and workload by about half.

\begin{figure}[htb]
  \centering
  \includegraphics[width=0.95\textwidth]{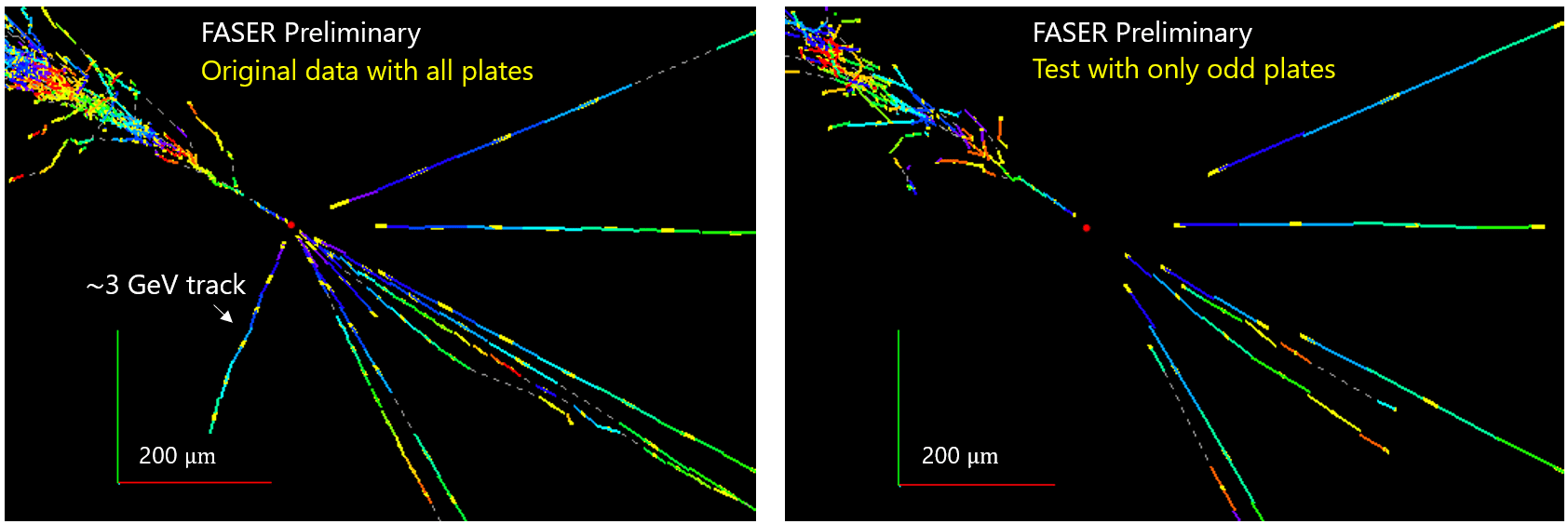}
  \caption{Event displays of a neutrino interaction vertex in FASER$\nu$ reconstructed with all plates (left) and with only odd plates (right).}
  \label{fig:FASERnu2_FASERnu_nue_odd_plates}
\end{figure}

\subsubsection{Sweeper Magnet}
\label{sec:sweeper_magnet}

The implementation of an effective sweeper magnet to reduce the background muon fluence in the FPF would be highly beneficial. Ongoing studies aim to evaluate the performance of a potential sweeper magnet to be installed in the LHC tunnel after the LOS exits the LHC magnet cryostats, but before it leaves the tunnel; see \cref{fig:sweepermagnet_concept}.

\begin{figure}[htbp]
  \centering
  \includegraphics[width=\textwidth]{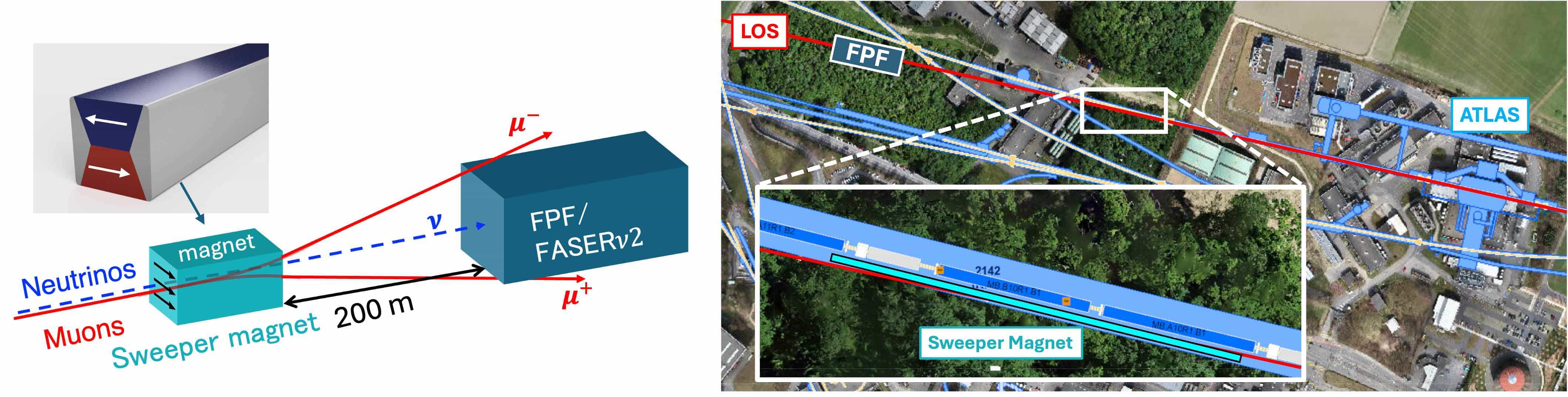}
  \caption{Left: Concept of the sweeper magnet to deflect muons. Right: Plan view of the LHC complex and the currently-considered location of the sweeper magnet. The LOS is shown as a red line, and the region after the LHC magnet cryostats, but still inside the tunnel, is shown in a zoomed-in view. From the CERN GIS portal.}
  \label{fig:sweepermagnet_concept}
\end{figure}

To obtain a first estimate of the effectiveness of this approach, a simplified \texttt{Geant4}-based simulation has been conducted using muons generated from \texttt{BDSIM}~\cite{Nevay:2018zhp} as input. A uniform magnetic field of 1 T was applied within a magnet structure—including the magnet core and its iron yoke—with an overall size of 40 cm $\times$ 40 cm in cross section and 40 meters in length, representing the sweeper magnet under consideration. Because muons reaching the detector are expected to have significant horizontal spread, the magnet length of 40 meters along the tunnel is chosen to effectively cover approximately 1 meter in the horizontal direction. The simulation does not include any detector response; instead, it tracks muons to the FPF location and evaluates their spatial and energy distributions at the location of the FASER$\nu$2 detector.

The current magnet concept consists of two identical magnets arranged vertically, one above and one below the LOS, with their magnetic fields oriented in opposite horizontal directions. Iron yokes are placed on the sides of this assembly, providing a return path for the magnetic flux and confining the field within the magnet structure.

The simulation results include a comparison of the expected muon rates in three scenarios: the current FASER Run 3 configuration, the anticipated rates in the FPF without the sweeper magnet, and the predicted rates in the FPF with the sweeper magnet installed. The muon rate for the FASER Run 3 scenario was calculated from the number of muon tracks measured with the FASER detector. The expected muon rate in the FPF without the sweeper magnet was obtained from \texttt{BDSIM} simulations of muon production and transport. The predicted rate for the FPF with the sweeper magnet was calculated by applying the reduction factor derived from the simplified \texttt{Geant4} simulations, which compare the number of muons reaching the FPF location with the magnet to the expected rate in the FPF without the magnet.

\begin{table}[tb]
  \centering
  \begin{tabular}{l | c}
  \hline \hline
  Scenario & Muon rate ($\mathrm{Hz/cm^2}$) \\
  \hline
  FASER Run 3                & 0.71 \\
  FPF without sweeper magnet & 0.82 \\
  FPF with sweeper magnet    & 0.18 \\
  \hline \hline
  \end{tabular}
  \caption{Comparison of expected muon rates at the detector location in different scenarios. The muon rates estimated using \texttt{BDSIM} simulations are used as input for this study.}
  \label{tab:muon_rates}
\end{table}

The results are shown in \cref{tab:muon_rates}.  They indicate that installing the sweeper magnet could reduce the muon rate at the FPF location to approximately 20\% of the expected rate without the magnet, and to about 25\% of the rate observed during FASER Run 3. 
Studies on the integration of a possible sweeper magnet in the LHC are ongoing.

\subsubsection{Emulsion Film Production and Scanning Systems}

The total emulsion film surface of the FASER$\nu$2 detector is $\sim$530 m$^2$/year. 
The emulsion film production and its readout will be conducted at facilities in Japan; see \cref{fig:ProductionScanning}.  The capacity of the film production facility~\cite{Rokujo:2024gqw} is 1200 m$^2$ per year. 
The Hyper Track Selector (HTS) system~\cite{Yoshimoto:2017ufm} can read out $\sim$0.5 m$^2$ per hour. The effective readout rate is $\sim$2400 m$^2$/year with HTS, and higher with the upgraded device described below. It will be possible to complete the readout of the data collected each year within a year using either of the systems.

\begin{figure}[htb]
  \centering
  \includegraphics[width=\textwidth]{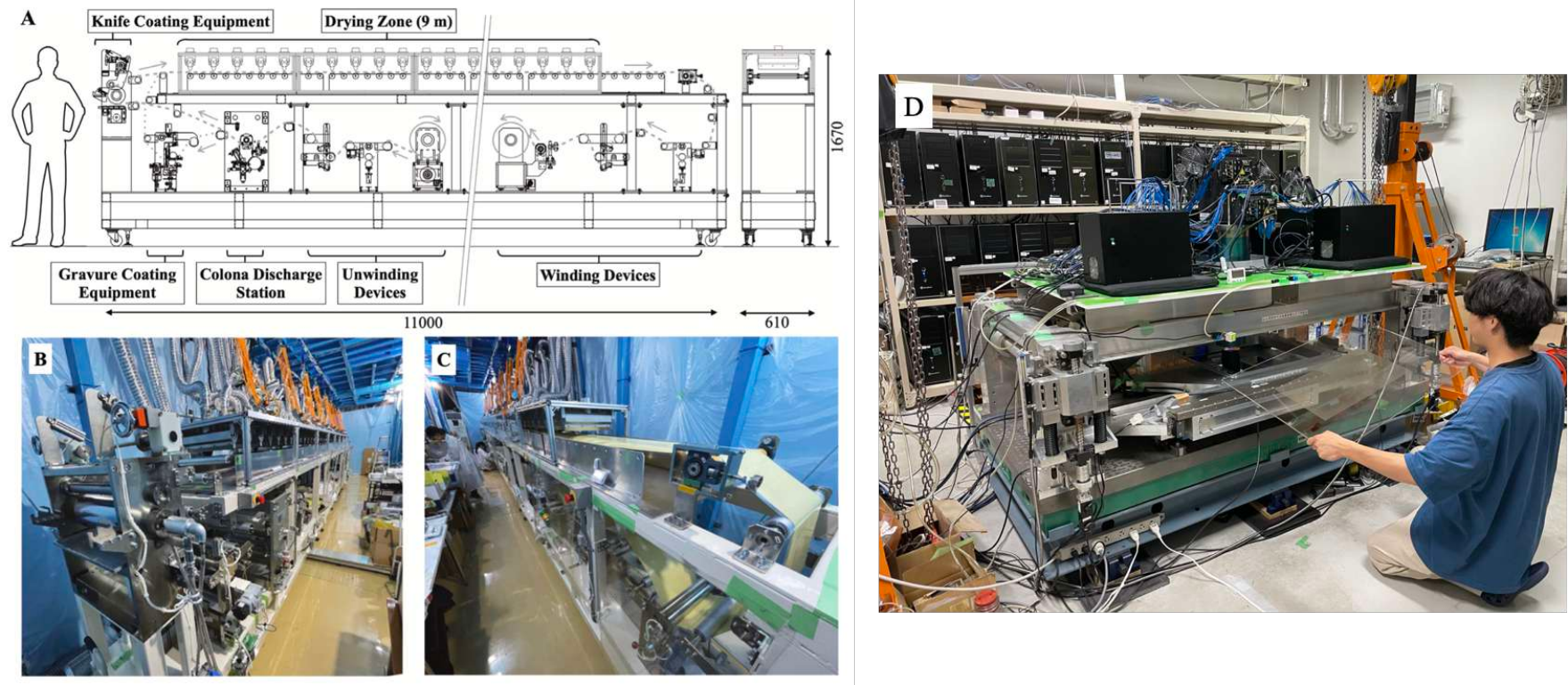}
  \caption{A: Overall view of the roll-to-roll emulsion film production system. B and C: Photographs of the system taken from the upstream and downstream sides, respectively~\cite{Rokujo:2024gqw}. D: A photograph of Hyper Track Selector-2 (HTS2).}
  \label{fig:ProductionScanning}
\end{figure}

\par
The nuclear emulsion gel, which is the raw material for the emulsion film, is developed and produced at Nagoya University. The large-scale gel production machine for mass production, which has been in operation since 2021, has undergone updates to the  manufacturing recipe and currently has the capacity to produce emulsion gel equivalent to 20 m$^2$ of film area per batch (1.5 days of operation). Additionally, the university has a small-scale gel production machine for R\&D purposes, which can produce a small amount of gel (0.3 m$^2$) per batch. This system can also be used to improve the emulsion gel  itself for specific experimental purposes. AgBr(I) crystals can be produced by adding silver nitrate solution and halogen solution to a gelatin solution by precisely controlling the addition rate, temperature, and stirring speed. The current crystal size for FASER$\nu$ is approximately 240 nm. The sensitivity and resolution for charged particles can be adjusted by changing the crystal size. For FASER$\nu$2, which will accumulate tracks at higher density, there is room to select the optimal emulsion gel by changing the crystal size, crystal shape, and amount of sensitiser added.

The coating and drying of the nuclear emulsion gel is performed mechanically using a roll-to-roll nuclear emulsion film coating device developed independently at Nagoya University. This device unrolls a roll of polystyrene film, continuously transports it through an emulsion coating head and a 9-meter-long drying section, and then rewinds it into a roll at the end, enabling stable, high-speed film production. In preparation for the start of the production of film for FASER$\nu$2, the development of methods and equipment for stable and labour-saving production with higher functionality (including a surface protection layer that greatly reduces the work required from development to data acquisition) and higher precision (improving the uniformity of coating, eliminating local distortion of the emulsion layer, and improving angle accuracy) are in progress. 
From 2022 to the present (Summer 2025), approximately 600 m$^2$ of double-sided coated emulsion film has been supplied to FASER$\nu$ without interruption. We expect that the facilities will continue to operate stably, and there will be no issues regarding film supply.
\par
Recently, operations have begun for the HTS upgrade device, HTS2, shown in \cref{fig:ProductionScanning} D. HTS2 features a wide field of view of 9.5 mm $\times$ 5.3 mm, which is twice that of HTS, and employs an array of 72 camera sensors to divide the image and achieve high-speed scanning. In addition to speed, efforts have been made to improve measurement accuracy. This includes synchronizing all sensor captures, measuring and correcting the vertical position difference of each sensor's focal plane, and introducing an accuracy-enhancing algorithm into the processing software. As a result, angle measurement accuracy has been improved to more than twice that of HTS within the angle range where $\tan \theta >  0.5$. Currently, HTS2 is operating at approximately twice the speed of HTS, and scanning of FASER$\nu$ film has begun. HTS2 is being improved in parallel with data acquisition work, and it is expected that the readout speed may be further improved by a factor of 2.5 through the introduction of a new image acquisition method. A further upgraded system called HTS3 is also being developed and is planned to be introduced for the readout of FASER$\nu$2.

\subsubsection{FASER$\nu$2 Summary}
Analysis methodologies dedicated to TeV neutrino interactions are currently being developed and tested in FASER$\nu$. These methods include momentum measurements using multiple Coulomb scattering information, electromagnetic shower reconstruction, and machine learning algorithms for neutrino energy reconstruction. First estimates of the efficiencies/performance for flavour-specific neutrino interactions have been obtained~\cite{FASER:2019dxq}, and work is ongoing to refine them.

FASER$\nu$2 has a clear and broad physics target, and the detector is based on a well-tested technology for tau neutrino and short-lived particle detection. 
The performance of FASER$\nu$2 is based on the experience of the FASER$\nu$ detector operating at the LHC. Studies of a possible sweeper magnet in the LHC tunnel have shown promising prospects to reduce the muon flux during HL-LHC operation by a factor of 5, making it a factor of 4 below what has been seen for FASER$\nu$ in LHC Run 3. While such a magnet would further improve the detector environment, the physics reach of FASER$\nu$2 remains well motivated even without the magnet. 
Some novel aspects of FASER$\nu$2, such as how the detector is assembled, have been studied in dedicated test beams with positive results. 
Further studies are being carried out to optimise the detector performance, the detector operational environment, and the installation scheme.

\subsection{FASER2 }

FASER2 is a large-volume detector comprised of a spectrometer, electromagnetic and hadronic calorimeters, veto detectors, and a muon detector, that is designed to be sensitive to a wide variety of models of BSM physics and to precisely reconstruct muons produced in upstream neutrino interactions. It builds on the experience gained from the successful operation of the existing FASER experiment~\cite{FASER:2022hcn}, a much smaller detector, which was constrained to be situated within a former LEP transfer tunnel. The FASER2 detector, specifically designed for the FPF facility, is much larger (by a factor of $\sim 600$ in decay volume) and includes new detector elements. It provides an increase in rates for various BSM signals of several orders of magnitude compared to FASER and can discover new particles that were previously out of reach, such as dark Higgs bosons, HNLs, and ALPs, as studied in Refs.~\cite{FASER:2018ceo,FASER:2020gpr,FASER:2018eoc}. In addition, FASER2 is uniquely sensitive to well-motivated BSM particles that cannot be discovered elsewhere, including, for example, inelastic DM particles in cosmologically-motivated regions of parameter space and quirks in mass ranges motivated by naturalness arguments.  For more details, see \cref{sec:darkmatter,sec:newparticles}.

In addition to the BSM case for FASER2, the SM neutrino program at the FPF will rely on the identification of muons from neutrino interactions and precise measurement of their momentum and charge. The FASER2 spectrometer will be required for these measurements for both FASER$\nu$2 and FLArE. 

The detector optimisation performed to date has been based on sensitivity to the range of physics signatures presented in \cref{tab:faser2_physics_benchmark}. Together these signatures inform the detector performance requirements that in turn guide the more detailed detector design.
The level of simulation performed was as appropriate to determine the performance characteristics of the detector, ranging from Monte-Carlo truth level studies to determine energy distributions and thresholds for the calorimeter, through to full \texttt{Geant4} simulations, where detector acceptance and interactions are required. 

\begin{table}[t]
  \setlist{nolistsep}
  \setlist[itemize]{topsep=0pt, leftmargin=*}
  \centering
  \begin{tabular}{l |c |l}
  \hline\hline
  \textbf{Physics Process} 
    & \textbf{Reconstructed Particles} 
    & \textbf{Detector Requirements} \\
  \hline\hline

  Dark Photon / Dark Higgs 
    & \begin{minipage}[t]{4.5cm} High energy $e^+e^-, \mu^+\mu^-, \text{hadrons}$ \end{minipage}  
    & \begin{minipage}[t]{6cm}
        \begin{itemize}[leftmargin=*]
          \item Tracker resolution
          \item Muon detector
          \item Magnet design
          \item Calorimeter resolution
          \item Hadron PID
        \end{itemize}
        \vspace{2mm}
      \end{minipage} \\ \hline

  ALPs 
    & \centering \begin{minipage}[t]{4.5cm} High energy $ \gamma\gamma$  \end{minipage}  
    & \begin{minipage}[t]{6cm}
        \begin{itemize}[leftmargin=*]
          \item Photon ID and separation
          \item Calorimeter resolution
        \end{itemize}
        \vspace{2mm}
      \end{minipage} \\ \hline

  Inelastic DM 
    & \centering \begin{minipage}[t]{4.5cm} Low energy photon or lepton pairs from $X_2 \to X_1 \gamma, \ell^+\ell^-$   \end{minipage}  
    & \begin{minipage}[t]{6cm}
        \begin{itemize}[leftmargin=*]
          \item Low energy $\gamma$ reconstruction
          \item Calorimeter resolution
          \item Tracker resolution
          \item Magnet design
          \item Dynamic range
        \end{itemize}
        \vspace{2mm}
      \end{minipage} \\ \hline

  Quirks 
    & \centering \begin{minipage}[t]{4.5cm} Slow/delayed charged particle tracks  \vspace{1mm} \end{minipage}  
    & \begin{minipage}[t]{6cm}
        \begin{itemize}[leftmargin=*]
          \item Timing resolution
        \end{itemize}
      \end{minipage} \\ \hline

  Neutrino Physics 
    & \centering \begin{minipage}[t]{4.5cm} $\mu$ from FLArE / FASER$\nu$2, $e$ from interactions in veto  \end{minipage} 
    & \begin{minipage}[t]{6cm}
        \begin{itemize}[leftmargin=*]
          \item Detector acceptance
          \item Good charge identification
          \item Momentum resolution
          \item Veto design
          \vspace{1mm}
        \end{itemize}
      \end{minipage} \\ \hline\hline

  \end{tabular}
  \caption{Table of benchmark physics processes used to optimize FASER2 detector performance~\cite{Salin:2927003}.}
  \label{tab:faser2_physics_benchmark}
\end{table}

\subsubsection{Design of the Experiment}

\begin{figure}[tb]
  \centering
  \includegraphics[width=0.99\textwidth]{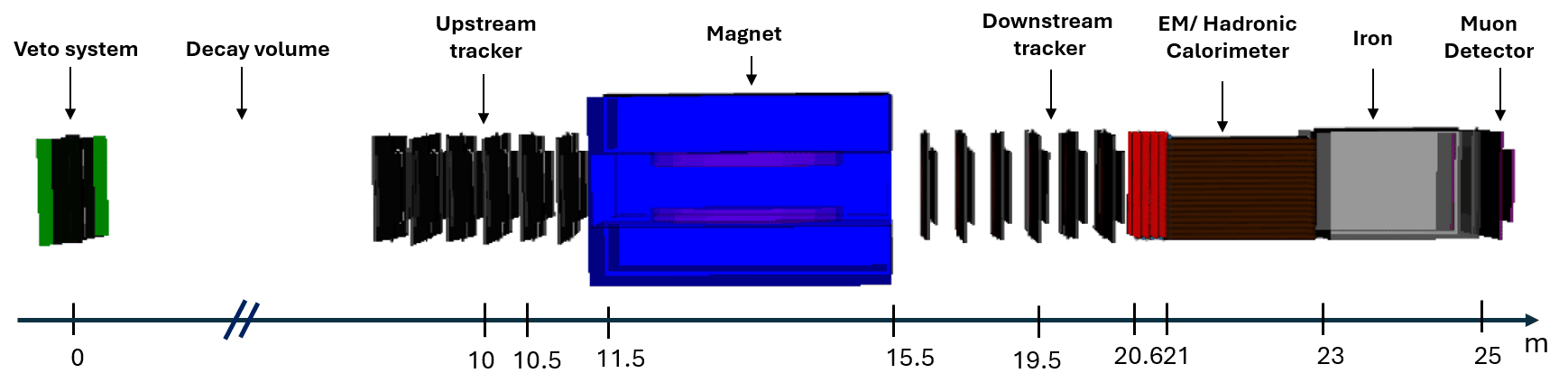} 
  \caption{Visualisation of the full FASER2 detector, showing the veto system, uninstrumented 10~m decay volume, tracker, magnet, electromagnetic calorimeter,  hadronic calorimeter, iron absorber, and muon detector. 
  \label{fig:FASER2-Design}}
\end{figure}

Figure~\ref{fig:FASER2-Design} shows a rendering of the \texttt{Geant4} model of the baseline FASER2 detector. The overall layout is largely driven by the spectrometer, which is itself constrained by considerations relating to deliverable and affordable magnet technology. This leads to a baseline detector configuration consisting of a spectrometer with a large-volume dipole magnet. The magnet has a rectangular aperture of 1~m in height and 3~m in width. This also defines the transverse size of the decay volume, which is the 10~m uninstrumented region upstream of the first tracking station (a $2.6 \times 1 \times 10$~m$^3$ cuboid) and downstream of the first veto station. Maximising the transverse size is a general design requirement driven by the need to have sufficient acceptance for BSM particles originating from heavy flavour decays and charged leptons arising from neutrino interactions in FLArE. 
Studies are ongoing as to the additional advantages of filling the decay volume with a low-mass container containing low-density gas such as helium, which would further reduce the backgrounds from neutrino interactions in that volume. 

\begin{figure}[H] 
  \centering
  \includegraphics[width=1.0\textwidth]{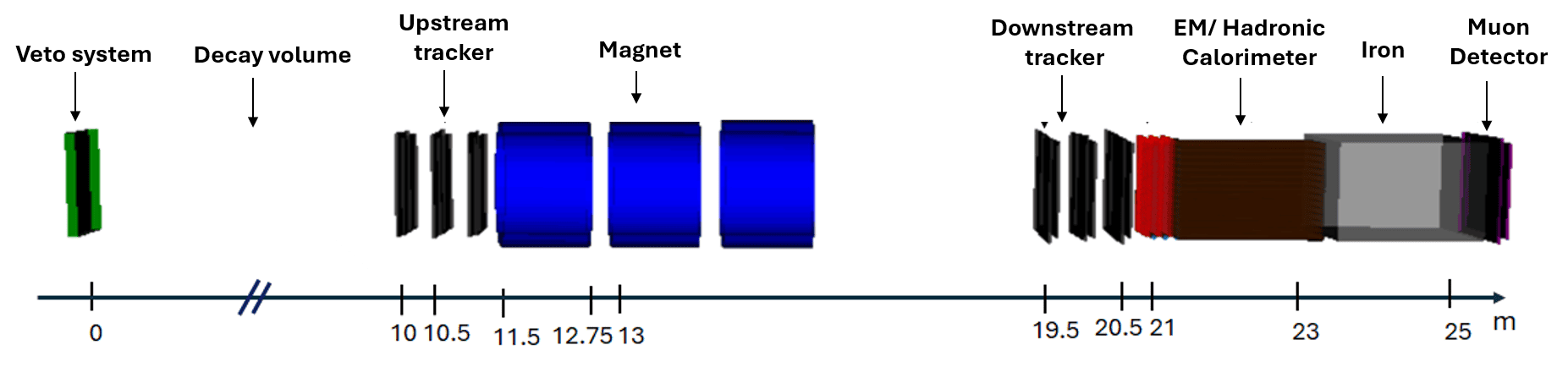} 
\caption{Alternative design with three ``crystal-puller'' magnets for the FASER2 detector, presented in \texttt{Geant4}. Here each module has a diameter of 1.6~m and an integrated magnetic field of 0.7 Tm. The design also illustrates a smaller number of tracking stations (6) compared to the baseline design. The rest of the design is similar to the baseline shown in \cref{fig:FASER2-Design}.}
  \label{fig:FASER2_indu} 
\end{figure}

For most FASER2 sub-detectors, a performant baseline is achievable from simpler well-understood detector technologies that will allow the major physics goals to be achieved. However, more advanced technologies are also under consideration to augment these baseline capabilities. Such augmentations are especially appealing in the case that they can come via existing R\&D activities, for example, in the context of future colliders, where FASER2 can act as a mid-term testbed.

\paragraph{Magnet.} The baseline integrated magnetic field is 2~Tm. This has been optimised based on simulations~\cite{Salin:2927003} that demonstrate that the required charged particle separation, momentum resolution, and charge identification are obtained for the BSM and neutrino programme, while keeping the field strength to an acceptable minimum to reduce cost. Superconducting magnet technology is required to maintain such a field strength across a large aperture. Investigations by KEK magnet experts, along with discussions with manufacturing experts at Toshiba in Japan and Tesla Engineering in the UK, have demonstrated that a dipole design is feasible at an acceptable cost ($\sim 4$ MCHF) and lead-time (3.5 years).  

Alternative magnet options are being investigated to make use of tooling and parts used for industrial crystal-puller magnets available commercially in the semiconductor industry which have a smaller, circular aperture (typically 1.6\,m diameter). An illustration of a design containing three such magnets can be found in \cref{fig:FASER2_indu}. The individual commercial crystal-puller magnets provide field-integrals that are a factor of several smaller than is required. Simulations~\cite{Salin:2927003} show that the field provided by several such magnets of this design would be sufficient to meet the physics requirements. 
A more practical design than daisy-chaining the commercial magnets is to produce a single cold-mass of the same diameter with the desired field-integral. This is expected to be possible with reuse of tooling and components from the commercial crystal-puller magnets. A project is ongoing to commission a Phase-1 design of such a magnet. The proposal considers a single canted-cosine-theta magnet, reducing mechanical and cryo engineering issues that would result from the three-magnet solution illustrated in \cref{fig:FASER2_indu}. This single-magnet design project will include its electrical, mechanical and magnetic performance. Separate studies are being undertaken for the transport and installation plans, mechanical envelope, and operation parameters. The indications are that a magnet of this type can be designed, constructed, and operated for a cost smaller than that of a large dipole.

\begin{figure}[tb]
  \centering
  \includegraphics[width=0.7\textwidth]{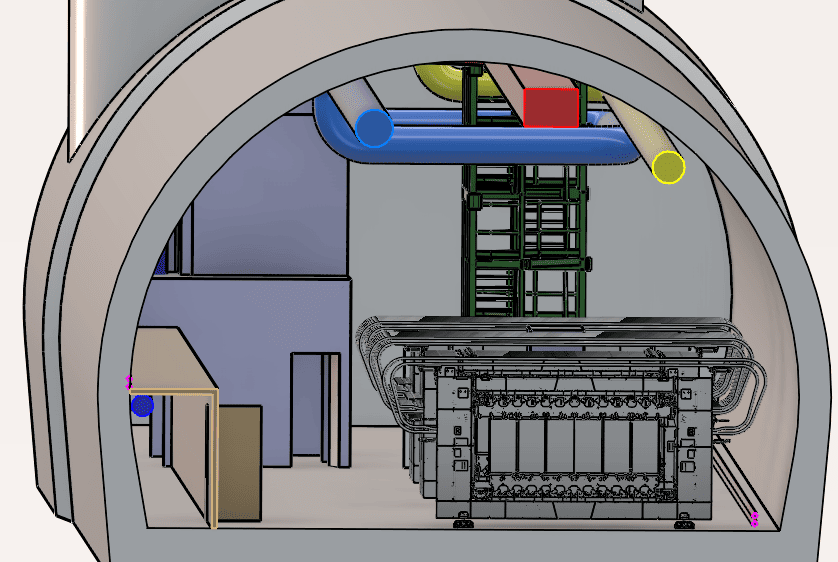} 
  \caption{CAD visualisation showing the FASER2 detector within the FPF cavern. The the SciFi tracking stations are visible in grey. The support structures and services are based on those of the LHCb SciFi detector.
  \label{fig:FASER2-cross-section}}
\end{figure}

\paragraph{Tracker.} The tracking detectors are foreseen to use a SiPM and scintillating fiber tracker technology, based on LHCb's SciFi detector~\cite{Hopchev:2017tee}, as shown in \cref{fig:FASER2-cross-section}. This technology gives sufficient spacial resolution at a significantly reduced cost compared to silicon detectors, as shown in \cref{fig:FASER2_Tracker_design}. Each layer has an active area of approximately $3~\text{m}\times 1~\text{m}$. It consists of vertical and horizontal fiber layers, with consecutive tracking stations rotated by about $1^\circ$ relative to each other to improve resolution. The fibers have a diameter of $250~\mu\text{m}$ and are arranged in mats of four fibers each. This design achieves a spatial resolution of around $80~\mu\text{m}$ and a hit detection efficiency exceeding 99\%.

Track-finding studies have been performed for different numbers of tracking stations in the range of 6 to 12, as illustrated in \cref{fig:FASER2-Design,fig:FASER2_indu}. In the absence of backgrounds, a 6-station configuration provides sufficient resolution and tracking capability and pointing resolution. The extent to which a larger number of tracking stations (up to 12) might be required to maintain performance in the presence of realistic backgrounds is under current study. 

\begin{figure}[thb] 
  \centering
  \includegraphics[trim=0 17mm 0 23mm, clip, width=0.7\textwidth]{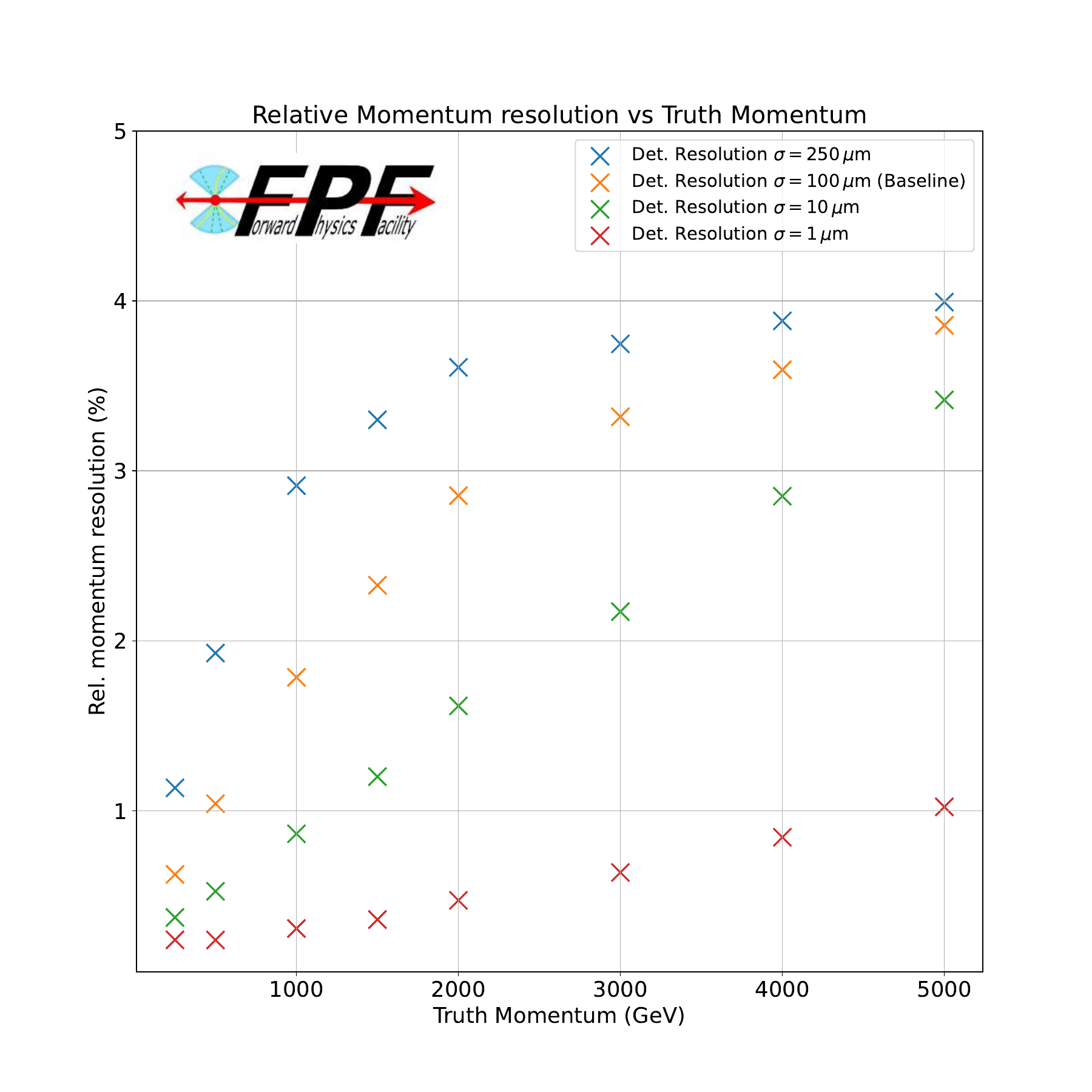} 
  \caption{Momentum resolution plots showing the track momentum reconstruction performance of trackers with detector resolution from 1 $\mu$m to 250 $\mu$m. The resolution of the LHCb-like SciFi detectors is about $100\,\mu m$.
  From Ref.~\cite{Salin:2927003}.}
  \label{fig:FASER2_Tracker_design} 
\end{figure}

Upgrade options include the use of silicon-based (rather than SciFi) tracking detectors for the interface between FASER2 and FASER$\nu$2, and for the first tracking station downstream of the decay volume. Possible augmentation utilising the LHCb MightyPix technology~\cite{Hennessy:2024jov} could provide improvement in particle separation power in both the first tracker layer and in the central region of the transverse plane, where the LLP energy is higher and decay products more collimated. 

\paragraph{Calorimeter.} A simple lead-scintillator calorimeter has been found to be sufficient for the reconstruction of energy deposits from electrons and hadronic decay products of LLPs. A more advanced calorimeter is also under study to be based on dual-readout calorimetry technology~\cite{Lee:2017xss,Antonello:2018sna}, especially for the central region. The calorimeter system is expected to contain a preshower to allow for particle identification, e.g. to identify photons from ALPs or iDM decays and separate them from possible backgrounds, and from neutrinos interacting in the calorimeter system. This builds upon experience of existing prototypes for future collider R\&D, but modifies them for the specific physics needs of FASER2: spatial resolution sufficient to identify particles at $\sim 1-10$~mm separation; good energy resolution; improved longitudinal segmentation with respect to FASER; and the capability to perform particle identification, separating, for example, electrons and pions.

\paragraph{Muon Detetcor.} The ability of the FPF to identify separately electrons and muons is required for signal characterisation, background suppression, and for the interface with FASER$\nu$2, as described in \cref{sec:fasernu2interface}. To achieve this, ${\cal O}(10)$ interaction lengths ($1.7~\textrm{m}$) of iron  will be placed after the calorimeter, with sufficient depth to absorb pions and other hadrons, followed by a detector for muon identification, for which additional SciFi planes could be used. 

\paragraph{Veto System.} Finally, the veto system will need to reject muon rates of roughly $20~\text{kHz}$ and backgrounds from their interactions. A scintillator-based design has proven sufficient for this purpose in FASER, and a similar approach is planned for FASER2. As noted in \cref{sec:physics}, such a veto would also serve as an additional target for neutrino measurements. Owing to its small optical thickness, it offers a unique opportunity to distinguish electron neutrinos from electron antineutrinos~\cite{Kling:2025lnt}. The target design will be re-optimised accordingly.

\paragraph{Readout.} The event rate and size are much lower than most LHC experiments, so the trigger needs are not expected to be a limiting issue. For instance, it is expected that it will be possible to significantly simplify the readout of the tracker, with respect to what is used in the LHCb SciFi detector.

\subsubsection{Performance Studies}

Various performance studies have been performed to assess different design considerations and technologies for FASER2. Metrics such as momentum resolution, LLP sensitivity, and geometrical acceptance of muons from neutrino interactions in FLArE have been studied both in terms of physics performance and the implied detector technology complexity and cost. Different simulation tools have been utilised for these studies:~the FORESEE~\cite{Kling:2021fwx} package is used for the simulation and event generation of LLP production from forward hadrons; the \texttt{Geant4}~\cite{Agostinelli:2002hh} simulation framework is used for material interactions and for the propagation of particles through a magnetic field in the LLP decay product separation studies; and the ACTS~\cite{Ai:2021ghi} tool is used for track reconstruction studies.

\paragraph{Momentum resolution.} An illustration of such studies is provided in the following for the expected momentum resolution. For the baseline detector outlined above, with an intrinsic resolution of 100~$\mu$m and 2 Tm integrated field strength, a muon momentum resolution of approximately 2(4)\% is achieved for 1(5)~TeV muons (\cref{fig:FASER2_Tracker_design}). This is expected to be sufficient for the physics goals of FASER2. Studies show the baseline design to be quite robust:~this performance is stable under a range of magnetic field strengths, and appreciable degradation only appears with a significantly worse intrinsic resolution. The momentum resolution was also studied as a function of the amount of material in each tracker layer, and only when approaching an interaction length is a significant loss in resolution observed. This study assumes perfect detector alignment; as is discussed below it is expected that the detector can be aligned with sufficient accuracy to give the required momentum resolution by using background muons. 

\begin{figure}
  \centering
  \includegraphics[scale=0.6]{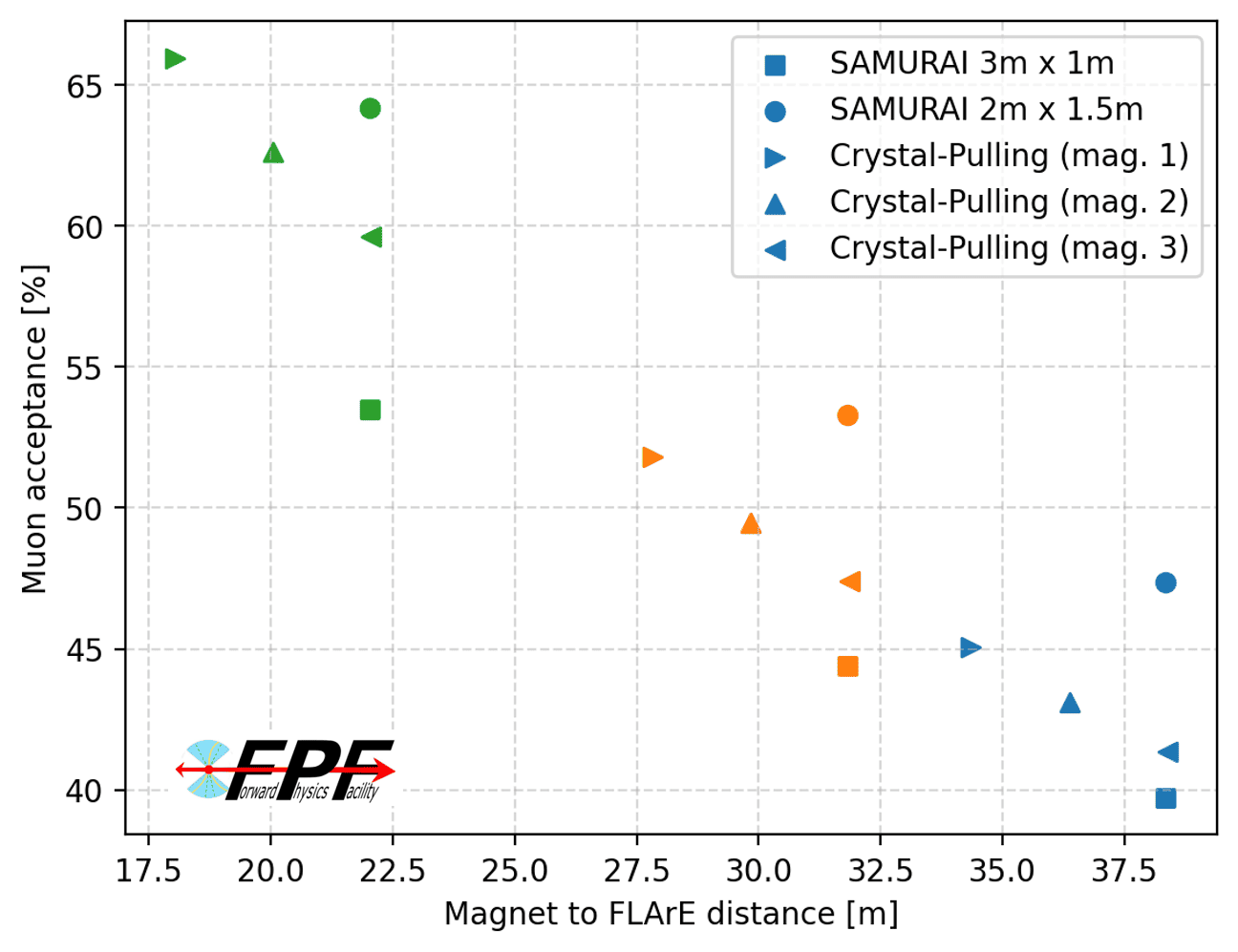} 
  \caption{Muon acceptance into the FASER2 magnets as a function of the distance between the magnets and the center of FLArE. The colours represent different configurations as described in Ref.~\cite{Salin:2927003}. In the legend `SAMURAI' refers to magnet designs similar to \cref{fig:FASER2-Design}, and `crystal-pulling' to designs similar to those illustrated in \cref{fig:FASER2_indu}.}
  \label{fig:FASER2_Muon_acceptance} 
\end{figure}

\paragraph{Acceptance from FLArE.} Since FASER2 provides the muon spectrometer for high-energy muons emitted from neutrino interactions in the FLArE experiment, the acceptance for such muons is of importance. Studies were performed using a sample of muons from \texttt{GENIE} simulation, involving a \texttt{Geant4} simulation in the FPF. The simulation tracked the trajectory of the muons resulting from $100~000$ $\nu_{\mu}$ CC interactions within the FLArE fiducial volume measuring 1m $\times$ 1m $\times$ 7m. The results~\cite{Salin:2927003} are summarised in \cref{fig:FASER2_Muon_acceptance} and show rather limited change in acceptance for the various magnet designs that were considered, demonstrating that all are viable options for this purpose. The largest sensitivity is set by the distance between the FLArE experiment and the magnet. A shorter distance could increase the solid angle subtended and improves acceptance to such muons, but at the cost of reducing the FASER2 decay volume.

\paragraph{Alignment.} Simulations were performed to reconstruct example translational misalignments of the tracking planes by finding the transformation parameters that best map the measured hit positions to the predicted hit positions.  The predicted hit positions come from simulating the propagation of the muon background through an aligned detector using the fitted momentum parameters and first tracking plane hit position given by ACTS.
Results are shown for 15 different simulated detectors, each with tracking plane misalignment parameters drawn from normal distributions with parameters specified in the relevant section. These detectors are fitted using 15 iterations of the ACTS fitting process and all of the available input data (150 000 propagated muons per iteration).
Taking $\sigma = 250$ $\mu$m in X and Y axis translation, these are reconstructed well (\cref{fig:tba_translation_plots}) with the maximum misalignment being approximately 10 $\mu$m after 15 iterations in all cases.
These studies demonstrate that the muon backgrounds should be sufficient to rapidly perform a sufficient alignment.

\begin{figure}
    \centering
    \includegraphics[width=0.99\textwidth]{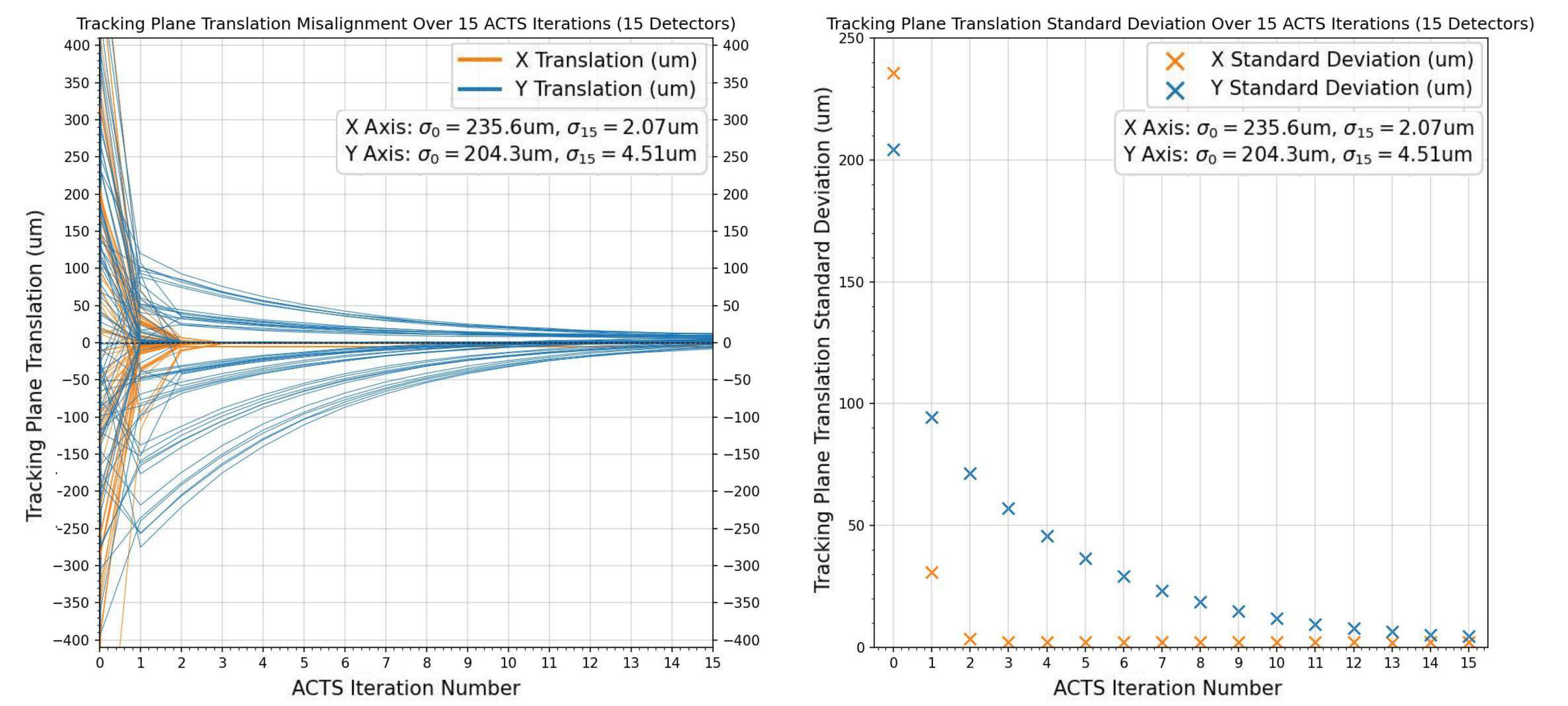}
    \caption{FASER2 alignment study results. Translation reconstruction value (left) and standard deviation (right) against ACTS iteration number (150 000 propagated muons per iteration). Each line in the left figure represents a single tracking plane. From Ref.~\cite{Salin:2927003}.}
    \label{fig:tba_translation_plots}
\end{figure}

 \begin{figure}
    \begin{center}
    \includegraphics[width=0.65\textwidth]{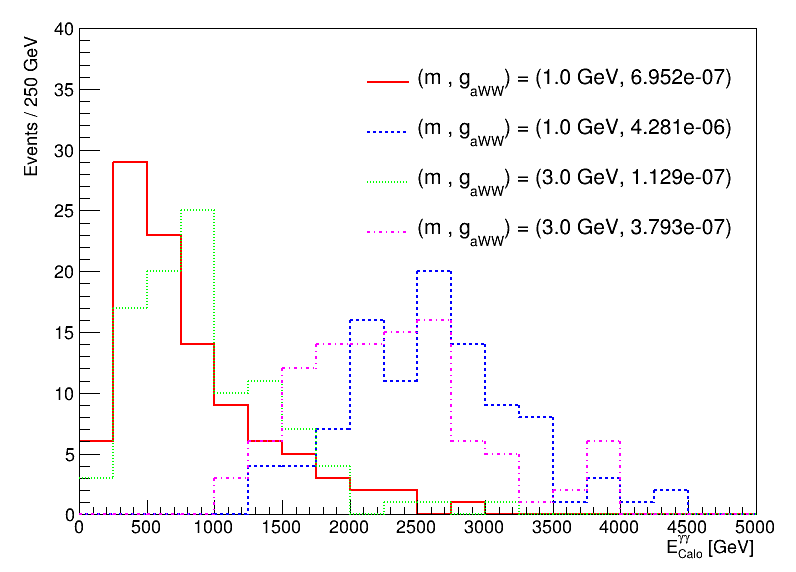}
    \caption{Truth-level calorimeter energy distributions of the diphoton system from the decay of an ALP for 4 chosen signal mass and lifetime points on the edge of the 90\% contour. From Ref.~\cite{Salin:2927003}.}
    \label{fig:ALPs_diphoton_energy}
    \end{center}

\end{figure}

\paragraph{Calorimetry.} While the final FASER2 calorimeter design proposal is currently under discussion, truth-level studies can be used to investigate the minimum calorimeter energy reconstruction requirements. Monte Carlo simulations (without material interactions or detector effects) were used to determine the typical energy distributions for models of interest. Figure~\ref{fig:ALPs_diphoton_energy} presents the diphoton energy spectrum from four selected ALP models on the edge of the expected FASER2 90\% exclusion contour. It is seen that the energy deposited is predominantly dependent upon the ALP's coupling, with lower couplings leading to a lower energy of the diphoton system. Reasonable sensitivity to the ALP models can be achieved provided that the chosen calorimeter design can reconstruct energy deposits of $\mathcal{O}(100)$\,GeV. Measuring deposits with these energies is easily achievable for the calorimeters under consideration, though design studies will be needed to ensure that such devices have sufficient linearity over a sufficient dynamic range to cover energy deposits in the GeV to TeV energy range.  

\paragraph{Timing.} Ref.~\cite{Feng:2024zgp} details the analysis strategy for discovering quirks through timing. Heavy particles typically have velocity much smaller than background muons. The resulting signal is hits in the scintillators that are either delayed (compared to the LHC clock) or slow (as they pass through the FPF detectors). 
Given a timing scintillator’s 500\,ps timing resolution, 
FASER2 is sufficient to probe quirk masses up to $\sim$1\,TeV, as shown in \cref{fig:BSM_newparticles}. This covers a range motivated by neutral naturalness solutions to the gauge hierarchy problem. 

\subsubsection{FASER2 Summary}

The FASER2 experiment will be essential to maximise the physics potential of the FPF. The baseline detector design has been optimised to obtain the required physics performance in an affordable way, but several systems could be upgraded to improve the performance at higher cost. Simulations have shown that the baseline detector has sufficient tracking performance, calorimeter performance, and integrated field strength to meet the physics goals. 
Given the importance of the FASER2 magnet in the design, significant work has been carried out to find a baseline solution for this, with a design study underway for a single-cold-mass canted-cosine-theta magnet based on standard parts and processes used in commercially-available magnet units.

\subsection{FORMOSA }

The high energy of the LHC beams can efficiently produce millicharged particles (mCPs) with masses up to $\sim 45$~GeV, and is capable of reaching masses up to $\sim 100$~GeV at reduced production rates. Taking advantage of this, the milliQan experiment~\cite{Ball:2016zrp,Alcott:2025rxn} has been running in the CMS service cavern at the LHC since 2018. A prototype version, demonstrated sensitivity to mCPs at a hadron collider using data collected during LHC Run~2~\cite{Ball:2020dnx}. In 2023, the completed milliQan ``bar" detector was installed for LHC Run 3~\cite{milliQan:2021lne}. Subsequently, the milliQan collaboration has analyzed 124.7~fb$^{-1}$ of $\sqrt{s} = 13.6$ TeV collision data and published the most stringent constraints to date for mCPs with charges $\leq0.24~\rm{e}$ and masses $\geq 0.45~\rm{GeV}$~\cite{Alcott:2025rxn}. 

The mass range for which the LHC provides unique mCP sensitivity ($\gtrsim$ GeV) is small compared to the LHC's centre-of-mass energy. Given this small mass compared to their transverse momentum, mCPs will be produced with an approximately flat distribution in rapidity~$y$, and thus, in pseudorapidity~$\eta$. Therefore, a mCP detector placed at $\eta \sim 7$ would be expected to see around a factor of 250 higher rate of mCPs compared to a detector of the same size at $\eta \sim 0$, as the forward detector intersects a much larger range in $\eta$, as well as having full azimuthal coverage.

\subsubsection{Design of the Experiment}

The FORMOSA experiment consists of a large array of rectangular plastic scintillator bars designed to be sensitive to signatures of millicharged particles produced along the axis of the detector. Similar to the milliQan experiment~\cite{Ball:2016zrp,Ball:2020dnx,milliQan:2021lne,Alcott:2025rxn}, plastic scintillator is chosen, as this provides the best combination of photon yield per unit length, response time, and cost~\cite{Adhikary:2024nlv}. The array will be oriented such that the long axis points at the ATLAS IP and will be located on the LOS. The array contains four longitudinal ``layers'' arranged to facilitate a 4-fold coincident signal for the detection of mCPs originating from the ATLAS IP. Each layer in turn contains 400 scintillator ``bars'', each measuring $5~\mathrm{cm} \times 5~\mathrm{cm} \times 100~\mathrm{cm}$,  in a $20\times20$ array. To maximise sensitivity to the smallest charges, each scintillator bar (made out of e.g., Eljen EJ-200~\cite{Eljen} or Saint-Gobain BC-408~\cite{SG}) is coupled to a high-gain photomultiplier tube (PMT) capable of efficiently reconstructing the waveform produced by a single photoelectron (PE), such as the Hamamatsu R7725~\cite{Hamamatsu}. 

To reduce random backgrounds, mCP signal candidates will be required to have a quadruple coincidence of hits with $\overline{N}_{\text{PE}} \ge 1$ within a 10-20 ns time window. The PMTs must therefore measure the timing of the scintillator photon pulse with a resolution of $\le5$ ns. The bars will be held in place by a steel frame~\cite{Adhikary:2024nlv}, resulting in a detector with dimensions $1.9~\mathrm{m} \times 1.9~\mathrm{m} \times 4.8~\mathrm{m}$. In addition to the scintillator bars, additional components will be installed to reduce or characterise certain types of backgrounds. Scintillator panels on the top and sides of the detector will provide the ability to tag and reject cosmic-ray muons and environmental radiation. Finally, segmented scintillator panels will be placed at the front and back of the detector to allow through-going muons to be identified and their path through the detector tracked. A conceptual design of the FORMOSA detector is shown in \cref{fig:FORMOSA_diagram}, while a more detailed description of the design is outlined in Ref.~\cite{Feng:2022inv}.

\begin{figure}[!ht]
  \centering
  \includegraphics[width=0.6\textwidth]{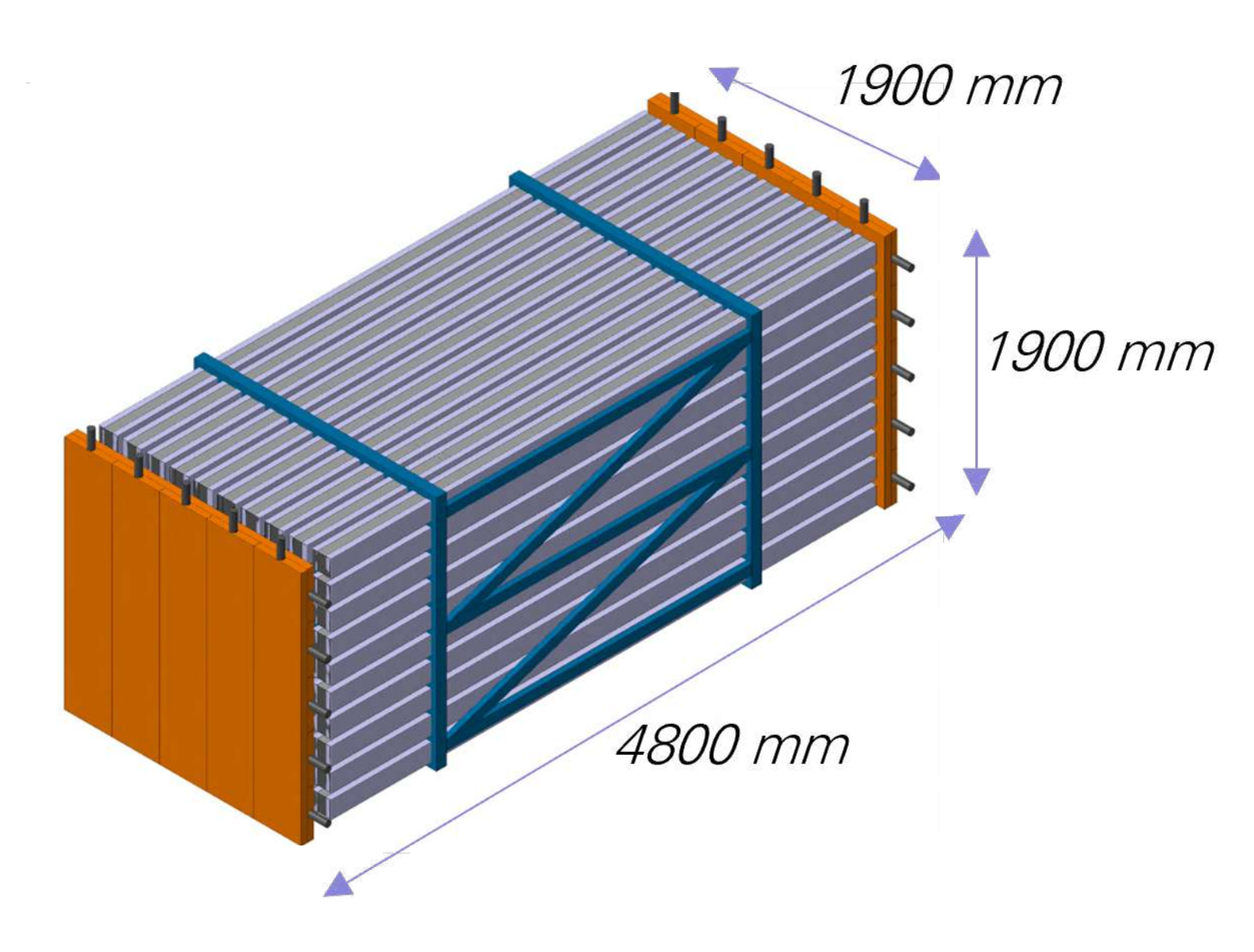}
  \caption{The FORMOSA detector design showing the supermodules (grey), which each hold $2\times2\times4$ scintillator bars, enclosed in a steel frame (blue). Segmented scintillator slabs are shown at the front and back of the detector (orange). For clarity, the top and side scintillator panels are not shown~\cite{Feng:2022inv}.
  \label{fig:FORMOSA_diagram}}
\end{figure}

To improve the sensitivity to particles with the lowest charges, the FORMOSA detector could be augmented by the inclusion of a high-performance non-organic scintillator such as $\mathrm{CeBr_3}$~\cite{BNC}. This material combines a fast response time and low internal radioactivity with a light yield factor $\sim 35$ times that of the same length of EJ-200. Due to the increased costs of such a material, this would be a smaller subdetector comprised of an array of up to $4\times4\times4$ bars of $\mathrm{CeBr_3}$, which would be installed in place of an equivalent array of plastic bars. 

The detector readout requires a resolution of $\sim 1\ \mathrm{mV}$ at a sampling rate of $\sim 1\ \mathrm{GS/s}$ to efficiently identify and trigger on single photo-electron signals (SPEs). This can be provided, for example, by the commercially-available CAENV1743 digitiser~\cite{CAENV1743}. Alternatively, a bespoke readout can be deployed at a significantly-reduced cost using the DRS4 chip~\cite{DRS4}. Such a design is being utilised by the SUBMET detector at J-PARC~\cite{Kim:2021eix}.

FORMOSA offers a unique opportunity as a dedicated experiment capable of detecting the weak signatures produced by mCPs. Its installation within the FPF enhances its search capability, given the large flux of mCPs in the very-forward region, and the fact that this region is shielded from most of the SM backgrounds produced at the IP and at the LHC.

\subsubsection{Sensitivity to Millicharged Particles} 
\label{sec:FORMOSA_sensitivity}

To evaluate the sensitivity of the FORMOSA detector, the signal contributions must be reliably simulated. Pair production of millicharged particles of a given mass and charge at the LHC is nearly model-independent. Every SM process that results in dilepton pairs through a virtual photon would, if kinematically allowed, also produce $\mathrm{mCP^+mCP^-}$ pairs with a cross section reduced by a factor of $(Q/e)^2$ and by mass-dependent factors that are well understood. Millicharged particles can also be produced through $Z$-boson couplings that depend on their hypercharge~\cite{Izaguirre:2015eya}. A full consideration of millicharged production mechanisms in the forward region of the LHC through pseudoscalar and vector meson decays has been carried out using Pythia~8~\cite{Skands:2014pea}, while heavier mCP production through Drell-Yan is simulated with MadGraph 5~\cite{Alwall:2011uj} and Pythia~8. For low-mass mCPs there will also be contributions from proton-bremsstrahlung, which provides a significant increase in the mCP production for masses below $\sim 1\ \mathrm{GeV}$. The size of the contribution from this process is currently under evaluation, and consequently, the expected sensitivities shown here are conservative.

\begin{figure}[!ht]
  \centering
  \includegraphics[width=0.5\textwidth]{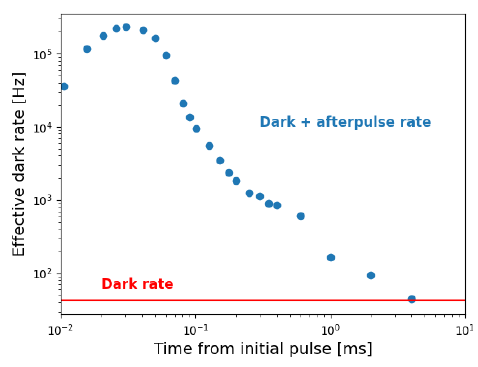}
  \caption{Effective increase in dark rate caused by afterpulses following a deposit similar in size to that expected from a through-going muon~\cite{Citron:2025kcy}. This is shown to drop below the targeted dark current rate of 1 kHz by $\sim$1 ms.}
  \label{fig:FORMOSA_afterpulsingR7725}
\end{figure}

The response of the FORMOSA detector to the signal deposits is simulated using \texttt{Geant4}. Backgrounds arise from three main processes: PMT dark rate, showers from cosmic-ray muons, and afterpulsing caused by beam muons.  Other backgrounds, such as secondaries produced at the IP or radiation produced at the LHC, are greatly attenuated by the inherent rock shielding around the FPF. Assuming a dark channel rate of 1~kHz (as measured for the R7725 PMTs in situ), the total background expected over the full HL-LHC running period for the coincidence of four layers within a $20\ \mathrm{ns}$ window is $\sim0.3$ events. The background from showers originated by cosmic-ray muons for a similar detector in the milliQan location ($\sim$70m underground) at the HL-LHC is projected to be $2\times 10^{-5}$ events in Ref.~\cite{milliQan:2021lne} (using a fully-calibrated \texttt{Geant4} simulation) after highly efficient signal selection; this background will be a similarly negligible contribution for the FORMOSA detector and can easily be studied in data taken while the beam is off.

Finally, bench tests strongly suggest that contributions from the afterpulses created by beam muons can be made negligible with a small deadtime after beam muons are identified passing through the detector:~$\sim 1\ \mathrm{ms}$ is enough time for a single PMT to recover (see \cref{fig:FORMOSA_afterpulsingR7725}), which corresponds to $2.5\%$ deadtime when considering the expected flux of muons traveling through each signal path at the FPF ($\sim 1\ \mathrm{Hz}/\mathrm{cm}^2$, see \cref{fig:mu-flux}).  In reality, since signal triggers require the coincidence of several channels, deadtime for signal triggers can be reduced further. All of the above are being studied in situ with the FORMOSA Demonstrator, as discussed in \cref{sec:FORMOSA_demonstrator}. The total background is therefore projected to be less than one event over the full HL-LHC data-taking period. Overall, one can safely assume a background of zero and 100\% detector efficiency with no visible impact on the sensitivity projections. The expected reach of the FORMOSA detector is shown in \cref{fig:FORMOSA_limit}, as well as the improvement possible with the inclusion of a $\mathrm{CeBr_3}$ subdetector.

\begin{figure}[!htb] 
  \centering
  \includegraphics[width=0.65\columnwidth]{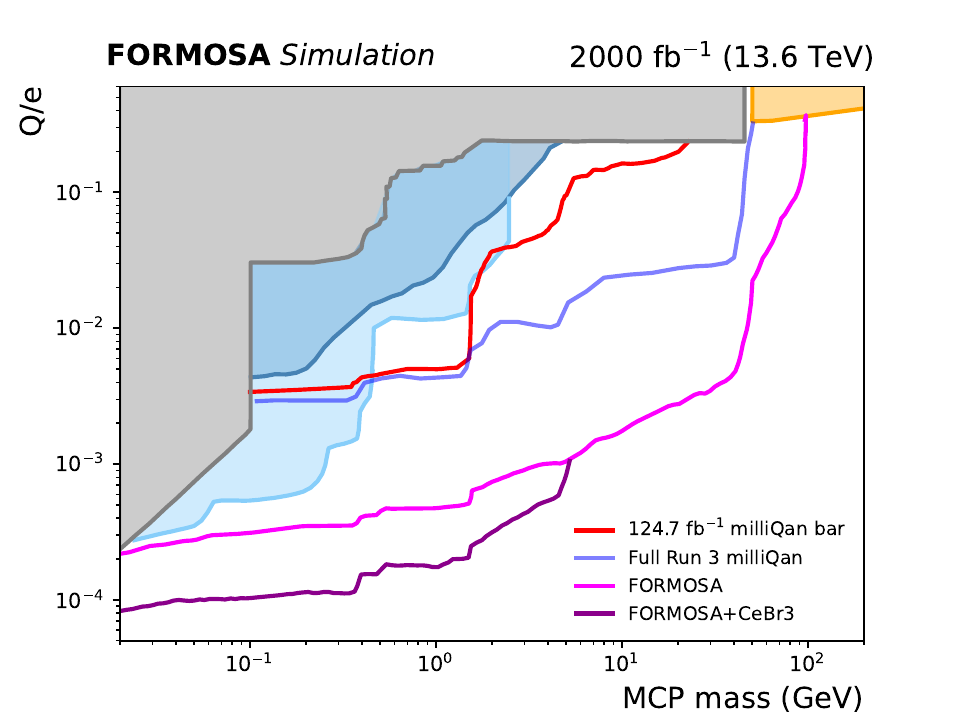}
  \caption{The exclusion limits projected to be achieved by FORMOSA with $2~\iab$ are compared to current bounds from ArgoNeuT, Sensei, CMS, and other experiments (shaded regions)~\cite{Plestid:2020kdm,Prinz:1998ua,Vogel:2013raa,Essig:2013lka,CMS:2012xi,CMS:2013czn,ArgoNeuT:2019ckq, Davidson:2000hf, Badertscher:2006fm,Ball:2020dnx,SENSEI:2023gie}, as well as current bounds determined with 124.7/fb of data and projected Run~3 sensitivity of the milliQan experiment~\cite{Alcott:2025rxn}.  Also shown is the potential sensitivity including a $4\times4\times4$ array subdetector comprised of $\mathrm{CeBr_3}$. Note that the FORMOSA signal simulation does not account for proton bremsstrahlung, which would increase mCP production for masses below $\sim1$~GeV. 
\label{fig:FORMOSA_limit}}
\end{figure}

\subsubsection{The FORMOSA Demonstrator}
\label{sec:FORMOSA_demonstrator}

The FORMOSA Demonstrator is a small-scale prototype of the detector described above that serves as a pathfinder experiment. It has been installed in the very forward region of the LHC in the UJ12 cavern, behind the FASER experiment, at a distance of $\sim480\ \mathrm{m}$ from the ATLAS IP (on the opposite side from where the FPF will be, see \cref{fig:FORMOSA_DemonstratorLocation}) to prove the feasibility of the design and data-taking strategy. The FORMOSA Demonstrator is comprised of a $2\times2\times4$ array of $5\ \mathrm{cm}\times 5\ \mathrm{cm}\times 80\ \mathrm{cm}$ scintillator bars. Scintillator panels are placed at the front and back of the bars for efficient tagging and rejection of beam muons. A diagram of the demonstrator is shown in \cref{fig:FORMOSA_DemonstratorDiagram}.

\begin{figure}[!htb] 
  \centering 
  \includegraphics[width=0.8\columnwidth]{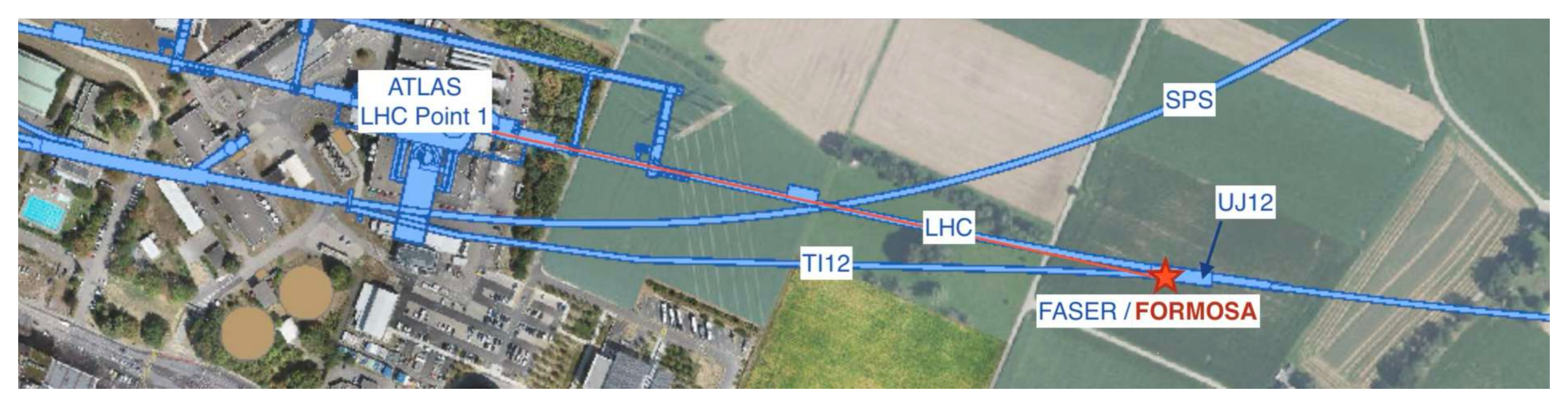}
  \caption{Diagram showing the location of UJ12, where the FORMOSA Demonstrator is installed.
  \label{fig:FORMOSA_DemonstratorLocation}}
\end{figure}

The UJ12 cavern is ideal for the installation and operation of the demonstrator as its location relative to the ATLAS's IP provides similar conditions to those expected at the FPF, although the limited size of the cavern sets a constraint allowing only for the installation of a reduced-size detector. A similar IP-generated-muon flux and comparable distance from the IP allows for operation of the demonstrator in realistic conditions. An important difference between the FPF and UJ12, however, resides in the proximity of the demonstrator to the LHC which results in additional background radiation produced by the circulation of the beams, resulting in a reduced physics sensitivity and calling for adapted DAQ strategies. 

\begin{figure}[!htb] 
  \centering 
  \includegraphics[width=0.65\columnwidth]{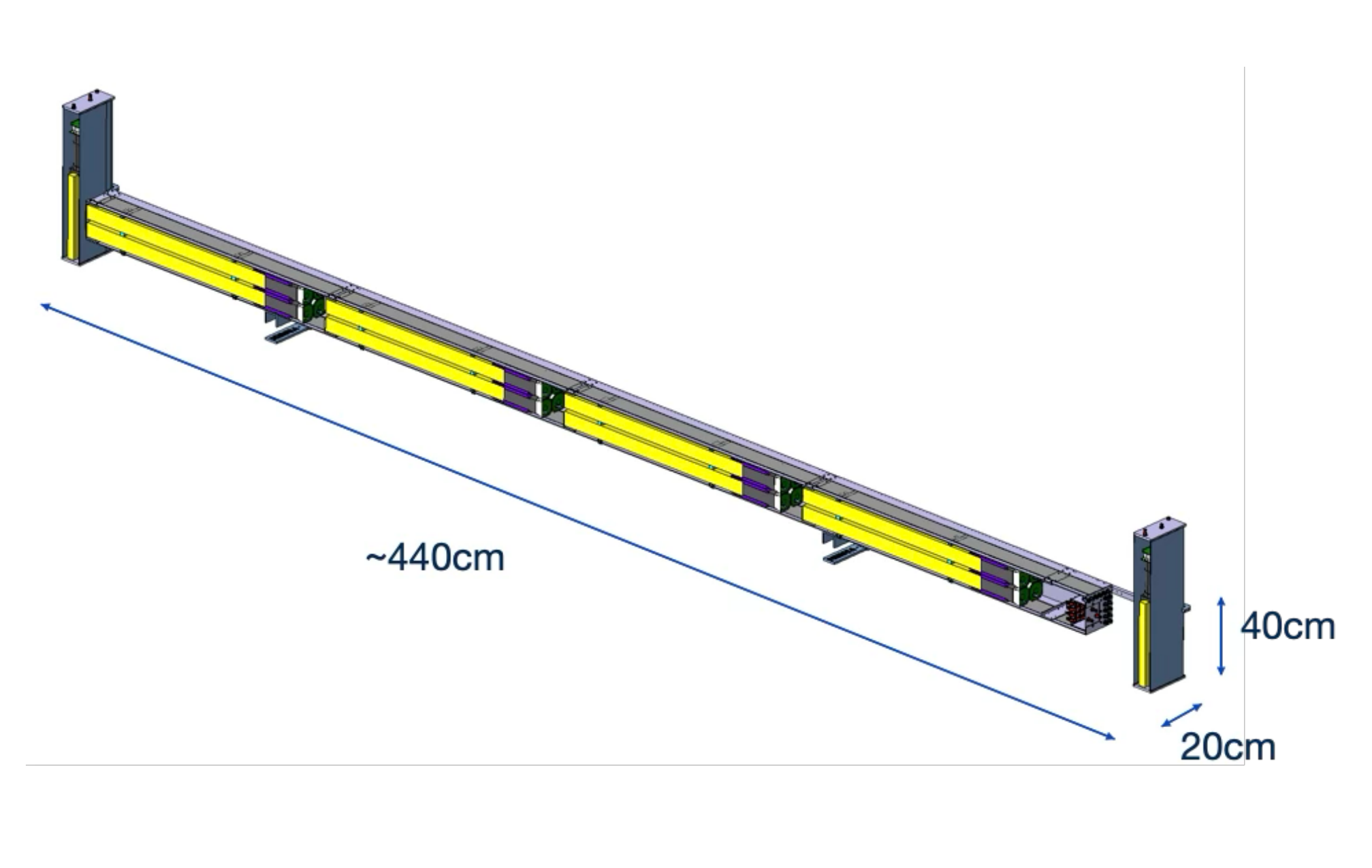}
  \caption{Diagram of the FORMOSA Demonstrator showing the bars and front and back panels.
  \label{fig:FORMOSA_DemonstratorDiagram}}
\end{figure}

The FORMOSA Demonstrator as shown in \cref{fig:FORMOSA_DemonstratorDiagram} was installed in February 2024 and was commissioned throughout the 2024 LHC Run~3 $pp$ collisions. In practice, the demonstrator aims to show that signal data can be efficiently collected, despite the challenging environment, to determine optimal selections for rejecting backgrounds from beam muon afterpulsing and beam radiation, and to measure any residual backgrounds after these selections. Figure~\ref{fig:FORMOSA_DAQValidation} shows the verification of the FORMOSA Demonstrator trigger rates compared to the ATLAS instantaneous luminosity with $pp$ collision data collected during LHC Run~3 in 2024, where a clear correlation between the signal triggers, muon tagging, and the LHC activity at the ATLAS IP can be observed.

\begin{figure}[!htb] 
  \centering \includegraphics[width=0.7\columnwidth]{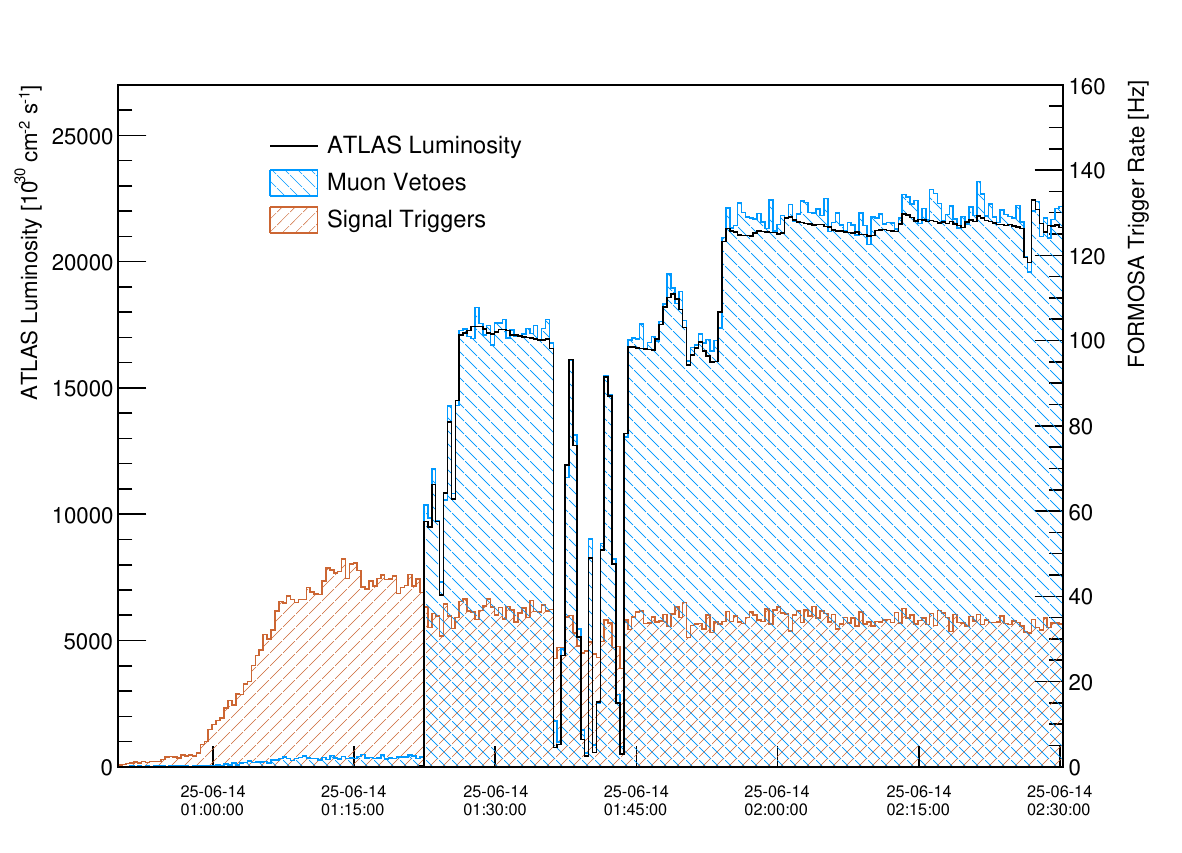}
  \caption{Verification of the FORMOSA Demonstrator trigger rates relative to ATLAS activity. The rate of signal triggers is shown to be stable during data-taking, while contributions from muons and beam-related radiation are effectively vetoed.
  \label{fig:FORMOSA_DAQValidation}}
\end{figure}

In late 2024 a $1\ \mathrm{cm}$ radius $\times~3\ \mathrm{cm}$ long cylindrical module of $\mathrm{CeBr_3}$ scintillator was added to the FORMOSA Demonstrator as a fifth layer to carry out preliminary performance studies of this scintillator material and to understand how it may be incorporated into the FORMOSA detector design. In February 2025, large side and top panels were installed to provide a fully hermetic coverage of the main part of the demonstrator and carry out background studies related to muon showers and beam radiation. The upgraded demonstrator is shown in \cref{fig:FORMOSA_DemonstratorPicture}. The background measurements will be completed with the data collected by the upgraded FORMOSA Demonstrator throughout 2025 LHC $pp$ collisions.

\begin{figure}[!htb] 
  \centering \includegraphics[width=0.4\columnwidth]{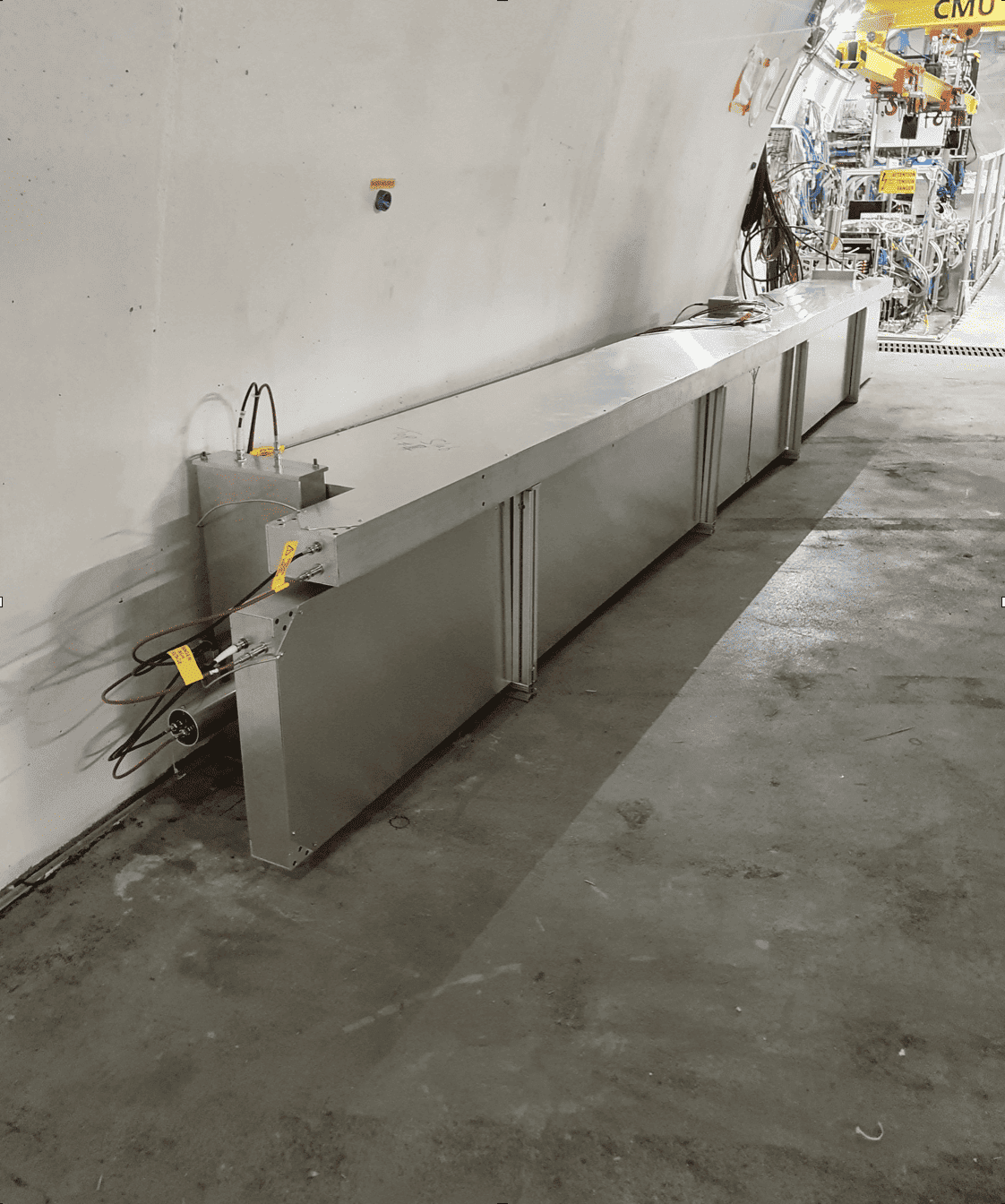}
  \caption{Photo of the upgraded FORMOSA Demonstrator in UJ12 showing hermetic veto panels. The $\mathrm{CeBr_3}$ module can be seen at the end.
  \label{fig:FORMOSA_DemonstratorPicture}}
\end{figure}

Studies around the timing of triggerable activity relative to the LHC clock and the orbit signal show that the active and logical timing resolution allows the finer structures, such as the LHC's bunch structure, to be discerned, which yields valuable information concerning the possible origin of a triggerable event and offline background rejection. Figure~\ref{fig:FORMOSA_Timing_LHCClock} shows the demonstrator's capability to identify possible background events as a function of the timing relative to the LHC clock. Figure~\ref{fig:FORMOSA_Timing_OrbitalCllock} shows the capability of the detector's timing resolution to classify and associate muon activity within the LHC's collision scheme. This capability will further reduce backgrounds, as signal signatures will be tightly peaked at the timing of the LHC collisions.

\begin{figure}[!htb] 
  \centering \includegraphics[width=0.5\columnwidth]{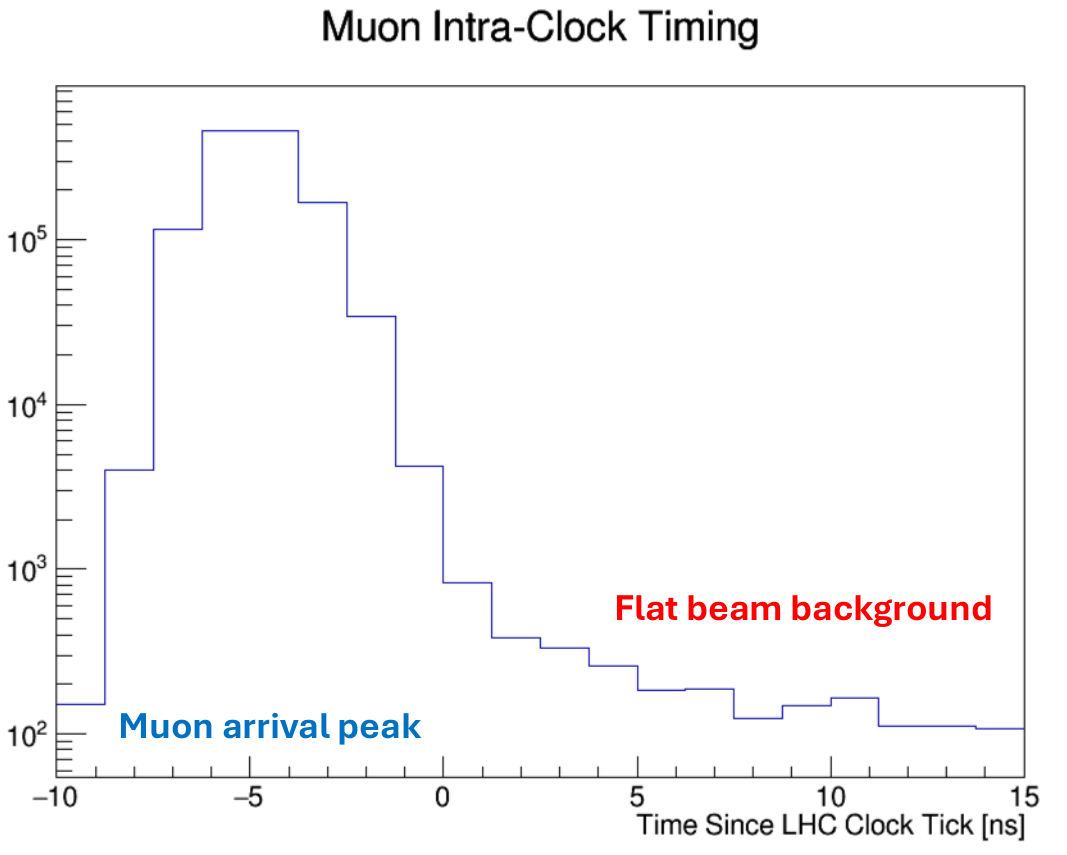}
  \caption{Categorisation of activity measured within the $25\ \mathrm{ns}$ period between bunches as given by the LHC clock. The plot shows the distribution of event counts relative to the LHC clock over a 14-hour-long LHC fill.}
  \label{fig:FORMOSA_Timing_LHCClock}
\end{figure}

\begin{figure}[!htb] 
  \centering \includegraphics[width=0.65\columnwidth]{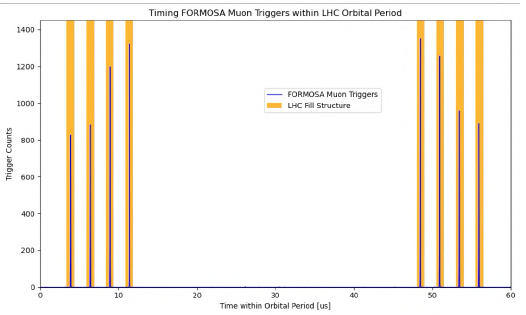}
  \caption{Measured muon activity relative to the LHC's orbit signal during a fill with 8 colliding bunches. The LHC's fill structure is depicted as yellow bands of the same height (i.e. no bunch luminosity is being displayed) centered at the timing of each collision and with exaggerated widths for the sake of visualisation.}
  \label{fig:FORMOSA_Timing_OrbitalCllock}
\end{figure}

Finally, it is worth noting that, while FORMOSA provides the leading sensitivity to mCPs at the FPF, synergy with the other experiments in the facility can extend the capabilities for mCP searches. For example, having FORMOSA provide trigger information to the FLArE experiment would allow looking for the faint tracks deposited by an mCP going through both experiments. This would allow the FPF experiments to further characterise potential mCP signatures.

\subsubsection{FORMOSA Summary}

The FORMOSA experiment at the HL-LHC will provide world-leading sensitivity for millicharged particles that cannot be achieved at any other facility. The experiment is fully costed, and a collaboration comprised of members with significant experience from past mCP experiments has been formed. The operation of the FORMOSA Demonstrator experiment at the LHC is proving the feasibility of such a detector, and we will be ready to construct the detector for LHC Run~4~\cite{Citron:2025kcy}.

\subsection{Simulation Framework}
\label{sec:simulation}

A custom simulation package, named \texttt{FPFSim}~\cite{FPFSim}, has been developed for the FPF and is now publicly available among other supporting packages in an online software repository on Github~\cite{FPFSoftware}.
\texttt{FPFSim} is a flexible \texttt{Geant4}-based toolkit that supports multiple event generator plugins and cavern and detector geometries with minimal external dependencies.
This framework allows a global optimisation of the FPF layout including studying the complementarity between the different detectors.
Its development has been instrumental to perform initial performance studies in the cavern for both FLArE~\cite{Vicenzi:2927282} and FASER2~\cite{Salin:2927003} and represents the starting point towards a unified infrastructure for simulation and reconstruction in the FPF.
While the package is continuously improved, it is fully functional and ready to support physics studies aimed at optimizing detector design and reconstruction strategies.

\begin{sidewaysfigure}[p]
  \centering
  \includegraphics[width=1.\textwidth]{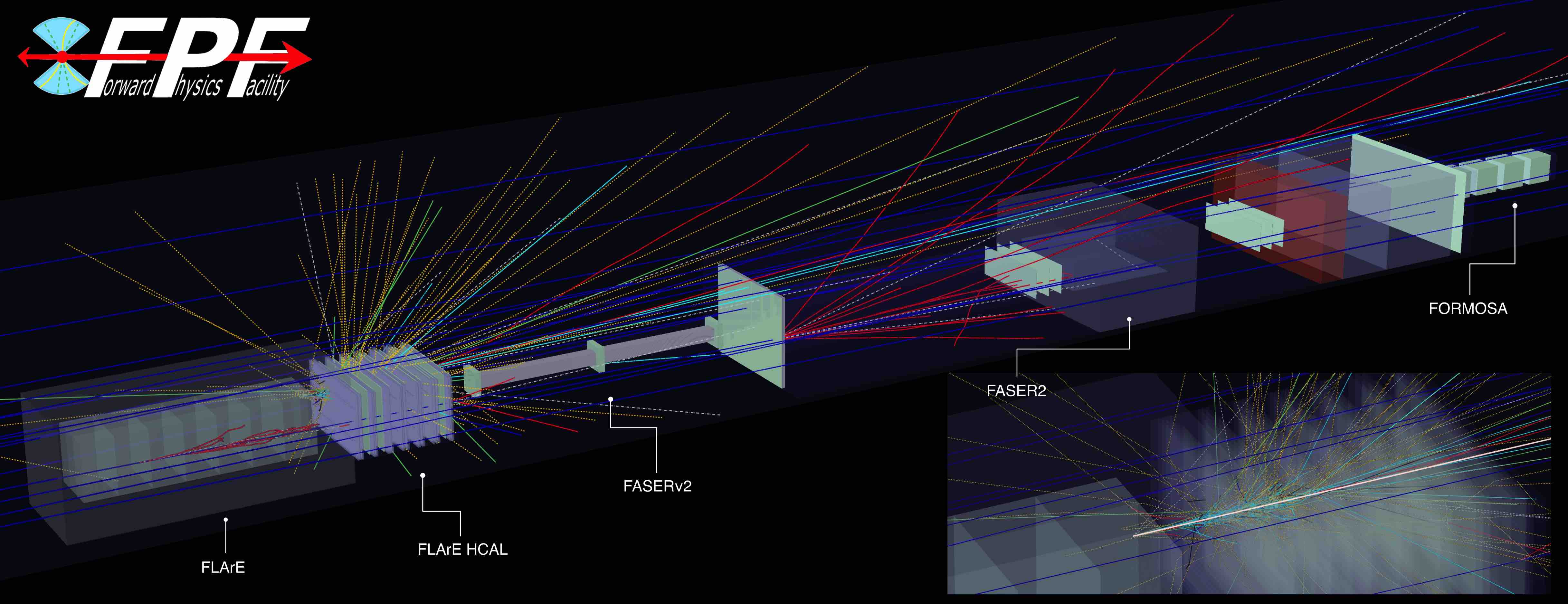}
  \caption{Example \texttt{FPFSim} event display showing a $1.4$ TeV $\nu_\mu$ CC interaction in FLArE and its HCAL. This interaction is overlayed with background muons equivalent to a full LHC orbit ($89~\mu s$) at the HL-LHC baseline luminosity of $5 \times 10^{34}$ cm$^{-2}$ s$^{-1}$. The insert shows a zoom on the signal interaction, highlighting in white the primary muon exiting from the neutrino vertex. Particles with kinetic energy $<60$ MeV are suppressed in the visualisation. Track colours represent different particle species: blue for $\mu^\pm$, red for $e^\pm$, orange (dashed) for $n$, cyan for $\pi^\pm$, magenta for $\pi^0$, grey (dashed) for $\gamma$, black for $p$ and green for everything else (including $\nu$).}
  \label{fig:FPFDisplay_example}
\end{sidewaysfigure}

\texttt{FPFSim} is designed with flexibility and modularity in mind. Detector geometries are defined in independent classes, and their parameters and positions inside the cavern can be fully controlled via macro options at runtime.
This makes it straightforward to explore and compare different layout scenarios.
Similarly, the toolkit itself has been designed for a staged simulation approach to be able to handle different detector technologies and future downstream reconstruction flows.
\texttt{FPFSim} simulates only the propagation and interactions of particles within the detectors and surrounding materials, and it provides their energy deposits in a common output format.
The detailed detector response simulation, such as the modeling of the charge and light transport in FLArE, is instead delegated to external downstream packages.
This separation is key for maintaining a clean interface between geometry-level simulation and detector-specific processing, ensuring that \texttt{FPFSim} remains lightweight, portable, and adaptable to the diverse FPF experiments.

A variety of physics event generators are already supported by \texttt{FPFSim}. Their integration is managed by reader classes that take the outputs of different generators as input.
For neutrino interactions, \texttt{FPFSim} supports the GENIE~\cite{Andreopoulos:2009rq} Monte Carlo generator, which is used to simulate interactions based on the expected FPF neutrino fluxes computed in Ref.~\cite{Kling:2021gos}. 
Its integration is performed via GENIE’s \texttt{gst} output format, which stores the kinematics of final-state particles in a plain ROOT tree.
This approach avoids any explicit dependencies on GENIE libraries, improving portability across computing environments.
\texttt{FPFSim} also supports inputs in the \texttt{HepMC} format, enabling integration with other generators such as FORESEE~\cite{Kling:2021fwx}. FORESEE is a numerical package to compute the sensitivity reach at the FPF for popular BSM models and can output simulated events in the \texttt{HepMC} format. This allows for the injection of LLP BSM events into \texttt{FPFSim}.
In addition to signal generators, a dedicated background generator has also been developed to simulate the radiation environment in the FPF cavern.
This is based on the detailed muon and neutron fluxes computed by the CERN FLUKA group (\cref{sec:muon_fluxes}).
These fluxes are stored in multidimensional histograms that preserve correlations between key quantities:~the particle’s entry position on the cavern front plane $(x,y)$, kinetic energy $E$, and direction.
For a given input time window corresponding to an integrated luminosity or a specific beam exposure, the expected number of background particles is sampled from these distributions. The resulting particles are injected into \texttt{FPFSim} and propagated through the hall and detector volumes. This background modeling infrastructure supports studies of cosmic-ray activity, detector occupancy, cavern-induced backgrounds, and the overlay of signal events based on realistic radiation conditions. 

Figure~\ref{fig:FPFDisplay_example} shows an event display produced with \texttt{FPFSim}, featuring a $1.4$ TeV $\nu_\mu$ CC interaction in the FLArE volume, which generates a shower in the downstream HCAL. This signal interaction is overlayed with background muons corresponding to one full LHC orbit ($89~\mu \text{s}$) at the HL-LHC baseline luminosity of $5 \times 10^{34}$ cm$^{-2}$ s$^{-1}$.
This example illustrates the complexity of the FPF environment and how \texttt{FPFSim} is mature enough to serve as a powerful tool for studying synergies between the detectors with full signal and background simulations.

\clearpage
\section{ Collaboration Structure and Project Planning } 
\label{sec:project}

In this section we examine the structure for the overall collaboration for the FPF. The structure of the collaboration and project planning are deeply linked, and therefore we provide some preliminary concepts which are likely to evolve.\medskip

\noindent \textbf{Governance models:}
The FPF and the forward physics detector collaborations will follow the best governance practices established by other major collaborations, such as ATLAS, CMS, DUNE, SBN at FNAL, and others.  The scale of the FPF enterprise is much smaller than many of these  collaborations, and so we will need to adjust the governance and coordination practices to be suitable.  
We will also look at other models for this coordination, such as the LHCb or R\&D collaborations at CERN.  
In general, FPF will follow the CERN rules and guidelines as outlined in the General Conditions for Experiments at CERN \cite{cernrules} and other relevant rules and procedure documents. 

An important difference between the major collaborations and the FPF community is the possibility of several independent FPF scientific collaborations using the same facility and sharing resources.  All of these collaborations will be international in nature.  Therefore, we are designing  a  structure that is composed of many independent self-governing collaborations that share a common resource, which is the facility. An outline of this structure is provided in \cref{subsec:collab}. Obviously, this structure will need to go through much review and input from  the participants, their funding sponsors, and the host laboratory, CERN.\medskip  

\noindent \textbf{Current community:} 
Three of the proposed experiments, FASER2, FASER$\nu$2, and FORMOSA, have pathfinder projects that are already installed and running at the LHC: FASER, FASER$\nu$, and milliQan, respectively. FASER has already placed world-leading bounds on dark photons~\cite{FASER:2023tle} and ALPs~\cite{FASER:2024bbl}, FASER$\nu$ has detected both electron and muon collider neutrinos for the first time~\cite{FASER:2023zcr} and measured their cross sections at TeV energies~\cite{FASER:2024hoe}, and milliQan has already placed world-leading bounds on mCPs~\cite{Ball:2020dnx}.  These results highlight the physics potential of the forward region, even with  modest luminosities and small experiments.  The collaborations behind the existing experiments are creating a community for this science that will serve all experiments in the FPF.  A brief description of the current collaborations is below.  

Currently, the broader FPF experimental and theoretical community is approximately 400 strong; this includes members of the collider community with interests in the scientific outcome of the FPF.  Please  see the list of contributors to Ref.~\cite{Feng:2022inv}.  Experimental scientists from this community  as well as new members are expected to form the collaborations needed for the FPF experiments.   The community is currently dominated by scientists from Europe, US, and Japan.  Discussions to obtain R\&D support from their respective agencies and institutions are in progress.    The international community is growing  with active interest from Korean and Indian colleagues.  The Indian community in particular has issued a new 10 year Mega-Science plan \cite{indiamegascience} with explicit assignment of resources to FPF in the next 10 years period.  \medskip   

\noindent \textbf{Coordination and growth:}
Along with the conceptual design report for the experiments and the facility, the community will propose a management   structure under the guidance of CERN. This structure will need to have a strong technical coordination team to construct the facility according to the scientific and engineering requirements and also to install the detectors. Each of the scientific collaborations will have representation in the technical coordination team and will provide scientific resources as needed. Each collaboration will also seek coordination and resolution of overlapping requirements. 

We expect the  scientific collaborations to  act independently to promote their respective  science and detectors, and seek collaborators from CERN and the broader international particle physics communities. CERN has a longstanding procedure for accommodating such new collaborators, mainly from Europe and North America.  For the FPF, we will attempt to broaden this to other parts of the world, especially in Asia, Africa,  and the Americas.  Such expansion will be welcome to inject new resources into this effort.  Fortunately, the current program for forward physics has created a core community that is centered around the experiments FASER, FASER$\nu$,  and milliQan.  This community is expected to lead the proposals for the FPF suite of experiments, but additional proposals that enhance the physics capabilities of the FPF are, of course, welcome;  they will be evaluated for science and feasibility. \medskip   

\noindent \textbf{Brief description of current experimental collaborations:} Below we make some remarks about the status of the collaborations for each of the constituent experiments along with what is needed for further expansion:  

\begin{itemize} 
\setlength\itemsep{-0.05in}

\item FLArE: FLArE is based on the liquid argon technology developed for the FNAL short baseline program, as well as DUNE. The FLArE collaboration will be based on the current  working groups, which have approximately $\sim$50 participants, equally divided between US and European collaborators.  The collaboration has  received support from a private foundation, and a US national laboratory-  (BNL-)directed R\&D program.  Because of the recent investment in DUNE prototypes, only limited and well-targeted R\&D is needed for FLArE. Specifically, the readout electronics and pixel readout will need optimisation for spatial resolution and dynamic range.  However, the majority of the design can be simply adapted from the DUNE ND-LAR design.  Furthermore, trigger strategies will need to be developed for the FLArE geometry.  At the moment, the collaboration has enough resources and person power to provide a physics proposal and a well-considered conceptual design.   A modest-sized international collaboration ($\sim$100 collaborators) with appropriate experience will have to be developed by the time of the technical design report in a few years.  

\item FASER$\nu$2:  The FASER$\nu$2 collaboration will largely be made up from the existing FASER$\nu$ experts who are part of the 110-person FASER Collaboration. The expertise on emulsion is mostly concentrated in Japan, where there is a strong tradition in using emulsion-based detectors for neutrino physics. Japan has the leading facilities for emulsion gel production and for the scanning of the emulsion films after exposure, and these will both be used in the FASER$\nu$2 operations.

\item FASER2: The current FASER Collaboration has more than 110 members from 28 institutions in 11 countries~\cite{FASERWebpage}. FASER is a magnetized spectrometer (using permanent dipole magnets) that is housed in an LHC service tunnel 500~m from ATLAS. It started taking data in Run 3.  The FASER Collaboration is expected to form the core of the FASER2 effort. FASER2 will require a much larger effort towards an appropriate spectrometer magnet and a larger tracking system that can handle the trigger rates from HL-LHC.  It also needs careful integration into the FPF hall and with the other experiments.  The Collaboration will need to expand to bring in the appropriate technical expertise and resources for the larger effort.     

\item FORMOSA: The FORMOSA Collaboration currently comprises 15 members from 8 institutions~\cite{FORMOSAWebpage}, and benefits significantly from the experience gained with the milliQan experiment. The FORMOSA concept is based on  well-known technologies that require limited R\&D and is focused on mCPs.  Alterations and improvements to the design to substantially improve the detector sensitivity, such as through the use of alternative scintillator material, are under study, and the collaboration is expected to grow accordingly.

\end{itemize}

\subsection{Collaboration Structure}
\label{subsec:collab}

The FPF collaboration will consist of a few hundred physicists from scores of  institutes, representing several  countries participating in several experiments.  Each of the experiments utilising the FPF infrastructure is expected to have its own collaboration structure and governance model.  These are expected to work independently, but coordinate their activities with the overall FPF collaboration.  

The joint technical, theoretical, and experimental activities of the FPF collaboration will be organised into working groups. Currently these working groups are self-organised and have studied the feasibility, cost, and logistical nature of the proposed FPF underground cavern, the designs and sensitivity of each of the proposed experiments, and the combined SM, BSM, and astroparticle physics potential of the FPF compared to existing and future facilities. Each working group is led by two conveners, and meets regularly to discuss progress. 

The work of the various groups is coordinated by a bi-weekly ``FPF conveners'' meeting. Common facility issues (for example, the radiation environment, the footprints of the experiments laid out in CAD, needed services, and other infrastructure), consistency of theoretical/experimental assumptions, funding opportunities, publications, and annual meetings are among the topics discussed. While individual technical choices are left to the leaders of the respective experiments, decisions on topics that impact the FPF as a whole are taken by consensus of the FPF conveners. The chair of the FPF conveners meeting serves as the principal contact person for the FPF community for all matters (with the exception of technical coordination  issues between the FPF and CERN that are fielded by Jamie Boyd). The FPF conveners and its chair are also the primary contacts for the LHCC, which will be the responsible CERN committee to review the project for approval and during all stages of construction and operation.
 
It is expected that the current informal structure will need to be strengthened by  rules of governance that are commonly utilised in typical particle physics experimental collaborations. The organisation and the rules will need to be flexible so that the FPF enterprise can transition from the current proposal to the funded and design stage to the construction and operation stage.  We have studied various other collaborations that employ large accelerator facilities and locate experimental programs in these facilities. Based on these we are considering a commonly used  set of  organisational bodies. When these bodies come  into practice, the current working groups will be subsumed into the working groups that will report to these bodies. We briefly list these groups and their proposed roles. When the institutional board forms, the detailed functioning of these bodies will be discussed.

\begin{itemize} 
\setlength\itemsep{-0.05in}
\item {FPF Program Oversight Board (FPF-OB).}
The FPF-OB will provide a key forum for cross-collaboration communication or MoU development on issues relevant to construction, commissioning, operations, data management, and analysis.  We expect that this board will be formed early in the process and that it will be critical during the formation and construction phase.  

\item {FPF Technical Coordination Board (FPF-TB).  } 
 The technical coordination board will be the main body where design, construction, and installation of systems for both the FPF infrastructure and the detector will be coordinated. 

\item {FPF Collaboration  Board (FPF-CB).} 
The FPF-CB will be the main governing body and will be the forum for program-wide communication on issues relevant to the program including procedures and policies covering joint aspects of operation, data sharing, data analysis, publications, etc. 

\item {FPF Joint Working Groups.} 
A set of FPF Joint Working Groups are needed to co-develop many key aspects of operations and physics analysis.  These working groups will be constituted by the FPF-CB and are expected to develop and change over time.  

\end{itemize}

\subsection{Schedule, Budget, and Technical Coordination} 
A very preliminary budget and schedule is being  assembled for the FPF facility and the component experiments; this has been discussed in successive FPF workshops~\cite{FPF5Meeting, FPF6Meeting, FPF7Meeting}.  The costs are in several separate groups, as indicated in \cref{costtab}.  For this report we provide new cost numbers compared to the US-based estimates in Ref.~\cite{usp5fpf}: the civil construction design has been improved, providing more space and creating a section of the tunnel for cryogenic infrastructure away from the experiments~\cite{PBCnote2}, and the overall concepts for installation of all detectors and facilities have been improved. 

The cost for the civil construction and the outfitting was provided by the CERN civil engineering group and the technical infrastructure groups, respectively.  They reviewed the  initial experimental  requirements for the needed location, underground space, and services, and performed  a Class 4  estimate~\cite{CEcosting}.  According to international standards of conventional construction, a Class 4 estimate has a range of $-30\%$ to $+50\%$ around the point estimate.  The outfitting includes electrical service, as well as safety, ventilation, transportation, and lift services that are needed for the facility.  Obviously, the facility costs depend on the experimental requirements, which are expected to evolve as we progress towards a technical design. 

The costs for the experimental program were assembled by the proponents.  These  core costs are shown in \cref{costtab} for the FPF experiments.  The costs for FASER2 are dominated by the proposed magnetic spectrometer systems. For FASER$\nu$2, the costs are dominated by the production and handling of emulsion.  The international division of scope for components for these projects is currently not well defined, and therefore these core costs are provided without labour, overhead, contingency, and additional factors that must be used for a full cost estimate according to the rules of each national sponsor.  

FLArE and FORMOSA have substantial US portions; US cost estimates tend to include preliminary estimates for engineering, management, labour, overhead, and contingency factors.  These are not included in \cref{costtab} so that uniform core costs can be presented for each experimental project.  

FORMOSA is a conventional plastic scintillator-based detector with PMT readout.  The estimate includes mostly off-the-shelf parts and conventional assembly.  The number presented is for the more expensive commercial option for the readout  electronics. The FLArE estimate is based on the DUNE ND design with some modifications and includes a scintillator/steel hadron calorimeter using the Baby-MIND detector as a model~\cite{Hallsjo:2018mmo}.   The design will require targeted R\&D for the TPC electronics, a sophisticated photon sensor system, trigger electronics, and clean assembly. Granular details for the FLArE and FORMOSA costs including other factors are available at a pre-conceptual level.  These costs will be refined and further improved as we proceed to the conceptual design. 

A few additional comments are necessary for the project costs: 

\vspace*{-0.1in}
\begin{itemize} 
\setlength\itemsep{-0.05in}

\item The cost for FASER$\nu$2 includes the cost of replacing the emulsion films 10 times. These costs could change over time or be absorbed in the costs of detector operations.  

\item The baseline  for FLArE is now a single-walled, foam-insulated cryostat that is opened on the side for installation. This design has been verified, but needs detailed reviews from laboratory experts.  The cost should be considered very preliminary; it represents a substantial savings compared to a membrane-style cryostat.  The costs presented are for a pixel-style readout, but a very  significant option for FLArE is the ARIADNE optical readout option.  The cost of this option is dominated by the Timpix3 camera readout, but could be lower than the pixel option.   

\item Upon consultation with CERN experts, some of the cryogenic infrastructure has been separated from the experimental costs and included in the upper portion of the table.   The proximity cryogenics, which includes circulation and purification systems, is included in the FLArE experimental costs. 

\item Transport services will be needed for installation of large pieces such as the FASER2 magnet and the FLArE cryostat.  In addition, services will be needed for transporting the emulsion detector periodically.  
The cryostat/cryogenics and additional infrastructure design and costs clearly need to be coordinated and shared with CERN. This process of coordination has started only recently.  

\item The cost for  experiments  does not include engineering, labour, project management, contingency, and   the research support that will be needed.  Obviously, for an effort of this size, considerable support will be needed by a collaboration for students, postdocs, travel, and R\&D.  This is not included in the table. We estimate the total size of the collaboration to range from 250 to 350 people engaged full-time with corresponding annual support from the national agencies. 
\end{itemize} 

\begin{table}[t]
  \centering
  \begin{tabular}{ l | l |l  }
  \hline \hline
  \ Component &  \ Approximate Cost  &  \ Comments \\ 
  \hline \hline
  \ {\bf Facility Costs} &   &  \\ 
  \ FPF civil construction &  \ 35.3 MCHF & \ Construction of shaft and cavern \\
  \ FPF outfitting costs & \ 10.0  MCHF &  \ Electrical, safety, and integration \\ 
  \ Cryogenic infrastructure  & \ 3.8 MCHF & \ Cryogen storage and cooling systems \\ 
  \ {\bf Total} & \ {\bf 49.1  MCHF} & \ Includes integration for infrastructure \\ 
  \hline 
  \ {\bf Experiment Costs} & &  \ Core costs only\\ 
  \ FASER2 & \  11.6 MCHF &  \ 3+3 tracker layers, SAMURAI-style  \\
  & & \ \ magnet, dual-readout calorimeter   \\ 
  \ FASER$\nu$2 & \  15.9 MCHF & \ Tungsten target, scanning system, \\
  & & \ \ emulsion films (10 replacements), interface detector \\ 
  \ FLArE  & \  10.8 MCHF & \ Cryostat,~proximity cryogenics, detectors  \\    
  \ FORMOSA & \ 2.3 MCHF  & \ Plastic scintillator, PMTs, readout   \\ 
  \ {\bf Total} & \ {\bf 40.6  MCHF} &  \   Core~cost~experimental program \\ 
  \hline \hline
  \end{tabular}
  \caption{Cost for components of the FPF and the experimental program.  Costs of the infrastructure at CERN are Class 4 estimates according to international standards; they have a range from $-30\%$ to $+50\%$. The costs for experimental components  are estimated as core costs, which consist of direct costs of materials and contracts only.  Each core cost was computed with conservative technical choices; as new ideas and designs are considered, the costs are expected to change.  }
  \label{costtab}  
\end{table} 

The cost estimate for the FPF and its experiments will be refined in successive stages and reviews, as normally done for large  acquisitions.  We expect the review process to be defined by CERN, as the host lab, and the leading funding organisations that will be involved, such as the UK-STFC, US DOE and NSF, and Japanese JSPS.  Clearly, additional steps are needed to better define the scope of the facility and the constituent physics experiments.  The FPF community will continue with its working group activities and the FPF workshops. Detailed simulation activities have commenced and have provided critical information on detector sizes and depths needed for good efficiency and energy containment for various types of neutrino interactions, as well as for sensitive searches for the many possible BSM scenarios.  The CERN accelerator and radiation protection groups have contributed immensely by providing detailed simulations of the muon rates.  These simulation activities will require appropriate levels of support to develop the detailed requirements needed for a conceptual design report.  

\cref{tab:profile} has the proposed approximate funding profile using the current understanding of the cost estimates for components.  In the following, we provide the constraints used for assembling this funding profile: 

\begin{itemize} 
\setlength\itemsep{-0.05in}

\item \cref{tab:profile} includes some milestones and the nominal HL-LHC schedule.  Any FPF construction must be coordinated with the HL-LHC, so that the civil construction and demands on personnel do not interfere with LHC operations. 
\item We present an ambitious and technically feasible timeline, that maximises the physics. Given the complexity of managing several experiments and many international funding sponsors, it will be challenging to keep this timeline.  However, the design can accommodate timeline shifts for some of the experiments without endangering the overall physics goals.
\item The CERN radiation protection group has concluded that the FPF can be accessed during LHC operations with appropriate controls for radiation safety.  This will allow detector installation to proceed during Run 4.  
\item It would be desirable  that detector construction, installation, and commissioning  happen before Run 4 concludes.  This would contribute significantly to the overall scientific productivity of the FPF and organisation of the FPF community.  
\item We assume that the funding profile for the CERN infrastructure will follow the appropriate profile to allow start of detector installation in the 2033 time frame. 
\item Full detector construction funding is assumed to start in 2027. However, critical development, such as the FASER2 magnet systems and FLArE cryostat, may require funding ahead of this date. Planning and integration of the FPF program will require excellent technical coordination with leadership from the host laboratories.  
\end{itemize} 

\begin{table}[t]
\scriptsize
\renewcommand{\arraystretch}{1.3}
\centering
\begin{tabular}{l|l|l|l|l|l|l|l|l|l}
\hline\hline
\begin{minipage}[t]{1.8cm}\raggedright\textbf{Year}\end{minipage} &
\begin{minipage}[t]{1.45cm}\raggedright 2025\end{minipage} &
\begin{minipage}[t]{1.45cm}\raggedright 2026\end{minipage} &
\begin{minipage}[t]{1.45cm}\raggedright 2027\end{minipage} &
\begin{minipage}[t]{1.45cm}\raggedright 2028\end{minipage} &
\begin{minipage}[t]{1.45cm}\raggedright 2029\end{minipage} &
\begin{minipage}[t]{1.45cm}\raggedright 2030\end{minipage} &
\begin{minipage}[t]{1.45cm}\raggedright 2031\end{minipage} &
\begin{minipage}[t]{1.45cm}\raggedright 2032\end{minipage} &
\begin{minipage}[t]{1.45cm}\raggedright 2033\end{minipage} \\
\hline

\begin{minipage}[t]{1.8cm}\raggedright\textbf{(HL-)LHC Nominal Schedule} \vspace{1mm}\end{minipage} &
\begin{minipage}[t]{1.45cm}\raggedright Run~3\end{minipage} &
\begin{minipage}[t]{1.45cm}\raggedright Run~3/LS-3\end{minipage} &
\begin{minipage}[t]{1.45cm}\raggedright LS~3\end{minipage} &
\begin{minipage}[t]{1.45cm}\raggedright LS~3\end{minipage} &
\begin{minipage}[t]{1.45cm}\raggedright LS~3\end{minipage} &
\begin{minipage}[t]{1.45cm}\raggedright LS~3/Run~4\end{minipage} &
\begin{minipage}[t]{1.45cm}\raggedright Run~4\end{minipage} &
\begin{minipage}[t]{1.45cm}\raggedright Run~4\end{minipage} &
\begin{minipage}[t]{1.45cm}\raggedright Run~4\end{minipage} \\
\hline

\begin{minipage}[t]{1.8cm}\raggedright\textbf{FPF Milestones}\end{minipage} &
\begin{minipage}[t]{1.45cm}\raggedright LOI and physics proposal\end{minipage} &
\begin{minipage}[t]{1.45cm}\raggedright R\&D and prototype detectors\end{minipage} &
\begin{minipage}[t]{1.45cm}\raggedright CDR, long lead items, magnet\end{minipage} &
\begin{minipage}[t]{1.45cm}\raggedright Start of civil construction\end{minipage} &
\begin{minipage}[t]{1.45cm}\raggedright TDR for detectors\end{minipage} &
\begin{minipage}[t]{1.45cm}\raggedright Detector construction start\end{minipage} &
\begin{minipage}[t]{1.45cm}\raggedright Major equipment acquisition\end{minipage} &
\begin{minipage}[t]{1.45cm}\raggedright End of civil construction,\\ install services\end{minipage} &
\begin{minipage}[t]{1.45cm}\raggedright Detector installation and commissioning \vspace{1mm}\end{minipage} \\
\hline

\begin{minipage}[t]{1.8cm}\raggedright\textbf{Experiment Core Costs (kCHF)}\end{minipage} &
\begin{minipage}[t]{1.45cm}\raggedright \textemdash\end{minipage} &
\begin{minipage}[t]{1.45cm}\raggedright 154\end{minipage} &
\begin{minipage}[t]{1.45cm}\raggedright 1275\end{minipage} &
\begin{minipage}[t]{1.45cm}\raggedright 3473\end{minipage} &
\begin{minipage}[t]{1.45cm}\raggedright 7257\end{minipage} &
\begin{minipage}[t]{1.45cm}\raggedright 11220\end{minipage} &
\begin{minipage}[t]{1.45cm}\raggedright 9503\end{minipage} &
\begin{minipage}[t]{1.45cm}\raggedright 6978\end{minipage} &
\begin{minipage}[t]{1.45cm}\raggedright 741\end{minipage} \\
\hline\hline
\end{tabular}
\caption{ Proposed funding profile for the FPF  experimental program using the core cost numbers  from \cref{costtab}. The infrastructure cost profile is being developed. This funding schedule assumes that all approvals can be obtained in a timely manner, and so it should be considered a technically feasible timeline. The approval and cost rules will be different for the various international sponsors who are proposed to contribute to this overall profile. Nevertheless, for the purpose of this illustration, the profile is shown in as-spent funds in a single currency. Obviously delays are expected and will need to be factored in as the project moves forward.}
\label{tab:profile}
\end{table}

\subsection{Community and Societal Benefits}

The FPF and its experimental program will continue until the end of the HL-LHC program into the early 2040's.  We have designed the facility and its program to be flexible, and we expect it to evolve over the period of a decade or longer after it is built.  Given its modest footprint, relatively low cost, and  ability to install and operate detectors independently of the LHC running, a continuous program of detector upgrades could be imagined as new ideas or new discoveries come forth. The novel forward geometry, and independence from LHC operations will allow a new vibrant scientific community to grow around the FPF program. The community  interest can be broadly divided into three parts, although with many overlapping topics.
\begin{itemize} 
\setlength\itemsep{-0.05in}
\item {Collider Physics:}  A large part of the FPF's scientific program directly impacts collider physics through its impact on precision measurements of fundamental electroweak parameters and contributions to constraining proton and nuclear structure functions. This provides a unique and attractive opportunity for ambitious and creative approaches for those interested in jointly analyzing the collider and FPF data sets. 

\item {Neutrino Physics and Astrophysics:} A new community that is expert in neutrino physics could develop at CERN,  dedicated to very high-energy neutrinos and the technological approaches for their measurements.  

\item {Beyond Standard Model Physics:} We have planned an initial suite of experiments that target a variety of benchmark BSM models.  As we learn to analyze the new FPF data, we expect new detector technologies to emerge. Entirely new sets of experimental plans might also need to be examined.  This is a very important opportunity for early career researchers who need leadership experience.   

\end{itemize} 

The new  community and its three parts, is likely to have several characteristics that should be attractive to CERN and the international high energy physics program.  First, the community will be driven by a set of scientific questions that are related to the LHC, but also independent, thus adding excitement to the LHC program. Given the novelty of the facility, we expect to attract early career theorists and experimentalists.  Junior physicists are drawn to the FPF because the physics goals are new and exciting, there is room for their creative and innovative ideas to have a real impact, and they can contribute at every stage of an experiment’s life cycle, from detector design to construction to analysis, in the time it takes to obtain a graduate degree. Given the younger profile, we also expect a much more energetic approach to detector development and R\&D.  The FPF offers the possibility of developing modest-sized detectors in a short period of time and utilising it for physics without the enormous challenges that are encountered in a large collaboration.  

Lastly, we emphasise that the field of high energy physics must have platforms where new leadership can be developed and expertise  retained.  Future very large projects cannot be undertaken without providing challenges to a new generation of scientists to prove their independence.  The FPF is an ideal platform for creating a new generation of ambitious physicists with the skill set to propose, construct, and execute sophisticated experimental projects. It also provides an avenue to attract scientists from nations that are new to   fundamental particle physics,  without the huge barriers that they would face in large collaborations.  

In summary, the FPF will generate a new, large, and diverse community of scientists who will be attracted for reasons of physics, detector development, and a chance at leadership.  They will enhance CERN as the flagship of international fundamental science and provide further support for the future of the particle physics enterprise.  

\clearpage 
\acknowledgements
We gratefully acknowledge the invaluable support of the CERN Physics Beyond Colliders (PBC) study group, under the excellent coordination of Gianluigi Arduini, Joerg Jaeckel, and Gunar Schnell and, previously, Mike Lamont and Claude Vallee.
PBC members who have directly contributed to technical studies related to the feasibility of the implementation of the FPF are:
\begin{itemize}
\setlength\itemsep{-0.05in}
    \item Civil Engineering studies: K. Balazs, T. Bud, J. Gall, D. Gamba, A. Navascues Cornago, J. Osborne
    \item FLUKA simulation studies: F. Cerutti, M. Sabat\'e Gilarte\
    \item Radiation Protection studies: L. Elie, A. Infantino
    \item Integration: J.-P. Corso, A. Magazinik, J. Prosic
    \item Cryogenics: J.~Bremer
    \item Safety: M.~Andreini, F.~Corsanego 
\end{itemize}
We further thank Anatoli Fedynitch and Felix Riehn for useful discussions and invaluable technical support.

The work of L.A.~Anchordoqui is supported by the U.S. National Science Foundation grant PHY-2412679.
The work of A.J.~Barr is funded in part through STFC grants ST/R002444/1 and ST/S000933/1, and through research funded by the John Fell Oxford University Press Research Fund.
The work of B.~Batell is supported by U.S.~Department of Energy grant DE–SC-0007914.
The work of J.~Bian, M.V.~Diwan, S.~Linden, M.~Vicenzi, and W.~Wu is supported in part by Heising-Simons Foundation Grant 2022-3319.
The work of M.~Citron is supported in part by U.S.~Department of Energy grant DE-SC0009999. 
I.~Coronado and D.~Soldin acknowledge the support and resources from the Center for High Performance Computing at the University of Utah. I.~Coronado is also supported by the Summer Undergraduate Research Fellowship Program at the Department of Physics \& Astronomy at the University of Utah.
The work of J.L.~Feng, M.~Fieg, F.~Kling, R.M.~Abraham, and T.~M\"akel\"a is supported in part by U.S.~National Science Foundation grants PHY-2111427, PHY-2210283, and PHY-2514888.
The work of J.L.~Feng is supported in part by Simons Investigator Award \#376204, Heising-Simons Foundation Grants 2019-1179 and 2020-1840, and Simons Foundation Grant 623683.  
The work of R.~Francener is supported by Conselho Nacional de Desenvolvimento Cient\'{\i}fico e Tecnol\'ogico (CNPq, Brazil), Grant 161770/2022-3.
The work of E. Hammou is supported by the European
Research Council under the European Union’s Horizon
2020 research and innovation Programme (grant agreement n.950246), and partially by the STFC consolidated
grant ST/T000694/1 and ST/X000664/1.
The work of C.S.~Hill is supported in part by U.S.~Department of Energy Grant DE-SC0011726.
The work of Y.S.~Jeong is supported in part by the National Research Foundation of Korea (NRF) grant funded by the Korea government through Ministry of Science and ICT Grant No. RS-2025-00555834. 
The work of F.~Kling is supported in part by the Deutsche Forschungsgemeinschaft under Germany's Excellence Strategy -- EXC 2121 Quantum Universe -- 390833306.
The work of J.~McFayden is supported by the Royal Society grant URF\textbackslash R1\textbackslash201519 and STFC grant ST/W000512/1.
The work of J.~Rojo is partially supported by NWO, the Dutch Research Council, and by the Netherlands eScience Center (NLeSC).
The work of S.~Trojanowski is supported by the National Science Centre, Poland, research grant No.~2021/42/E/ST2/00031. 
The work of K.~Watanabe is supported by JSPS KAKENHI Grant No. JP25K07286.
The work at BNL is under U.S.~Department of Energy contract No.~DE-SC-0012704.  

\clearpage
\appendix 

\section{Comparison between different Luminosities}

In this appendix, we compare the projected physics sensitivity of the FPF experiments operating at the HL-LHC with integrated luminosities of $2~\iab$ and $3~\iab$.

In \cref{sec:qcd}, we discussed how differential neutrino DIS measurements at the FPF can be used to constrain  proton structure and reduce production cross section uncertainties for many processes probed at ATLAS and CMS during the HL-LHC era, such as Higgs and electroweak boson production. In the upper panels of \cref{fig:lumi_pdf}, we show the fractional uncertainties of the valence up, valence down, and strange quark PDFs in the PDF4LHC21 baseline scenario (blue), compared to the results with FPF data at the HL-LHC for $2~\iab$ (red) and $3~\iab$ (light red). The corresponding impact on PDF uncertainties for HL-LHC measurements is shown in the bottom panel. We see that the FPF improves the PDF uncertainties for both $2~\iab$ and $3~\iab$ of integrated luminosity, with the difference between the two being relatively small.

\begin{figure}[bth]
  \centering
  \includegraphics[width=.32\textwidth]{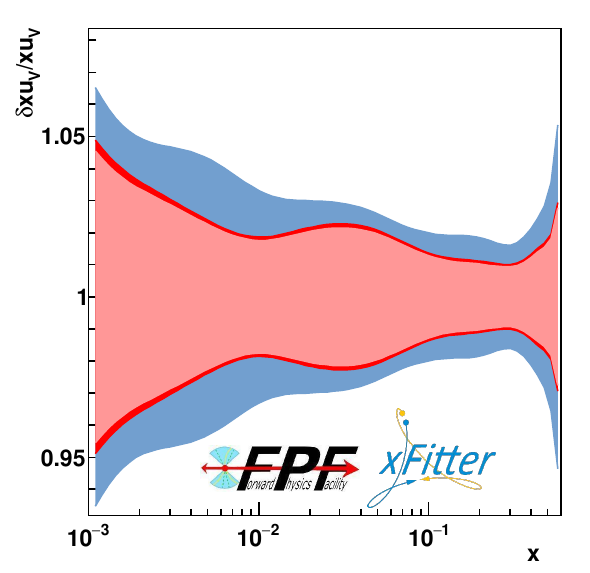}
  \includegraphics[width=.32\textwidth]{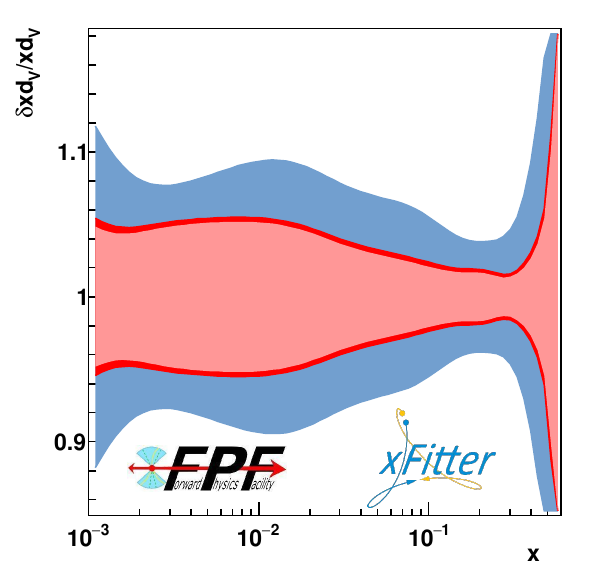}
  \includegraphics[width=.32\textwidth]{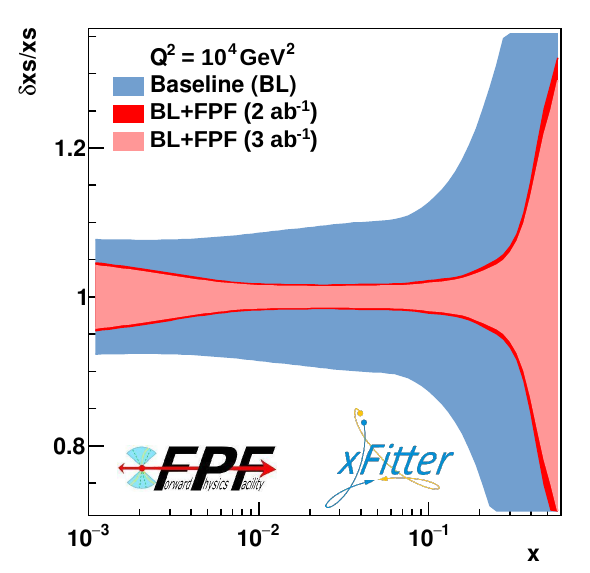}
  \includegraphics[width=.49\textwidth]{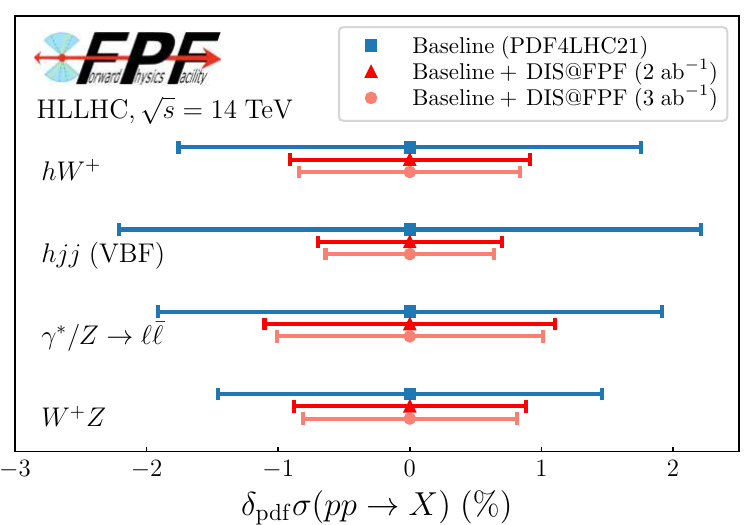}
  \caption{\textbf{Sensitivity Projections for Proton Structure and its Implications.} In the upper panels, we show the projected fractional uncertainties for the valence up, valence down, and strange quark PDF (also see \cref{fig:FPF_PDFs})). The lower panels show the resulting impact of these measurements on PDF uncertainties in HL-LHC cross sections predictions for a number of Higgs and electroweak processes (also see \cref{fig:PDF_applications}). Results are shown for the PDF4LHC21 baseline scenario and compared to the results with  FPF at the HL-LHC with 2~ab$^{-1}$ and 3~ab$^{-1}$ of integrated luminosity.}
  \label{fig:lumi_pdf}
\end{figure}

In \cref{fig:lumi_bsm}, we display the projected sensitivities for various searches for new particles at different FPF experiments. The upper panels show the search for decays of long-lived excited states in inelastic dark matter models at FASER2 as considered in \cref{sec:darkmatter}. The lower panels show the searches for scattering of millicharged particles at FORMOSA and FLARE, and for a muon-philic scalar at FASER$\nu$2. In all panels, we show the reach for $2~\iab$ and $3~\iab$ of integrated luminosity. We find that the sensitivity reach remains largely unchanged. This can be understood as follows: in the short-lifetime limit (large couplings) of decay searches, the reach is limited not by luminosity but by the probability of particles to reach the detector which changes exponentially in this regime. At low couplings, the event rate in most searches scales with the fourth power of the coupling (two powers from production and two from decay or interaction). Increasing the luminosity from $2~\iab$ to $3~\iab$ therefore extends the reach in coupling by a factor $(2/3)^{1/4} \sim 1.1$, corresponding to roughly a 10\% increase in sensitivity. A notable exception is the search for millicharged particles at low masses, where the event rate scales as the tenth power of the coupling, further reducing the sensitivity to luminosity.

\begin{figure}[thb]
  \centering
  \includegraphics[width=.49\textwidth]{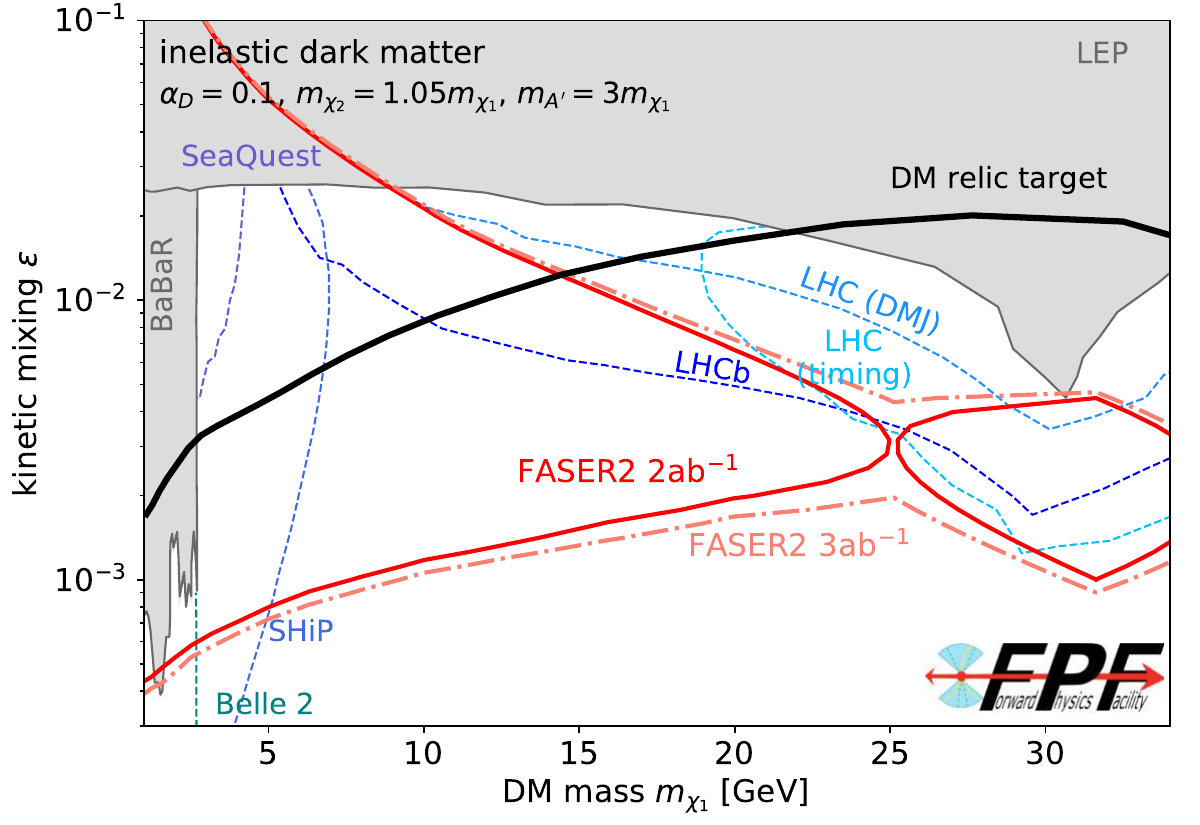}
  \includegraphics[width=.49\textwidth]{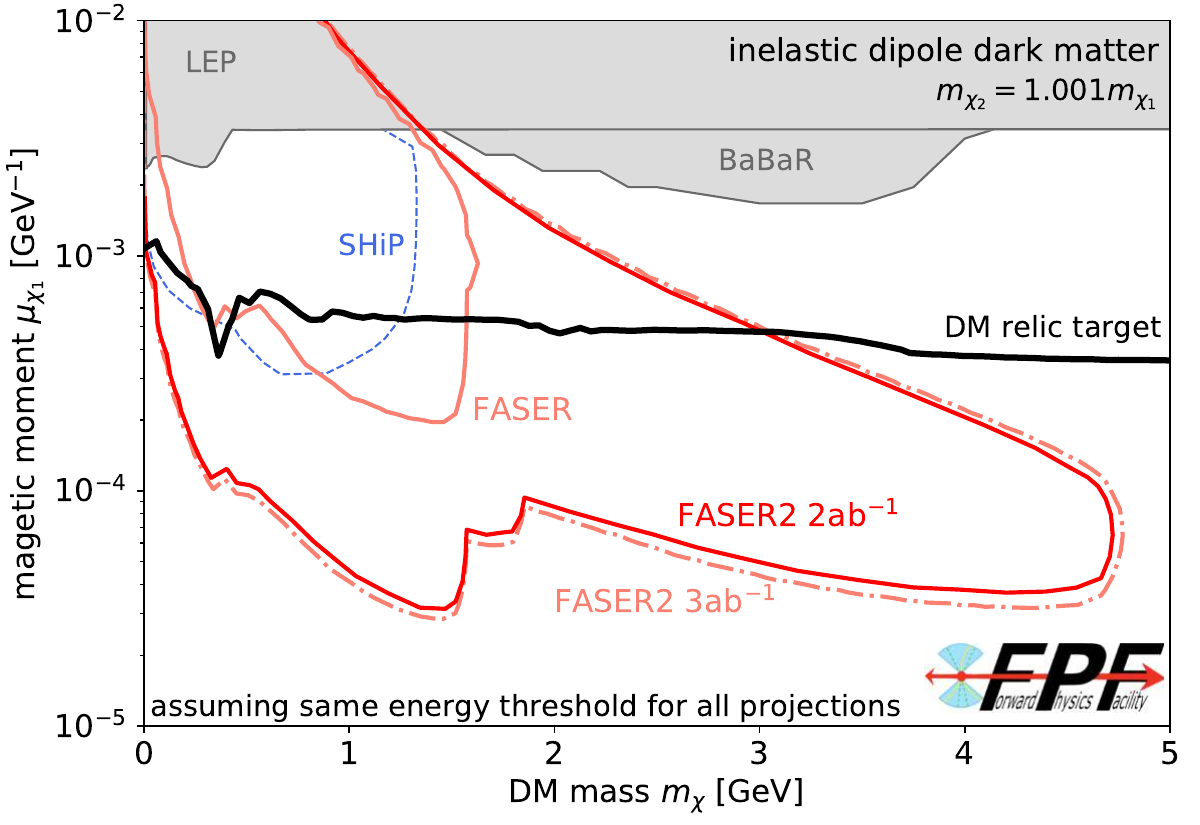}
  \includegraphics[width=.49\textwidth]{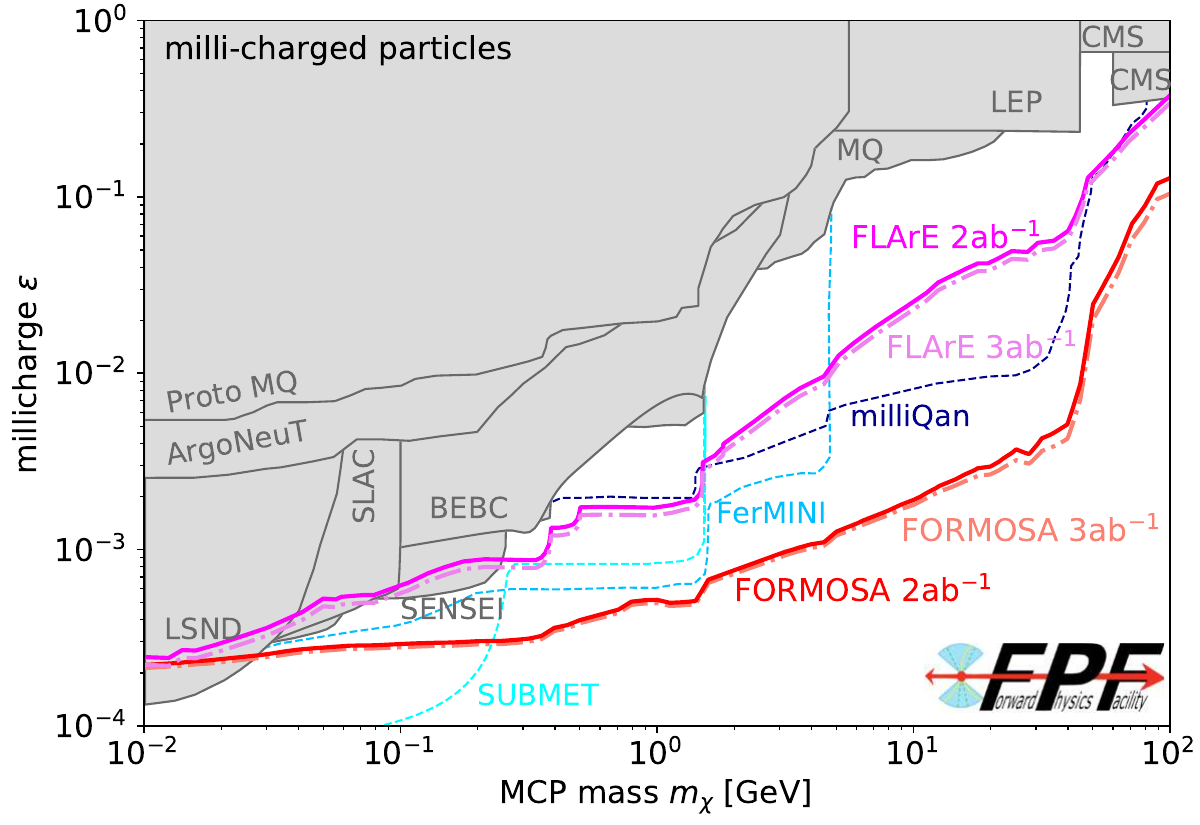}
  \includegraphics[width=.49\textwidth]{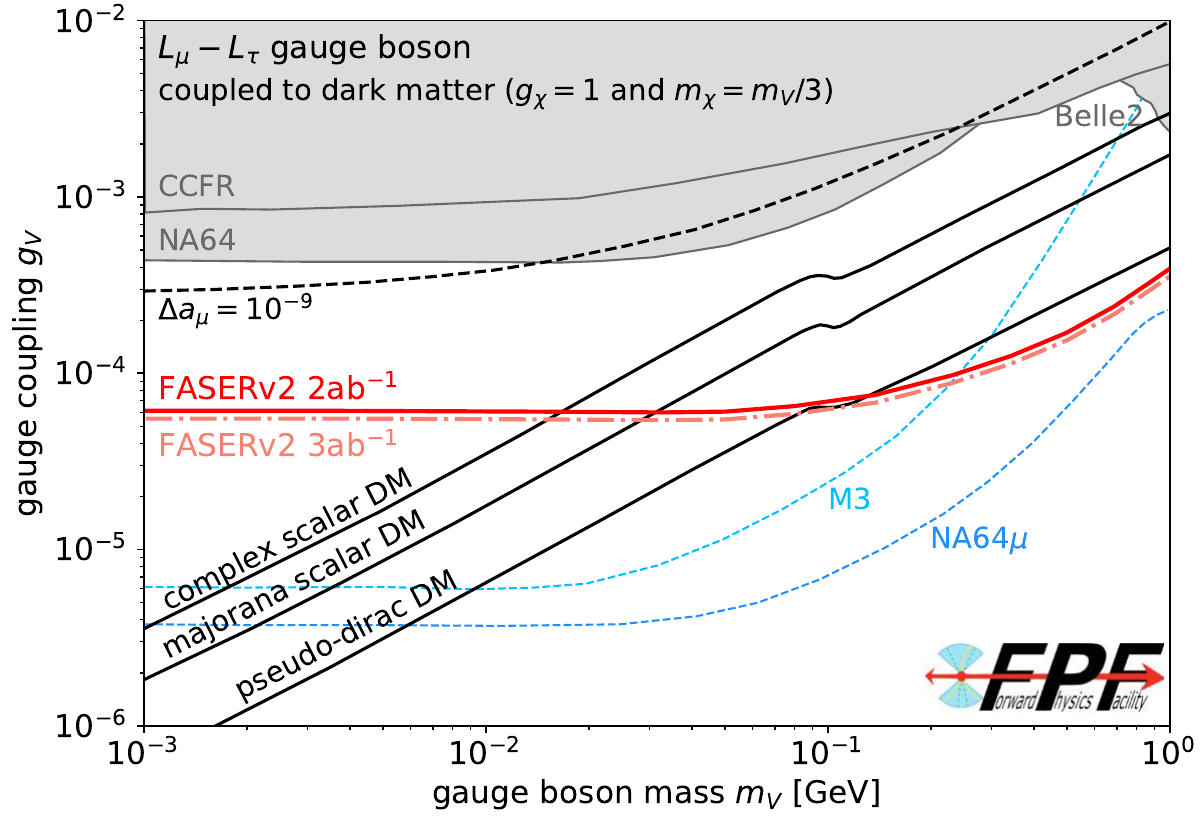}
  \caption{\textbf{Sensitivity Projections for BSM Models at different Luminosities.} We show the projected sensitivity for the dark photon mediated inelastic dark matter model at FASER2 (upper left, also see left panel of \cref{fig:inelasticDarkMatter}), the dipole portal mediated inelastic dark matter model at FASER2 (upper right, also see right panel of \cref{fig:inelasticDarkMatter}),  millicharged particles at FORMOSA and FLArE  (lower left, also see left panel of \cref{fig:BSM_newparticles}), 
  and muon-philic force carriers at FASER$\nu$2 (lower right, also see left panel of \cref{fig:BSM_muons}). Solid and dash-dotted lines indicate the reach with 2~ab$^{-1}$ and 3~ab$^{-1}$ of integrated luminosity, respectively. }
  \label{fig:lumi_bsm}
\end{figure}

Overall, we find only small differences between projections performed with integrated luminosities of $2~\iab$ and $3~\iab$. This demonstrates that an FPF operating only during LHC Run~5 would be nearly as impactful as one running throughout Runs~4 and~5. The examples shown here illustrate this point for representative cases, but the conclusion holds broadly across essentially all other analyses. Consequently, projected FPF results obtained in the literature for $3~\iab$ also hold when considering only the integrated luminosity of LHC Run~5, $2~\iab$.

\clearpage
\bibliography{references}

\end{document}